%% file: shellmodelPR.tex
\documentclass[final,3p,times,authoryear]{elsarticle}

\usepackage[dvipdfm]{hyperref}
\usepackage {amsmath}
\usepackage {version}
\usepackage{color}
\usepackage{rotating}
\usepackage{setspace}
\usepackage{natbib}
\usepackage{ulem}

\providecommand\bb{\boldsymbol{\rm b}}
\providecommand\bu{\boldsymbol{\rm u}}
\providecommand\e{{\rm e}}
\newcommand{\Pm}{\mathrm{Pm}}
\newcommand{\Rm}{\mathrm{Rm}}
\newcommand{\Rey}{\mathrm{Re}}
\newcommand{\Rh}{{\rm R_H}}
\newcommand{\tp}{{\rm P}}
\newcommand{\ttor}{{\rm T}}

\newcommand{\bB}{\textbf{B}}
\newcommand{\bD}{\textbf{D}}

\newcommand{\bF}{\textbf{F}}
\newcommand{\bG}{\textbf{G}}
\newcommand{\PP}{\textbf{\textsl{P}}}
\newcommand{\bQ}{\textbf{Q}}
\newcommand{\bR}{\textbf{R}}
\newcommand{\bS}{\textbf{S}}
\newcommand{\bU}{\textbf{U}}

\newcommand{\bW}{\textbf{W}}
\newcommand{\bX}{\textbf{X}}
\newcommand{\bY}{\textbf{Y}}
\newcommand{\bZ}{\textbf{Z}}
\newcommand{\ba}{\textbf{a}}
\newcommand{\bc}{\textbf{c}}
\newcommand{\bff}{\textbf{f}}
\newcommand{\be}{\textbf{e}}
\newcommand{\bh}{\textbf{h}}
\newcommand{\bk}{\textbf{k}}
\newcommand{\bl}{\textbf{l}}
\newcommand{\bp}{\textbf{p}}
\newcommand{\bq}{\textbf{q}}
\newcommand{\br}{\textbf{r}}
\newcommand{\bs}{\textbf{s}}

\newcommand{\bx}{\textbf{x}}
\newcommand{\by}{\textbf{y}}
\newcommand{\bz}{\textbf{z}}
\newcommand{\bom}{\boldsymbol{\omega}}
\newcommand{\bOm}{\boldsymbol{\Omega}}
\newcommand{\Complex}{\mathbb{C}}
\renewcommand{\i}{\text{i}}
\renewcommand{\imath}{\text{i}}

\usepackage{amssymb}

\journal{Physics Reports}

\begin{document}

\begin{frontmatter}



\title{Shell Models of Magnetohydrodynamic Turbulence}


\author[label1]{Franck Plunian}
\address[label1]{ISTerre, CNRS, Universit\'e Joseph Fourier, Grenoble, France}
\author[label2,label3]{Rodion Stepanov}
\author[label2,label3]{Peter Frick}
\address[label2]{Institute of Continuous Media Mechanics of RAS, Perm, Russia}
\address[label3]{Perm National Research Polytechnical University, Russia}

\begin{abstract}

Shell models of hydrodynamic turbulence originated in the seventies. Their main aim was to describe the statistics of homogeneous and isotropic turbulence in spectral space, using a simple set of ordinary differential equations. In the eighties, shell models of magnetohydrodynamic (MHD) turbulence emerged based on the same principles as their hydrodynamic counter-part but also incorporating interactions between magnetic and velocity fields.
In recent years, significant improvements have been made such as the inclusion of
non-local interactions and appropriate definitions for helicities.
Though shell models cannot account for the spatial complexity of MHD turbulence, their dynamics are not over simplified and do reflect those of real MHD turbulence including intermittency or chaotic reversals of large-scale modes. Furthermore, these models use realistic values for dimensionless parameters (high kinetic and magnetic Reynolds numbers, low or high magnetic Prandtl number) allowing extended inertial range and accurate dissipation rate. 
Using modern computers it is difficult to attain an inertial range of three decades
with direct numerical simulations, whereas eight are possible using shell models.

In this review we set up a general mathematical framework allowing the description of any MHD shell model. The variety of the latter, with their advantages and weaknesses, is introduced. Finally we consider a number of applications, dealing with free-decaying MHD turbulence, dynamo action, Alfv\'en waves and the Hall effect.

\end{abstract}

\end{frontmatter}
\newpage
\tableofcontents
\newpage

\begin{table}
\begin{center}
\begin{tabular}{|@{\hspace{0.cm}}l@{\hspace{0.2cm}}l|@{}}
\hline
Variables	&	$\bu$, $\bb$, $p$, $\bz^{\pm}$, $\ba$, $u^{\pm}$, $b^{\pm}$			\\*[0cm]
External magnetic field, forcing, vorticity &	$\bb_0$, $\bff$, $\bom$		\\*[0cm]
Wave numbers &	$\bk$, $\bp$, $\bq$		\\*[0cm]
Energies, helicities, enstrophy, squared magnetic potential	&	$E^u$, $E^b$, $H^u$, $H^b$, $H^c$, $\Xi$, $A$\\*[0cm]
Complex conjugation, complex conjugate of $z$ & $c.c.$,	$z^*$					\\*[0cm]
Dimensionless numbers &	$\Rey$, $\Rm$, $\Pm$, $\Rh$		\\*[0cm]
Viscosity, diffusivity, density & $\nu$, $\eta$, $\rho$ \\*[0cm]
Injection rate of energy, kinetic helicity, cross helicity, magnetic helicity & $\epsilon$, $\zeta $, $\chi$, $\xi$ \\*[0cm]
Viscous, Joule dissipation rate & $\epsilon_{\nu}$, $\epsilon_{\eta}$ \\*[0cm]
Rotation rate, Alfv\'en wave velocity & $\Omega$, $b_0$ \\*[0cm]
Characteristic time scales & $t_{NL}$, $t_{\Omega}$, $t_{A}$, $t_{\nu}$, $t_{\eta}$ \\*[0cm]
Characteristic length scales & $l_F,l_{\nu}$, $l_{\eta}$ \\*[0cm]
Characteristic velocity and magnetic fluctuations at scale $l$ & $u_l, b_l$ \\*[0cm]
Velocity and magnetic structure functions, scaling exponent & $S_p^u(l), S_p^b(l), \zeta_p$ \\*[0cm]
Specific wave number modulus & $k_F$, $k_{\nu}$, $k_{\eta}$, $k_{\perp}$, $k_{\parallel}$  \\*[0cm]
Shell common ratio & $\lambda$ \\*[0cm]
Kinematic growthrate & $\Gamma_{\rm kin}$ \\*[0cm]
Shell variables & $U_n$, $B_n$, $Z_n^{\pm}$, $U_n^{\pm}$, $B_n^{\pm}$ \\*[0cm]
Shell wave number, forcing & $k_n$, $F_n$  \\*[0cm]
Quadratic functions &$\bQ(\bX), \bW(\bX,\bY), \widetilde{\bW}(\bX,\bY)$ \\*[0cm]
Quadratic quantities in shell $n$ & $E^U_n$, $E^B_n$, $H^U_n$, $H^B_n$, $H^C_n$, $\Xi_n$ $A_n$ \\*[0cm]
Energy transfer  & $T^{XY}_{i}, T^{XY}_{ij}$  \\*[0cm]
Energy flux & $\Pi^{X<}_{Y<}, \Pi^{X<}_{Y>}, \Pi^{X>}_{Y>}$  \\*[0cm]
Mode-to-mode energy transfer & $S^{XY}(i|j|k)$  \\*[0cm]
Non-locality parameter & $\gamma$ \\*[0cm]
\hline
	\end{tabular}	
	\caption{Table of notations.}
	\label{notations}
	\end{center}
\end{table}
\newpage
\input{Introduction}

\input{MHD}
\input{Shell}

\input{Dynamo}	
\input{Applications}

\input{Conclusion}

\input{Appendix}




\newpage

\bibliographystyle{elsarticle-harv}
\bibliography{ref}







\end{document}

%% file: Introduction.tex
\section{Introduction}
\label{s1}

In astrophysical objects most fluids are electrically conducting and generally exhibit highly turbulent motion due to the large dimensions involved \citep{Schekochihin2007}. Such magnetohydrodynamic (MHD) turbulence is at the heart of the dynamo action generating magnetic fields in planets, stars and galaxies  \citep{Brandenburg2005,Tobias2011}.
Dynamo action has been the object of several experiments \citep{Gailitis2000,Stieglitz2001,Shew2005,Monchaux2007, Spence2007,Nataf2008,Frick2010b} and is suspected to occur in nuclear reactors cooled with liquid sodium \citep{Plunian1999}.
MHD turbulence is also responsible for the propagation of Alfv\'en waves \citep{Alfven1942} in the presence of an external magnetic field as in e.g. the solar wind. Such waves can be reproduced in MHD experiments \citep{Alboussiere2011} and measured in plasma tokamaks \citep{Gekelman1999}.
Complementary to observation and experiment, direct numerical simulations aimed at reproducing the finest details of MHD turbulence have been performed \citep{Muller2003}. However, one serious difficulty faced by simulations is that the processes involved are
strongly non-linear implying, for example, that the energy is transferred over an extended range of scales \citep{Verma2004}. This range is several orders of magnitude larger than what is attainable with
present day or, indeed, projected computers.
In this respect shell models are of primary importance in building up our understanding.
First introduced to deal with hydrodynamic (HD) turbulence, shell models have now been generalized to MHD, leading to interlocked progresses
of both types of model.
We will now summarize the evolution of these ideas.

\citet{Obukhov1971}
introduced a multilevel
system of non-linear triplets to mimic the energy transport, in the spirit of the Richardson-Kolmogorov
scenario for the energy cascade in HD turbulence.
This idea has been successfully developed by his team \citep{Gledzer1973,Desnianskii1974,Glukhovskii1975,Gledzer1981}.
At the same time
\citet{Lorenz1971} started from
the full Fourier representation of the Navier-Stokes equations. Aiming at studying the statistical properties of turbulent flow with limited computer facilities, he
reduced the set of equations to a ``very low order model".
Though both approaches were different, \citet{Lorenz1972} and \citet{Gledzer1973} eventually derived the same shell model of HD turbulence
\footnote{First denoted ``cascade" or ``scalar" models,
such models have been called ``shell" in the beginning of the nineties.}.
They both applied the conservation of kinetic energy and enstrophy
 with the description of atmospheric turbulence statistics in mind.
We note that these two quantities, kinetic energy and enstrophy, are positive definite and so
are rather straightforward to define in the framework of shell models, explaining why shell models of 2D-turbulence were preferred at that time.
It took a further 20 years \citep{Kadanoff1995} to identify and adequately describe kinetic helicity (not sign definite) in shell models, leading to 3D-turbulence modeling (for which kinetic helicity, instead of enstrophy, is ideally conserved).

Other low order models of turbulence appearing in the seventies were all based on the same principle:
the division of
isotropic spectral space into a set of concentric shells using only one variable per shell to characterize velocity fluctuations. The main difference between the models were the degree of locality between two interacting shells, each shell interacting either with one first neighbor \citep{Obukhov1971,Desnianskii1974,Bell1978,Kerr1978} - with only one quadratic invariant (kinetic energy) - or two first neighbors \citep{Lorenz1972,Gledzer1973,Glukhovskii1975,Gledzer1979} - with two quadratic invariants (kinetic energy and enstrophy).

In the next decade, \citet{Zimin1981} introduced the so-called ``hierarchical model of turbulence" with self-similar functions localized in both physical and Fourier spaces
\footnote{
In terms of contemporary scientific language, these functions would be called wavelets, as discussed by \citet{Frick1993a}.}. Projecting the Navier-Stokes equations on this base of functions, he obtained a set of
ordinary differential equations organized in a hierarchical tree. By reducing this hierarchical tree to one vertical chain, \cite{Frick1983} \footnote{In papers, translated from Russian journals, Peter Frick was spelt as Frik P.G. and we keep each time the spelling from the cited paper.} constructed a shell model for 2D-turbulence - with two quadratic invariants (kinetic energy and enstrophy) - including not only local interactions as in \citet{Lorenz1972} and \citet{Gledzer1973} but also non-local interactions.

From the very first numerical simulations of the shell model equations, it was clear that the Kolmogorov solution (or Kraichnan's in 2D) gave an unstable fixed point, and that a Kolmogorov spectrum of energy could be obtained only by averaging over time, as expected in real turbulence. However, such models failed to show any chaotic behavior.
The link to intermittency was still missing, until the first MHD shell models were derived \citep{Frick1984, Gloaguen1985}. Indeed, by doubling the degrees of freedom (adding the magnetic field), chaotic behavior was obtained.
A similar effect had also been observed with temperature in shell models of convective turbulence \citep{Frick1986,Frick1987}.
Applying this idea to HD turbulence, \citet{Yamada1987,Yamada1988} used a velocity with two \textit{real} components instead of only one, and obtained solutions showing chaotic behavior. With such \textit{complex} velocity they found intermittency statistics in excellent agreement with real HD turbulence.

In the following years such models of HD turbulence aroused wide interest \citep{Pisarenko1993,Carbone1994,Biferale1995a,Kadanoff1995,Frick1995}.
A spurious regularity in the spectral properties (a three-shell periodicity)
identified by \citet{Biferale1993} was corrected either by using a slightly different model \citep{L'Vov1998} or by considering a velocity with
three \textit{real} components per shell instead of two \citep{Aurell1994}.

After the identification of kinetic helicity by \citet{Kadanoff1995}, the first model of 3D MHD turbulence was derived \citep{Brandenburg1996,Basu1998,Frick1998} with total energy, magnetic and cross helicities as quadratic invariants. Meanwhile, new shell models for HD turbulence were elaborated in which the velocity is projected onto helical modes \citep{Zimin1995,Benzi1996a}. In such \textit{helical} models the helicity is not correlated with the kinetic energy, contrary to the other models. This gives rise to important differences when dealing with kinetic helicity in HD turbulence \citep{Stepanov2009,Lessinnes2011}. It is also suitable to study magnetic and cross helicities in MHD turbulence \citep{Frick2010}.

Within the last ten years more complex shell models have been elaborated
to account for characteristics peculiar to MHD turbulence. These models include non-local interactions, directly within triads \citep{Plunian2007} or with the help of multi-scale models \citep{Frick2006}, anisotropy \citep{Nigro2004}, and the Hall effect \citep{Frick2003}.

Our review is organized as follows. After a short description of MHD turbulence in Sec.~\ref{s2},
a general framework is introduced in Sec.~\ref{s3} providing a description of various shell models derived so far. This is followed in Sec.~\ref{s4} by a review of results obtained for different applications. For the sake of clarity, HD shell models will often be introduced before MHD shell models. However, we will focus on MHD applications only. For HD applications the reader can refer to reviews by e.g. \citet{Bohr1998}, \citet{Biferale2003}, \citet{Frick2003a}, \citet{Ditlevsen2011}.
A list of notations that are used in the review is given in Table \ref{notations}.

%% file: MHD.tex
\section{MHD turbulence}
\label{s2}
In this section we simply review some background information necessary to
address the next sections.
For deeper knowledge the reader can refer to reference books
on HD \citep{Frisch1995,Lesieur1997} and MHD \citep{Moreau1990,Davidson2001,Biskamp2003} turbulence.

\subsection{Physical space}
\subsubsection{MHD equations}
\label{s2:MHDequations}

The incompressible MHD equations that govern the time evolution of the velocity $\bu$ and the magnetic induction $\bb$
are
\begin{eqnarray}
\left(\partial_t-\nu \nabla^2 \right)  \bu &=&- (\bu\cdot \nabla) \bu + (\bb \cdot \nabla) \bb -\nabla p + \bff, \qquad \nabla \cdot \bu = 0,  \label{MHDU}\\
\left(\partial_t-\eta \nabla^2 \right) \bb &=&- (\bu\cdot \nabla) \bb + (\bb \cdot \nabla)  \bu ,    \qquad \nabla \cdot \bb =0,
\label{MHDB}
\end{eqnarray}
where $\nu$ is the viscosity, $\eta$ the magnetic diffusivity, $p$ the total pressure (including the magnetic pressure) and $\bff$ the flow forcing, normalized by the fluid density $\rho$. These equations are derived from the Navier-Stokes equations supplemented by the Lorentz force, and Maxwell equations for which advantage has been taken of fast charge redistribution commonly found in liquid metals.
The magnetic induction has been normalized by $\sqrt{4 \pi \rho}$ such that $\bb$ is given in units of velocity.
Here we are interested only in fluctuation, assuming that $\bu$ and $\bb$ average to zero in space and time.

The non-linear terms on the r.h.s. (right hand side) redistribute kinetic and magnetic energies among the full range of scales
from the largest, defined by the system boundaries, to the smallest at which the total energy dissipates.
Different kinds of helicities are also transferred.
Such transfers are called \textit{direct} or \textit{inverse}, depending on whether they occur towards smaller or larger scales.
They are also described as \textit{local} or \textit{non-local} depending on whether they occur between neighboring scales or not.

We speak of \textit{forced} or \textit{decaying} turbulence depending on whether $\bff$ is different from or equal to zero and \textit{dynamo action} when the magnetic energy does not decay in time, meaning that the energy transfer from kinetic to magnetic is sufficient to compensate for the magnetic dissipation.

In the presence of an external magnetic field $\bb_0$ 
(having a velocity dimension as it is normalized by $\sqrt{4\pi \rho}$), 
Eqs.~(\ref{MHDU}-\ref{MHDB}) can be rewritten by replacing $\bb$ by $\bb + \bb_0$.
Introducing the so-called \textit{Els\"{a}sser variables} defined as
\begin{equation}
\label{elsasser}
\bf z^{\pm}= \bf u \pm \bf b,
\end{equation}
the MHD equations become
\begin{equation}
\label{elsasser2}
\partial _t \bz^{\pm} + (\bz^{\mp}\cdot \nabla)\bz^{\pm}= \pm(\bb_0\cdot \nabla)\bz^{\pm} - \nabla p + r^{+}\nabla^2\bz^{\pm}+ r^{-}\nabla^2\bz^{\mp} + \bff^{\pm},
\end{equation}
where $r^{\pm}=\frac{1}{2}(\nu \pm \eta)$ and $\bb_0$ is assumed to be independent of space coordinates.
Provided $\bb_0$ is sufficiently strong, Eq.~(\ref{elsasser2}) can be linearized and thus becomes a wave equation the solutions of which are the so-called
Alfv\'en waves \citep{Alfven1942}.
This set of equations is only symmetric for $\Pm=1$. Taking $\Pm=1$ has the effect of suppressing the reflection of Alfv\'{e}n waves at the walls \citep{Schaeffer2012}.

\subsubsection{Quadratic invariants}
\label{s2:conservative}

In MHD  three integral quantities play a special role: the total energy, the cross helicity and the magnetic helicity.
The total energy, $E$, is the sum of the kinetic energy $E^u$ and the magnetic energy $E^b$,
\begin{equation}
	E=E^u+E^b, \quad E^u=\frac{1}{2} \int_V \bu^2 dV, \quad E^b=\frac{1}{2} \int_V \bb^2 dV,
	\label{totalenergy}
\end{equation}
where $V$ is the volume of integration.
The cross helicity $H^c$ and magnetic helicity $H^b$ are defined as
\begin{equation}
	H^c=\int_V \bu \cdot \bb \;dV, \quad H^b=\int_V \ba \cdot \bb \;dV,
	\label{defhelicities}
\end{equation}
where $\ba$ is the vector potential satisfying $\nabla \times \ba = \bb$.
The absolute value of cross helicity  is maximal if $\bu$ and $\bb$ are aligned (and zero if they are perpendicular).\\

The quadratic quantities $E, H^c$ and $H^b$ are called \textit{invariant} because 
in the ideal limit of a non-viscous and non-diffusive fluid $\nu=\eta=0$, and in the absence of 
forcing and an external magnetic field ($\bff=\bb_0=0$), they  are conserved in time,
\begin{equation}
	d_t E = d_t H^c = d_t H^b = 0.
	\label{MHDconservation}
\end{equation}
The first two conservation laws
can be shown from Eqs.~(\ref{MHDU}-\ref{MHDB}) provided appropriate boundary conditions are used while the third conservation law is obtained from Eq.~(\ref{MHDB}).
We can also show that the first two conservation laws are equivalent using the following property
\begin{equation}
	\int_V  \by \cdot (\bx\cdot \nabla) \by  \; dV =0,
	\label{advectionproperty}
\end{equation}
which is satisfied for any divergence-free vectors $\bx$ and $\by$. Then the conservation of total energy and cross helicity take the following forms
\begin{eqnarray}
	\int_V \bu\cdot (\bb\cdot \nabla) \bb \; dV + \int_V \bb\cdot (\bb\cdot \nabla) \bu \; dV &=& 0 \label{Consenergy}\\
	\int_V \bb\cdot (\bu\cdot \nabla) \bu \; dV + \int_V \bu\cdot (\bu\cdot \nabla) \bb \; dV &=& 0. \label{Conscross}
\end{eqnarray}
Exchanging $\bu$ and $\bb$ does not change $E$ and $H^c$. However, Eqs.~(\ref{Consenergy}) and (\ref{Conscross}) are exchanged, showing the equivalence of the two conservation laws.

We note that the kinetic helicity, defined by
\begin{equation}
	H^u=\int_V \bu\cdot \bom dV,
\end{equation}
where $\bom = \nabla \times \bu$ is the vorticity, is not conserved in MHD.
Hence
\begin{equation}
	d_t H^u=2\int_V \bb\cdot \nabla \times (\bb \times\bom)\; dV,
	\label{helicityderivative}
\end{equation}
indicating that kinetic helicity can be created or suppressed by the interaction between the vorticity and the magnetic field.
For $\bb=0$ the kinetic energy
and helicity are conserved.
In the presence of an external magnetic field $\bb_0$, the magnetic helicity is not conserved anymore.

Finally, in 2D HD turbulence the kinetic helicity is always zero. Instead enstrophy
\begin{equation}
	\Xi=\frac{1}{2}\int_V \bom^2 dV
\end{equation}
is conserved along with the kinetic energy.

In 2D MHD turbulence magnetic helicity is always  constant. Instead the square potential
\begin{equation}
	A=\frac{1}{2}\int_V \ba^2 dV
\end{equation}
is conserved together with the total energy and cross helicity.

\subsubsection{Dimensionless parameters}
\label{s2:dimension}

The Reynolds number is defined as
\begin{equation}
	\Rey=u_l l/\nu,
\label{re}
\end{equation}
where $u_l$ is a characteristic velocity and $l$ is a characteristic scale.
Putting $\bb=0$ in Eq.~(\ref{MHDU}), $\Rey$ shows how strong the non-linear interactions are, compared with viscous dissipation.

In MHD the magnetic Reynolds number is defined as
\begin{equation}
 \Rm=u_l l/\eta.
\label{rm}
\end{equation}
From Eq.~(\ref{MHDB}) $\Rm$ shows how strong the non-linear interactions are,  compared with magnetic dissipation.

The ratio of both numbers is called the magnetic Prandtl number
\begin{equation}
	\Pm = \Rm / \Rey = \nu / \eta
\end{equation}
and depends only on the fluid properties.
The dimensionless form of Eqs. (\ref{MHDU}-\ref{MHDB}) is obtained by replacing $\nu$ and $\eta$ by $\Rey^{-1}$ and $\Rm^{-1}$ respectively.

Fully developed turbulence implies high $\Rey$ (greater than $10^3$ at the largest scale). Dynamo action requires $\Rm > 1$ at the scale where the magnetic energy grows. This corresponds to typical $\Rm$ of $10$ to $10^3$ when $\Rm$ is calculated at the largest scale of the system. The latter condition is difficult to meet experimentally with liquid metals.
Indeed the power necessary to run a liquid metal experiment increases as $\Rm^{3}$ \citep{Petrelis2001},
demanding a considerable effort to reach $\Rm > 10$ at the largest scale.
Liquid sodium is usually used for its high conductivity, and for its density about unity. With $\Pm\approx 10^{-5}$,
reaching $\Rm \approx 10^2$ would require $\Rey \approx 10^{7}$.

\begin{figure}[ht]
\centering
\includegraphics[width=0.9\textwidth]{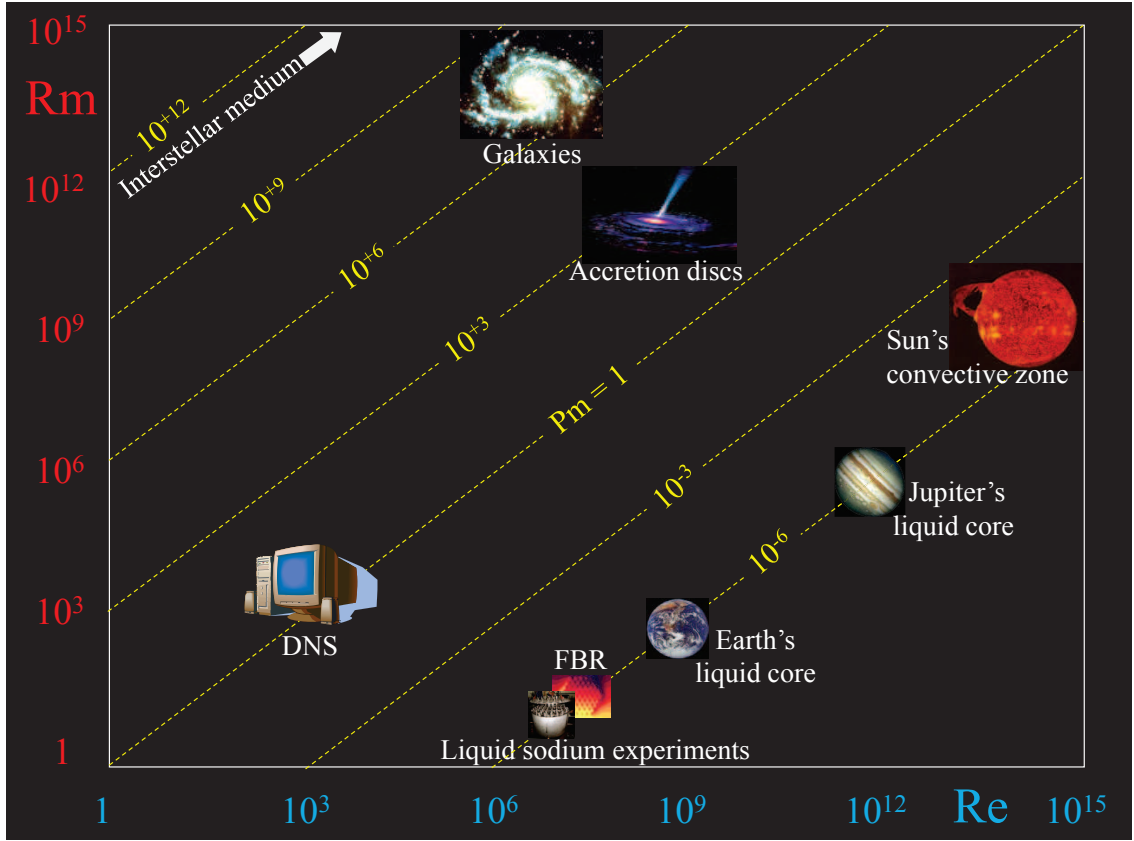}
\caption{Map of typical objects in the plane ($\Rey$,$\Rm$). DNS stands for Direct Numerical Simulation and FBR for Fast Breeder Reactor (cooled with liquid sodium). Yellow dashed lines are $\Pm$ isolines.}
\label{ReRm}
\end{figure}

In Fig.~\ref{ReRm} a few typical objects are placed on the map ($\Rey$,$\Rm$). Among them liquid metal experiments, fast breeder reactors, the Earth's core, Jupiter's core and the Sun's convective zone correspond to $\Pm \sim 10^{-5}$ to $10^{-6}$. We note that 
such low values for $\Pm$ and also realistic $\Rey$ values remain beyond the limits of current
direct numerical simulations \citep{Sakuraba2009,Uritsky2010}.

\subsubsection{Homogeneity and isotropy}
\label{s2:homogeneity}

Two assumptions are usually made in order to obtain theoretical predictions for both HD and MHD turbulence. The first assumes \textit{homogeneity}, meaning that
the statistical quantities derived from the flow and magnetic fields are invariant under translation in physical space.
This assumption fails to predict, for example, the effect of boundary layers.
The second assumes \textit{isotropy}, meaning that the statistical quantities are independent of direction.
In principle isotropy is broken as soon as a sufficiently strong external magnetic field or rotation is applied.

Most 3D shell models are based upon this double assumption of homogeneity and isotropy.
However, several models have been developed to relax the assumption of isotropy
in the context of Alfv\'{e}n waves (see Sec~\ref{S4:Alfven}).

\subsubsection{Isotropic phenomenology}
\label{s2:phenomenology}

In his paper \citet{Kolmogorov1941} introduced the structure function for the velocity field
\begin{equation}
	S^u_p(l)=\langle|u_l|^p\rangle
\label{sfunc}
\end{equation}
where $u_l$ is the longitudinal velocity increment
\begin{equation}
	u_l=\left[\bu(\bx+\bl)-\bu(\bx)\right]\cdot \bl/l,
	\label{deltaul}
\end{equation}
and $\langle \; \rangle$ denotes ensemble averaging.
In HD turbulence, the power $\epsilon$ which drives the flow at forcing scale $l_F$, is transferred towards smaller scales and, in a stationary state, is equal to the energy dissipation rate $\epsilon_{\nu}$ \citep{Kolmogorov1941,Obukhov1941}.
This \textit{direct} energy cascade occurs within the \textit{inertial range} corresponding to scales $l$ such that $l_{\nu} < l < l_F$, where
$l_{\nu}$ is the viscous scale below which viscous dissipation dominates.
In such an inertial range, assuming isotropy, a simple dimensional analysis leads to the estimation
\begin{equation}
	u_l \propto (\epsilon l)^{1/3}.
	\label{Scalinglaw}
\end{equation}
This corresponds to an energy transfer rate $u_l^2 / t_{NL} \propto \epsilon$, where $t_{NL}=l/ u_l$ is the eddy turn-over time.
Then for any $p$,
\begin{equation}
S^u_p(l)\propto (\epsilon l)^{p/3}
\label{zetak41}.
\end{equation}
The viscous scale $l_{\nu}$ is estimated by assuming that the power $\epsilon$, which is injected into the fluid at some forcing scale, is subsequently dissipated by viscosity at scale $l_{\nu}$, corresponding to
$\epsilon \approx \nu u^2_{l_{\nu}}/l_\nu^2$. This leads to
\begin{equation}
l_{\nu} \propto \epsilon^{-1/4} \nu^{3/4}.	
\label{eqlnu}
\end{equation}
In MHD turbulence at high $\Rm$, the estimate of $l_{\eta}$ depends on $\Pm$.
For $\Pm \le 1$
applying the same type of phenomenology as above for $\bu$ and $\bb$, we estimate the viscous and ohmic scales
to be
\begin{equation}
l_{\nu} \propto \epsilon_{\nu}^{-1/4} \nu^{3/4}, \quad \quad l_{\eta}\propto\epsilon_{\eta}^{-1/4}\eta^{3/4},
\end{equation}
where $\epsilon_{\nu}$ and $\epsilon_{\eta}$ are the fractions of $\epsilon$ that correspond to the viscous and the ohmic dissipation respectively ($\epsilon_{\nu}+\epsilon_{\eta}=\epsilon$) . The ratio of these two scales is
\begin{equation}
\frac{l_{\eta}}{ l_{\nu}} \propto \left(\frac{\epsilon_{\nu}}{\epsilon_{\eta}}\right)^{1/4}\Pm^{-3/4}.	
\end{equation}
For $\Pm\ge 1$, assuming that the magnetic energy is produced by the velocity shear which is
maximum at scale $l_{\nu}$, the scale at which the magnetic energy dissipates is given by
$l_{\nu}^{-1}u(l_{\nu}) \propto \eta l_{\eta}^{-2}$, leading to
\begin{equation}
\frac{l_{\eta}}{ l_{\nu}} \propto \Pm^{-1/2}.	
\end{equation}

This immediately shows the difference between low-$\Pm$ MHD turbulence, as in liquid metal,
high-$\Pm$ MHD turbulence, as in interstellar medium, and $\Pm\approx 1$, as in direct numerical calculations.
For $\Pm \ll 1$ magnetic energy dissipation occurs at a much larger scale than that of the kinetic energy, and vice-versa for $\Pm \gg 1$. For $\Pm\approx 1$, an example is given in Fig.\ref{MHDspectra_Carati}, both kinetic and magnetic energies dissipate at about the same scale.

The ratio $\epsilon_{\nu}/\epsilon_{\eta}$, about which little is known, 
depends on the level of the magnetic energy compared to the level of the kinetic energy.
In general we have 
$\epsilon_{\nu} \gg \epsilon_{\eta}$ in experiments and
$\epsilon_{\nu} \le \epsilon_{\eta}$ in real astrophysical objects. 
In MHD shell models, $\epsilon_{\nu}/ \epsilon_{\eta} \approx 1/10 $
for $\Pm\approx 10^{-5}$, leading to $l_{\eta}/ l_{\nu} \approx 10^{3.5}$ \citep{Plunian2010}.

The structure functions for the magnetic field are defined similarly to Eq.~(\ref{sfunc}), as
\begin{equation}
	S^b_p(l)=\langle|b_l|^p\rangle.
\label{sfuncb}
\end{equation}
Assuming a Kolmogorov scaling law, given by Eq.~(\ref{Scalinglaw}), for both magnetic and velocity fields leads to the same scaling for both structure functions  $S^u_p\propto S^b_p\propto l^{p/3}$.

In the presence of an external magnetic field $\bb_0$, a different mechanism of energy transfer occurs due to the interaction of Alfv\'en waves \citep{Iroshnikov1963,Kraichnan1965}.
Indeed, such an applied field leads to an additional time scale $t_A=l/ b_0$.
Provided $b_0$ is sufficiently strong, $t_A$ can be shorter than the eddy turnover time $l/u_l $.
Then energy transfer occurs on the Alfv\'{e}n time scale,
leading to
\begin{equation}
	u_l \propto b_l \propto(\epsilon b_0)^{1/4} l^{1/4},
	\label{irkrul}
\end{equation}
and onto
\begin{equation}
S^u_p(l) \propto S^b_p(l)\propto (\epsilon b_0)^{p/4}l^{p/4} .
\label{irkr}
\end{equation}
Deviation from isotropy leads to other time scales and scaling laws (see Sec.~\ref{s2:b0ne0}).
\subsubsection{Intermittency}
\label{s2:intermitency}

If structure functions obtained from HD experimental measurements exhibit clear scaling laws,
their slopes, however, clearly deviate from $p/3$ as $p$ is increased.
This is interpreted as the signature of intermittent events, like bursts, which are not captured by the self-similarity
assumption of the \cite{Kolmogorov1941} theory.
Such intermittency is quantified by the scaling exponent $\zeta_p$ such that
\begin{equation}
	S_p(l)\propto l^{\zeta_p}.
\end{equation}
Various models of intermittency have been proposed aiming at an analytical formula for the scaling exponent $\zeta_p$ \citep{Frisch1995}. Here we draw attention to the elegant parameter-free formula of \citet{She1994}
\begin{equation}
\zeta^{HD}_p = {p}/{9} + 2\bigl( {1 - ( {{2}/{3}})^{{p}/{3}}} \bigr),
\label{sheleveque}
\end{equation}
which gives $\zeta^{HD}_3 = 1$, consistent with the Kolmogorov ``4/5" law.
It is based on \\
(i) the Kolmogorov refined similarity hypothesis: $\langle u_l^p\rangle \propto \langle\epsilon_l^{p/3}\rangle l^{p/3}$
where $\epsilon_l$ is the energy dissipation averaged over a volume $l^3$, \\
(ii) log-Poisson statistics for the dissipation rate fluctuations, \\
(iii) one-dimensional (filament-like) form of the ultimate dissipative structures.

The scaling exponent given by Eq.~(\ref{sheleveque}) is in excellent agreement with experimental measurements of isotropic HD turbulence.
There is, however, good reason to expect that Eq.~(\ref{sheleveque}) is not valid in MHD turbulence, even if both magnetic and velocity fields satisfy the same Kolmogorov scaling law given by Eq.~(\ref{Scalinglaw}).
The difference comes from the different nature of the ultimate dissipative structures, which might be sheet-like rather than filament-like. Thus, after replacing hypothesis (iii) of She-Leveque by\\ 
(iv) the ultimate dissipative structure is two-dimensional (sheet-like),\\
 \citet{Horbury1997} and \citet{Mueller2000} proposed a MHD version of Eq.~(\ref{sheleveque})
for $S_p^{z^{\pm}}(l)$
\begin{equation}
\zeta^{MHD}_p = {p}/{9} + 1 - ( {{1}/{3}})^{{p}/{3}},
\label{shelevequeMHD}
\end{equation}
which again gives $\zeta^{MHD}_3 = 1$.

When applied to Alfv\'en wave turbulence not only must (iii) be replaced by (iv), but (i) must also be replaced by the Iroshnikov-Kraichan relation\\
(v) $\langle u_l^p\rangle \propto \langle b_l^p\rangle \propto \langle(\epsilon_l b_0)^{p/4}\rangle l^{p/4}$.\\
Subsequently  \citet{Grauer1994} and \citet{Politano1995} developed the Alfv\'enic version of Eq.~(\ref{sheleveque})
\begin{equation}
\zeta^{IK}_p = {p}/{8} + 1 - ( {{1}/{2}})^{{p}/{4}},
\label{shelevequeMHDb}
\end{equation}
with $\zeta^{IK}_4 = 1$.

In experiments and/or numerical simulations, accurate measurements of scaling exponents are needed in order
to discriminate between the three formulas given above by Eqs.~(\ref{sheleveque}-\ref{shelevequeMHDb}).
When dealing with high order structure functions, which is necessary for discrimination, it is even hard to identify the appropriate range of scales which can be used for the determination of the scaling laws. In this respect, significant progress has been made by \citet{ESS} who discovered the concept of {\it Extended Self-Similarity} (ESS) while calculating high order structure functions from wind tunnel experimental results. The ESS takes advantage of the Kolmogorov ``4/5" law
\begin{equation}
\langle u_l^3\rangle = -\frac{4}{5} \epsilon l ,
\label{K4over5}
\end{equation}
so  providing an exact linear relation between the third order structure function $\langle u_l^3\rangle$ and the scale $l$ within the inertial range.
Instead of plotting the structure function $\langle u_l^p\rangle$ versus $l$, leading to a power scaling
$l^{\zeta_p}$, they plotted $\langle u_l^p\rangle$ versus $\langle u_l^3\rangle$. As expected they found that the power scaled as
$\langle u_l^p\rangle\propto \langle u_l^3\rangle^{\zeta_p/\zeta_3}$. Furthermore, they discovered that this scaling held for a range of scales $l$ much larger (both in small and large scale directions) than that for which $\langle u_l^p\rangle\propto l^{\zeta_p}$
holds.
\citet{ESS} therefore claimed an extended self-similarity range of scales. In addition, they found that by using this method the accuracy in the estimate of the scaling exponents was much improved. ESS was tested and used for the measurement of scaling exponents in a variety of turbulent flow conditions, including MHD turbulence \citep{Rowlands2005}.

\subsection{Fourier space}
\subsubsection{Triads}
\label{s2:triads}
Assuming triply periodic boundary conditions in a cube of volume $L^3$, both fields, flow and induction, can be expanded into discrete Fourier series:
\begin{equation}
	\bu(\bx)=\sum_{\bk}\bu(\bk)\e^{\i\bk\cdot\bx}, \qquad \bb(\bx)=\sum_{\bk}\bb(\bk)\e^{\i\bk\cdot\bx}, \qquad \bk \in \frac{2\pi}{L}\mathbb{Z}^3
	\label{Fourierub}
\end{equation}
where $\bu(\bk)$ and $\bb(\bk)$ are the \textit{complex} Fourier coefficients defined as
\begin{equation}
	\bu(\bk)=\frac{1}{L^3}\int\bu(\bx)\e^{-\i\bk\cdot\bx}d\bx^3, \qquad \bb(\bk)=\frac{1}{L^3}\int\bb(\bx)\e^{-\i\bk\cdot\bx}d\bx^3.
\end{equation}
The conditions $\bu(-\bk)=\bu^*(\bk)$ and $\bb(-\bk)=\bb^*(\bk)$
must be satisfied in order that $\bu(\bx)$ and $\bb(\bx)$ be \textit{real} vectors. In Fourier space the divergence-free form of both $\bu(\bx)$ and $\bb(\bx)$ is given by
\begin{equation}
	\bk \cdot \bu(\bk) = 0,  \qquad \bk \cdot \bb(\bk) = 0,
	\label{Fouriersolenoidality}
\end{equation}
indicating that both fields are perpendicular to $\bk$. Similarly, both Navier-Stokes and induction equations can be projected onto a plane perpendicular to $\bk$ in Fourier space, making it possible to remove the pressure terms without loss of generality. These equations are \citep{Biskamp2003}
\begin{eqnarray}
\left(\partial_t+\nu k^2 \right)	\bu(\bk) &=& -\i\PP(\bk)\cdot\sum_{\substack{\bp,\bq\\\bk+\bp+\bq=0}}\left(\left(\bu^*(\bq)\cdot\bk\right)\bu^*(\bp)-\left(\bb^*(\bq)\cdot\bk\right)\bb^*(\bp)\right)
 + \PP(\bk) \cdot \bff(\bk),\label{Fourieru}\\
\left(\partial_t+\eta k^2 \right)	\bb(\bk) &=& -\i\PP(\bk)\cdot\sum_{\substack{\bp,\bq\\\bk+\bp+\bq=0}}\left(\left(\bu^*(\bq)\cdot\bk\right)\bb^*(\bp)-\left(\bb^*(\bq)\cdot\bk\right)\bu^*(\bp)\right)
\label{Fourierb}
\end{eqnarray}
where $\PP(\bk)$ is the operator  defined by the matrix $\textsl{P}_{ij}=\delta_{ij}-\frac{k_i k_j}{k^2}$
and corresponds to the projection of $\bx$ on the plane perpendicular to $\bk$.
Only the subset of wave numbers $\bk$, $\bp$ and $\bq$ satisfying $\bk+\bp+\bq=0$, interact together. Such a triad is illustrated in Fig.~\ref{figtri}.

It can be shown that all the quadratic invariants introduced above are also conserved within each triad.
Hence we can define energy and helicity transfer only between three modes belonging to the same triad.
The formalism for mode-to-mode energy transfer in MHD turbulence has been developed in detail  by \citet{Verma2004} and can be generalized to helicity  (cross or magnetic) transfer. This formalism will be transposed to shell models in Sec.~\ref{s3}.

\subsubsection{Spectra}
\label{s2:spectra}

The Fourier spectra of the quadratic quantities introduced in Sec.~\ref{s2:conservative} are
\begin{eqnarray}
  E^u(\bk)&=& \frac{1}{2}\bu(\bk)\cdot\bu^*(\bk), \label{psd1}\\
  E^b(\bk)&=& \frac{1}{2}\bb(\bk)\cdot\bb^*(\bk),  \label{psd2}\\
	H^b(\bk)&=&\frac{1}{2}\left(\frac{\i}{k^2}(\bk \times \bb(\bk)) \cdot \bb^*(\bk)+c.c.\right) ,  \label{psd3}\\
	H^c(\bk)&=&\frac{1}{2}\left(\bu(\bk) \cdot \bb^*(\bk)+c.c.\right),  \label{psd4}\\
	H^u(\bk)&=&\frac{1}{2}\left(\i(\bk \times \bu(\bk)) \cdot \bu^*(\bk)+c.c.\right).  \label{psd5}
\end{eqnarray}
where $c.c.$ means the complex conjugate.
Their power density spectra
are defined in their integral form
\begin{equation}
	X(k)=\int_{|\bk|=k} X(\bk) d \bk,
\end{equation}
or discrete form
\begin{equation}
	X(k_n)=\sum_{k_n \le|\bk|< k_{n+1}} X(\bk),
\end{equation}
where $X$ denotes any of the above quadratic quantities.
In addition the following conditions are satisfied \citep{Frisch1975}
\begin{equation}
	|H^u(k)|\le k E^u(k), \qquad |H^b(k)|\le k^{-1} E^b(k), \qquad |H^c(k)|\le \left(E^u(k)E^b(k)\right)^{1/2}.
	\label{Realizability}
\end{equation}
Of course all these quantities depend strongly on time. However, we can look for states which are statistically stationary or at least with a time dependency much larger than that of the time scale of the fluctuations (e.g. in free-decaying turbulence).
Not only energies but also helicities are expected to cascade. From
absolute equilibrium distributions, which depend only on the quadratic invariants, \cite{Frisch1975} found that the cascade is \textit{direct} for the total energy and cross helicity, and \textit{inverse} for the magnetic helicity. This makes a striking difference with the HD case for which, using the same method, \citet{Kraichnan1973} found that both the kinetic energy and helicity cascades are \textit{direct}.
The results also depend on whether the MHD turbulence is forced or freely decaying , with or without the presence of an external magnetic field $\bb_0$.

\begin{figure}[ht]
\centering
\includegraphics[width=0.5\textwidth]{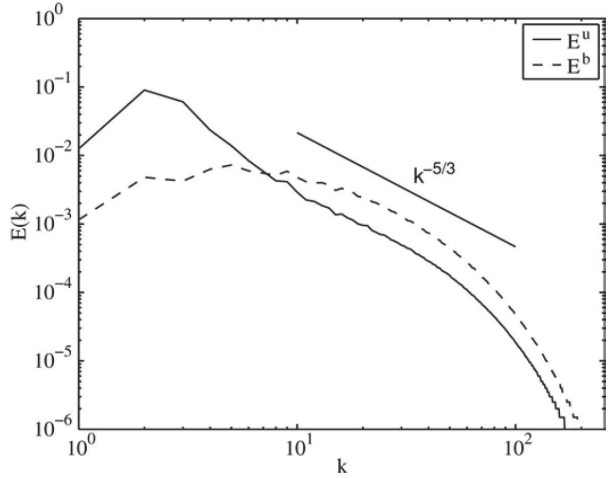}
\caption{Plots of kinetic and magnetic energy density spectra $E^u(k)$ and $E^b(k)$, for Pm=1, obtained from a $512^3$ DNS, for forced MHD turbulence. From \citet{Carati2006}.}
\label{MHDspectra_Carati}
\end{figure}

An example of forced MHD dynamo turbulence for $\bb_0=0$ is given in Fig.~\ref{MHDspectra_Carati}. 
From dimensional arguments, if the velocity and magnetic field increments satisfy $u_l \propto b_l \propto l^\alpha$, then the corresponding spectral energy densities satisfy $E^u(k)\propto E^b(k) \propto k^{-2\alpha -1}$. 
Assuming $u_l \propto b_l \propto (\epsilon l)^{1/3}$, this leads to the famous ``-5/3" Kolmogorov scaling law
\begin{equation}
E^u(k)\propto E^b(k) \propto \epsilon^{2/3} k^{-5/3},
\label{kol41}
\end{equation}
for both kinetic and magnetic energy density spectra.
HD turbulence experiments clearly demonstrate the existence of such a scaling law
over more than three decades \citep{Saddoughi1994,Pope2000}. This is also observed
in DNS over more than one decade \citep{Gotoh2002}.
In MHD turbulence there is, however, not such a clear inertial range
for the magnetic energy density spectrum, as depicted in  Fig. \ref{MHDspectra_Carati}. Even the kinetic energy inertial range is rather short, less than one decade in Fig.~\ref{MHDspectra_Carati}, making it difficult to identify a clear scaling law.
Short spectra are due to limited numerical resolution.
Presumably future higher resolution will give rise to wider inertial range.
The shape of the magnetic spectrum also depends on forcing. In particular if the forcing is helical
the spectrum can be peaked at the largest possible scale \citep{Brandenburg2001,Brandenburg2009}.
Such large-scale magnetic field generation by the small-scale MHD turbulence corresponds to the so-called $\alpha$-effect \citep{Krause1980}.
For free-decaying MHD turbulence and $\Pm=1$ \citet{Mueller2000} and \citet{Mueller2005} found clear Kolmogorov scaling laws.

\subsubsection{Spectra in presence of an external magnetic field}
\label{s2:b0ne0}

\begin{figure}[ht]
\centering
\includegraphics[width=0.4\textwidth]{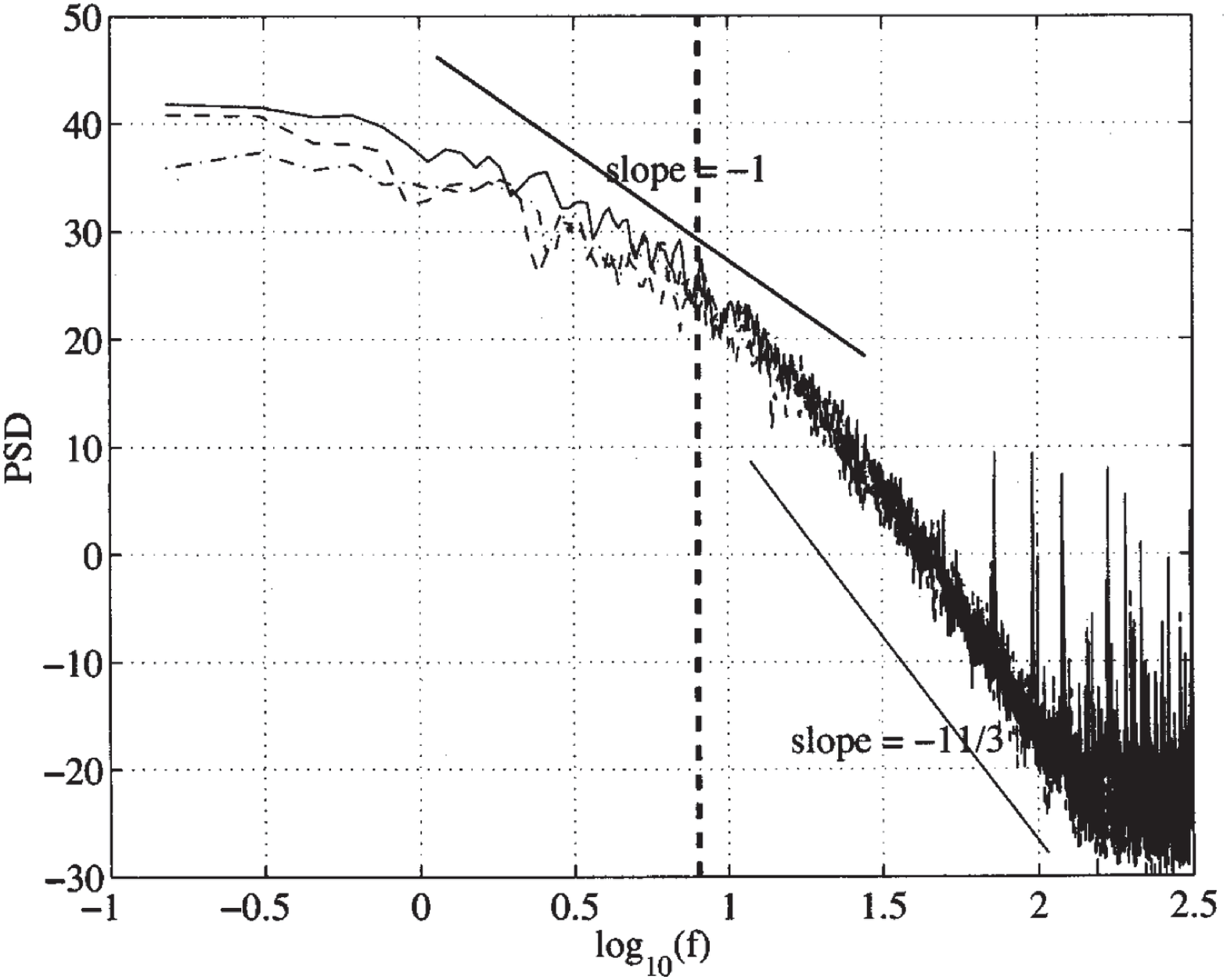}
\includegraphics[width=0.5\textwidth]{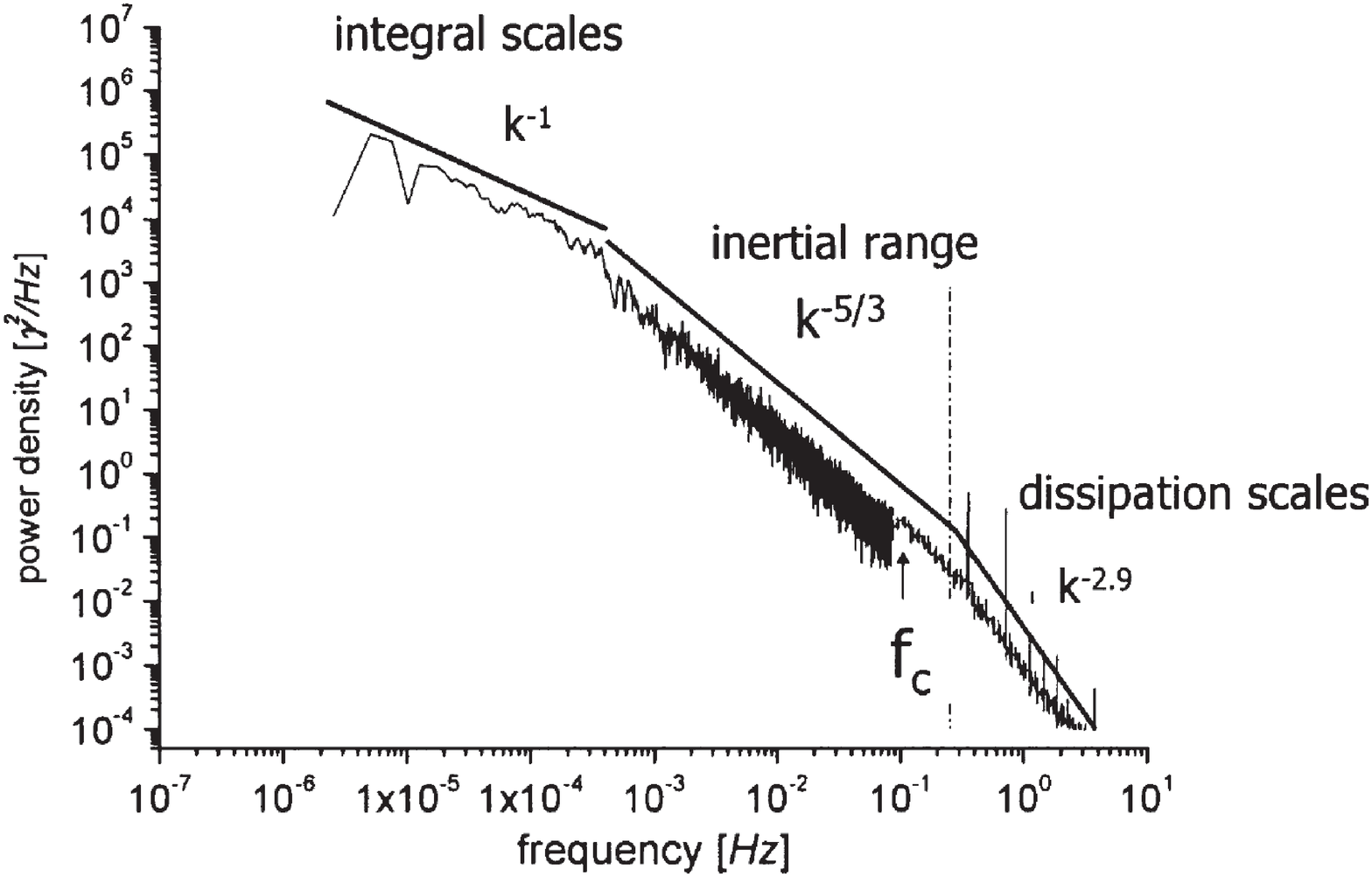}
\caption{Magnetic energy spectra of MHD turbulence embedded in an external magnetic field. Left panel: spectrum obtained from the VKS experiment (adapted from \citet{Bourgoin2002}). The dashed line corresponds to the motor frequency (8 Hz). Right panel: composite solar wind spectrum. From \citet{Bruno2005}.}
\label{MHDspectra-exp-sol}
\end{figure}

As mentioned in Sec.~\ref{s2:phenomenology}, the presence of an external magnetic field $\bb_0$ changes the energy transfer which occurs at the Alfv\'en time scale $t_A$ rather than at the eddy turn-over time scale $t_{NL}$, provided $b_0=|\bb_0|$ is strong enough. 
Assuming isotropy, the following spectra are expected \citep{Iroshnikov1963,Kraichnan1965}
  \begin{equation}
E^u(k) \propto E^b(k) \propto (b_0 \epsilon)^{1/2} k^{-3/2}.
\label{mhd 3_2_law}
\end{equation}
Unfortunately, in MHD turbulence experiments \citep{Odier1998,Alemany2000,Forest2007} the values of $\Rm$ which are possible to achieve are too low (less than 10 at the largest scale) to observe a sufficiently wide magnetic inertial range.
In the left panel of Fig.~\ref{MHDspectra-exp-sol},
a typical magnetic energy density spectrum is plotted versus frequency \citep{Bourgoin2002}.
Two slopes are observed, -1 and -11/3, neither of which can be attributed to an inertial range. The first slope is not easy to understand. On the other hand the second slope can be justified as follows \citep{Golitsyn1960,Moffatt1961}. First we have to assume that
the Taylor hypothesis applies in order to interpret the frequency as a wave number.
The induction equation~(\ref{MHDB}), replacing $\bb$ by $\bb + \bb_0$, where $\bb_0$ is taken to be independent of any spatial coordinate, and assuming $|\bb| \ll b_0$ (low $\Rm$), implies
\begin{equation}
\left(\partial_t-\eta \nabla^2 \right) \bb = (\bb_0 \cdot \nabla)  \bu.
\label{MHDB0}
\end{equation}
In a stationary statistic state
the induction term $(\bb_0 \cdot \nabla)  \bu$ is balanced by the dissipation term $\eta \nabla^2 \bb$, implying  $b_0 k u(k) \sim \eta k^2 b(k)$. Now assuming that the turbulent velocity obeys the Kolmogorov scaling law~(\ref{Scalinglaw}), we find
\begin{equation}
E^b(k) \propto b_0^2 \epsilon^{2/3} \eta^{-2} k^{-11/3}.
\label{11_3_law}
\end{equation}
Applying Taylor hypothesis, Eq.~(\ref{11_3_law}) implies that $E^b(k) \propto f^{-11/3}$.
We note that such line of argument assumes non-local interactions between the applied field $\bb_0$ and the small-scale turbulence.

If the IK scaling law~(\ref{mhd 3_2_law}) cannot be tested against experiments, at least it can be compared with observations.
The right panel of Fig.~\ref{MHDspectra-exp-sol} shows the magnetic energy density spectrum measured in the solar wind. The corresponding MHD turbulence subjected to the strong magnetic field $\bb_0$ emanating from the Sun, shows three successive slopes, again $-1$ at large scales,
$-5/3$ in the inertial range, and $-2.9$ at the smallest scales. An interesting point here is that an inertial Kolmogorov scaling law $f^{-5/3}$ is obtained instead of the IK scaling law $f^{-3/2}$ (for a discussion of the two other slopes see e.g. \citet{Verdini2012a} and
\citet{Howes2011}).

In fact anisotropy plays a crucial role in Alfv\'en wave turbulence
\citep{Goldreich1995}, leading to
modified definitions for both the Alfv\'en time
$t_A\propto(k_{\parallel}b_0)^{-1}$ and the eddy turn-over time $t_{NL}\propto (k_{\perp}b(k_{\perp}))^{-1}$, where the subscripts $\parallel$ and $\perp$ denote the directions
parallel and perpendicular to the applied field $\bb_0$. Two regimes are possible, depending whether $t_A \ll t_{NL}$ or $t_A \approx t_{NL}$. They are denoted by \textit{weak} and \textit{strong} turbulent regimes respectively.
In the \textit{weak} regime, on the basis of resonant three-wave interactions,
\citet{Galtier2000} found a cascade restricted to the perpendicular plane, with $E^b(k_{\perp})\propto k_{\perp}^{-2}$. This has been
numerically confirmed \citep{Boldyrev2009}.
In the \textit{strong} regime the cascade occurs in both perpendicular and parallel directions.
Provided the critical balance $t_A = t_{NL}$ is satisfied, the magnetic energy spectrum is now expected to satisfy  $E^b(k_{\perp})\propto k_{\perp}^{-5/3}$ and
$E^b(k_{\parallel})\propto k_{\parallel}^{-2}$ \citep{Goldreich1995}. This seems to be well supported by solar wind measurements \citep{Horbury2008}, but still lacks numerical confirmation. Instead simulations give an energy spectrum $E^b(k_{\perp})\propto k_{\perp}^{-3/2}$ \citep{Mueller2005,Mason2008}. It has been suggested that such discrepancy is due to the dominance of the one Els\"asser variable on the other \citep{Boldyrev2006}. However, recent results based on shell models in the perpendicular direction (Sec.~\ref{S4:Alfven}) manage to reproduce the transition between \textit{weak} and \textit{strong} turbulence for a ratio $t_A / t_{NL}$ varying from 0 to 1 \citep{Verdini2012b}.

\subsubsection{Transfer functions}
\label{s2:transfers}
From Eqs.~(\ref{Fourieru}-\ref{Fourierb}) and following \citet{Verma2004}, the time evolution of the Fourier modes of the kinetic and magnetic energies  $E^u(\bk)$ and $E^b(\bk)$ is given by
\begin{eqnarray}
(	\partial_t  + 2\nu k^2 )E^u(\bk) &=& \frac{1}{2}\sum_{\substack{\bp,\bq\\\bk+\bp+\bq=0}}\left(S^{uu}(\bk|\bp,\bq)+S^{ub}(\bk|\bp,\bq)\right) + \Re\left\{\bff(\bk)\cdot\bu^*(\bk)\right\}\label{SuuSub}\\
(	\partial_t  + 2\eta k^2 )E^b(\bk) &=& \frac{1}{2}\sum_{\substack{\bp,\bq\\\bk+\bp+\bq=0}}\left(S^{bu}(\bk|\bp,\bq)+S^{bb}(\bk|\bp,\bq)\right),\label{SbuSbb}
\end{eqnarray}
where each $S^{xy}(\bk|\bp,\bq)$ term represents the energy transfer rate from the modes $\bp$ and $\bq$ of field $\by$,
 into the mode $\bk$ of field $\bx$. They are defined as
\begin{eqnarray}
	S^{uu}(\bk|\bp,\bq)&=&S^{uu}(\bk|\bp|\bq)+S^{uu}(\bk|\bq|\bp) \\
	S^{ub}(\bk|\bp,\bq)&=&S^{ub}(\bk|\bp|\bq)+S^{ub}(\bk|\bq|\bp) \\
	S^{bu}(\bk|\bp,\bq)&=&S^{bu}(\bk|\bp|\bq)+S^{bu}(\bk|\bq|\bp) \\
	S^{bb}(\bk|\bp,\bq)&=&S^{bb}(\bk|\bp|\bq)+S^{bb}(\bk|\bq|\bp),
\end{eqnarray}
with
\begin{eqnarray}
	S^{uu}(\bk|\bp|\bq)&=&-\Im{\left\{\left[\bk\cdot\bu(\bq)\right]\left[\bu(\bk)\cdot\bu(\bp)\right]
	                                  \right\}} \\
	S^{ub}(\bk|\bp|\bq)&=&+\Im{\left\{\left[\bk\cdot\bb(\bq)\right]\left[\bu(\bk)\cdot\bb(\bp)\right]
	                                  \right\}} \\
	S^{bu}(\bk|\bp|\bq)&=&+\Im{\left\{\left[\bk\cdot\bb(\bq)\right]\left[\bb(\bk)\cdot\bu(\bp)\right]
	                                  \right\}} \\
	S^{bb}(\bk|\bp|\bq)&=&-\Im{\left\{\left[\bk\cdot\bu(\bq)\right]\left[\bb(\bk)\cdot\bb(\bp)\right]
	                                  \right\}},
\end{eqnarray}
and where the terms $S^{xy}(\bk|\bq|\bp)$ are obtained from $S^{xy}(\bk|\bp|\bq)$ by exchanging $\bp$ and $\bq$.
Each $S^{xy}(\bk|\bp|\bq)$ term represents the mode-to-mode energy transfer rate from the mode $\bp$ of field $\by$
 into the mode $\bk$ of field $\bx$, with the mode $\bq$ acting as a mediator.

Another way to write Eqs.~(\ref{SuuSub}-\ref{SbuSbb}) is
\begin{eqnarray}
(	\partial_t  + 2\nu k^2 )E^u(\bk) &=& T^{uu}(\bk)+T^{ub}(\bk) +\Re\left\{\bff(\bk)\cdot\bu^*(\bk)\right\}\label{TuuTub}\\
(	\partial_t  + 2\eta k^2 )E^b(\bk) &=& T^{bu}(\bk)+T^{bb}(\bk),\label{TbuTbb}
\end{eqnarray}
where the quantities $T^{xy}(\bk)$ are interpreted as the energy transfer rate from all modes of the $\by$-field into the $\bk$ mode of the $\bx$-field. They are defined as
\begin{equation}
	T^{xy}(\bk)=\frac{1}{2}\sum_{\substack{\bp,\bq\\\bk+\bp+\bq=0}} S^{xy}(\bk|\bp,\bq).
\end{equation}

\begin{figure}[ht]
\centering
\includegraphics[width=0.4\textwidth]{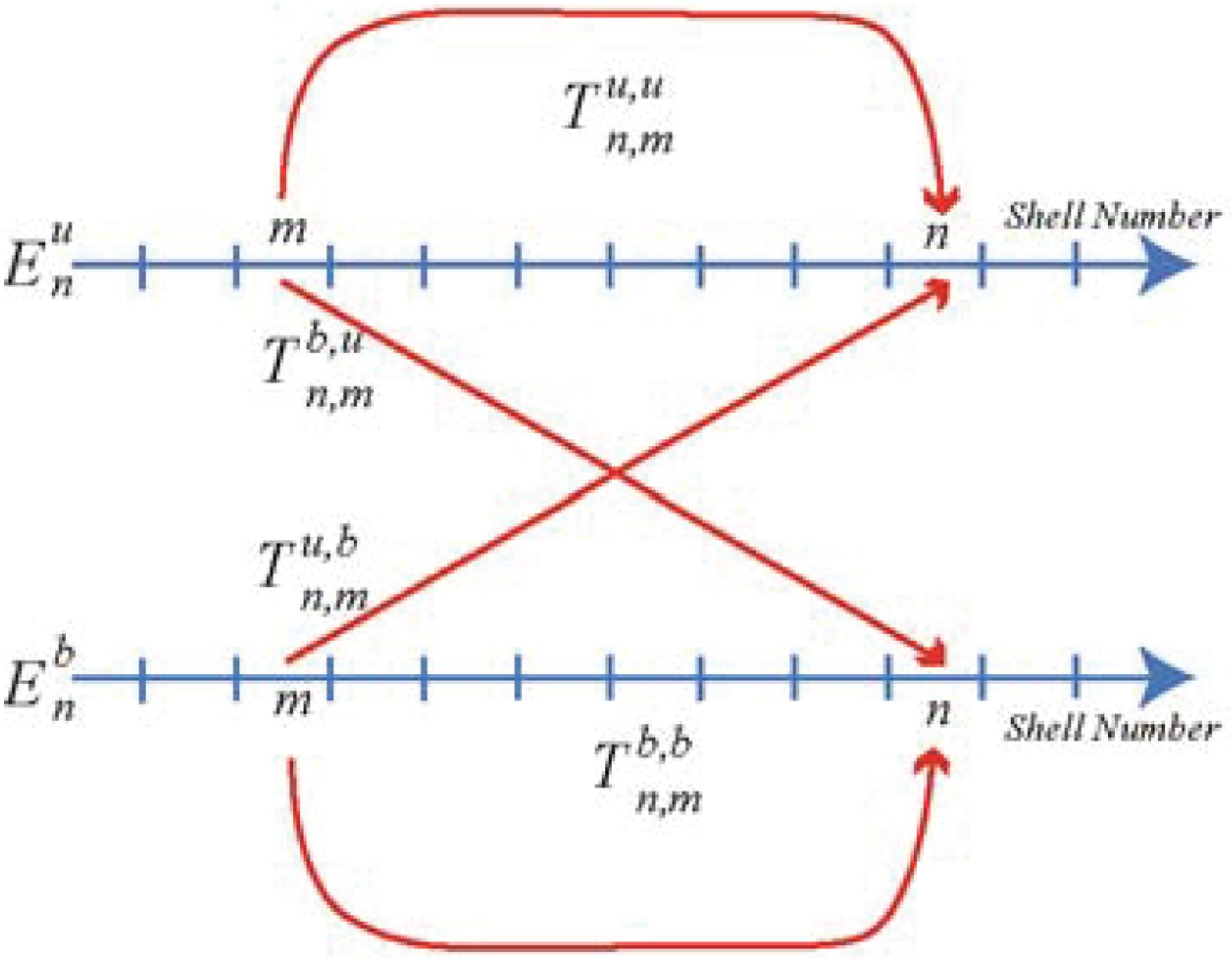}
\includegraphics[width=0.59\textwidth]{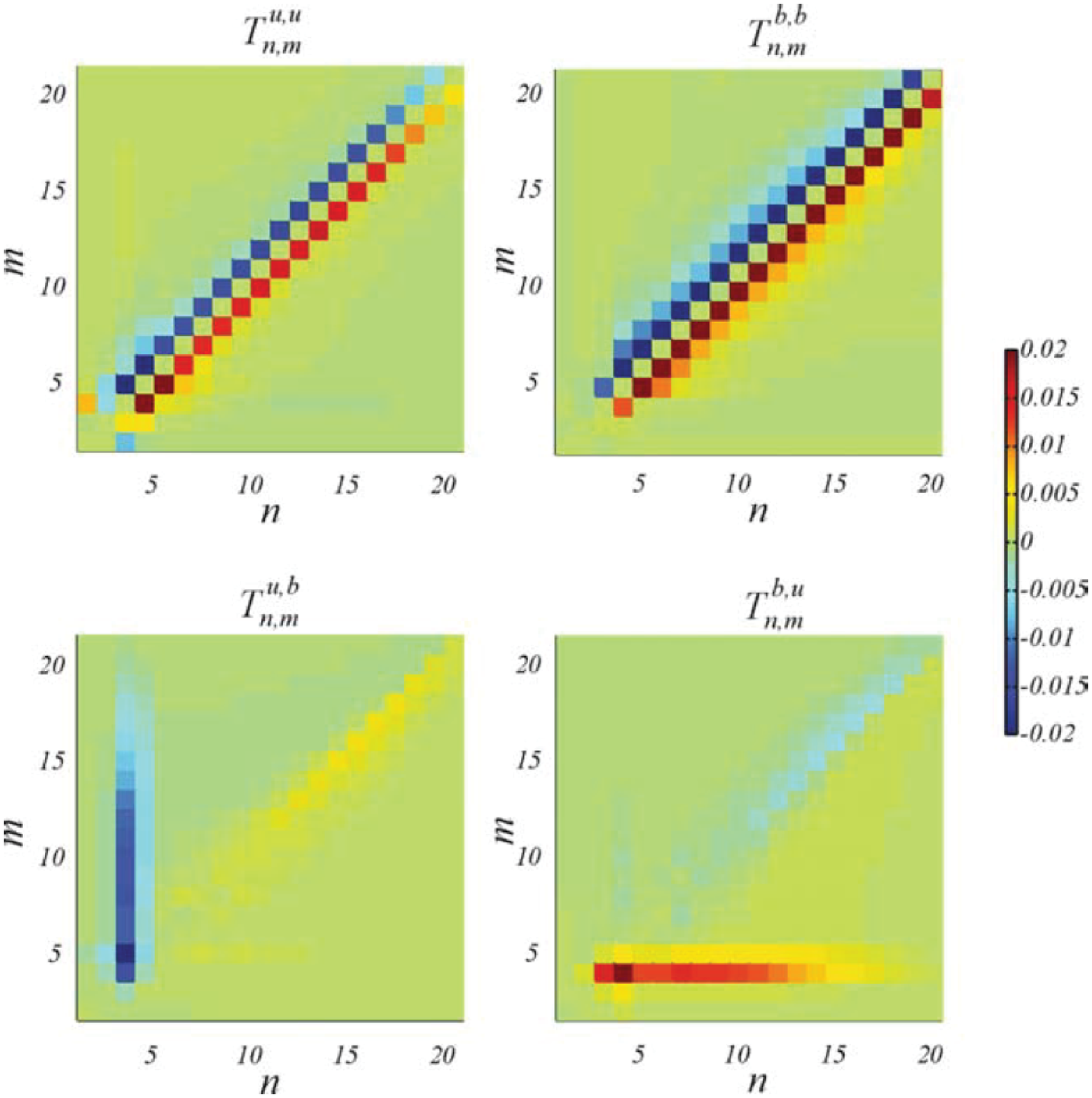}
\caption{Left panel: schematic representation of the shell-to-shell energy transfers between velocity and magnetic fields. Right panel: shell-to-shell energy transfers $T^{xy}_{nm}$ from $\by$ in shell $m$ to 
$\bx$ in shell $n$. From \citet{Carati2006}.}
\label{Figuretransfers}
\end{figure}

The Fourier space is divided into spherical shells $a_n$ that contain
all wave vectors $\bk$ such that
$k_{n}\le|\bk|< k_{n+1}$. The energy transfer rate from
the shell $m$ of field $\by$, to the shell $n$ of field $\bx$ is given by
\begin{equation}
	T^{xy}_{nm}=\sum_{\substack{\bk\in a_n,\bp\in a_m\\\bk+\bp+\bq=0}} S^{\bx\by}(\bk|\bp|\bq).
	\label{defTxynm}
\end{equation}
Defining the kinetic and magnetic energies in shell $n$ as
\begin{equation}
	E_n^u=\frac{1}{2}\sum_{\bk\in a_n}|\bu(\bk)|^2, \qquad E_n^b=\frac{1}{2}\sum_{\bk\in a_n}|\bb(\bk)|^2,
\end{equation}
we obtain
\begin{eqnarray}
\partial_t  E^u_n &=& \sum_{m} \left(T^{uu}_{nm} + T^{ub}_{nm}\right) -D^u_n + F^u_n \label{dtEun}\\
\partial_t  E^b_n &=& \sum_{m} \left(T^{bu}_{nm} + T^{bb}_{nm}\right) -D^b_n,\label{dtEbn}
\end{eqnarray}
with
\begin{equation}
	D^u_n=\nu \sum_{\bk\in a_n}k^2|\bu(\bk)|^2, \qquad D^b_n=\eta \sum_{\bk\in a_n}k^2|\bb(\bk)|^2, \qquad F^u_n= \sum_{\bk\in a_n}\Re\left\{\bff(\bk) \cdot \bu^*(\bk)\right\}.
\end{equation}
Similar definitions of shell-to-shell energy transfers were used in \citet{Alexakis2005a} and \citet{Alexakis2005}.
A schematic representation of the shell-to-shell energy transfers is given in Fig.~\ref{Figuretransfers} (left),
together with in Fig.~\ref{Figuretransfers} (right) the results obtained from the same DNS used to produce the energy spectrum of Fig.~\ref{MHDspectra_Carati}. The two top figures in Fig.~\ref{Figuretransfers} (right) show that the $u$-to-$u$ and $b$-to-$b$ transfers are local and forward. On the other hand the two bottom figures show that the $b$-to-$u$ and $u$-to-$b$ transfers are non-local, with a strong contribution coming from the velocity forcing shell to all magnetic shells.
We stress that in DNS the sequence of $k_n$ is chosen to be arithmetic in contrast to \textit{shell models} in which the sequence is geometric.

Extending the previous definitions to fluxes, we can separate Fourier space into two parts, inside and outside a sphere of radius $k_n$. We define four fluxes from $\by$ to $\bx$, from the inside / outside of the $\by$-sphere to the inside / outside of the $\bx$-sphere,
\begin{eqnarray}
  \Pi^{x< y<}_n&=&\sum_{\substack{|\bk|<k_n, |\bp|<k_n\\\bk+\bp+\bq=0}} S^{xy}(\bk|\bp|\bq)\label{flux1}\\
	\Pi^{x> y<}_n&=&\sum_{\substack{|\bk|>k_n, |\bp|<k_n\\\bk+\bp+\bq=0}} S^{xy}(\bk|\bp|\bq)\\
	\Pi^{x< y>}_n&=&\sum_{\substack{|\bk|<k_n, |\bp|>k_n\\\bk+\bp+\bq=0}} S^{xy}(\bk|\bp|\bq)\\
	\Pi^{x> y>}_n&=&\sum_{\substack{|\bk|>k_n, |\bp|>k_n\\\bk+\bp+\bq=0}} S^{xy}(\bk|\bp|\bq).\label{flux4}
\end{eqnarray}

We note that spherical shell transfers are rather unsuited to anisotropic turbulence,
e.g. with a strong $\bb_0$ \citep{Teaca2009a}.
\citet{Alexakis2007} introduced cylindrical shells concentric with the direction of $\bb_0$, and plane layers perpendicular to the latter, leading to transfer maps between cylindrical shells on the one hand and parallel planes on the other hand. \citet{Teaca2009b} introduced ring-to-ring transfers by dividing each spherical shell into rings. This has the advantage of showing how the transfers are distributed with the angle between the ring and the direction of $\bb_0$.

Helicity transfer can also be defined in a way similar to that used for energy transfer.
Starting from Eqs.~(\ref{Fourieru}-\ref{Fourierb}), the time evolution of the cross helicity $H^c(\bk)$
and magnetic helicity $H^b(\bk)$ is
\begin{eqnarray}
(	\partial_t  + (\nu+\eta) k^2 )H^c(\bk) &=& \frac{1}{2}\sum_{\substack{\bp,\bq\\\bk+\bp+\bq=0}}S^c(\bk|\bp,\bq) + \Re\left\{\bff(\bk)\cdot\bb^*(\bk)\right\},\label{Hcevolution}\\
(	\partial_t  + 2\eta k^2 )H^b(\bk) &=& \frac{1}{2}\sum_{\substack{\bp,\bq\\\bk+\bp+\bq=0}}S^b(\bk|\bp,\bq),\label{Hbevolution}
\end{eqnarray}
where $S^c(\bk|\bp,\bq)$ and $S^b(\bk|\bp,\bq)$ are respectively the cross helicity and magnetic helicity transfer rates from modes $\bp$ and $\bq$,
to mode $\bk$. They are defined as
\begin{eqnarray}
	S^c(\bk|\bp,\bq)&=&S^c(\bk|\bp|\bq)+S^c(\bk|\bq|\bp),\\
	S^b(\bk|\bp,\bq)&=&S^b(\bk|\bp|\bq)+S^b(\bk|\bq|\bp),
\end{eqnarray}
where $S^c(\bk|\bp|\bq)$ and $S^b(\bk|\bp|\bq)$ are the transfer rates from mode $\bp$
to mode $\bk$, with the mode $\bq$ acting as a mediator.
Hence
\begin{eqnarray}	 S^c(\bk|\bp|\bq)&=&\Im\left\{-\left[\bu(\bq)\cdot\bk\right]\left[\bu(\bp)\cdot\bb(\bk)+\bb(\bp)\cdot\bu(\bk)\right]+\left[\bb(\bq)\cdot\bk\right]\left[\bu(\bp)\cdot\bu(\bk)+\bb(\bp)\cdot\bb(\bk)\right]\right\}, \\
S^b(\bk|\bp|\bq)&=&2\Re\left\{
 \bu(\bq)\cdot\left[\bb(\bp)\times \bb(\bk)\right]
\right\}.
\end{eqnarray}
We note that $S^c(\bk|\bp|\bq)=-S^c(\bp|\bk|\bq)$ and $S^b(\bk|\bp|\bq)=-S^b(\bp|\bk|\bq)$, meaning that the transfers from $\bp$-to-$\bk$ and $\bk$-to-$\bp$ are opposite, as expected from mode-to-mode transfers.

Defining the transfer rates from the shell $m$ to the shell $n$, as
\begin{equation}
	T^{c}_{nm}=\sum_{\substack{\bk\in a_n,\bp\in a_m\\\bq=-(\bk+\bp)}} S^{c}(\bk|\bp|\bq), \qquad
	T^{b}_{nm}=\sum_{\substack{\bk\in a_n,\bp\in a_m\\\bq=-(\bk+\bp)}} S^{b}(\bk|\bp|\bq),
\end{equation}
the helicities in shell $n$ as
\begin{equation}
	H_n^c=\sum_{\bk\in a_n}H^c(\bk), \qquad H_n^b=\sum_{\bk\in a_n}H^b(\bk),
\end{equation}
gives
\begin{eqnarray}
\partial_t  H^c_n &=& \sum_{m} T^{c}_{nm}  -D^c_n + F^b_n \label{dtHcn},\\
\partial_t  H^b_n &=& \sum_{m} T^{b}_{nm}  -D^b_n,  \label{dtHbn}
\end{eqnarray}
with
\begin{equation}
	D^c_n=(\nu+\eta) \sum_{\bk\in a_n}k^2H^c(\bk), \qquad 
	D^b_n=2\eta \sum_{\bk\in a_n}k^2H^b(\bk), \qquad 
	F^b_n= \sum_{\bk\in a_n}\Re\left\{\bff(\bk) \cdot \bb^*(\bk)\right\}.
\end{equation}
The inverse cascade of magnetic helicity has been studied numerically by \citet{Alexakis2006}, on the basis of a similar shell-to-shell formalism.

\subsubsection{Helical decomposition}
\label{Helical decomposition}
Following the approach presented by \citet{Waleffe1992} for HD turbulence, we introduce, in Fourier space, a base of polarized helical waves $\textbf{h}^{\pm}$ defined as the eigenvectors of the curl operator \citep{Craya1958,Herring1974,Cambon1989},
\begin{equation}
	\i \bk \times \bh^{\pm} = \pm k \bh^{\pm}.
	\label{roteig}
\end{equation}
Note that the helical vectors $\bh^{\pm}(\bk)$ are defined up to an arbitrary rotation of axis $\bk$. \citet{Waleffe1992} suggests taking
\begin{equation}
\bh^{\pm}(\bk) = \bu_2(\bk) \pm \i\bu_1(\bk)
\end{equation}
with $\bu_1(\bk) = (\bz_{\bk}\times \bk)/|(\bz_{\bk}\times \bk)|$ and
$\bu_2(\bk)= \bu_1(\bk) \times \bk /k$, where $\bz_{\bk}$
is an arbitrary vector that, in general, may depend on $\bk$, though it is not proportional to $\bk$.
\citet{Lessinnes2009a} extended this approach to MHD with the following line of argument.

\begin{figure}[ht]
\begin{center}
\includegraphics[width=0.3\textwidth]{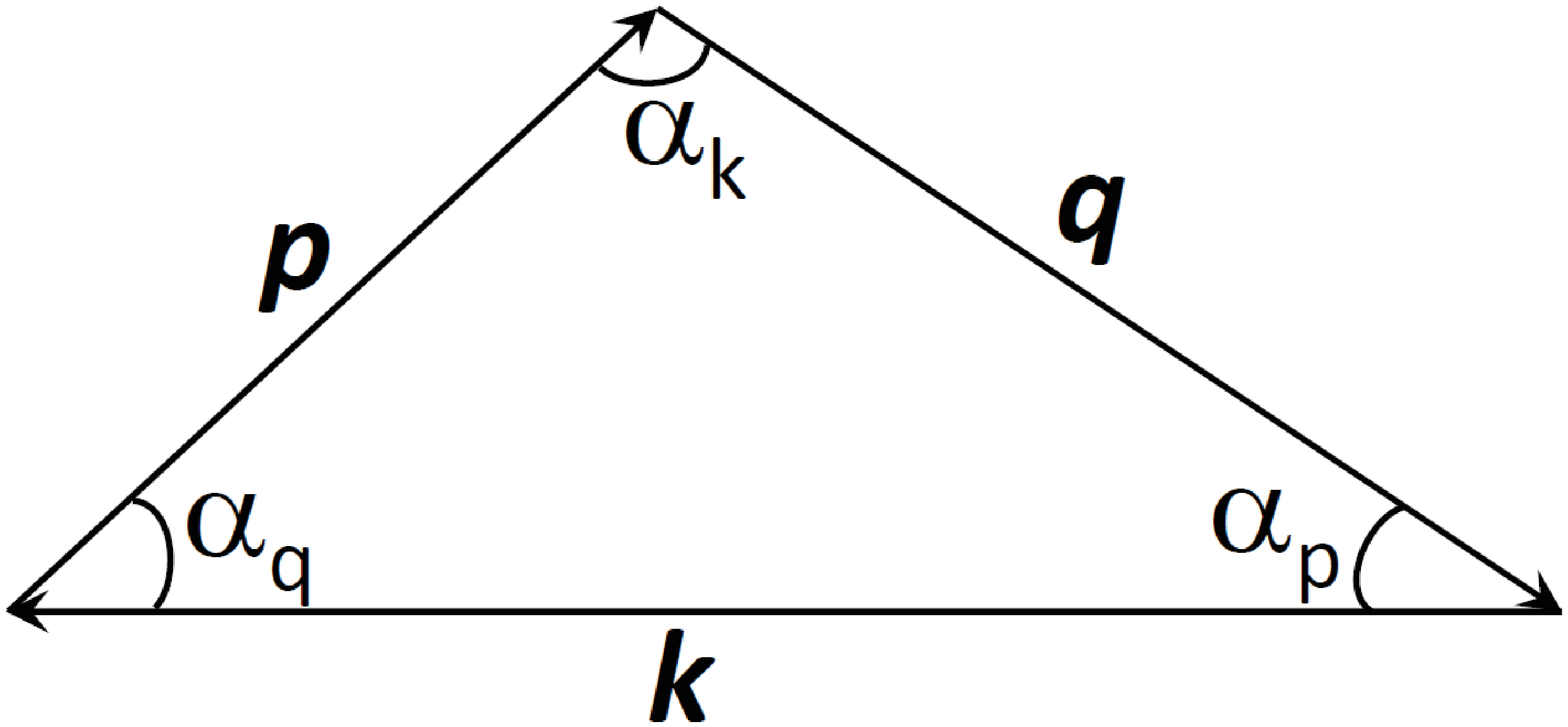}
\caption{Representation of the triad formed by the wave vectors $\bk$, $\bp$ and $\bq$.}
\label{figtri}
\end{center}
\end{figure}

The Fourier modes of both fields, velocity and magnetic, are expanded on that helical base
\begin{eqnarray}
\bu(\bk)&=&u^+(\bk)\, \bh^+ +u^-(\bk)\, \bh^-,\label{helu}\\
\bb(\bk)&=&b^+(\bk)\, \bh^+ +b^-(\bk)\, \bh^-,\label{helb}
\end{eqnarray}
leading to energies and cross helicity expressions
\begin{eqnarray}
	E^u(k)&=&\frac{1}{2}\left(|u^+(\bk)|^2 + |u^-(\bk)|^2\right), \quad \quad \quad \label{EUwal}\\
  E^b(k)&=&\frac{1}{2}\left(|b^+(\bk)|^2 + |b^-(\bk)|^2\right), \quad \quad \quad \label{EBwal}\\
  H^c(k)&=&\frac{1}{2}\left(u^+(\bk) {b^+}^*(\bk) + u^-(\bk) {b^-}^*(\bk)+ c.c.\right). \quad \quad \quad \label{HCwal}
\end{eqnarray}
The vorticity $\boldsymbol{\omega}$ and potential vector $\ba$ (with an appropriate gauge) can also be
expanded on the same base
\begin{eqnarray}
\bom(\bk)&=&k\left(u^+(\bk)\, \bh^+ -u^-(\bk)\, \bh^-\right),\\
\ba(\bk)&=&k^{-1}\left(b^+(\bk)\, \bh^+ -b^-(\bk)\, \bh^-\right),
\end{eqnarray}
leading to the kinetic and magnetic helicities
\begin{eqnarray}
  H^u(k)&=&k\left(|u^+(\bk)|^2-|u^-(\bk)|^2 \right) \label{HUwal}\\
  H^b(k)&=&k^{-1}\left(|b^+(\bk)|^2-|b^-(\bk)|^2 \right), \quad \quad \quad \label{HBwal}\\
\end{eqnarray}
enstrophy and square potential
\begin{eqnarray}
  \Xi(k)&=&\frac{1}{2}k^2\left(|u^+(\bk)|^2+|u^-(\bk)|^2 \right) \label{Enstrophywal}\\
  A(k)&=&\frac{1}{2}k^{-2}\left(|b^+(\bk)|^2+|b^-(\bk)|^2 \right). \quad \quad \quad \label{Awal}\\
\end{eqnarray}
Replacing the expressions~(\ref{helu}-\ref{helb}) for $\bu$ and $\bb$ in Eqs.~(\ref{Fourieru}-\ref{Fourierb}), and projecting onto the helical base $\bh^{s_k} (\bk)$ ($s_k=\pm 1$) lead to the following system
\begin{eqnarray}
\left(\partial_t+\nu k^2\right)  u^{s_k}(\bk)  &=& \frac{1}{2}
\sum_{\substack{\bp,\bq\\\bk+\bp+\bq=0}} \sum_{s_p, s_q} \left(s_p p-s_q q\right)\, g\, \left(u^{s_p}(\bp) u^{s_q}(\bq)-b^{s_p}(\bp) b^{s_q}(\bq)\right)^*   + f^{s_k}(\bk),\label{MHDhelU}\\
\left (\partial_t+\eta k^2\right)  b^{s_k}(\bk) &=&
-\frac{1}{2} \sum_{\substack{\bp,\bq\\\bk+\bp+\bq=0}}
\sum_{s_p, s_q} s_k\, k\, g  \left(u^{s_p}(\bp) b^{s_q}(\bq)-b^{s_p}(\bp) u^{s_q}(\bq)\right)^*
\label{MHDhelB},
\end{eqnarray}
where $g$, a function of $\bk$, $\bp$, $\bq$, $s_k$, $s_p$ and $s_q$, is defined as
\begin{align}
g(\bk,\bp,\bq, s_k, s_p, s_q) &\equiv -\frac 1 {\bh^{s_k}(\bk)^* \cdot  \bh^{s_k}(\bk)} (\bh^{s_k}(\bk)^* \times \bh^{s_p}(\bp)^*)\cdot \bh^{s_q}(\bq)^*.\label{MHDgdef}
\end{align}
Considering a single triadic interaction $\bk+\bp+\bq=0$, it is not necessary to introduce an arbitrary unit vector $\bz_{\bk}$ to define the unit vectors $\bu_1$ and $\bu_2$. Indeed, there is a natural direction which is represented by the unit vector perpendicular to the plane of the triad:
\begin{equation}
\text{\boldmath $\lambda$} = (\bk \times \bp) /|\bk \times \bp|=(\bp \times \bq) /|\bp \times \bq|=(\bq \times \bk) /|\bq \times \bk|\,.
\end{equation}
A second unit vector $\text{\boldmath $\mu$}_{\bk}=\bk \times \text{\boldmath $\lambda$}/k$ is introduced and the helical vectors are then defined as
\begin{equation}
\bh^{s_k} (\bk) = \e^{\i s_k \varphi_{\bk}}\ \left(\text{\boldmath $\lambda$}+\i\, s_k\, \text{\boldmath $\mu$}_{\bk}\right)\,.\label{hlam}
\end{equation}
The angle $\varphi_{\bk}$ defines the rotation around $\bk$ needed to transform the base $(\text{\boldmath $\mu$}_{\bk},\text{\boldmath $\lambda$})$ onto the base $(\bu_1(\bk),\bu_2(\bk))$. Since the base $(\text{\boldmath $\mu$}_{\bk},\text{\boldmath $\lambda$})$ depends on the triad, the angle $\varphi_{\bk}$ is also a function of $(\bk, \bp, \bq)$. The coupling constant for this triad then simply reduces to
\begin{eqnarray}
g(\bk,\bp,\bq, s_k, s_p, s_q)&=& -
\e^{-\i(s_k \varphi_{\bk}+s_p \varphi_{\bp}+s_q \varphi_{\bq})}\, s_k\, s_p\, s_q\ (s_k \sin\alpha_k + s_p \sin \alpha_p + s_q \sin
\alpha_q)
\label{gwal1}\,, \\
&=&-\i\ \e^{-\i\Phi_{kpq}(s_k,s_p,s_q)}\ (s_k \sin\alpha_k + s_p \sin \alpha_p + s_q \sin
\alpha_q)\label{gwal2}
\end{eqnarray}
where the phase $\Phi_{kpq}(s_k,s_p,s_q)=s_k (\varphi_{\bk}+\pi/2)+s_p (\varphi_{\bp}+\pi/2)+s_q (\varphi_{\bq}+\pi/2))$, and $\alpha_k$, $\alpha_p$ and $\alpha_q$ are the triad angles (see Fig.~\ref{figtri}). 

The sines are defined analytically as:
\begin{equation}
\sin\alpha_k=\displaystyle{\frac{Q}{2\, p\, q}}\hspace{7truemm}
\sin\alpha_p=\displaystyle{\frac{Q}{2\, k\, q}}\hspace{7truemm}
\sin\alpha_q=\displaystyle{\frac{Q}{2\, k\, p}}\label{angletriad}\,,
\end{equation}
where $Q=\sqrt{2\,k^2\,p^2+2\,q^2\,p^2+2\,q^2\,k^2-k^4-q^4-p^4}$. The expression~(\ref{gwal2}) shows that $g$ depends on the shape of the triangle formed by the triad but not on its scale. In the ideal limit ($\nu=\eta=0$) and in the absence of external forcing, the triadic dynamical system obtained by neglecting all the interactions with wave vectors different from $\bk$, $\bp$ or $\bq$ is given by
\begin{equation}
\renewcommand{\arraystretch}{1.4}
\left\{
\begin{array}{ccccc}
d_t u^{s_k}(\bk)&=& g(\bk,\bp,\bq, s_k, s_p, s_q) &(s_p p-s_q q) & (u^{s_p}(\bp)\,u^{s_q}(\bq)-b^{s_p}(\bp)\,b^{s_q}(\bq))^*,\\
d_t u^{s_p}(\bp)&=& g(\bk,\bp,\bq, s_k, s_p, s_q) &(s_q q-s_k k) & (u^{s_q}(\bq)\,u^{s_k}(\bk)-b^{s_q}(\bq)\,b^{s_k}(\bk))^*,\\
d_t u^{s_q}(\bq)&=& g(\bk,\bp,\bq, s_k, s_p, s_q) &(s_k k-s_p p) & (u^{s_k}(\bk)\,u^{s_p}(\bp)-b^{s_k}(\bk)\,b^{s_p}(\bp))^*,\\
d_t b^{s_k}(\bk)&=& g(\bk,\bp,\bq, s_k, s_p, s_q) &  (-s_k k)  &   (u^{s_p}(\bp)\,b^{s_q}(\bq)-b^{s_p}(\bp)\,u^{s_q}(\bq))^*,\\
d_t b^{s_p}(\bp)&=& g(\bk,\bp,\bq, s_k, s_p, s_q) & (-s_p p)   &   (u^{s_q}(\bq)\,b^{s_k}(\bk)-u^{s_q}(\bq)\,b^{s_k}(\bk))^*,\\
d_t b^{s_q}(\bq)&=& g(\bk,\bp,\bq, s_k, s_p, s_q) & (-s_q q)   &   (u^{s_k}(\bk)\,b^{s_p}(\bp)-u^{s_k}(\bk)\,b^{s_p}(\bp))^*.
\end{array}\right .\label{MHDtriad}
\renewcommand{\arraystretch}{1}
\end{equation}
This dynamical system couples six \textit{complex} variables. The geometric and scale independent $g$ factor is the same in all equations. The second prefactors in Eqs.~(\ref{MHDtriad}) only depend on the wave numbers of the triad or, more specifically, on the eigenvalues of the curl operator. The nature of interactions in Eqs.~(\ref{MHDtriad}) is obviously affected by the values of the parameters $s_k=\pm1$, $s_p=\pm1$ and $s_q=\pm1$ (eight possible choices). However, the structure of the system is unchanged if all the signs of $s_k,s_p$ and $s_q$ are reversed. Therefore, there are only four different types of interaction.
In each triad ($\bk, \bp, \bq$), Eqs.~(\ref{MHDtriad}) automatically conserve the total energy $E^u(\bk)+E^b(\bk) + E^u(\bp)+E^b(\bp)+E^u(\bq)+E^b(\bq)$, the magnetic helicity $H^b(\bk)+H^b(\bp)+H^b(\bq)$, and cross helicity $H^c(\bk)+H^c(\bp)+H^c(\bq)$. This is also true for kinetic energy and helicity in HD.

Such helical decomposition turns out to be useful in the derivation of shell models of turbulence, mainly because the kinetic and magnetic helicities are then unambiguously defined.
 

%% file: Shell.tex
\section{Derivation of MHD shell models}
\label{s3}

\subsection{Principles and generic equations}

\subsubsection{The HD GOY model as a first example}
\label{s3:GOY}
\begin{figure}[ht]
\centering
\includegraphics[width=0.3\textwidth]{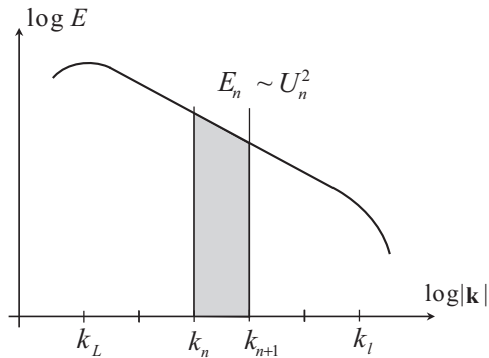}
\caption{Illustration of the shell $n$ (in gray).}
\label{f3:shells}
\end{figure}

Shell models have been elaborated keeping in mind the spectral representation of the Navier-Stokes equations.
The Fourier space is divided into spherical shells, defined by
\begin{equation}
	k_n \le | \bk|<k_{n+1}
\end{equation}
for which an illustration is given in Fig.~\ref{f3:shells}.

The sequence of $k_n$ is chosen to be geometric with the common ratio  $\lambda$. Therefore $k_n=k_0 \lambda^n$ and the kinetic energy in a given shell $n$ is given by
\begin{equation}
	E^U_n = \int_{k_n}^{k_{n+1}} E^u(k) dk.
\end{equation}

Next we introduce a \textit{complex} quantity $U_n$, such that $|U_n|^2/2$
 characterizes the kinetic energy $E^U_n$ in shell $n$. This quantity $U_n$ depends on time only and is interpreted
as a typical velocity fluctuation in shell $n$, a kind of collective variable for all fluctuations $\bu$ whose Fourier images belong to shell $n$.
As turbulence is assumed to be homogeneous and isotropic, all directions are equivalent and so all directional information is lost.

The simplest HD shell model consists of a system of ordinary differential equations of the form
\begin{equation}
d_t U_n = Q_n -\nu k_n^2 U_n + F_n, \qquad n \in\left\{1,\cdots,N\right\},
\label{shell_hydro}
\end{equation}
which mimics, to a degree, the original Navier-Stokes equations.
The term $-\nu k_n^2 U_n$ corresponds to the viscous dissipation of $U_n$ in shell $n$. The last (\textit{complex}) term $F_n$ corresponds to the forcing applied in shell $n$. In general, $F_n$ is applied only in one shell. However, nothing prevents from applying $F_n$ in several shells in order to control, for example, helicity injection in addition to energy injection (see Sec.~\ref{s4:forcing}).

Without the $Q_n$ term, Eq.~(\ref{shell_hydro}) is a simple diffusive equation with each shell being independent of the others. The $Q_n$ term mimics the non-linear interactions within triads. A variety of shell models can be derived depending
on the choice of $Q_n$. As an introductory example we choose $Q_n$ to have the form of the so-called GOY model \citep{Gledzer1973,Yamada1987}
\begin{equation}
	Q_n = \i k_n \left[a U_{n-2}U_{n-1} + b U_{n-1}U_{n+1} + c U_{n+1}U_{n+2} \right]^*
\label{QnGOY}
\end{equation}
where $a$, $b$ and $c$ are \textit{real} coefficients.
The expression of $Q_n$ is inspired by the Fourier form of the non-linear terms in Eq.~(\ref{Fourieru}).
We keep only the transfers within the subset of triads defined by $(k_{n-2},k_{n-1},k_{n})$, $(k_{n-1},k_{n},k_{n+1})$ and $(k_{n},k_{n+1},k_{n+2})$.

To obtain a model of 3D turbulence the coefficients $a$, $b$ and $c$ are derived writing that kinetic energy $E^U$ and helicity $H^U$ are ideally conserved. In this model the latter quantities are defined as
\begin{equation}
	E^U=\frac{1}{2}\sum\left|U_n\right|^2, \qquad  H^U=\frac{1}{2}\sum(-1)^nk_n\left|U_n\right|^2.
\end{equation}
In the absence of forcing and for $\nu=0$, the equations $d_t E^U=0$ and $d_t H^U=0$ take the form
\begin{equation}
\sum Q_n U_n^* + c.c.=0,  \qquad  \sum (-1)^n k_n Q_n U_n^* + c.c.=0
\end{equation}
where $c.c.$ denotes the complex conjugate.
This leads to
\begin{eqnarray}
\i\sum k_n\left[a\Delta_{n-1}+b\Delta_n+c\Delta_{n+1}\right] + c.c.&=&0 \label{GOYsum}\\
\i\sum (-1)^n k_n^2\left[a\Delta_{n-1}+b\Delta_n+c\Delta_{n+1}\right] + c.c.&=&0
\label{GOYsum1}
\end{eqnarray}
where $\Delta_n=U_{n-1}^*U_n^*U_{n+1}^*$.
With appropriate subscript changes Eqs.~(\ref{GOYsum}-\ref{GOYsum1}) are satisfied if and only if
\begin{eqnarray}
ak_{n+1}+bk_n+ck_{n-1}&=&0\\
a k_{n+1}^2 - b k_{n}^2 + c k_{n-1}^2&=&0.
\end{eqnarray}
Replacing $k_n$ by $k_0\lambda^n$ leads to
\begin{equation}
	a=-c/\lambda^3, \qquad b=-c(\lambda-1) / \lambda^2.
\end{equation}
Time and viscosity can be renormalized respectively by $c$ and $c^{-1}$, leading to an $c$-independent problem.
Taking $\lambda=2$ Eq.~(\ref{shell_hydro}) becomes
\begin{equation}
d_t U_n = \i k_n \left[U_{n+1}U_{n+2} -\frac{1}{4} U_{n-1}U_{n+1} -\frac{1}{8} U_{n-1}U_{n-2} \right]^* -\nu k_n^2 U_n + F_n, \qquad n \in\left\{1,\cdots,N\right\},
\label{shell_hydro2}
\end{equation}
which is the GOY model for 3D HD turbulence.

A GOY model can also be derived for 2D HD turbulence writing that enstrophy (defined in Eq.~(\ref{enstrophy})) instead of helicity is conserved. Taking $\lambda=2$ Eq.~(\ref{shell_hydro}) becomes
\begin{equation}
d_t U_n = \i k_n \left[U_{n+1}U_{n+2} -\frac{5}{8} U_{n-1}U_{n+1} +\frac{1}{16} U_{n-1}U_{n-2} \right]^* -\nu k_n^2 U_n + F_n, \qquad n \in\left\{1,\cdots,N\right\}.
\label{shell_hydro3}
\end{equation}

\subsubsection{General formalism of MHD shell models}
\label{s3:general formalism}
Any shell model can be recast within a general formalism such as the one introduced in \citet{Lessinnes2009} and \citet{Lessinnes2010}.
This formalism has the advantage of enhancing the link to the original MHD equations and of clarifying the differences between the variety of shell models elaborated so far.
The equations are given by
\begin{eqnarray}
   d_t \bU &=& \bQ(\bU)-\bQ(\bB) -   \nu \bD(\bU) + \bF, \label{shell_NS}\\
   d_t \bB &=& \bW(\bU,\bB)-\bW(\bB,\bU) -   \eta \bD(\bB),   \label{shell_induction}
\end{eqnarray}
where $\bU$ and $\bB$ are vectors in space $\Complex^N$, $N$ being the total number of shells,
\begin{equation}
	\bU=(U_1, U_2, \cdots, U_N), \quad \quad \bB=(B_1, B_2, \cdots, B_N).
\end{equation}
The coordinates $U_n$ and $B_n$ thus correspond to the velocity and magnetic fluctuations in shell $n$.

The linear operator $\bD$ is defined as
\begin{equation}
	\bD(\bX)=(k_1^2 X_1, k_2^2 X_2, \cdots, k_N^2 X_N).
\end{equation}
The vector $\bF$ stands for forcing and has non-zero coordinates only in shells which experience forcing.
The operators $\bQ$ and $\bW$ stand for the non-linear terms in Eqs.~(\ref{Fourieru}-\ref{Fourierb}).
The general expressions of the $n$-th coordinate of $\bQ$ and $\bW$ are assumed to be of the form
\begin{eqnarray}
	Q_n(\bX)&=&    k_n\sum_{i,j=1}^N a_{nij}^{Q} X_i X_j   + b_{nij}^{Q} X_i^* X_j + c_{nij}^{Q} X_i X_j^* + d_{nij}^{Q} X_i^* X_j^*.
	\label{Qnform}\\
	W_n(\bX,\bY)&=& k_n \sum_{i,j=1}^N a_{nij}^{W} X_i Y_j + b_{nij}^{W} X_i^* Y_j + c_{nij}^{W} X_i Y_j^* + d_{nij}^{W} X_i^* Y_j^*.
	\label{Wnform}
\end{eqnarray}

As an example, in the GOY model $a_{nij}^{Q}=b_{nij}^{Q}=c_{nij}^{Q}=0$, leading to the general form $Q_n(\bX)=k_n\sum d_{nij}^{Q} X_i^* X_j^*$. This choice is arbitrary.

To determine the remaining coefficients some additional criteria have to be applied. These are:
\begin{enumerate}
	\item the number of variables per shell,
	\item the number of shells interacting with a given shell $n$,
	\item the locality of the interactions between the shell $n$ and the other shells,
	\item the conservation laws,
	\item the symmetries coming from Eqs.~(\ref{MHDU}-\ref{MHDB}).
\end{enumerate}

In most MHD shell models, the variables in each shell $n$ are $U_n$ and $B_n$. However, in helical models such as the one presented in Sec.~\ref{s3:Helical mode decomposition}, the number of variables is doubled to $U_n^+$, $U_n^-$, $B_n^+$ and $B_n^-$.

In shell model terminology we distinguish the first-neighbor models from two-first-neighbor models, and the local from non-local models, depending on the interacting triads.
\begin{itemize}
	\item In L1-models (local, two feet in the same shell, the third foot in a neighboring shell), the shell $n$ is involved in triads $(n-1,n,n)$ and $(n,n+1,n+1)$, or $(n-1,n-1,n)$ and $(n,n,n+1)$, or all the four.
	\item In L2-models (local, three feet in three neighboring shells), the shell $n$ is involved in the triads $(n-2,n-1,n)$, $(n-1,n,n+1)$ and $(n,n+1,n+2)$. Clearly the GOY model introduced above is a L2-model.
	\item In N1-models (non-local, two feet in the same shell, the third foot in an inner shell), the shell $n$ is involved in triads $(n-m,n,n)$ and $(n,n+m,n+m)$.
  \item In N2-models (non-local, two feet in two neighbor shells, the third foot in an inner shell), the shell $n$ is involved in triads $(n-m-1,n-1,n)$, $(n-m,n,n+1)$ and $(n,n+m,n+m+1)$.
\end{itemize}

\begin{figure}[ht]
\centering
\includegraphics[width=0.4\textwidth]{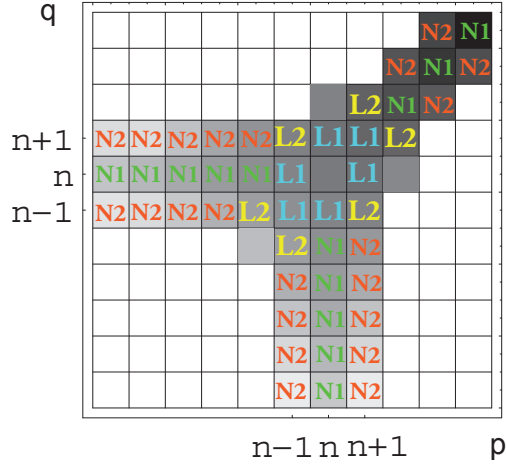}
\caption{Interactions between three modes belonging to shells $n,p$ and $q$. The graph is given in the map $(p,q)$. The labels are defined in the text and refer to different types of shell models. The gray intensity is proportional to the probability of interaction for $\lambda=\lambda_g$, the darker the higher the probability. White corresponds to zero probability.
Adapted from \citet{Plunian2007}.}
\label{interacprob}
\end{figure}
This classification is illustrated in Fig.~\ref{interacprob} where the interactions between three modes belonging to shells $n,p$ and $q$ are reported in the map $(p,q)$.
The possible interacting triads are dictated by the geometry. As mentioned in Sec.~\ref{s2:triads}, a triad is defined by three wave vectors $\bk, \bp, \bq$ such that $\bk+\bp+\bq=0$ and therefore define a triangle as in Fig.~\ref{figtri}. Here each vector belongs to a shell. For a common ratio $\lambda=2$, triads of the form $(n,n,n+m)$ with $m> 1$ are therefore impossible, as it would require a triangle with two identical sides and a third side larger than the sum of the two others. Of course the possible interactions depend on $\lambda$.
However, only the interactions corresponding to L1, L2, N1 and N2 are kept in shell models.
The gray squares represent the relative probability of interactions between three shells calculated for 
$\lambda=\lambda_g$, where $\lambda_g=(1+\sqrt{5})/2$ is the golden number. In this case we see in Fig.~\ref{interacprob} that using a shell model four possible interactions are ignored $(n,n,n), (n,n-2,n-2), (n,n,n+2), (n,n+2,n)$.
However, compared to the case $\lambda=2$ in which $(n,n,n)$ is also ignored and $(n,n-2,n-2), (n,n,n+2), (n,n+2,n)$ do not exist, the results do not change significantly.\\

The kinetic and magnetic energies $E^U$ and $E^B$, and the cross helicity $H^C$ are defined by
\begin{equation}
	E^U= \frac{1}{2}\bU^2, \quad E^B= \frac{1}{2}\bB^2, \quad H^C= \bU\cdot\bB,
	\label{conservative quantities}
\end{equation}
with the scalar product
\begin{equation}
	\bX\cdot\bY= \frac{1}{2}\sum_{i=1}^N(X_i Y_i^* + Y_i X_i^*).
	\label{scalar product}
\end{equation}
Note that $E^U$ is homogeneous to the square of a velocity. Therefore an energy spectrum $E^U_n=\frac{1}{2}U_n^2$
obeying to a scaling law $E^U_n \propto k_n^{\alpha}$ would give an energy density scaling law $E^u(k)\propto k^{\alpha -1}$.
Typically the -5/3 Kolmogorov scaling law for $E^u(k)$ corresponds to $E^U_n \propto k_n^{-2/3}$.\\

The conservation of total energy and cross helicity leads to
\begin{eqnarray}
   \left[\bQ(\bU)-\bQ(\bB)\right]\cdot\bU &+&  \left[\bW(\bU,\bB)-\bW(\bB,\bU)\right]\cdot\bB =  0, \label{energy conservation}\\
   \left[\bQ(\bU)-\bQ(\bB)\right]\cdot\bB &+&  \left[\bW(\bU,\bB)-\bW(\bB,\bU)\right]\cdot\bU =  0. \label{cross helicity conservation}
\end{eqnarray}
These equations must be satisfied for any vectors $\bU$ and $\bB$. In particular they must be satisfied when $\bU$ and $\bB$ are exchanged. This shows that both relations conservation of total energy
and cross helicity are equivalent, in agreement with Sec.~\ref{s2:conservative}.

Eqs.~(\ref{energy conservation}-\ref{cross helicity conservation}) must also be satisfied for $\bB=0$ (or $\bU=0$) implying that
\begin{equation}
   \bQ(\bX)\cdot\bX = 0, \label{X conservation}\\
\end{equation}
for any variable $\bX$.

Since a curl operator is included in the definition of the potential vector, the latter is not trivially defined in shell models. Therefore such a general framework fails to provide a general equation for the conservation of magnetic helicity in MHD or kinetic helicity in HD turbulence. 
Actually the definition of the latter quantities
depends on the type of the shell model used, helical or non helical. It is postponed to Sec.~\ref{section local}, \ref{s3:nonlocal} and \ref{s3:HM}.
The enstrophy $\Xi$ and squared magnetic potential $A$ are more easy to define as, contrary to helicity, they are always positive in any shell
\begin{equation}
	\Xi = \frac{1}{2}\sum_n k_n^2 |U_n|^2, \qquad A = \frac{1}{2}\sum_n k_n^{-2} |B_n|^2.
	\label{enstrophy}
\end{equation}

We note that 
the necessity of having the same structure L1, L2, N1 or N2, for $\bQ$ and $\bW$ follows from the conservation laws.\\

At this stage we have all the information necessary to start elaborating a shell model. It is a matter of choosing between the several possibilities mentioned earlier. However, we note that contrary to $\bQ$, the operator $\bW$ is not uniquely defined. Indeed $\bW(\bX,\bY)$ can be replaced by any operator of the form
\begin{equation}
	\widetilde{\bW}(\bX,\bY)= \frac{w}{w-1}\bW(\bX,\bY)+\frac{1}{w-1}\bW(\bY,\bX)
\label{mu}
\end{equation}
where $w$ is a scalar quantity,
without changing Eq.~(\ref{shell_induction}), or the conservation laws. Indeed it is easy to show that
\begin{equation}
	\widetilde{\bW}(\bX,\bY)-\widetilde{\bW}(\bY,\bX)=\bW(\bX,\bY)-\bW(\bY,\bX).
\end{equation}
In other words $\widetilde{\bW}(\bX,\bY)$ corresponds to a combination of
$(\bx\cdot \nabla)\by$ and $(\by\cdot \nabla)\bx$.
Though this does not change the shell model results in terms of $\bU$ and $\bB$,
it may change their analysis (post-processing) in terms of energy transfer and shell-to-shell exchange.

Following \citet{Lessinnes2009} we arbitrary make $\widetilde{\bW}(\bX,\bY)$
correspond to $-(\bx\cdot \nabla)\by$. From Eq.~(\ref{advectionproperty}) this implies that
\begin{equation}
	\widetilde{\bW}(\bX,\bY)\cdot \bY= 0,
	\label{Wtilde}
\end{equation}
and uniquely determines the value of $w$ in Eq.~(\ref{mu}).

So at this stage Eqs.~(\ref{shell_NS}-\ref{shell_induction}) can be rewritten in the form
\begin{eqnarray}
   d_t \bU &=& \bQ(\bU)-\bQ(\bB) -   \nu \bD(\bU) + \bF, \label{shell_NS21}\\
   d_t \bB &=& \widetilde{\bW}(\bU,\bB)-\widetilde{\bW}(\bB,\bU) -   \eta \bD(\bB),   \label{shell_induction21}
\end{eqnarray}
with
\begin{eqnarray}
   \bQ(\bX)\cdot \bX &=& 0, \label{X1}\\
   \widetilde{\bW}(\bX,\bY)\cdot \bY &=& 0 ,\label{X2}\\
   \bQ(\bX)\cdot \bY + \widetilde{\bW}(\bX,\bY)\cdot \bX &=& 0. \label{X3}
\end{eqnarray}
Now
Eqs.~(\ref{X2}-\ref{X3}) imply
\begin{equation}
	\widetilde{\bW}(\bX,\bX)= \bQ(\bX), \label{X4}
\end{equation}
which again is present in the original Navier-Stokes and induction equations.
To show that Eq.~(\ref{X4}) is satisfied, we start
by replacing $\bY$ by $\bX+\bY$
in Eq.~(\ref{X2}).
Assuming that $\widetilde{\bW}$ is a bilinear operator,

\begin{equation}
	\widetilde{\bW}(\bX,\bX)\cdot \bX +
	\widetilde{\bW}(\bX,\bX)\cdot \bY +
	\widetilde{\bW}(\bX,\bY)\cdot \bX +
	\widetilde{\bW}(\bX,\bY)\cdot \bY = 0\label{X5}
\end{equation}
which, from Eq.~(\ref{X2}) again, simplifies to
\begin{equation}
	\widetilde{\bW}(\bX,\bX)\cdot \bY +
	\widetilde{\bW}(\bX,\bY)\cdot \bX = 0.
	\label{X5b}
\end{equation}
From Eq.~(\ref{X3}), $\widetilde{\bW}(\bX,\bY)\cdot \bX$ can be replaced by
$-\bQ(\bX)\cdot \bY$ in Eq.~(\ref{X5}), leading to
\begin{equation}
	\left(\widetilde{\bW}(\bX,\bX)-\bQ(\bX)\right)\cdot \bY=0. \label{X6}
\end{equation}
As Eq.~(\ref{X6}) must be satisfied for any $\bY$, it implies Eq.~(\ref{X4}).

Finally Eqs.~(\ref{shell_NS}-\ref{shell_induction}) can be rewritten in the form
\begin{eqnarray}
   d_t \bU &=& \widetilde{\bW}(\bU,\bU)-\widetilde{\bW}(\bB,\bB) -   \nu \bD(\bU) + \bF, \label{shell_NS2}\\
   d_t \bB &=& \widetilde{\bW}(\bU,\bB)-\widetilde{\bW}(\bB,\bU) -   \eta \bD(\bB),   \label{shell_induction2}
\end{eqnarray}
with
\begin{equation}
	\widetilde{\bW}(\bX,\bY)\cdot \bY = 0 \label{relation1}
\end{equation}
being the only condition required to be satisfied so that both total energy and cross helicity are conserved along with
the symmetries of the non-linear operators in the original Navier-Stokes and induction equations.
In particular Eqs.~(\ref{X1}) and (\ref{X3}) are automatically satisfied,
using again the bilinear property of $\widetilde{\bW}$ to show Eq.~(\ref{X3}).

Note that changing $\bU$ to $K\bU$, $\bB$ to $K\bB$ and $\widetilde{\bW}$ to $K^{-1}\widetilde{\bW}$,
where $K$ is scalar quantity,
does not change the system of Eqs.~(\ref{shell_NS2}-\ref{shell_induction2}).

Similar to $\bW$, $\widetilde{\bW}$ takes the general form
\begin{equation}
	\widetilde{W}_n(\bX,\bY)=\sum_{i,j=1}^N a_{nij}^{\widetilde{W}} X_i Y_j + b_{nij}^{\widetilde{W}} X_i^* Y_j + c_{nij}^{\widetilde{W}} X_i Y_j^* + d_{nij}^{\widetilde{W}} X_i^* Y_j^*.
	\label{Wntildeform1}
\end{equation}

As the velocity $\bu$ and induction $\bb$ are divergence-free, the Navier-Stokes and induction equations
satisfy Liouville's theorem in the ideal limit $\nu=\eta=0$. For shell models Liouville's theorem
gives
\begin{equation}
	\sum_n \frac{\partial}{\partial U_n}\left(\frac{d U_n}{dt}\right) + \frac{\partial}{\partial B_n}\left(\frac{d B_n}{dt}\right)
	+ \frac{\partial}{\partial U^*_n}\left(\frac{d U^*_n}{dt}\right) + \frac{\partial}{\partial B^*_n}\left(\frac{d B^*_n}{dt}\right) = 0.
	\label{e3:Liouville}
\end{equation}

Finally, the magnetic helicity which has not been considered so far will give rise to an important additional constraint on  shell models designed for 3D MHD turbulence.

\subsubsection{Energy Flux}
\label{s3:fluxes}

\begin{figure}[ht]
\centering
\includegraphics[width=0.35\textwidth]{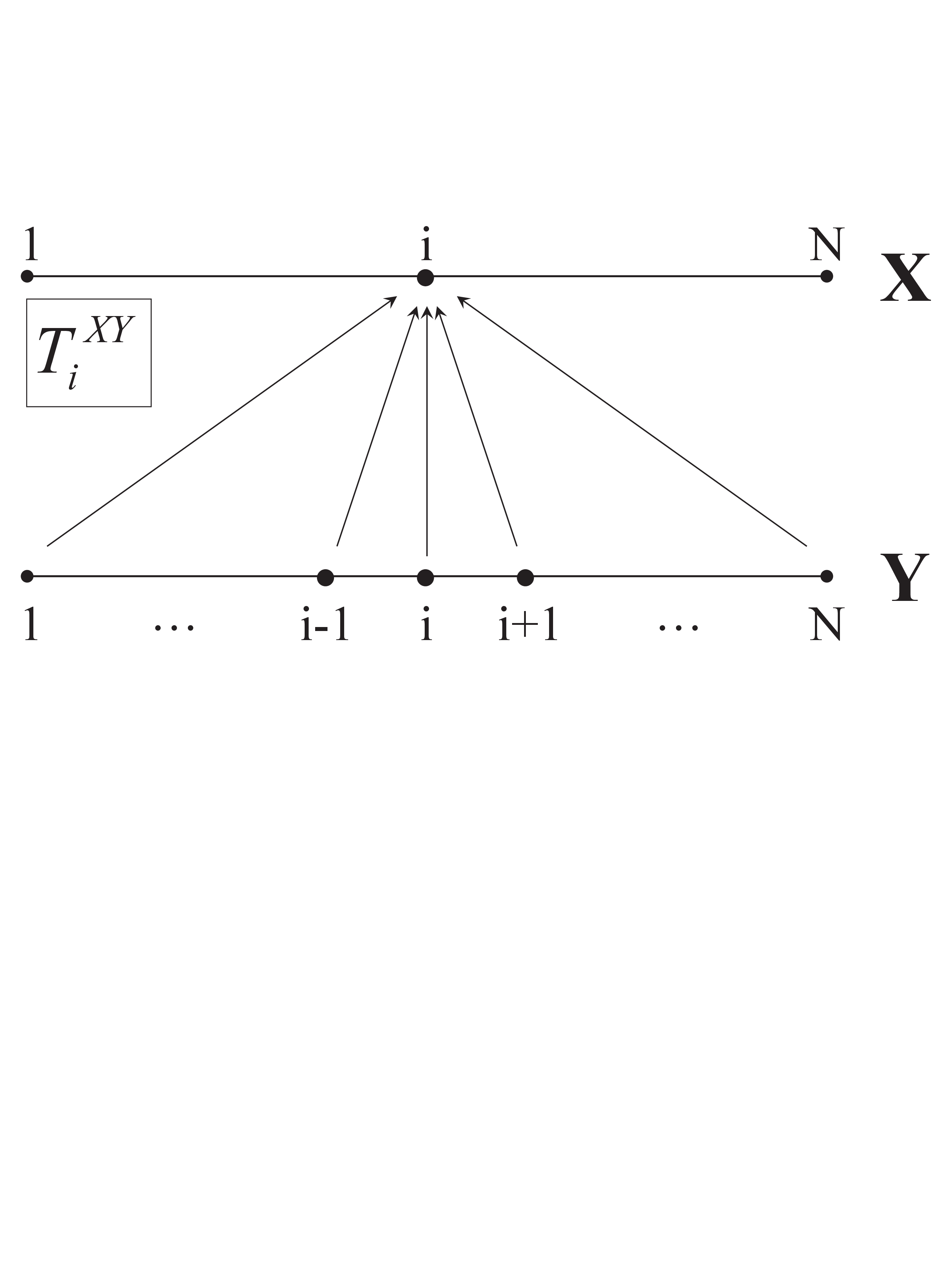} \qquad
\includegraphics[width=0.35\textwidth]{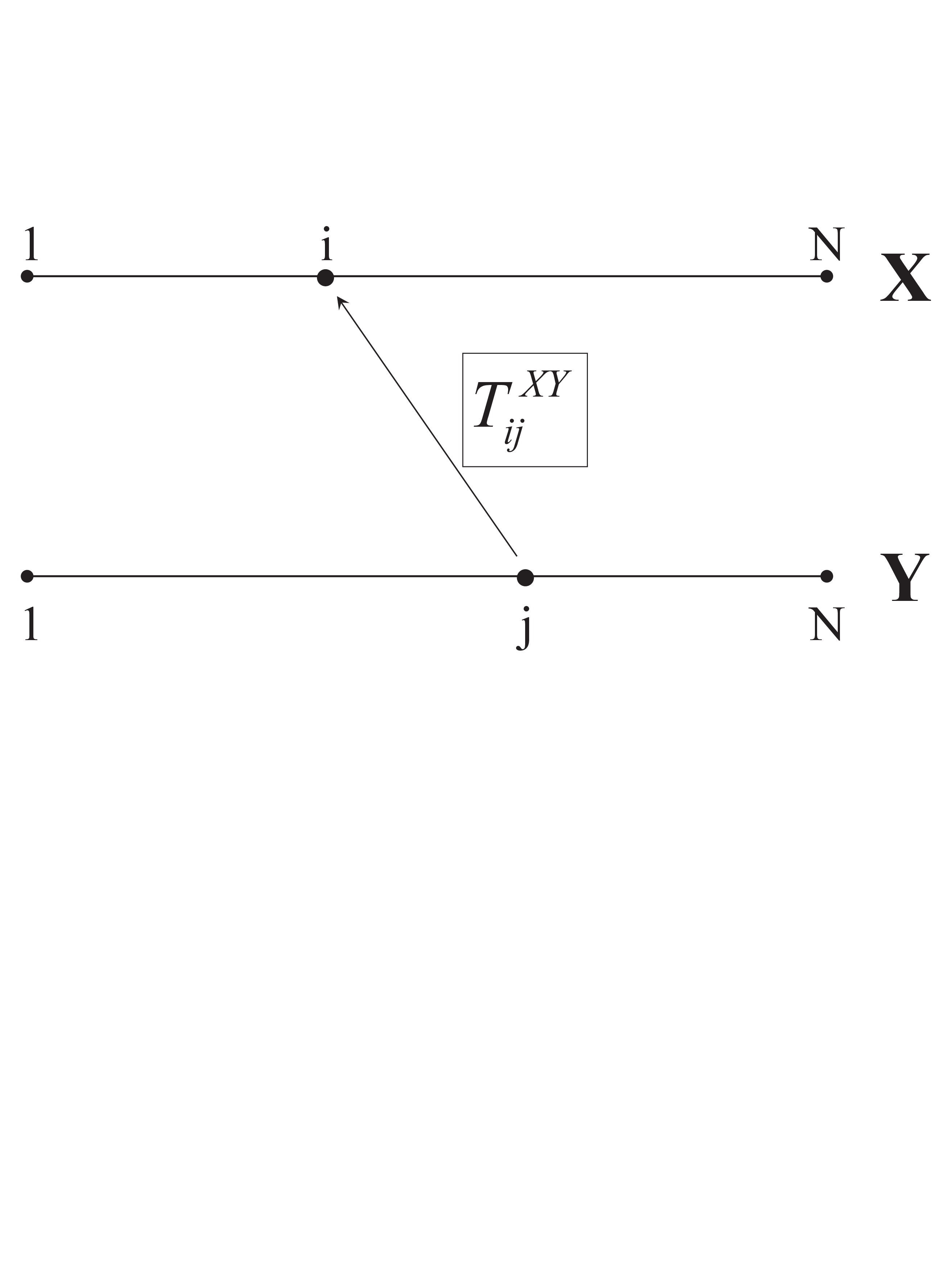}
\caption{Illustrations of the energy rate flowing into the $i$-th shell of $\bX$, coming from
either all shells of $\bY$ (left) or only one shell $j$ of $\bY$ (right).}
\label{TXY}
\end{figure}

Starting from Eqs.~(\ref{shell_NS2}-\ref{shell_induction2}) we obtain the kinetic and magnetic energy equations in any shell $i$
\begin{eqnarray}
    d_t U_i^2/2 &=& T_i^{UU} + T_i^{UB} -  \nu k_i^2 U_i^2 + \bF\cdot \bU_i, \label{shell_NS3}\\
    d_t B_i^2/2 &=& T_i^{BB} + T_i^{BU}   -  \eta k_i^2 B_i^2,   \label{shell_induction3}
\end{eqnarray}
where
\begin{eqnarray}
    &T_i^{UU}=\widetilde{\bW}(\bU,\bU)\cdot \bU_i  ,& T_i^{UB} = -\widetilde{\bW}(\bB,\bB)\cdot \bU_i, \label{T1}\\
    &T_i^{BB}=\widetilde{\bW}(\bU,\bB)\cdot \bB_i  ,& T_i^{BU} = -\widetilde{\bW}(\bB,\bU)\cdot \bB_i, \label{T2}
\end{eqnarray}
and $\bX_i = (0,\cdots,0,X_i,0, \cdots,0)$.
The quantities $T_i^{XY}$ are interpreted as the energy rate flowing from all shells of the $\bY$-field into the $i$-th shell of the $\bX$-field. They are illustrated in Fig.~\ref{TXY}.

Eqs.~(\ref{X5b}) and (\ref{relation1})  imply respectively
$\sum\limits_{i=1}^N T_i^{UB} + \sum\limits_{i=1}^N T_i^{BU}=0$ and
$\sum\limits_{i=1}^N T_i^{UU} = \sum\limits_{i=1}^N T_i^{BB}=0$.\\

Going one step further we introduce the quantity $T_{ij}^{XY}$, the shell-to-shell energy exchange rate from shell $j$ of the $\bY$-field into shell $i$ of the $\bX$-field, defined as
\begin{eqnarray}
    &T_{ij}^{UU}=\widetilde{\bW}(\bU,\bU_j)\cdot \bU_i  ,& T_{ij}^{UB} = -\widetilde{\bW}(\bB,\bB_j)\cdot \bU_i, \label{T3}\\
    &T_{ij}^{BB}=\widetilde{\bW}(\bU,\bB_j)\cdot \bB_i  ,& T_{ij}^{BU} = -\widetilde{\bW}(\bB,\bU_j)\cdot \bB_i, \label{T4}
\end{eqnarray}
implying that
\begin{equation}
	T_i^{XY}=\sum_{j=1}^N T_{ij}^{XY},
\end{equation}
that is illustrated in Fig.~\ref{TXY}.
Then the energy exchange rate from $\bY_j$ to $\bX_i$ must be opposite to that from $\bX_i$ to $\bY_j$:
\begin{equation}
	T_{ij}^{XY} = - T_{ji}^{YX}.
	\label{TijTji}
\end{equation}
This implies the conservation of total energy (\ref{energy conservation}), the relation (\ref{X conservation})
for $\bX=\bU$, and Eq.~(\ref{Wtilde}). The notation $T_{ij}^{XY}$ is chosen in agreement with $T_{nm}^{xy}$
given in Eq.~(\ref{defTxynm}).

In order to calculate energy fluxes we introduce the following vectors
\begin{eqnarray}
    \bX_n^< &=& (X_1, \cdots, X_n, 0, \cdots, 0), \label{Xninf}\\
    \bX_n^> &=& (0, \cdots, 0, X_{n+1}, \cdots, X_N), \label{Xnsup}\\
\end{eqnarray}
with $\bX=\bX_n^<+\bX_n^>$.
The corresponding energies are defined in the same manner as in Eq.~(\ref{conservative quantities}),
\begin{equation}
	E_n^{X^{<}}=\frac{1}{2}(\bX_n^<)^2, \qquad E_n^{X^{>}}=\frac{1}{2}(\bX_n^>)^2.
\end{equation}
Then from Eqs.~(\ref{shell_NS3}-\ref{shell_induction3}),
\begin{eqnarray}
    d_t E_n^{U^{<}} &=& \sum_{i=1}^n \left(T_i^{UU} + T_i^{UB}\right) -  \nu \bD(\bU)\cdot \bU_n^< + \bF\cdot \bU_n^<, \label{shell_NS31}\\
    d_t E_n^{B^{<}} &=& \sum_{i=1}^n \left(T_i^{BB} + T_i^{BU}\right)   -  \eta \bD(\bB)\cdot \bB_n^<,   \label{shell_induction31}
\end{eqnarray}
with $\bD(\bX)\cdot \bX_n^<=\sum\limits_{i=1}^nk_i^2 X_i^2$.
The quantity
$\sum\limits_{i=1}^n T_i^{XY}$ defines the flux from $\bY$ (in all shells) to $\bX_n^<$. It is denoted by
\begin{equation}
	\Pi_{X^<}^{Y} \equiv \sum_{i=1}^n T_i^{XY}. \label{flux21}
\end{equation}
We also define
\begin{eqnarray}
	\Pi_{X^<}^{Y^<}&\equiv&\sum_{i=1}^n \sum_{j=1}^n T_{ij}^{XY}, \qquad
	\Pi_{X^<}^{Y^>}\equiv\sum_{i=1}^n \sum_{j=n+1}^N T_{ij}^{XY},\nonumber \\
	\Pi_{X^>}^{Y^<}&\equiv&\sum_{i=n+1}^N \sum_{j=1}^n T_{ij}^{XY}, \qquad
	\Pi_{X^>}^{Y^>}\equiv\sum_{i=n+1}^N \sum_{j=n+1}^N T_{ij}^{XY}, \label{flux22}
\end{eqnarray}
which is the shell model counterpart of Eqs.~(\ref{flux1}-\ref{flux4}).
In the flux notation given in Eqs.~(\ref{flux21}-\ref{flux22}) the subscript $n$ has been dropped for convenience,
e.g. $\Pi_{X^<}^{Y^<}$ must be understood as  $\Pi_{X_n^<}^{Y_n^<}$.
We have
\begin{eqnarray}
	\Pi_{X^<}^{Y} &=& \Pi_{X^<}^{Y^<}+\Pi_{X^<}^{Y^>}, \qquad \Pi_X^{Y^<} = \Pi_{X^<}^{Y^<}+\Pi_{X^>}^{Y^<} \nonumber \\
	\Pi_{X^>}^{Y} &=& \Pi_{X^>}^{Y^<}+\Pi_{X^>}^{Y^>}, \qquad \Pi_X^{Y^>} = \Pi_{X^<}^{Y^>}+\Pi_{X^>}^{Y^>}.
\end{eqnarray}
Using Eq.~(\ref{TijTji}) we can show that
\begin{eqnarray}
	\Pi_{X^<}^{X^<}&=&\Pi_{X^>}^{X^>}=0, \nonumber \\
	\Pi_{X^<}^{X^>}&=&-\Pi_{X^>}^{X^<}, \qquad
	\Pi_{X^<}^{Y^<}=-\Pi_{Y^<}^{X^<}, \nonumber \\
  \Pi_{X^<}^{Y^>}&=&-\Pi_{Y^>}^{X^<}, \qquad
  \Pi_{X^>}^{Y^>}=-\Pi_{Y^>}^{X^>}.
  \label{Eq_fluxes}
\end{eqnarray}
\begin{figure}[ht]
\centering
\includegraphics[width=0.3\textwidth]{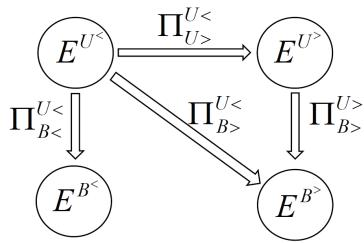}
\caption{Illustration of some energy fluxes. The others can be found with the help of Eqs.~(\ref{Eq_fluxes}).}
\label{PiXY}
\end{figure}

These equations, illustrated in Fig.~\ref{PiXY}, imply the following expressions for the energy equations
\begin{eqnarray}
    d_t E_n^{U^{<}} &=& \Pi_{U^<}^{U^>} + \Pi_{U^<}^{B^<} + \Pi_{U^<}^{B^>} -  \nu \bD(\bU)\cdot \bU_n^< + \bF\cdot \bU_n^<, \label{shell_NS32}\\
    d_t E_n^{B^{<}} &=& \Pi_{B^<}^{B^>} + \Pi_{B^<}^{U^<} + \Pi_{B^<}^{U^>}   -  \eta \bD(\bB)\cdot \bB_n^<,   \label{shell_induction32}
\end{eqnarray}
where
\begin{eqnarray}
	\Pi_{U^<}^{U^>}&=&\widetilde{\bW}(\bU,\bU_n^>)\cdot \bU_n^<, \qquad
	\Pi_{U^<}^{B^<}=-\widetilde{\bW}(\bB,\bB_n^<)\cdot \bU_n^<, \nonumber\\
	\Pi_{U^<}^{B^>}&=&-\widetilde{\bW}(\bB,\bB_n^>)\cdot \bU_n^<, \qquad
	\Pi_{B^<}^{B^>}=\widetilde{\bW}(\bU,\bB_n^>)\cdot \bB_n^<, \nonumber\\
	\Pi_{B^<}^{U^<}&=&-\widetilde{\bW}(\bB,\bU_n^<)\cdot \bB_n^<, \qquad
	\Pi_{B^<}^{U^>}=-\widetilde{\bW}(\bB,\bU_n^>)\cdot \bB_n^<.
\end{eqnarray}

The evolution of kinetic and magnetic energies in shell $i$ has the form
\begin{eqnarray}
    d_t U_i^2/2 &=& \sum_{j,k=1}^N \frac{1}{2}S^{UU}(i|j,k) + \sum_{j,k=1}^N \frac{1}{2}S^{UB}(i|j,k) -  \nu k_i^2 U_i^2 + \bF\cdot \bU_i, \label{shell_NS4}\\
    d_t B_i^2/2 &=& \sum_{j,k=1}^N \frac{1}{2}S^{BB}(i|j,k) + \sum_{j,k=1}^N \frac{1}{2}S^{BU}(i|j,k) -  \eta k_i^2 B_i^2,   \label{shell_induction4}
\end{eqnarray}
with
\begin{eqnarray}
    S^{UU}(i|j,k)=&  \widetilde{\bW}(\bU_k,\bU_j)\cdot \bU_i &+ \widetilde{\bW}(\bU_j,\bU_k)\cdot \bU_i, \nonumber \\
    S^{UB}(i|j,k)=& -\widetilde{\bW}(\bB_k,\bB_j)\cdot \bU_i &- \widetilde{\bW}(\bB_j,\bB_k)\cdot \bU_i, \nonumber  \\
    S^{BB}(i|j,k)=&  \widetilde{\bW}(\bU_k,\bB_j)\cdot \bB_i &+ \widetilde{\bW}(\bU_j,\bB_k)\cdot \bB_i, \nonumber  \\
    S^{BU}(i|j,k)=& -\widetilde{\bW}(\bB_k,\bU_j)\cdot \bB_i &- \widetilde{\bW}(\bB_j,\bU_k)\cdot \bB_i.
\end{eqnarray}
Each $S^{XY}(i|j,k)$ term represents the transfer rate from the modes $j$ and $k$ of field $\bY$ into the mode $i$ of field $\bX$.

\begin{figure}[ht]
\centering
\includegraphics[width=0.45\textwidth]{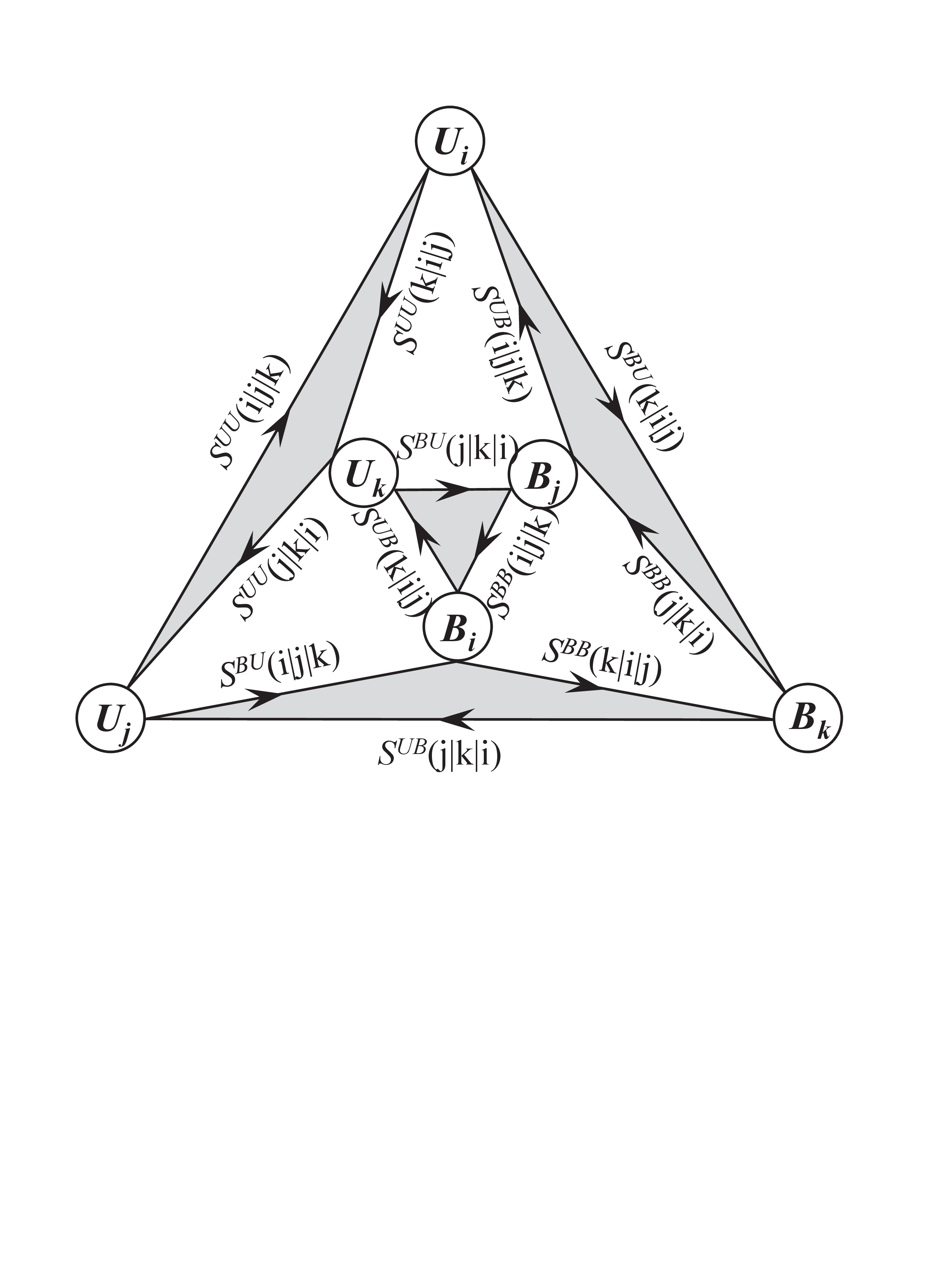}
\caption{Illustration of the mode-to-mode energy transfers. The gray triangles correspond to the interacting triads.}
\label{SXY}
\end{figure}

We also define the following quantities
\begin{eqnarray}
    S^{UU}(i|j|k)=&  &\widetilde{\bW}(\bU_k,\bU_j)\cdot \bU_i, \nonumber  \\
    S^{UB}(i|j|k)=& -&\widetilde{\bW}(\bB_k,\bB_j)\cdot \bU_i, \nonumber  \\
    S^{BB}(i|j|k)=&  &\widetilde{\bW}(\bU_k,\bB_j)\cdot \bB_i,  \nonumber \\
    S^{BU}(i|j|k)=& -&\widetilde{\bW}(\bB_k,\bU_j)\cdot \bB_i,
\end{eqnarray}
 where $S^{XY}(i|j|k)$ is the mode-to-mode energy transfer rate from the mode $j$ of field $\bY$
 to the mode $i$ of field $\bX$, with the mode $k$ acting as a mediator.
 Within one triad $(i,j,k)$ the related interactions are illustrated in Fig.~\ref{SXY}.
 The following relation applies
\begin{equation}
T_{ij}^{XY}=\sum_k	S^{XY}(i|j|k).
\end{equation}
MHD fluxes were introduced in \citet{Stepanov2006a} and later corrected in \citet{Plunian2007} and \citet{Lessinnes2009}.

\subsection{Local models}
\label{section local}
\subsubsection{L1-models (local, first-neighbor)}
In \ref{s7:L1}, we give the form of all possible L1-models obeying to the general requirements given by Eqs.~(\ref{relation1}) and (\ref{Wntildeform1}).
They were already introduced in the seminal paper by \citet{Gloaguen1985}.
Two L1-models have been studied in detail, one by \citet{Gloaguen1985} and subsequent authors, the other by \citet{Biskamp1994}.

\begin{itemize}
\item
The model investigated by \citet{Gloaguen1985} corresponds to
\begin{equation}
\widetilde{W}_n(\bX,\bY)=k_n \left[C_1(X_{n - 1}Y_{n-1} - \lambda X_{n}Y_{n +1} )+C_2 (X_{n}Y_{n - 1} - \lambda X_{n+1} Y_{n +1})\right], \label{gloagen}
\end{equation}
where $C_1$ and $C_2$ are \textit{real} parameters, and $\bX$ and $\bY$ are \textit{real} variables. The set of equations for $\bU$ and $\bB$ is as follows
\begin{eqnarray}
d_t U_n &=& C_1 \left(k_n U_{n-1}^2 - k_{n+1}U_{n}U_{n+1} - k_n B_{n-1}^2 + k_{n+1}B_{n}B_{n+1}\right)\nonumber\\
                  &+&  C_2 \left(k_n U_{n-1}U_{n} - k_{n+1}U_{n+1}^2 - k_n B_{n-1}B_{n} + k_{n+1}B_{n+1}^2\right)
 -\nu k_n^2 U_n +  F_n,\label{e3:gloag_u}\\
d_t B_n &=& C_1 k_{n+1} \left(U_{n+1}B_n- U_{n}B_{n+1}\right)
                  +  C_2 k_n     \left(U_{n}B_{n-1} - U_{n-1}B_{n} \right)
 -\eta k_n^2 B_n.
\label{e3:gloag_b}
\end{eqnarray}

In addition to total energy and cross helicity, provided
$C_1 + \lambda^q C_2=0$, this model has another conserved quantity,
$\sum k_n^{q}B_n$. However, as it is not quadratic it cannot be considered as an analog of magnetic helicity (which is conserved in ideal 3D MHD) or as a squared magnetic potential  (which is conserved in ideal 2D MHD). So it has no real meaning.
For $\bB=0$, the \citet{Bell1978} model is recovered, which in turn gives \citet{Obukhov1971} model if $C_1=0$, and \citet{Desnianskii1974} model if $C_2=0$.

In the dissipationless limit ($\nu=\eta=0$) and for an infinite number of shells, the system of Eqs.~(\ref{e3:gloag_u}-\ref{e3:gloag_b}) has the Kolmogorov stationary solution
\begin{equation}
B_n \sim U_n \sim k_n^{-1/3}
\label{kolm_sol}
\end{equation}
for any value of the ratio $c=C_1/C_2$. However, \citet{Grappin1986} found
three kinds of attractors depending on the value of $c$. For $c\geq2$ the system tends to a supercorrelated state, for which the attractor is reduced to a fixed point $\bB=\pm \bU$,
characterized by a total absence of spectral energy transfer and very steep energy spectra.
For $0.5\leq c\leq1$, the attractor is a non-magnetic ($\bB=0$) stable fixed point. Finally, for $c<0.3$ the attractor has a high dimension, with a chaotic solution characterized by an extended inertial range and equipartition between kinetic and magnetic energies.
Therefore only the latter range of $c$ is of interest for MHD turbulence as it reproduces the expected chaotic behavior of real turbulence. Finally, for $c=0.01$, \citet{Grappin1986} found a Lyapunov dimension for the attractor which is consistent with the standard Kolmogorov HD turbulence, rather than the Kraichnan MHD \citep{Kraichnan1965}, phenomenology . The latter scenario is lost due the absence of non-local interactions in the model.

In its HD version, i.e. for a model reduced to Eq.~(\ref{e3:gloag_u}) with $\bB=0$, \citet{Dombre1998} found a solution for $\bU$ which again depends on $c$. A stable fixed point is found for $c >0.55$ (including Desnianskii-Novikov's model for $c\rightarrow \infty$), and chaotic solutions for $c < 0.536$ (including Obukhov's model for $c=0$). The transition between both regimes corresponds to a succession of Hopf bifurcations.\\

MHD intermittency has been studied by \citet{Carbone1994a,Carbone1994} for $c=0.01$ and $N=19$. Solving the system of Eqs.~(\ref{e3:gloag_u}-\ref{e3:gloag_b}) expressed in terms of Els\"{a}sser variables
$Z_n^{\pm}=U_n \pm B_n$,
 he calculated the $p$-th-order structure functions $S_n^{\pm}(p) = \langle|Z_n^{\pm}|^p\rangle$. For scales belonging to the inertial range,
he found a scaling exponent $\zeta_p$ such that $S_n^{\pm}(p) \propto k_n^{-\zeta_p}$.
For a range of scales much larger than the inertial range, he
found that the structure functions satisfy the relation $S_n^{\pm}(p) \propto S_n^{\pm}(3) ^{\zeta_p/\zeta_3}$, thus confirming that the concept of extended self-similarity applies to MHD.
\citet{Carbone1994} also compared these scaling exponents to
those obtained from solar wind measurements, by the Voyager 2 satellite at 8.5 AU \citep{Burlaga1991}. As shown in Fig.\ref{carbone_solarwind}, the shell model results lie inside the error bars of the observed data.
For completeness we note that \citet{Carbone1995} showed that the solutions of Eqs.~(\ref{e3:gloag_u}-\ref{e3:gloag_b})
are sign-singular, meaning that their sign reverses continuously on arbitrary finer time scales, similarly to some signed measurements in turbulence and fast dynamos \citep{Ott1992}.

\begin{figure}[ht]
\centering
\includegraphics[width=0.45\textwidth]{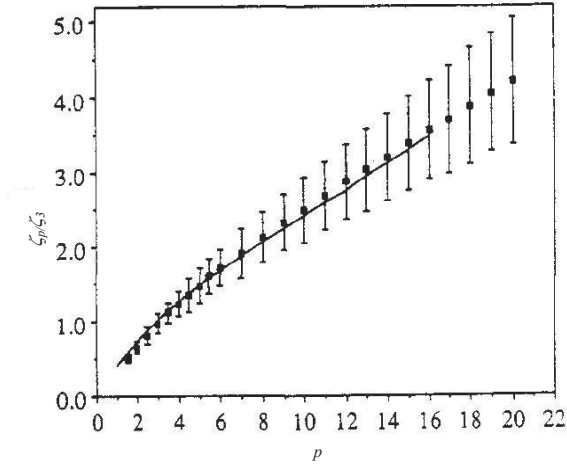}
\caption{Scaling exponents $\zeta_p/\zeta_3$ versus $p$ obtained from the L1-model (\ref{gloagen}) (full line), and from Voyager 2 solar wind data analysis \citep{Burlaga1991} (points with error bars). Adapted from \cite{Carbone1994}.}
\label{carbone_solarwind}
\end{figure}

\item
The model investigated by \citet{Biskamp1994} corresponds to
\begin{equation}
\widetilde{W}_n(\bX,\bY)=k_n \left[C_1(X^*_{n - 1}Y^*_{n-1} - \lambda X^*_{n}Y^*_{n +1} )+C_2 (X^*_{n}Y^*_{n - 1} - \lambda X^*_{n+1} Y^*_{n +1})\right], \label{gloagencomplex}
\end{equation}
and gives
\begin{eqnarray}
d_t U_n &=& C_1 \left(k_n U_{n-1}^{*\;2} - k_{n+1}U_{n}^*U_{n+1}^* - k_n B_{n-1}^{*\;2} + k_{n+1}B_{n}^*B_{n+1}^*\right)\nonumber\\
                  &+&  C_2 \left(k_n U_{n-1}^*U_{n}^* - k_{n+1}U_{n+1}^{*\;2} - k_n B_{n-1}^*B_{n}^* + k_{n+1}B_{n+1}^{*\;2}\right)
 -\nu k_n^2 U_n +  F_n,\label{e3:bisk_u}\\
d_t B_n &=& C_1 k_{n+1} \left(U_{n+1}^*B_n^*- U_{n}^*B_{n+1}^*\right)
                  +  C_2 k_n     \left(U_{n}^*B_{n-1}^* - U_{n-1}^*B_{n}^* \right)
 -\eta k_n^2 B_n,
\label{e3:bisk_b}
\end{eqnarray}
where $\bU$ and $\bB$ are \textit{complex} vectors.
Again, depending on the ratio $c=C_1/C_2$, \citet{Biskamp1994} studied the dynamics of the solutions of Eqs.~(\ref{e3:bisk_u}-\ref{e3:bisk_b}) for both HD and MHD turbulence.

For HD turbulence, chaotic Kolmogorov solutions were found only for $|c|\ll 1$. This confirms the argument made by \citet{Gloaguen1985} that $|c|\ll 1$ is more representative of incompressible turbulence, since
the $C_1$-terms referring only to flat triads make a negligible contribution to the non-linear interactions.

For MHD turbulence \citet{Biskamp1994} calculated the structure functions scaling exponents of $\bZ^{\pm}$, $\bU$ and $\bB$ for $c=-1.33$ and $c=10^{-2}$. He found the same scaling exponents for all variables (within the error bars), but the results depend on $c$. For $c =-1.33$ he found stronger multifractal behavior (stronger deviation from the Kolmogorov scaling exponent $\zeta_p=p/3$) while for $c=10^{-2}$ he found  scaling exponents that were more compatible with experimental observations and DNS.
\citet{Biskamp1994} also studied the effect of an externally applied magnetic field.
\end{itemize}

It is remarkable that the peak in popularity of L1-models reached at the beginning of the nineties was mainly due to the \citet{Gloaguen1985} model which, in contrast to the HD models of \citet{Obukhov1971} and \citet{Desnianskii1974}, demonstrated not only chaotic behavior but also intermittency.
This led \citet{Brandenburg1992} to generalize this model to MHD Boussinesq convection by adding an equation for temperature fluctuation $\theta$. For $\bB=0$ he found (for a range of parameter $c$) the scaling laws $E_u(k)\sim k^{-11/5}$ and $E_{\theta}(k)\sim k^{-7/5}$ \citep{Obukhov1959,Bolgiano1959}.
For $\bB\neq 0$ he found Kolmogorov spectra for the three fields $E_{\theta}(k)\sim E_u(k)\sim E_b(k)\sim k^{-5/3}$.
\citet{Geertsema1992} introduced a vectorial three-component version of the \citet{Gloaguen1985} model in order to evaluate the turbulent stress tensor in a differentially rotating disk. This model will be discussed in Sec.~\ref{s5:transport coeff}.\\

A generic problem encountered with the previous models is that there is an insufficient number of quadratic invariants.
In the HD models introduced by \citet{Obukhov1971} or \citet{Desnianskii1974} only the kinetic energy is conserved 
bringing the number of quadratic invariants to one, instead of two
in real HD turbulence. 
In the MHD models introduced by \citet{Gloaguen1985} or \citet{Biskamp1994} only the total energy and cross helicity are conserved,
bringing the number of quadratic invariants to two, instead of three in real MHD turbulence.
Such a problem is due to a too simplistic expression of $\widetilde{W}_n(\bX,\bY)$.
For example the  \citet{Biskamp1994} model corresponds to only the four first terms in Eq.~(\ref{L1complex}) among the twenty possible terms (assuming $A_{10}=0)$. 
However as shown in \ref{s3:HM1} one additional quadratic invariant can be obtained taking more terms in Eq.~(\ref{L1complex}). Such a model is called helical and will be detailed in Sec.~\ref{s3:HM1}. It is called helical for at least two reasons. First it allows to have the kinetic helicity in HD or magnetic helicity in MHD as an additional quadratic invariant. Second such a model can be interpreted in a framework of helical mode decomposition.

\subsubsection{L2-models (local, two-first-neighbor)}

Another way to introduce an additional quadratic invariant is to increase the number of interacting triads, with interactions between two-first-neighbors instead of just first-neighbors.
This led \citet{Gledzer1973} to propose a ``System of hydrodynamic type admitting two quadratic integrals of motion" (his paper's title). 
In our notation it is a L2-model.
It is remarkable that at about the same time \citet{Lorenz1972} derived exactly the same model, starting from the Navier-Stokes equations. Both authors imposed enstrophy conservation in addition to the kinetic energy, thus the model is relevant to 2D-turbulence only.

Following a different approach, \citet{Frick1983} elaborated a non-local L2-model with again kinetic energy and enstrophy conservation. Applying the same method \citet{Frick1984} also derived the first MHD shell model, using the square of the magnetic potential as the third quadratic invariant. This model was thus relevant to 2D MHD turbulence. In our notation both models are N2-models.
For this reason they will be discussed in Sec.~\ref{s3:nonlocal} when dealing with non-local models.

Finally,
\citet{Brandenburg1996} introduced the first 3D MHD L2-model using magnetic helicity as the third quadratic invariant (see also \citet{Basu1998,Frick1998}).\\


The L2-model which has received most attention is undoubtedly the GOY model, named
	 after \citet{Gledzer1973}, \citet{Yamada1987}\footnote{Unfairly ignoring Lorenz's contribution.}. Among all possible L2-models given in \ref{s7:L2} it corresponds to
\begin{equation}
	\widetilde{W}_n(\bX,\bY)=\i k_n \left[
    C_1 (X^*_{n-2}    Y^*_{n-1}  -  \lambda X^*_{n-1}   Y^*_{n+1} )
 +  C_2 (X^*_{n-1}      Y^*_{n-2}   -  \lambda^2 X^*_{n+1} Y^*_{n+2} )
 +  C_3 (X^*_{n+1}      Y^*_{n-1}   -  \lambda X^*_{n+2} Y^*_{n+1} )\right].
 \label{GOY_general}
\end{equation}
The HD version of this model became popular because of its relevance to real turbulence in terms of high-order structure functions. In Sec.~\ref{s3:N2models} we will also discuss another L2-model, called the Sabra model\footnote{The name Sabra model was introduced by \citet{L'Vov1998} in the context of shell models, presumably as an insider joke: Sabra denotes a Jewish person born in Israel, while Goy denotes a non-Jewish person.}, introduced by \citet{L'Vov1998} as an ``improved" version of the GOY model. It is an improvement in so far as some spurious correlations existing between different shells in the GOY model, are suppressed with the Sabra model. Such spurious correlations do not exist in the MHD version of the GOY model, both models giving the same results.

\begin{itemize}
\item

To derive the GOY model for HD turbulence, we write $\widetilde{W}_n(\bX,\bX)$ in the form
\begin{equation}
	\widetilde{W}_n(\bX,\bX)=-\i k_n \lambda(C_3 + \lambda C_2)\left[
	 X^*_{n+1} X^*_{n+2}
	-\frac{\varepsilon}{\lambda} X^*_{n-1} X^*_{n+1}
	-\frac{1-\varepsilon}{\lambda^2} X^*_{n-1} X^*_{n-2}\right],
	\label{e3:GOY1}
\end{equation}
where $\varepsilon = (C_3 - \lambda C_1) / (C_3 + \lambda C_2)$.
With an appropriate renormalization of $\bU$ and $\widetilde{\bW}$ in Eq.~(\ref{shell_NS2}), the term $- \lambda(C_3 + \lambda C_2)$ can be taken equal to unity,
leading to \citep{Biferale1995a}
\begin{equation}
d_t U_n= \i k_n \left[U^*_{n+1} U^*_{n+2}
	-\frac{\varepsilon}{\lambda} U^*_{n-1} U^*_{n+1}
	-\frac{1-\varepsilon}{\lambda^2} U^*_{n-1} U^*_{n-2}\right] -\nu k_n^2 U_n + F_n.
	\label{GOY_epsilon}
\end{equation}
\citet{Kadanoff1995} showed that, in addition to the kinetic energy, this system has the following quadratic invariant
\begin{equation}
H=\sum_n |U_n|^2(\varepsilon -1)^{-n}.
\label{geneen}
\end{equation}

Depending on $\varepsilon$, different physical interpretations can be given to $H$ \citep{Frick1995}:
\begin{itemize}
	\item
For $\varepsilon=1-\lambda^{-1}$,
\begin{equation}
	H=\sum (-1)^n k_n |U_n|^2
	\label{e3:kin-helicity}
\end{equation}
which is analogous to kinetic helicity \citep{Kadanoff1995}.
This corresponds to the GOY model for 3D HD turbulence introduced in Sec.~\ref{s3:GOY}.
It has been studied numerically \citep{Jensen1991,Pisarenko1993}
and analytically using a closure model \citep{Benzi1993}.
\cite{Frick1995} showed that the GOY model displays the same intermittency as 3D turbulence provided $\varepsilon \approx 1-\lambda^{-1}$.
\item
For $\varepsilon=1$, all cascades are impossible.
\item
For $\varepsilon=1+\lambda^{-2}$,
\begin{equation}
	H=\sum k_n^2 |U_n|^2
\end{equation}
which is analogous to enstrophy \citep{Lorenz1972,Gledzer1973}. The
two stable solutions are $U_n\propto k_n^{-1/3}$ and $U_n\propto k_n^{-1}$,
leading to spectral properties analogous to 2D turbulence. The energy spectrum density should be
made of two power laws depending on whether $k$ is smaller or larger than $k_F$, the forcing wave-number.
\begin{itemize}
	\item For $k\le k_F$, $E(k)\propto k^{-5/3}$ with an inverse energy cascade.
	\item For $k\ge k_F$, $E(k)\propto k^{-3}$ with a direct cascade of enstrophy.
\end{itemize}
However such a 2D HD shell model is not able to show a true inverse energy cascade \citep{Aurell1994,Ditlevsen2011} contrary to its 2D MHD analog (see below).

\item
For $\varepsilon=2$ the only
quadratic invariant is the kinetic energy, leading to a Kolmogorov fixed point.

\item
For $\varepsilon=1+\lambda$,
\begin{equation}
	H=\sum k_n^{-1} |U_n|^2,
\end{equation}
which is the dimensional equivalent of the ``action", a hidden integral of motion in 3D turbulence
written in Clebsch variables \citep{Yakhot1993}.
\end{itemize}

For $0<\varepsilon<1$ and $\lambda=2$, a numerical study of the transition to chaos was performed by \citet{Biferale1995a}.
They found a Kolmogorov stable fixed point solution for $\varepsilon<0.3843$.
For $\varepsilon=0.3843$ the solution becomes unstable via a Hopf
bifurcation. For larger values of $\varepsilon$, the system evolves towards a chaotic state following a Ruelle-Takens
scenario. For $\varepsilon > 0.3953$ the dynamics are intermittent with a positive Lyapunov exponent. This regime is
characterized by a strange attractor remaining close to the Kolmogorov unstable fixed point.

\item
In MHD, taking $\widetilde{W}_n(\bX,\bY)$ defined by Eq.~(\ref{GOY_general}),
and provided
\begin{equation}
	C_1 = \frac{-1 + (1-\varepsilon) + (1-\varepsilon)^2}{1 + (1-\varepsilon) + (1-\varepsilon)^2} C_2, \qquad
	C_3 = \lambda \frac{1 + (1-\varepsilon) - (1-\varepsilon)^2}{1 + (1-\varepsilon) + (1-\varepsilon)^2} C_2,
	\label{e3:C1C3}
\end{equation}
we can show that Eq.~(\ref{shell_induction2}) has an additional quadratic invariant
\begin{equation}
	I=\sum_n(\varepsilon-1)^n|B_n|^2.
\label{thirdint}
\end{equation}
Using the same renormalization applied to the HD case,
and thus omitting the term $-\lambda(C_3 + \lambda C_2)$ in front of $\widetilde{W}_n(\bX,\bY)$, the latter becomes
\begin{eqnarray}
\label{GOY_general2}
	\widetilde{W}_n(\bX,\bY)
	&=&\frac{\i k_n}{2} \left[ \right. X^*_{n+1} Y^*_{n+2} + X^*_{n+2} Y^*_{n+1}
	+ \frac{(1-\varepsilon)^2}{2-\varepsilon}(X^*_{n+1} Y^*_{n+2} - X^*_{n+2} Y^*_{n+1})\nonumber \\
	&-& \frac{\varepsilon}{\lambda} (X^*_{n-1}   Y^*_{n+1} + X^*_{n+1}      Y^*_{n-1})
	+ \frac{1-\varepsilon}{\lambda(2-\varepsilon)} (X^*_{n-1}   Y^*_{n+1} - X^*_{n+1}      Y^*_{n-1}) \\ 
	&-& \frac{1-\varepsilon}{\lambda^2}(X^*_{n-2}    Y^*_{n-1} + X^*_{n-1}      Y^*_{n-2})
	+ \frac{1}{\lambda^2(2-\varepsilon)}(X^*_{n-2}    Y^*_{n-1} - X^*_{n-1}      Y^*_{n-2})\left. \right].\nonumber
\end{eqnarray}
It leads to the system \citep{Frick1998,Antonov2000}:
 \begin{eqnarray}
d_t U_n &=& \i k_n \Bigl \lbrace (U_{n+1}^*U_{n+2}^* - B_{n+1}^*B_{n+2}^* )
- \frac{\varepsilon}{\lambda} (U_{n-1}^*U_{n+1}^* - B_{n-1}^*B_{n+1}^* )
-\frac{(1-\varepsilon)}{\lambda^2} (U_{n-2}^*U_{n-1}^* - B_{n-2}^*B_{n-1}^* )\Bigl \rbrace -\nu k_n^2 U_n +  F_n,
\label{e3:goy_u}\\
d_t B_n &=&
\frac{\i k_n}{2-\varepsilon} \Bigl \lbrace (1-\varepsilon)^2(U_{n+1}^*B_{n+2}^*
- B_{n+1}^*U_{n+2}^* )
+ \frac{(1-\varepsilon)}{\lambda} (U_{n-1}^*B_{n+1}^* -
 B_{n-1}^*U_{n+1}^* )
+ \frac{1}{\lambda^2} (U_{n-2}^*B_{n-1}^*
- B_{n-2}^*U_{n-1}^* )\Bigl \rbrace -\eta k_n^2 B_n.\nonumber\\
\label{e3:goy_b}
\end{eqnarray}
Similar to the quantity $H$ in HD, different physical interpretations can be given to $I$ depending on $\varepsilon$ :
\begin{itemize}
	\item
	For $\varepsilon=1-\lambda^{-1}$,
	
\begin{equation}
	I=\sum_n(-1)^n k_n^{-1} |B_n|^2,
	\label{w3hb}
\end{equation}
which is analogous to magnetic helicity \citep{Brandenburg1996}
and, contrary to the choice of $\varepsilon=1+\lambda^{-1}$ made by \citet{Biskamp1994}, leads to
an unsigned quantity as expected for magnetic helicity. This corresponds to the GOY model for 3D MHD turbulence.

\item
For $\varepsilon=1+\lambda^{-2}$,
\begin{equation}
	I=\sum_n k_n^{-2} |B_n|^2,
\end{equation}
which is analogous to the square of the potential vector. This corresponds to the GOY model for 2D MHD turbulence.

\item
For $\varepsilon=2$,
\begin{equation}
	I=\sum_n |B_n|^2,
\end{equation}
which is twice the magnetic energy. As total energy is conserved this corresponds to separate conservation of both kinetic and magnetic energies, which has no obvious physical meaning.
\end{itemize}

\citet{Antonov2000} solved Eqs.~(\ref{e3:goy_u}-\ref{e3:goy_b}) for $\varepsilon \in[-10,10]$, $\lambda=2$,
$\nu=\eta=10^{-9}$,
 and for a stationary forcing applied at shell $n=0$. They found that in contrast to the HD case
there are no stable solutions and the behavior is always stochastic over the whole range of $\varepsilon$.
Kinetic and magnetic spectra are shown in Fig.~\ref{antonov} for different values of $\varepsilon$.
For $\varepsilon\le 1.01$ (left column) a small-scale dynamo action occurs with near equipartition between kinetic and magnetic energies and approximately Kolmogorov spectral slopes.
For $\varepsilon\ge 1.1$, the magnetic energy at large scales is depleted (right column). On increasing $\varepsilon$ the peak of the magnetic spectrum moves to smaller scales until it reaches the dissipation scale for $\varepsilon\approx 1.7$.
For $\varepsilon \in[1.7, 2]$ the magnetic spectrum is much lower than the kinetic spectrum, but the dynamo still occurs. For $\varepsilon>2$ the small-scale dynamo is lost. However, the magnetic energy at the largest scales slowly grows. 
To understand why it is so, we can write $I$ in the form
$I=\sum k_n^d |B_n|^2$. Then taking $\varepsilon>2$ corresponds with having $d>0$, $I$ becoming a magnetic analog of generalized enstrophy. Similar to 2D HD turbulence, energy can only be transferred towards large scales (inverse cascade)
while $I$ is transferred towards small scales (direct cascade).

\begin{figure}[ht]
\centering
\includegraphics[width=0.5\textwidth]{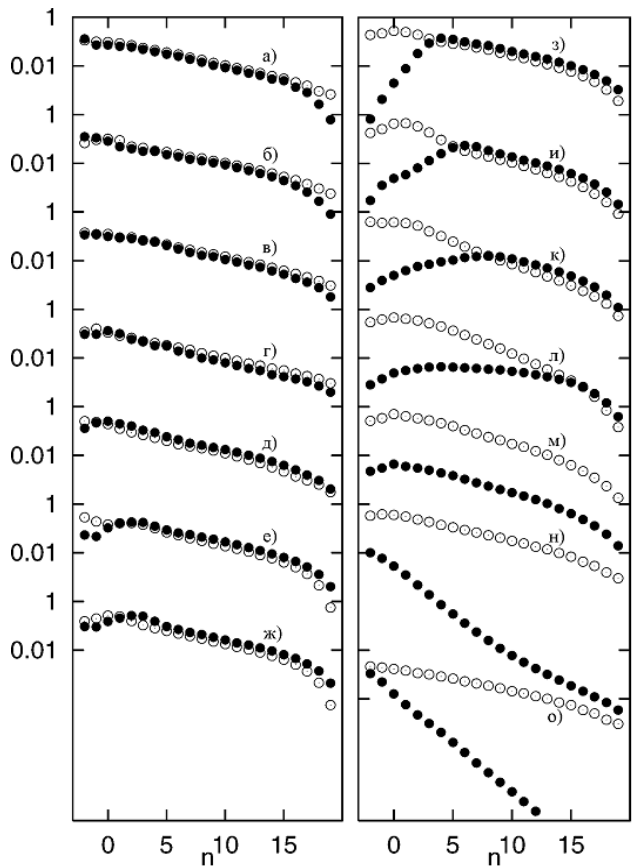}
\caption{Kinetic (white) and magnetic (black) shell energies vs the shell number $n$ for different values of the parameter $\varepsilon$. Left(up-down) $\varepsilon=-5, -2, -0.5, 0, 0.5, 0.99, 1.01 $.  Right (up-down) $\varepsilon = 1.1, 1.3, 1.5, 1.7, 1.99, 3, 5$. From \cite{Antonov2000}.}
\label{antonov}
\end{figure}

\citet{Frick1998} investigated the model (\ref{e3:goy_u}-\ref{e3:goy_b}) in detail  for $\varepsilon=1 -\lambda^{-1}$ and $\varepsilon=1+\lambda^{-2}$, corresponding respectively to 3D and 2D MHD turbulence, with $I$ being respectively the magnetic helicity and the square of the magnetic potential.
In free decaying turbulence (without forcing), they found dynamo action in the 3D case, and magnetic decay in the 2D case (as expected from antidynamo theorems). In 3D (Fig.\ref{fig_decay}-left) the magnetic energy reaches equipartition after 20 turn-over times.
In the kinematic approximation, $\widetilde{\bW}(\bB,\bB)=0$, the growth of magnetic energy is unbounded as expected from kinematic dynamo action. In 2D (Fig.\ref{fig_decay}-right) the magnetic energy for both non-linear and kinematic cases grows up to a level of about 1/100 of kinetic energy and slowly decays on a dissipation time scale.

In free decaying turbulence, \citet{Frick1998} also tested the analogy between magnetic energy in 2D MHD turbulence and
temperature gradients \citep{Zeldovich1956}. They considered a GOY model for temperature fluctuations $\theta_n$,
with $\sum \theta_n^2$ as an additional ideal invariant.
For both 2D and 3D turbulence they found that the thermal energy decays smoothly, while the temperature gradients exhibit temporal
growth followed by decay similar to the magnetic energy in 2D MHD turbulence.

\begin{figure}[ht]
\begin{center}
\includegraphics[width=0.4\textwidth]{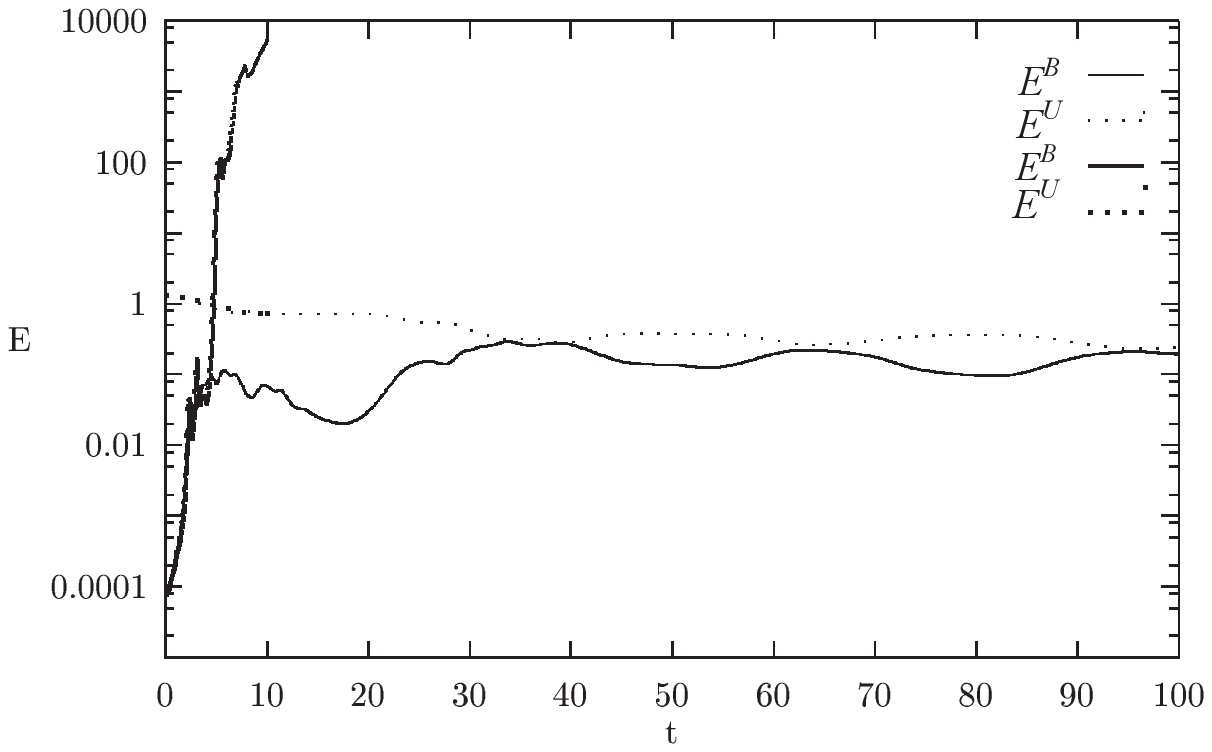}
\includegraphics[width=0.4\textwidth]{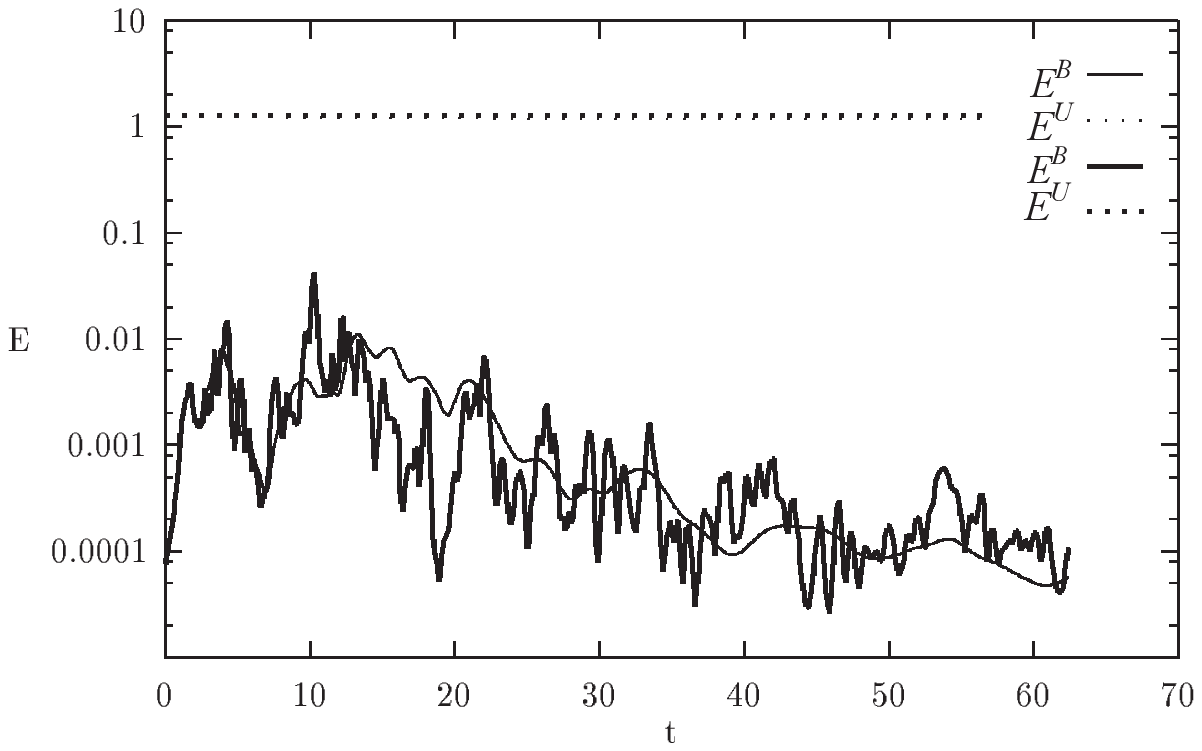}
\end{center}
\caption{Free decaying 3D (left) and 2D (right) MHD turbulence:
 kinetic $E^U$ and magnetic $E^B$ energies
versus time. Thick lines corresponds to kinematic simulations imposing $\widetilde{\bW}(\bB,\bB)=0$. From \cite{Frick1998}.}
\label{fig_decay}
\end{figure}

Forced MHD turbulence has been tested against the concept of extended self-similarity \citep{Basu1998},
while the scaling properties and long-time behavior of the solutions
have been studied in detail \citep{Frick1998}. In particular, it was shown that the model given by Eqs.~(\ref{e3:goy_u}-\ref{e3:goy_b}) is rather sensitive to the dynamics of the applied forcing. For a constant forcing \citet{Giuliani1998} found that
after some time a supercorrelation state $\bB=\pm \bU$ appears which is spurious and has no real physical interpretation. 
Instead of applying a forcing \citet{Frick2000} imposed the modulus of the 
\textit{complex} velocity at shell $n=0$ letting its phase evolving freely.
They found different time stages with both low or high correlations between $\bU$ and $\bB$.
Using forcing with a random \textit{complex} phase suppresses the problem of supercorrelation state \citep{Stepanov2006a}.

\end{itemize}

\subsection{Non-local models}
\label{s3:nonlocal}
Some time after the emergence of shell models in Moscow, mainly in the wake of \citet{Obukhov1971}, another approach, the so-called hierarchical approach, was developed in Perm (Russia) by \citet{Zimin1981}. The idea was to model HD turbulence as a network of vortices with a double repartition in both physical and Fourier spaces. Under the assumptions of homogeneity and isotropy, it is possible to calculate the Reynolds tensor for each interacting triad of vortices, and derive a system for the intensity of each vortex still depending on space and scale. Averaging in space over all vortices having the same scale leads to a non-local shell model.
\citet{Frick1983} found an original way of enabling such a model to conserve two ideal invariants. Considering kinetic energy and enstrophy conservation, he elaborated a non-local shell model of 2D-turbulence. Applying the same method to MHD, with conservation of total energy, cross helicity and the square of magnetic potential, \citet{Frick1984} developed a non-local shell model for 2D MHD turbulence. Both shell models are N2 models. Completing the work started about 15 years earlier by \citet{Zimin1981}, \citet{Zimin1995} derived an N1-model of 3D HD turbulence, which satisfies both kinetic energy and helicity requirements (though helicity is not mentioned in their work). The hierarchical approach followed by \citet{Frick1983} is detailed in Sec.~\ref{nonlocal} along with the model derived by \citet{Zimin1995}.

Another way to derive a non-local shell model is to begin directly from the shell model structure given by Eqs.~(\ref{Wntildeform1}), including all possible non-local interactions.
The general shape of $\widetilde{\bW}$ for N1 and N2 models is given respectively  in \ref{s7:N1}
and \ref{s7:N2}. One example of an N2-model
is presented in Sec.~\ref{s3:N2models}. Such direct derivation does not provide unique definitions for the non-local coefficients. One free parameter, directly related to the strength of the non-local interactions, remains. This free parameter can be estimated 
either from phenomenological argument or using the hierarchical approach.

\subsubsection{N2-model derived using a hierarchical approach}
\label{nonlocal}
Following \citet{Frick1983}, we consider a network of parallel 2D-vortices, depending on the horizontal coordinates only, with velocities perpendicular to the third direction ${\bf e_z}$.
Each vortex is denoted by two integers $n$ and $N$. The first integer $n$ indicates the shell to which the vortex wave number belongs. As with shell models we consider shells obeying a geometric sequence, here with $\lambda=2$ as their common ratio.

As two different vortices may belong to the same shell in Fourier space, a second integer $N$ is necessary to differentiate the vortices in real space \footnote{For convenience, and in this section only, $N$ denotes the second integer and not the maximum number of shells.}.
Then we can write the velocity field as
\begin{equation}
\label{eq614}
\bu(t,x,y) =\sum_{n,N}U_{nN}(t) \; \bu_{nN}(\br-\br_{nN}),
\end{equation}
where $U_{nN}(t) $  is the amplitude of the vortex $nN$,  $\br_{nN} $ is the position of the vortex center and $\bu_{nN}$ is defined by its Fourier coefficients
\begin{eqnarray}
\label{bas_vf}
 \hat{\bu}_{nN}(\bk)&=& \i\;\frac{2^{1-n}}{\sqrt{3\pi}}\frac{\bk\times\be_z}{k^2} e^{-\i\bk \cdot\br_{nN}},\quad \mbox{for} \quad k_n \le \vert \bk \vert < k _{n + 1}, \quad \mbox{with} \quad k_n =\pi 2^n,  \\
 \hat{\bu}_{nN}(\bk) &=& 0 \quad\quad\quad\quad \mbox{outside the shell.} \nonumber
\end{eqnarray}

\begin{figure}[ht]
\begin{center}
\includegraphics[width=0.4\textwidth]{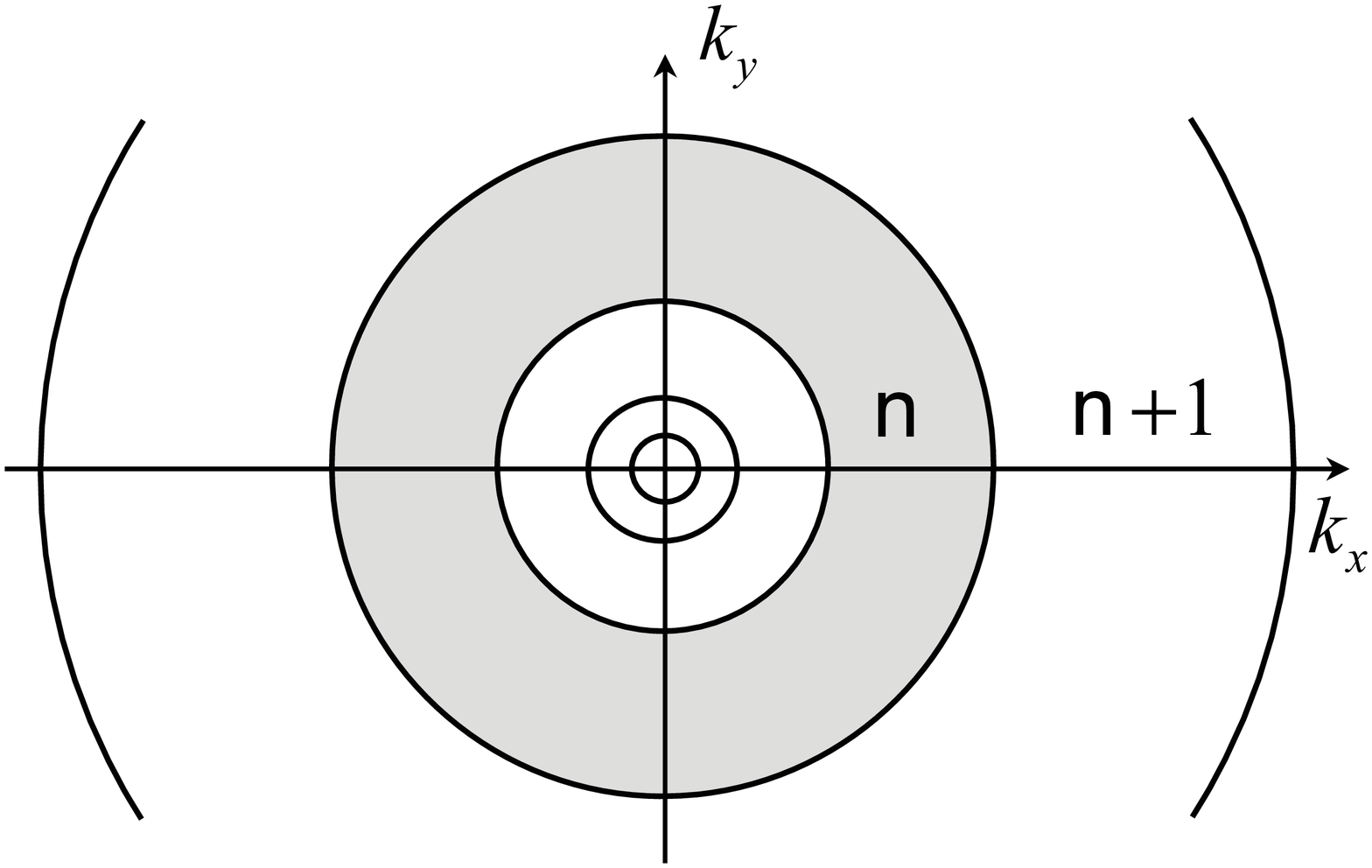}
\includegraphics[width=0.4\textwidth]{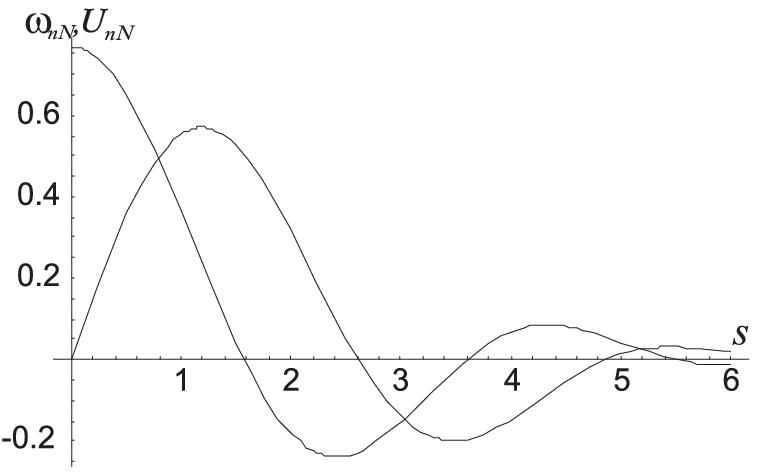}
\end{center}
\caption{Left: illustration of logarithmic shells in Fourier space, where the gray annulus corresponding to shell $n$. Right: functions $u_{nN}(s)$ and ${\omega}_{nN}(s)$ for $n=0$.}
\label{shells-func}
\end{figure}

Taking the inverse Fourier transform of $\hat{\bu}_{nN}$, and calculating the vorticity intensity ${\omega}_{nN}=(\nabla \times \bu_{nN})\cdot \be_z$, we find
\begin{align}
\label{eq618}
\bu_{nN} (\br - \br_{nN} ) &= (\be_z\times \bs)(3\pi)^{-1/2}
\left( {{J_0 (2s) - J_0 (s)}}\right)/{s^2},\\
\label{eq619}
\omega _{nN} ({\bf r} - {\bf r}_{nN} ) &= 2^{ - n} ({\pi
}/{3})^{1/2} \left( {{2J_1 (2s) - J_1 (s)}} \right)/{s},
\end{align}
where $\bs = \pi 2^n( {\bf r} - {\bf r}_{nN} ) $, $s=|\bs|$,
and $J_0(s)$ and $J_1 (s)$ are the zero and first order Bessel functions.
Fig.~\ref{shells-func} shows
both functions,
$U_{nN}$ (defined such that $\bu_{nN}= (\be_z\times \bs) U_{nN}$)
and $\omega _{nN}$,
along with the corresponding shell $n$ (in gray) in which
$\hat{\bu}_{nN}(\bk)$ is non zero.
It can be shown that the density of vortices (number of vortices divided by the shell surface) increases with $n$ as
$2^{2n}$.

All functions $\hat{\bu}_{nN}(\bk)$ and $\hat{\bu}_{mM}(\bk)$ are orthogonal provided $n\ne m$ (functions of different scale do not overlap in Fourier space). This implies that
 $\bu_{nN}$ and $\bu_{mM}$ are also orthogonal. In contrast, $\bu_{nN}$ and $\bu_{nM}$ are not necessarily orthogonal, which will have several consequences as discussed below.

By replacing the expression for $\bu$ given by Eqs.~(\ref{eq614}) and (\ref{eq618}) in the Navier-Stokes equations, we find that amplitude $U_{nN}$ has the form
\begin{equation}
\label{eq634}
d_t U_{nN} = \sum_{p,P} {\sum_{q,Q}
{R_{nNpPqQ} U_{pP} U_{qQ} - \nu k_n^2U_{nN} } } .
\end{equation}
where non-linear interactions are given by
\begin{equation}
\label{eq629}
R_{nNpPqQ} =\int \bu _{nN}\cdot\left[ (\bu_{pP} \cdot \nabla) \bu_{qQ}\right] d\br .
\end{equation}
From Eq.~(\ref{eq629}) we obtain the exact relations
\begin{equation}
R_{qQpPnN}= - R_{nNpPqQ}, \qquad R_{pPqQnN}= - R_{nNqQpP}, \qquad R_{pPnNqQ}= - R_{qQnNpP},
\label{eq638}
\end{equation}
which imply energy conservation, where energy is defined as $\sum\limits_{n,N}U_{nN}^2/2$.
We note that this definition of energy is not exact due to the fact that the $\bu_{nN}$ functions are not orthogonal with respect to $N$ \citep{Frick1983}.
Such a hierarchical model has been used to study enstrophy intermittency in 2D turbulence \citep{Aurell1994a}.

Now the aim is to reduce Eq.~(\ref{eq634}) to a shell model of the form
\begin{equation}
\label{eq635}
d_t U_{n} = \sum_{p,q}
{T_{npq} U_{p} U_{q} - \nu k_n^2 U_{n} }  ,
\end{equation}
where the coefficients $T_{npq}$ need to be defined.

We introduce the vectors $\bR_{npq}$, which we assume, as in Eq.~(\ref{eq638}), satisfy
\begin{equation}
\bR_{qpn}= - \bR_{npq}, \qquad \bR_{pqn}= - \bR_{nqp}, \qquad \bR_{pnq}= - \bR_{qnp}.
\label{eq638b}
\end{equation}
Their moduli are defined to be the root-mean-square value of $R_{nNpPqQ}$, calculated for all possible positions of any three interacting vortices $nN$, $pP$ and $qQ$,
\begin{equation}
|\bR_{npq}|^2 = \int \int R_{nNpPqQ}^2 d{\bf r}_{pP}d{\bf r}_{qQ}.
\label{eq637}
\end{equation}
\citet{Frick1983} introduced the correlation between all triads of vortices belonging to shells
$n,p$ and $q$
\begin{equation}
\cos \theta_{npq} = \frac{\int \int R_{nNpPqQ} R_{nNqQpP} d{\bf r}_{pP}d{\bf r}_{qQ}}{|\bR_{npq}|^2|\bR_{nqp}|^2},
\label{eq639}
\end{equation}
and found that $\theta_{npq}+\theta_{qnp}+\theta_{pqn}=2\pi$,
and $\theta_{nqp}+\theta_{qpn}+\theta_{pnq}=2\pi$.
Together with relation (\ref{eq638b}) this shows that the six vectors $\bR_{npq}, \bR_{nqp}, \bR_{pnq}, \bR_{pqn}, \bR_{qpn}, \bR_{qnp}$ are coplanar.
The angles $\theta_{ijk}$ uniquely define the mutual positions of these vectors, but not their absolute positions as the set of vectors can be rotated by any angle. This degree of freedom corresponds to some arbitrary coefficient that will be set to unity after renormalization of the equations.

Now we define the coefficients $T_{npq}$ as
\begin{equation}
T_{npq}=(\be\cdot{\bf R}_{npq}+\be\cdot{\bf R}_{nqp}) ,
\label{eq642}
\end{equation}
where $\be$ is a unit vector.
To determine the direction of $\be$, \citet{Frick1983} considered enstrophy conservation, relevant to 2D turbulence. From the enstrophy equation he introduced the non-linear interactions
\begin{equation}
\label{eq640}
S_{nNpPqQ} = \int \omega _{nN} (\bu_{pP} {\bf \nabla})\omega _{qQ} d\br ,
\end{equation}
and the six vectors $\bS_{npq}, \bS_{nqp}, \bS_{pnq}, \bS_{pqn}, \bS_{qpn}, \bS_{qnp}$.
The $\bS$-vectors are found to be coplanar with the $\bR$-vectors and the
$T_{npq}$ coefficients are given by
\begin{equation}
T_{npq}=({\be}\cdot{\bf S}_{npq}+{\be}\cdot{\bf S}_{nqp}).
\label{eq642b}
\end{equation}
There is, however, only one possible choice for $\be$ such that both
definitions (\ref{eq642}) and (\ref{eq642b}) give the same value for $T_{npq}$ .
For this direction of $\be$, both energy and enstrophy are ideally conserved.

The corresponding shell model has the form
\begin{eqnarray}
d_t {U}_n + \nu k_n^2U_n =   \sum_{m>0}[ T_{n,n-m-1,n-1}U_{n -m- 1}U_{n -1}
+ T_{n,n-m,n+1}U_{n -m}U_{n +1}  +T_{n,n+m,n+m+1}U_{n+m}U_{n +m+1} ] + F_n.
\label{eq641}
\end{eqnarray}
The $T_{0pq}$ coefficients were directly estimated by \citet{Frick1983} for $\lambda=2$, their numerical values are given in Table \ref{table_fr83}. The other $T_{npq}$ coefficients can be determined by applying the formula
\begin{equation}
	T_{npq}=k_n T_{0,p-n,q-n}.
	\label{relation_hierarchical}
\end{equation}
\begin{table}
\begin{center}
\begin{tabular}
{|c|c|c|c|c|c|c|c|c|}
\hline $q\backslash p$& $-4$ & $-3$ & $-2$ & $-1$ & 0& 1& 2& 3 \\
\hline 4& & & & & & & &
0.155 \\
\hline 3& & & & & & & 0.242& 0
 \\
\hline 2& & & & & & 0.431& 0 &
 \\
\hline 1& ${-0.0088}$& ${-0.0257}$& ${-0.0796}$& ${-0.269}$& 0 & 0 & &
 \\
\hline 0& 0 & 0  & 0 & 0 & 0 & & &
 \\
\hline 1& 0.0032& 0.0096& 0.0269& 0 & & & &
 \\
\hline 2& & & & & & & &
 \\
\hline
\end{tabular}
\caption{Numerical values of the $T_{0pq}$ coefficients \citep{Frick1983}.}
\label{table_fr83}
\end{center}
\end{table}
In our classification, the shell model (\ref{eq641}) is an N2-model. The corresponding function $\widetilde{W}(\bX,\bY)$ is given by Eq.~(\ref{N2complex}) with \textit{real} variables and coefficients
\begin{eqnarray}
\widetilde{W}_n(\bX,\bY)&=& k_n\sum_{m >0}\left[ \right.
    C_{1} (X_{n-m-1}    Y_{n-1}  -  \lambda X_{n-m}   Y_{n+1} )
 +  C_{2} (X_{n-1}      Y_{n-m-1}   -  \lambda^{m+1} X_{n+m} Y_{n+m+1} )
 +  C_{3} (X_{n+1}      Y_{n-m}   -  \lambda^m X_{n+m+1} Y_{n+m} ), \nonumber \\
\label{N2hierachical}
\end{eqnarray}
where $C_1, C_2$ and $C_3$ depend on $m$ only.
Taking $\lambda=2$ and identifying the different terms in Eqs.~(\ref{eq641}) and (\ref{N2hierachical})
we find that the $T_{npq}$ coefficients must satisfy the relation
\begin{equation}
	T_{n,n-m,n+1} + 2 T_{n,n-m-1,n-1} + 2^{-m} T_{n,n+m,n+m+1} =0
	\label{energy_hierarchical}
\end{equation}
which corresponds to energy conservation.
The relation corresponding to enstrophy conservation is
\begin{equation}
	T_{n,n-m,n+1} + 2^{-1} T_{n,n-m-1,n-1} + 2^{m} T_{n,n+m,n+m+1} =0.
	\label{enstrophy_hierarchical}
\end{equation}
Finally, from Eqs.~(\ref{relation_hierarchical}), (\ref{energy_hierarchical}) and (\ref{enstrophy_hierarchical}), we have
\begin{eqnarray}
	T_{n,n-m,n+1} &=& k_nT_{0,-m,1}\\
	T_{n,n-m-1,n-1} &=& (1-2^{-2m})(2^{-2m-1}-2)^{-1} k_n  T_{0,-m,1}\\
	T_{n,n+m,n+m+1} &=& 3(2^{-m}-2^{2+m})^{-1} k_n  T_{0,-m,1}.
\end{eqnarray}
Putting $m=1$ in Eq.~(\ref{eq641}) gives an L2-model with local interactions only.
\citet{Frick1983} found that
such local interactions provide about 35 \% of the total energy transfer and 25\% of the total enstrophy transfer, showing the relative importance of non-local transfers.

The generalization of model (\ref{eq641}) to MHD 2D turbulence, with conservation of total energy, cross helicity and square of magnetic potential \citep{Frick1984} is given by
\begin{eqnarray}
d_t {U}_n + \nu k_n^2U_n &=&   \sum_{m>0}[ T_{n,n-m-1,n-1}(U_{n -m- 1}U_{n -1} - B_{n -m- 1}B_{n -1})
+ T_{n,n-m,n+1}(U_{n -m}U_{n +1} - B_{n -m}B_{n +1}) \nonumber \\
&+& T_{n,n+m,n+m+1}(U_{n+m}U_{n +m+1} - B_{n +m}B_{n +m+1}) ] + F_n ,
\label{frick21}\\
d_t {B}_n+ \eta k_n^2B_n &=&  \sum_{m>0}[ M_{n,n-m-1,n-1}(U_{n -m- 1}B_{n -1} - B_{n -m- 1}U_{n -1})
+ M_{n,n-m,n+1}(U_{n -m}B_{n +1} - B_{n -m}U_{n +1}) \nonumber \\
&+&M_{n,n+m,n+m+1}(U_{n+m}B_{n +m+1} - B_{n +m}U_{n +m+1}) ], \label{frick22}
\end{eqnarray}
with
\begin{eqnarray}
M_{n,n-m-1,n-1}=\frac{4}{1-2^{-2m}} T_{n,n-m-1,n-1}, \quad M_{n,n-m,n+1}=\frac{1}{4-2^{-2m}} T_{n,n-m,n+1}, \quad
M_{n,n+m,n+m+1}=\frac{1}{2^{2m}(2^2-1)} T_{n,n+m,n+m+1}.
\label{M_to_T_2D}
\end{eqnarray}

The solutions of Eqs.~(\ref{frick21}-\ref{frick22}) reproduce the expected properties of 2D MHD turbulence: a direct kinetic energy cascade and growth of enstrophy.
In addition, the magnetic energy does not grow, meaning that a 2D dynamo is impossible. However, there is an inverse cascade of the square of the potential vector. This implies that the magnetic energy spectrum becomes steeper in time, with the energy concentrated at the largest scales. The role of non-local interactions was not studied in this work.

The hierarchical approach above has been applied to various problems: passive scalar in 2D turbulence and 2D turbulent convection \citep{Frick1986}, quasi-2D convective turbulence in a thin vertical layer \citep{Barannikov1988}, quasi-2D turbulent convection in a layer \citep{Aristov1989,Aristov1990}  or in a rotating system \citep{Aristov1988b}.

It is worth mentioning the study by \citet{Aristov1988} in which the hierarchical approach was used to model quasi-2D turbulent flow in a thin layer of an electrically conducting fluid heated from below. The
layer was rotated, between two solid horizontal boundaries in a vertical applied magnetic field.
For strong rotation and a weak magnetic field, the dimensionless quasi-2D equations for the 2D-fluctuations of the velocity $\bu$, magnetic field $\bb$ and temperature $\theta$ are given by
\begin{eqnarray}
\label{eq730}
\partial _t {\bf u} + ({\bf u}\cdot{\bf\nabla}){\bf u} - ({\bf b}\cdot{\bf\nabla}){\bf b} + ({\bf \tau}\cdot{\bf\nabla}){\bf \tau}&=&
 - {\bf\nabla} p + \nu \nabla^2{\bf u}-\mu {\bf u},\\
\label{eq731}
\partial_t{\bf b}+({\bf u}\cdot{\bf\nabla}){\bf b}-({\bf b}\cdot{\bf\nabla}){\bf u}&=&\eta \nabla^2{\bf b},\\
\label{eq732}
\partial_t{\theta}+({\bf u}\cdot{\bf\nabla}){\theta}&=&\kappa\nabla^2 \theta,\\
\label{eq733}
{\bf\nabla}\cdot{\bf u} &=& 0 , \qquad \bf\nabla\cdot{\bf b} = 0,
\end{eqnarray}
where ${\bf \tau}=\left(\partial_x \theta,\partial_y \theta\right)$ is the temperature gradient, $\mu$ characterizes the viscous friction at the horizontal boundaries,
and $\nu, \eta$ and $\kappa$ the viscosity, magnetic and thermal diffusivities.
The set of ideal quadratic invariants is then
\begin{eqnarray}
\label{eq734}
I_1=\langle{\bf u}^2\rangle+\langle{\bf b}^2\rangle-\langle{\bf \tau}^2\rangle, \qquad I_2= \langle{\bf u\cdot b}\rangle-\langle{\bf u \cdot\tau}\rangle, \qquad I_3=\langle\ba^2\rangle, \qquad I_4=\langle\theta^2\rangle,
\end{eqnarray}
where $\ba$ is the magnetic potential.
In the limit $\theta \to 0$ this set of invariants reduces to total energy, cross helicity and the square of magnetic potential as expected in 2D MHD turbulence.
However, because in its general form Eq.~(\ref{eq734}) includes two subtractions in $I_1$ and
$I_2$, a variety of scenarios are possible.
For example, both $\langle{\bb}^2\rangle$ and $\langle{\bf \tau}^2\rangle$ may grow simultaneously without changing $I_1$.

\begin{figure}[ht]
\begin{center}
\includegraphics[width=0.75\textwidth]{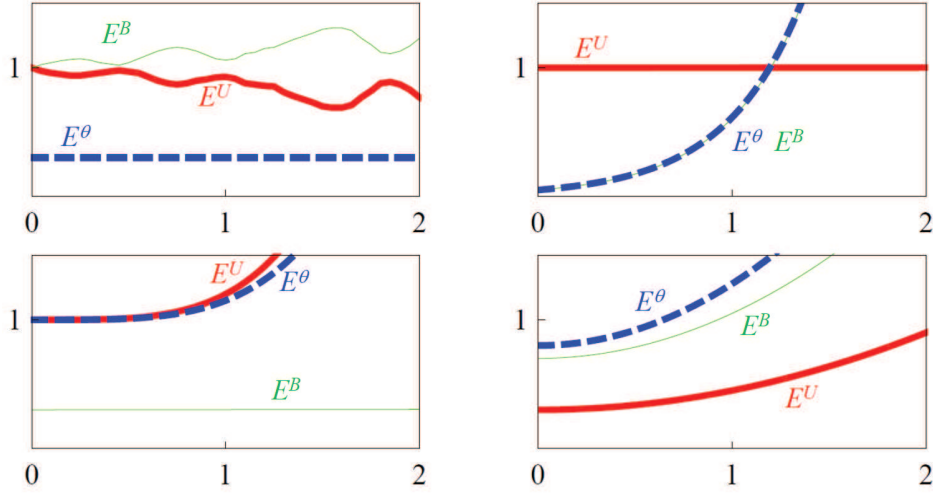}
\end{center}
\caption{Time evolution of kinetic energy (thick-red), magnetic energy (thin-green) and square of temperature gradients (dashed-blue) obtained from the shell model Eqs~(\ref{eq735}-\ref{eq737}), corresponding to quasi-2D turbulence in a thin rotating layer of conducting fluids, for four different initial conditions and ideal or non-ideal cases. Adapted from \cite{Aristov1988}.}
\label{fig_aristov}
\end{figure}

The shell model corresponding to Eqs.~(\ref{eq730}-\ref{eq732}) is described by
\begin{align}
\label{eq735}
d_t U_n = &\sum\limits_{p,q} {T_{npq} [U_p U_q -B_p B_q + \theta_p \theta_q]- \nu k_n^2 U_n-\mu U_n }, \\
\label{eq736}
d_t B_n = &\sum\limits_{p,q} {M_{npq} [U_p B_q -B_p U_q]- \eta k_n^2 B_n}, \\
\label{eq737}
d_t \theta_n = &\sum\limits_{p,q} {M_{npq} [U_p \theta_q -\theta_p U_q]- \kappa k_n^2 \theta_n}.
\end{align}
Some solutions are shown in Fig.~\ref{fig_aristov} for different initial conditions.
Only the bottom-right figure corresponds to a non-ideal case. The three others correspond to $\nu=\eta=\kappa=0$. If at $t=0$ the temperature gradient is weak, top-left, the sum of kinetic and magnetic energies is constant, as it should be in ideal MHD.
If at $t=0$ both the magnetic energy and temperature gradients are weak, top-right, they can grow without limit. If at $t=0$ the magnetic field is weak, bottom-left, the kinetic energy and temperature gradients grow simultaneously.
On bottom-right, it is shown that even with non-zero dissipations, growth of the three quantities is still possible. In this case the 2D anti-dynamo theorem does not apply. Because the temperature gradient cannot reach an infinite value, growth will eventually saturate.

Originally \citet{Zimin1981} introduced the hierarchical model for 3D HD turbulence, with
\begin{equation}
\label{eq714}
{\bf u}(t,x,y,z) =\sum \limits_{nN\nu}U_{nN\nu}(t){\bf u}_{nN\nu}({\bf r}-{\bf r}_{nN}),
\end{equation}
where the third index $\nu$ defines the orientation in space of the vortex, and can be equal to 1,2 or 3, denoting one of the three perpendicular directions of the 3D space. The shells are now defined in 3D Fourier space, and the base of function in the 3D space is
\begin{equation}
\label{eq718}
{\bf u}_{nN\nu} ({\bf r} - {\bf r}_{nN} ) = C ({\bf s}\times {\bf e_\nu})
\left( \sin 2s -2s \cos 2s -\sin s+s \cos s\right)/{s^3}.
\end{equation}
The density of vortices now increases with $n$ as $2^{3n}$.

From this base of function a new shell model
for 3D turbulence can be constructed following the approach of \citet{Frick1983} but with helicity as a second ideal invariant. To our knowledge this remains to be done. An attempt has been made by \citet{Shaid1987} but enstrophy was still used as the second ideal invariant.

Also using the base of function given by Eq.~(\ref{eq718}), \citet{Zimin1995}  obtained an N1-model given by
\begin{eqnarray}
d_t {U}_n + \nu k_n^2U_n =  k_n \sum_{m\ge-1}\Lambda_m[\lambda^{-5m/2}U_{n -m}U_{n}
- \lambda^{-3m/2}U_{n +m}^2 ] .
\label{eq741}
\end{eqnarray}
where $\Lambda_{-1}=0.387$, $\Lambda_{m\geq 1}=2.19$ and $\Lambda_0=0$.
Only kinetic energy is conserved in this model.
However, one remarkable feature is that the infra-red scaling laws are well reproduced.
A generalization of this model to MHD is proposed in Sec.~\ref{s3:HM}.

\subsubsection{Non-local version of the Sabra model}
\label{s3:N2models}
The non-local version of the Sabra model for 3D MHD turbulence has been introduced by \citet{Plunian2007}.
It corresponds to
\begin{eqnarray}
	\widetilde{W}_n(\bX,\bY)
	=\frac{\i k_n}{2}\sum_{m\ge1} \Lambda_m \left[ \right.
	\lambda^m(1+\lambda)( X^*_{n+m} Y_{n+m+1} + X_{n+m+1} Y^*_{n+m})
	&+& (-1)^{m+1}(X^*_{n+m} Y_{n+m+1} - X_{n+m+1} Y^*_{n+m})\nonumber\\
	- \lambda(1 -(-\lambda)^{-m-1}) (X^*_{n-m}   Y_{n+1} + X_{n+1} Y^*_{n-m})
	&+& (X^*_{n-m}   Y_{n+1} - X_{n+1} Y^*_{n-m}) \nonumber \\
	+ \lambda^{-1}(1 -(-\lambda)^{-m})(X_{n-1} Y_{n-m-1} + X_{n-m-1} Y_{n-1})
	&+& (X_{n-1}  Y_{n-m-1} - X_{n-m-1} Y_{n-1})\left. \right],
 \label{SabraW}
\end{eqnarray}
leading to the system 
\begin{eqnarray}
d_t U_n &=& \i \sum_{m\ge 1} \Lambda_m \left[ \right.
    (k_{n+m}+k_{n+m+1})( U^*_{n+m} U_{n+m+1} - B^*_{n+m} B_{n+m+1})
	- (k_{n+1}+(-1)^{m}k_{n-m}) (U^*_{n-m} U_{n+1} - B^*_{n-m} B_{n+1}) \nonumber \\
 &+&(k_{n-1}+(-1)^{m+1}k_{n-m-1})(U_{n-1} U_{n-m-1} - B_{n-1} B_{n-m-1}) \left. \right]
  - \nu k_n^2 U_n +  F_n,
\label{e3:sabra_u}\\
d_t B_n &=&
\i k_n\sum_{m\ge1} \Lambda_m \left[ \right.
	(-1)^{m+1}(U^*_{n+m} B_{n+m+1} - B^*_{n+m} U_{n+m+1}) \nonumber \\
	&+& U^*_{n-m}   B_{n+1}
- B^*_{n-m} U_{n+1} + U_{n-1}  B_{n-m-1} - B_{n-1}  U_{n-m-1}\left. \right] -\eta k_n^2 B_n,
\label{e3:sabra_b}
\end{eqnarray}
where $\Lambda_m$ is now some arbitrary parameter depending on $m$.
This model conserves total energy, cross helicity and magnetic helicity, the latter being defined by Eq.~(\ref{w3hb}). The choice of interacting triads is not arbitrary and corresponds to all possible triads
(Sec.~\ref{s3:general formalism}).

The Sabra model for 3D MHD turbulence corresponds to $m=1$ and $\Lambda_1=(\lambda+\lambda^2)^{-1}$,
\begin{eqnarray}
d_t U_n &=& \i k_n \left[ \right.
    U^*_{n+1} U_{n+2} - B^*_{n+1} B_{n+2}
	- \frac{\lambda -1}{\lambda^2} (U^*_{n-1} U_{n+1} - B^*_{n-1} B_{n+1}) \nonumber \\
 &+&\frac{1}{\lambda^3}(U_{n-1} U_{n-2} - B_{n-1} B_{n-2}) \left. \right]
  - \nu k_n^2 U_n +  F_n,
\label{e3:sabraL2_u}\\
d_t B_n &=&
\frac{\i k_n}{\lambda+\lambda^2} \left[ \right.
	U^*_{n+1} B_{n+2} - B^*_{n+1} U_{n+2}
	+ U^*_{n-1}   B_{n+1} - B^*_{n-1} U_{n+1}
	+ U_{n-1}  B_{n-2} - B_{n-1}  U_{n-2}\left. \right]
 -\eta k_n^2 B_n. \nonumber \\
\label{e3:sabraL2_b}
\end{eqnarray}

For $\bB=0$,
the HD model introduced by \citet{L'Vov1998} is refound.
It conserves kinetic energy and helicity, the latter being defined by Eq.~(\ref{e3:kin-helicity}).

In contrast with its local version, the system of Eqs.~(\ref{e3:sabra_u}-\ref{e3:sabra_b}) is not uniquely defined as it depends on the parameter $\Lambda_m$.
\citet{Plunian2007} suggested that $\Lambda_m= \lambda^{\gamma(m-1)}/\lambda(\lambda +1)$
with $\gamma < 0$. 
The Sabra model corresponds to $\gamma \rightarrow -\infty$.

Assuming isotropy, \citet{Plunian2007} estimated $\gamma$ that accounts for the number of all possible triads between three shells. They found $\gamma=-7/2$
for N2-models and $\gamma=-5/2$ for N1-models, the latter being consistent with the helical model derived by \citet{Zimin1995}.

\begin{figure}[ht]
\begin{center}
\includegraphics[width=0.4\textwidth,angle=0]{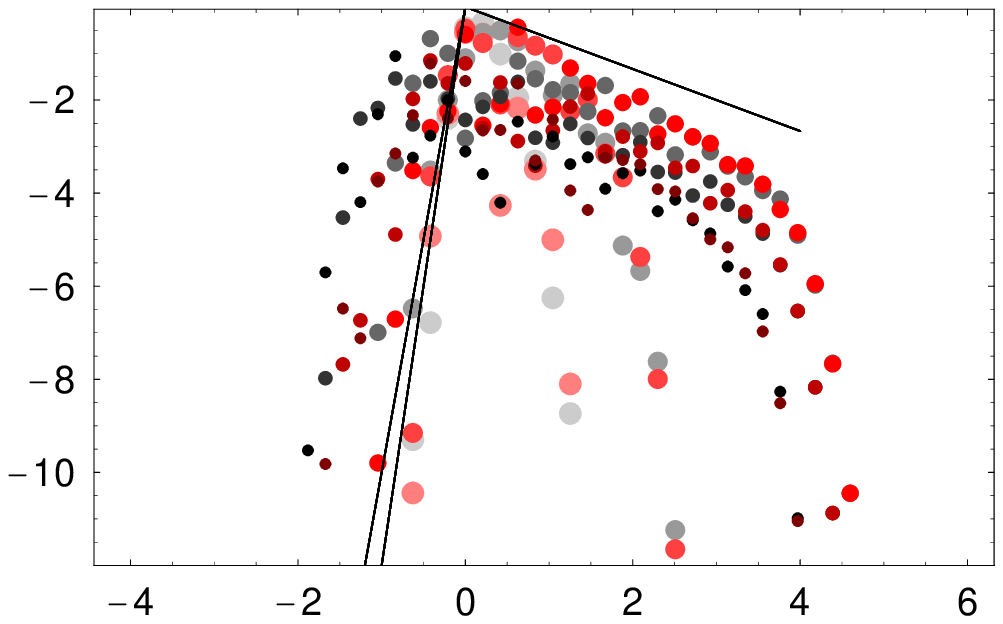}
\includegraphics[width=0.4\textwidth,angle=0]{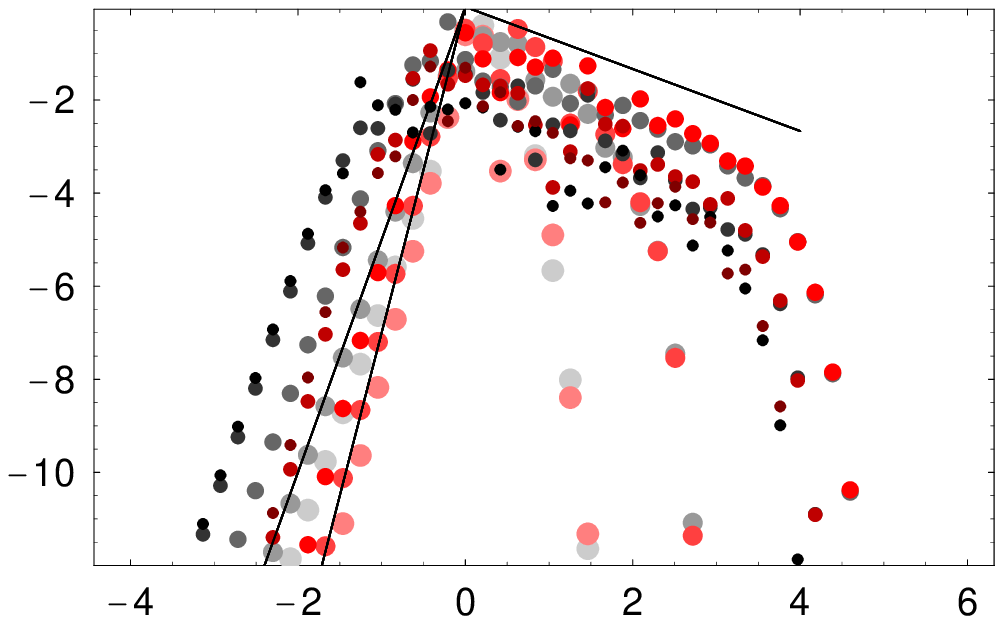}
\includegraphics[width=0.4\textwidth,angle=0]{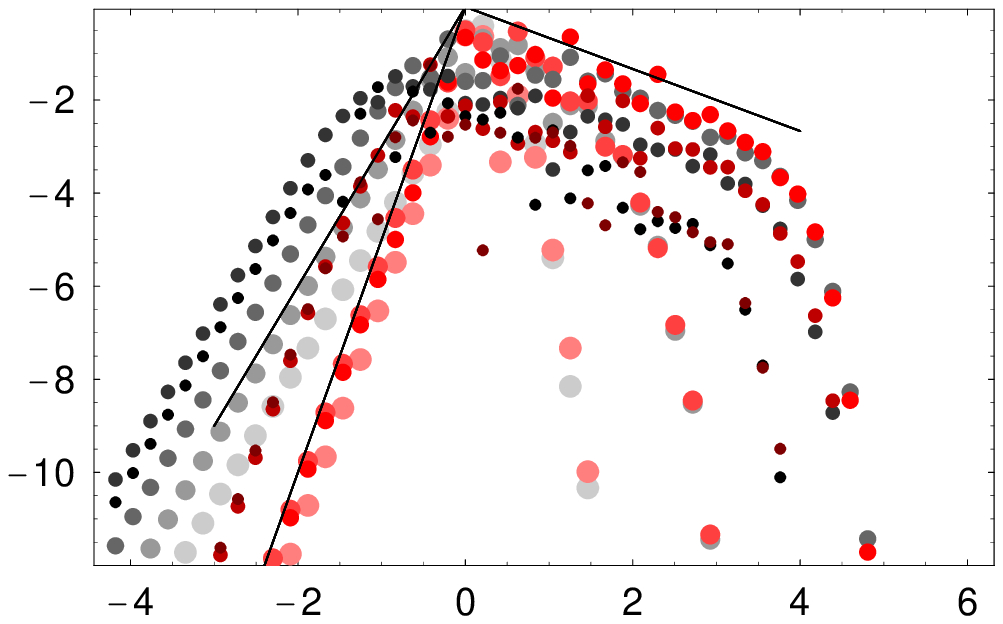}
\includegraphics[width=0.4\textwidth,angle=0]{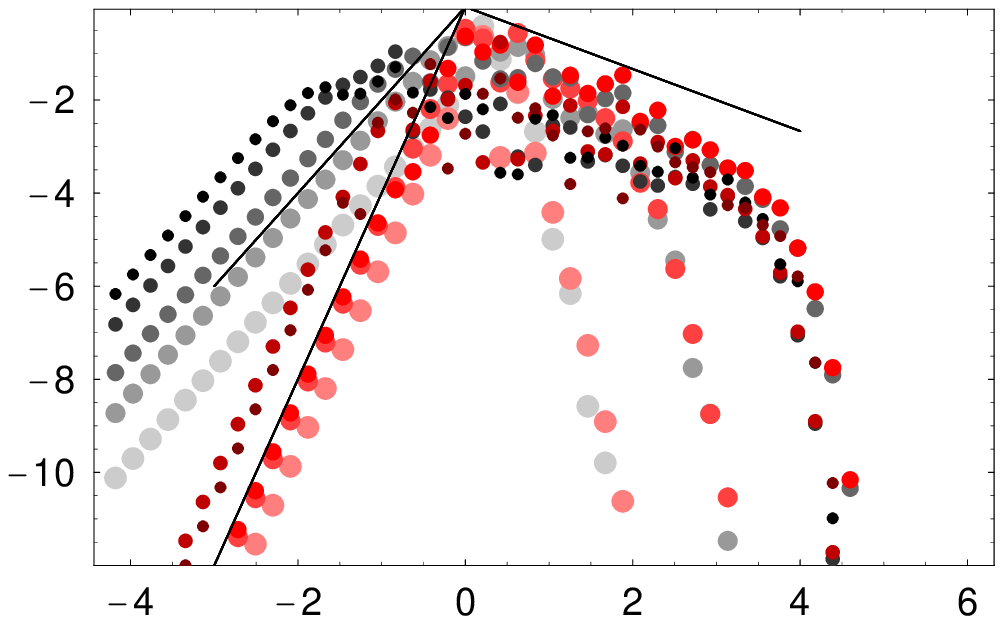}
\end{center}
\caption{Free decaying MHD turbulence for $\nu=10^{-6}$ and $\Pm=1$, with (top-left) local model,
(top-right) $\gamma=-2.5$, (bottom-left) $\gamma=-1.5$ and (bottom-right) $\gamma=-1$.
Gray corresponds to the kinetic energy, red to the magnetic energy. Increasing time is indicated with spots of lighter intensities. Adapted from \citet{Plunian2007}.}
\label{freedecay}
\end{figure}

For free decaying turbulence, \citet{Plunian2007} found that
$\gamma$ does not significantly change the slopes
of the kinetic and magnetic energy spectra in the inertial range, both compare well with Kolmogorov scaling $|U_n|^2\propto|B_n|^2\propto k_n^{-2/3}$
(corresponding to a $k^{-5/3}$ energy density spectrum).
On the other hand, $\gamma$ has a strong effect on the slopes of the infra-red spectrum, as shown in Fig.~\ref{freedecay}. It corresponds to the following dependency
\begin{equation}
	U_n^2 \propto k_n^{-2\gamma}, \qquad B_{n}^2 \propto k_{n}^{-2\gamma +2}.
\end{equation}

Different values of $\gamma$ may account for the presence of different infrared mechanisms in real 3D HD and MHD turbulence. We note that $\gamma = -5/2$ is in agreement with the arguments of \citet{Batchelor1948} for HD, and \citet{Pouquet1976} for MHD.
However, other values of $\gamma$ could also be satisfactory \citep{Saffman1967,Lesieur1997,Fournier1982}.
We note that the Sabra model given by Eqs.~(\ref{e3:sabraL2_u}-\ref{e3:sabraL2_b}), cannot give a realistic slope for the infra-red spectrum, emphasizing the importance of including non-local interactions.
Finally, for forced MHD turbulence (dynamo action) there are strong phenomenological arguments for taking $\gamma=-1$
as explained in Sec.~\ref{s4:forcing}.

\subsection{Helical models}
\label{s3:HM}

The kinetic and magnetic helicities given by Eqs.~(\ref{e3:kin-helicity}) and
(\ref{w3hb}),
have the particularity of being strongly correlated to, respectively, the kinetic and magnetic energies. In particular, as noted by \citet{Biferale1995} for the HD case, the kinetic helicity ``presents an asymmetry between odd and even shells that does not have any counterpart in physical flows''. The same remark applies to magnetic helicity.
In order to circumvent this problem, \citet{Biferale1995} introduced another type of shell model based on the decomposition into helical modes in the manner described by \citet{Waleffe1992}. At about the same time, \citet{Zimin1995} also introduced a helical, non-local shell model. The generalization of these helical shell models to MHD is given below.

In order to account for the helical decomposition of the velocity and the magnetic field
given in Eqs.~(\ref{helu}-\ref{helb}), we double the number of variables per shell,
$U_n^+, U_n^-, B_n^+$ and $B_n^-$.
The kinetic and magnetic energies, the cross helicity, and the kinetic and magnetic helicities  are now
\begin{eqnarray}
	E^U &=& \frac{1}{2}\sum_n\left(|U_n^+|^2+|U_n^-|^2\right) \label{EU_helical}\\
	E^B &=& \frac{1}{2}\sum_n\left(|B_n^+|^2+|B_n^-|^2\right) \\
	H^C &=& \frac{1}{2}\sum_n\left(U_n^+{B_n^+}^* +U_n^-{B_n^-}^* + c.c.\right) \\
	H^U &=& \sum_n k_n \left(|U_n^+|^2-|U_n^-|^2\right) \label{HU_helical}\\
	H^B &=& \sum_n k_n^{-1}\left(|B_n^+|^2-|B_n^-|^2\right). \label{HB_helical}
\end{eqnarray}
The enstrophy and square potential become
\begin{eqnarray}
	\Xi &=& \frac{1}{2}\sum_nk_n^2\left(|U_n^+|^2+|U_n^-|^2\right) \label{Enstrophy_helical}\\
	A &=& \frac{1}{2}\sum_nk_n^{-2}\left(|B_n^+|^2+|B_n^-|^2\right). \label{A_helical}
\end{eqnarray}
The general formalism derived in Sec.~\ref{s3:general formalism}
and given by Eqs.~(\ref{shell_NS2}-\ref{shell_induction2}) still applies,
defining any vector $\bX$ as
\begin{equation}
	\bX=(X_1^+, X_2^+, \cdots, X_N^+,X_1^-, X_2^-, \cdots, X_N^-).
	\label{defX}
\end{equation}
Applying this definition to $\bU$, $\bB$ and $\widetilde{\bW}$, we can rewrite Eqs.~(\ref{shell_NS2}-\ref{shell_induction2}) in the form
\begin{eqnarray}
   d_t U_n^{\pm} &=& \widetilde{W}_n^{\pm}(\bU,\bU)-\widetilde{W}_n^{\pm}(\bB,\bB) -   \nu k_n^2 U_n^{\pm} + F_n^{\pm}, \label{hel_NSpm}\\
   d_t B_n^{\pm} &=& \widetilde{W}_n^{\pm}(\bU,\bB)-\widetilde{W}_n^{\pm}(\bB,\bU) -   \eta k_n^2 B_n^{\pm}, \label{hel_Inducpm}
\end{eqnarray}
and Eq.~(\ref{relation1}) as
\begin{equation}
	\sum_n \widetilde{W}_n^+(\bX,\bY){Y_n^+}^* + \widetilde{W}_n^-(\bX,\bY){Y_n^-}^* + c.c.= 0. \label{relation1hel}
\end{equation}

In principle shell-to-shell and mode-to-mode energy transfer and flux can also be obtained using the definition of $\bX$ given in Eq.~(\ref{defX}). Transfer involves not only the velocity and the magnetic field but also both helical modes. Instead of just four, they are now sixteen possible transfers, $U^{\pm}$-to-$U^{\pm}$, $U^{\pm}$-to-$B^{\pm}$, $B^{\pm}$-to-$U^{\pm}$ and $B^{\pm}$-to-$B^{\pm}$.

We note that in each shell $n$, only the helical modes $U_n^{\pm}$
and $B_n^{\pm}$ are defined, from which the energies and helicities are calculated.
However if $U_n^{\pm}$ and $B_n^{\pm}$ are taken to be \textit{real}, then new \textit{complex} variables $U_n$ and  $B_n$ can be defined
\begin{equation}
	U_n = \left(U_n^+\;\e^{\i\pi/4} + U_n^-\;\e^{-\i\pi/4}\right) \qquad
	B_n = \left(B_n^+\;\e^{\i\pi/4} + B_n^-\;\e^{-\i\pi/4}\right),
	\label{UBhel}
\end{equation}
with energies and helicities
\begin{eqnarray}
	E^U &=& \frac{1}{2}\sum_n|U_n|^2 \label{Kin_energy}\\
	E^B &=& \frac{1}{2}\sum_n|B_n|^2 \\
	H^C &=& \frac{1}{2}\sum_n\left(U_n{B_n}^* + c.c.\right) \\
	H^U &=& \frac{\i}{2}\sum_nk_n\left({U_n^*}^2-U_n^2\right) \label{Kin_helicity}\\
	H^B &=& \frac{\i}{2}\sum_nk_n^{-1}\left({B_n^*}^2-B_n^2\right), \label{HB_helicalH1}
\end{eqnarray}
enstrophy and square potential
\begin{eqnarray}
	\Xi &=& \frac{1}{2}\sum_nk_n^2|U_n|^2 \\
	A &=& \frac{1}{2}\sum_nk_n^{-2}|B_n|^2.
\end{eqnarray}
The equivalence between Eqs.~(\ref{EU_helical}-\ref{HB_helical}) and Eqs.~(\ref{Enstrophy_helical}-\ref{A_helical}) is straightforward.\\

We call the H1-model a first-neighbor model for which
magnetic helicity as defined in Eq.~(\ref{HB_helical}) or (\ref{HB_helicalH1}) is conserved.
Similarly we call the H2-model a two-first neighbor model with the same quadratic invariant.
In Sec.~\ref{s3:HM1} we present the general expression of H1-models derived in terms of \textit{complex} variables $U_n$ and $B_n$
(\textit{real} $U_n^{\pm}$ and $B_n^{\pm}$).
In Sec.~\ref{s3:Helical mode decomposition} several H2-models 
in terms of \textit{complex} variables $U_n^{\pm}$ and $B_n^{\pm}$
are presented.


\subsubsection{H1-models (helical, two feet in the same shell)}
\label{s3:HM1}
A helical version of the HD model elaborated by \citet{Zimin1995} was first used to study 3D HD turbulence by \citet{Melander1997} (see also \citet{Melander2002,Melander2003,Melander2006}). The model is based on \textit{real} variables $U_n^+$ and $U_n^-$ and is both helical and non-local.
Equations were given for variables $S_n= (U_n^+ +U_n^-)$ and $D_n=(U_n^+ - U_n^-)$ and after appropriate renormalization we obtain
\begin{eqnarray}
d_t S_n&=&k_n \sum_{\substack{ m\ge-1\\ m\ne0}} \Lambda_m\left(S_n S_{n-m}-\lambda^{-m}D_n D_{n-m}-\lambda^m S^2_{n+m}+\lambda^m D^2_{n+m}\right)-\nu k_n^2 S_n +F_n^S,\nonumber\\
d_t D_n&=&k_n \sum_{\substack{m\ge-1\\m\ne0}} \Lambda_m\left(\lambda^{-m} S_n D_{n-m}-D_n S_{n-m}\right)-\nu k_n^{2} D_n + F_n^D.
\label{HDZimin}
\end{eqnarray}
where  $\Lambda_m$ is again an arbitrary parameter depending on $m$ as in model (\ref{SabraW}).
Using a hierarchical approach, Zimin found that
$\Lambda_m=\lambda^{-5|m+1/2|/2}$. However, any other function of $m$ does not change the conservation of kinetic energy and helicity.
This model was solved mainly for free decaying turbulence ($F_n^S=F_n^D=0$).
\citet{Melander1997} found that
any solution starting from a point close to the Kolmogorov stable manifold in phase space,
misses equilibrium and enters a chaotic
regime characterized by an exponential growth in helicity as the solution diverges from the equilibrium. Erratic
fluctuations of helicity and their dissipation imply intermittent energy decay.
Early enstrophy divergence was studied by \citet{Melander2002} and the transient laws were investigated by \citet{Melander2003}.

With 
\begin{equation}
	U_n=\i S_n \pm D_n,
\end{equation}
\cite{Stepanov2009} rewrote the
local version of model (\ref{HDZimin}) in a complex form. After appropriate renormalization
\begin{eqnarray}
	d_t U_n +\nu k_n^2 U_n&=& \i k_n [U_{n-1}^2+{U_{n-1}^*}^2
	+\lambda U_n^*(U_{n+1}- U_{n+1}^*)
	-\lambda^2 U_n(U_{n+1}+ U_{n+1}^*)]\nonumber\\
	&-&c \i k_n [U_n(U_{n-1}+U_{n-1}^*)
	+ \lambda U_n^*( U_{n-1}^* - U_{n-1})
	-\lambda^2 (U_{n+1}^2 + {U_{n+1}^*}^2)] + F_n,
	\label{helical_U}
\end{eqnarray}
where $c$ is a free parameter.
The kinetic energy and helicity defined by Eqs.~(\ref{Kin_energy}) and (\ref{Kin_helicity}) are ideally conserved whatever the value of $c$.
\cite{Stepanov2009} chose
$c=\lambda^{-5/2}$ in order that model (\ref{helical_U}) becomes equivalent
to the local part of the \citet{Zimin1995} model.
In terms of $S_n$ and $D_n$,
the kinetic energy and helicity are given by
\begin{eqnarray}
  E^U &=& \frac{1}{2}\sum_n(S_n^2+D_n^2) \label{Kin_energy_SD}\\
	H^U &=& \pm2\sum_nk_nS_nD_n. \label{Kin_helicity_SD}
\end{eqnarray}

The version of model (\ref{helical_U}) corresponding to 3D MHD turbulence is given by \citep{Mizeva2009}
\begin{eqnarray}
\widetilde{W}_n(\bX,\bY)&=&\i k_n \left[
 (X_{n-1}Y_{n-1}+X^*_{n-1}Y^*_{n-1})
-\lambda X_n^*Y_{n+1}^*
-\frac{\lambda^2}{2}(X_nY_{n+1}+X_{n+1}Y_n)\right.\nonumber\\
&&\;\; \left.-\frac{\lambda}{2}(X_{n-1}^*Y_{n-1}-X_{n-1}Y^*_{n-1})
+\lambda X_n^*Y_{n+1}
-\frac{\lambda^2}{2}(X_nY_{n+1}^*+X_{n+1}^*Y_n) \right]\nonumber\\
&-& \i c k_{n}\left[\frac{1}{2}(X_{n-1}Y_{n}+X_{n}Y_{n-1})\right.
+\lambda X_n^*Y_{n-1}^*
-\lambda^2(X_{n+1}Y_{n+1}+X_{n+1}^*Y_{n+1}^*)\nonumber\\
&&\qquad\left.+\frac{1}{2}(X_{n}Y_{n-1}^*+X_{n-1}^*Y_{n})
-\lambda X_n^*Y_{n-1}+
\frac{\lambda}{2}(X_{n+1}^*Y_{n+1}-X_{n+1}Y_{n+1}^*)
  \right],
\label{L1MHD}
\end{eqnarray}
for which the magnetic helicity (\ref{HB_helicalH1}) is conserved in addition to total energy and cross helicity. This model was used to study MHD helical turbulence \citep{Mizeva2009,Frick2010}, including global rotation and an external magnetic field \citep{Plunian2010}. The results will be detailed in Sec.~\ref{s6:rotation}.

\subsubsection{H2-models (helical, two feet in two neighboring shells, third foot in a smaller shell)}
\label{s3:Helical mode decomposition}

Expanding the velocity Fourier modes onto a base of polarized helical waves as described in Sec.~\ref{Helical decomposition}, \citet{Biferale1995} introduced four independent helical shell models for 3D HD turbulence.
Following \citet{Lessinnes2011} these can be written in a compact form
\begin{eqnarray}
	d_t U_n^{\pm} +\nu k_n^2 U_n^{\pm}
	= \i k_n \left[(s_1\lambda-s_2\lambda^{2})U_{n+1}^{\pm s_1}U_{n+2}^{\pm s_2}
	+(s_2\lambda-\lambda^{-1}) U_{n-1}^{\pm s_1}U_{n+1}^{\pm s_1s_2}
	+(\lambda^{-2}-s_1\lambda^{-1})U_{n-2}^{\pm s_2}U_{n-1}^{\pm s_1s_2}\right]^* +F_n^{\pm}\nonumber\\
	\label{helical_Upm}
\end{eqnarray}
where $U_n^{\pm}$ are \textit{complex} variables.
The four independent models correspond to $(s_1,s_2)\in\{(1,1),(-1,1),(1,-1),(-1,-1)\}$.
The kinetic energy and helicity defined by Eqs.~(\ref{EU_helical}) and (\ref{HU_helical}) are ideally conserved.
\begin{itemize}
	\item
For $(s_1,s_2)=(-1,1)$ Eq.~(\ref{helical_Upm}) consists of two independent sets of equations for $\bU=(U_1^+,U_2^-,U_3^+,\cdots,U_{2n-1}^+,U_{2n}^-)$
and
$\bU=(U_1^-,U_2^+,U_3^-,\cdots,U_{2n-1}^-,U_{2n}^+)$, each of which corresponds to the GOY model given by Eq.~(\ref{GOY_epsilon}) with $\varepsilon=1-\lambda^{-1}$ \citep{Biferale1995}.
\item
For $(s_1,s_2)=(1,1)$ Eq.~(\ref{helical_Upm}) consists of two independent sets of equations for
$\bU=(U_1^+,U_2^+,\cdots,U_{2n-1}^+,U_{2n}^+)$
and
$\bU=(U_1^-,U_2^-,\cdots,U_{2n-1}^-,U_{2n}^-)$
each of which conserving separately a positive-definite quantity.
This led \citet{Benzi1996} to conclude that such a model ``is equivalent to two uncorrelated GOY models for 2D turbulence''.
\item
For $(s_1,s_2)=(-1,-1)$ after \citet{Benzi1996} the model ``may present a significant backward energy transfer, which leads to possibly strong deviations from the Kolmogorov scaling".
\item
For $(s_1,s_2)=(1,-1)$ the model shows intermittent statistics
in excellent agreement with the Navier-Stokes equations \citep{Benzi1996}.
It is given by
\begin{equation}
	d_t U_n^{\pm} +\nu k_n^2 U_n^{\pm}
	= \i  \left[k_{n+1}(1+\lambda)U_{n+1}^{\pm}U_{n+2}^{\mp}
	-k_n(\lambda+\lambda^{-1}) U_{n-1}^{\pm}U_{n+1}^{\mp}
	+k_{n-1}(\lambda^{-1}-1)U_{n-2}^{\mp}U_{n-1}^{\mp}\right]^* +F_n^{\pm}.
	\label{helical_Umodel31}
\end{equation}	
	Another version of this model has also been formulated by \citet{Chen2003}, in the spirit of the Sabra model
\begin{equation}
	d_t U_n^{\pm} +\nu k_n^2 U_n^{\pm}
	= \i \left[k_{n+1}(1+\lambda)(U_{n+1}^{\pm})^*U_{n+2}^{\mp}
	-k_n(\lambda+\lambda^{-1}) (U_{n-1}^{\pm})^*U_{n+1}^{\mp}
	-k_{n-1}(\lambda^{-1}-1)U_{n-2}^{\mp}U_{n-1}^{\mp}\right] +F_n^{\pm}.
	\label{helical_Umodel32}
\end{equation}
\end{itemize}
The interest of helical rather than non-helical models has been clearly demonstrated by the study of the
 helicity flux in 3D HD turbulence.
Using the GOY model, \citet{Ditlevsen2001,Ditlevsen2001a} found a range of scales, within the inertial range, for which the helicity spectral flux is exponentially divergent in $k$ due to dissipation (see also \citet{Ditlevsen2011}). This implies that even if the flow is helical at large scales, there is a subdomain of the inertial range for which the flow is non-helical. This picture contrasts strongly with that arising due to the
energy cascade occurring all along the inertial range for which dissipation takes place only at scales smaller than the Kolmogorov scale.
The physical reason why helicity and energy do not dissipate on the same scale was not clear until helical models came into being.

First, using both a H1 \citep{Stepanov2009} and H2 \citep{Chen2003} models, the helicity spectral flux has been found to be constant all along the inertial range without the divergence in $k$ previously found with the GOY model. Apparently both types of model, non-helical and helical, were giving contradictory results.

Then using an H2-model corresponding to $(s_1,s_2)=(1,-1)$, \citet{Lessinnes2011} calculated the helicity spectral flux  for both $+$ and $-$ helical modes and found a subdomain of the inertial range in which both helicity flux spectra diverge in $k$ due to dissipation in agreement with the GOY model. However, as both fluxes have opposite signs they also found that their sum, which is the flux of total helicity, has a constant spectrum all along the inertial range. In the GOY model two helical modes cannot be present in the same shell
(see the case $(s_1,s_2)=(-1,1)$ above)
and therefore they cannot be summed. This clearly shows the superiority of helical shell models in dealing with kinetic helicity in 3D HD turbulence, and presumably with magnetic helicity in 3D MHD turbulence.

The helical shell model of MHD turbulence elaborated by \citet{Lessinnes2009a} is a linear combination of the four previous submodels $(s_1,s_2)=(\pm1,\pm1)$, and includes an estimate of the weighting to be given to each submodel. In fact it is not clear whether such a combination is appropriate, as only the model for $(s_1,s_2)=(1,-1)$ gives a relevant helicity spectral flux for 3D HD turbulence.
Instead it might be more appropriate to consider the MHD version of models (\ref{helical_Umodel31}) or (\ref{helical_Umodel32}). Their respective expressions in terms of $\widetilde{W}^{\pm}(\bX,\bY)$ are
\begin{eqnarray}
	\widetilde{W}_n^{\pm}(\bX,\bY) = \frac{\i}{2} \left\{
	 k_{n+1}(1+\lambda)(X_{n+1}^{\pm}Y_{n+2}^{\mp}+Y_{n+1}^{\pm}X_{n+2}^{\mp})\right.
	 &-&\lambda^{-1}k_{n+1}(X_{n+1}^{\pm}Y_{n+2}^{\mp}-Y_{n+1}^{\pm}X_{n+2}^{\mp})\nonumber\\
	 -k_{n}(\lambda^{-1}+\lambda)(X_{n-1}^{\pm}Y_{n+1}^{\mp}+Y_{n-1}^{\pm}X_{n+1}^{\mp})
	&+&k_{n}(X_{n-1}^{\pm}Y_{n+1}^{\mp}-Y_{n-1}^{\pm}X_{n+1}^{\mp})\nonumber\\
	 +k_{n-1}(\lambda^{-1}-1)(X_{n-2}^{\mp}Y_{n-1}^{\mp}+Y_{n-2}^{\mp}X_{n-1}^{\mp})
	 &+&\left.\lambda k_{n-1}(X_{n-2}^{\mp}Y_{n-1}^{\mp}-Y_{n-2}^{\mp}X_{n-1}^{\mp})\right\}^*,
	\label{HelMHDGOY}
\end{eqnarray}
and
\begin{eqnarray}
	\widetilde{W}_n^{\pm}(\bX,\bY) = \frac{\i}{2} \left\{
	k_{n+1}(1+\lambda)(X_{n+1}^{\pm *}Y_{n+2}^{\mp}+Y_{n+1}^{\pm *}X_{n+2}^{\mp})\right.
	&-&\lambda^{-1}k_{n+1}(X_{n+1}^{\pm *}Y_{n+2}^{\mp}-Y_{n+1}^{\pm *}X_{n+2}^{\mp})\nonumber\\
	-k_{n}(\lambda^{-1}+\lambda)(X_{n-1}^{\pm *}Y_{n+1}^{\mp}+Y_{n-1}^{\pm *}X_{n+1}^{\mp})
	&+&k_{n}(X_{n-1}^{\pm *}Y_{n+1}^{\mp}-Y_{n-1}^{\pm *}X_{n+1}^{\mp})\nonumber\\
	 -k_{n-1}(\lambda^{-1}-1)(X_{n-2}^{\mp}Y_{n-1}^{\mp}+Y_{n-2}^{\mp}X_{n-1}^{\mp})
	 &-&\left.\lambda k_{n-1}(X_{n-2}^{\mp}Y_{n-1}^{\mp}-Y_{n-2}^{\mp}X_{n-1}^{\mp})\right\}.
	\label{HelMHDSABRA}
\end{eqnarray}
In both models total energy, cross helicity and magnetic helicity are
ideally conserved.

%% file: Dynamo.tex
\section{Applications of MHD shell models}
\label{s4}
With the exception of models (\ref{HelMHDGOY}) and (\ref{HelMHDSABRA}) which are introduced here for the first time, the other shell models introduced in Sec.~\ref{s3} have been used to investigate the statistical properties of forced and free-decaying MHD turbulence (Sec.~\ref{s4:forced turbulence} and ~\ref{s4:free decaying}). They have also been the starting point for more complex models, like multi-scale dynamo models, Alf\'en-wave models or Hall-effect models (Sec.~\ref{s5}, ~\ref{s6:rotalfven}
and ~\ref{s7}).

\subsection{Forced MHD turbulence}
\label{s4:forced turbulence}
\subsubsection{Forcing}
\label{s4:forcing}
In forced MHD turbulence, choosing an appropriate forcing $\bF$ is of course essential. It is generally applied at the largest scales of the system.  The two main features of a forcing are its time dependency and the range of scales over which it is applied.

\begin{figure}[ht]
\begin{center}
\includegraphics[width=0.43\textwidth]{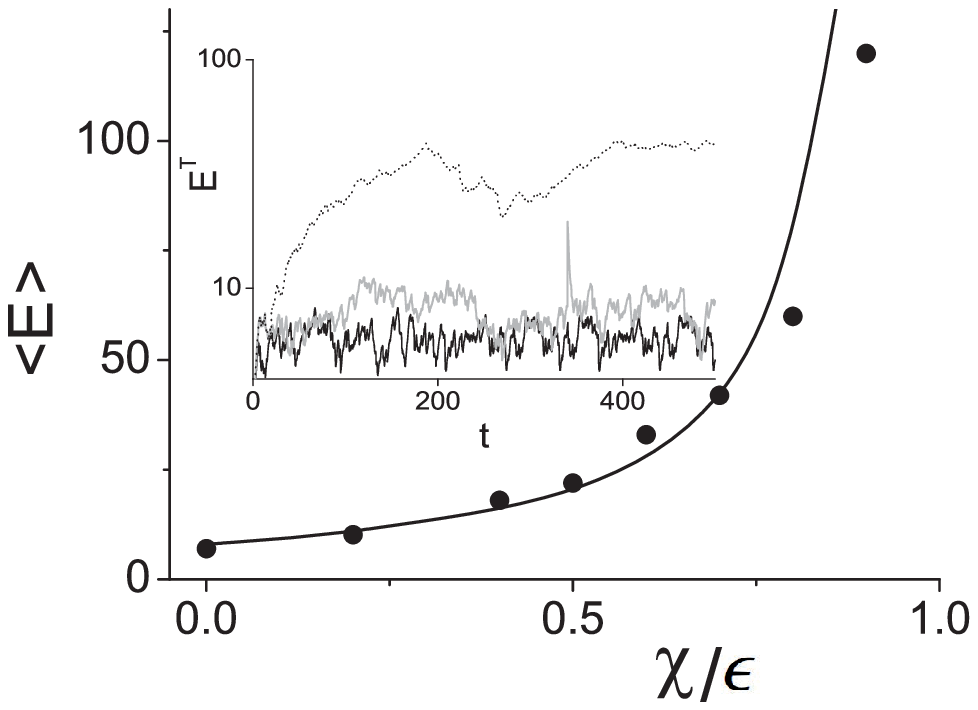}
\includegraphics[width=0.49\textwidth]{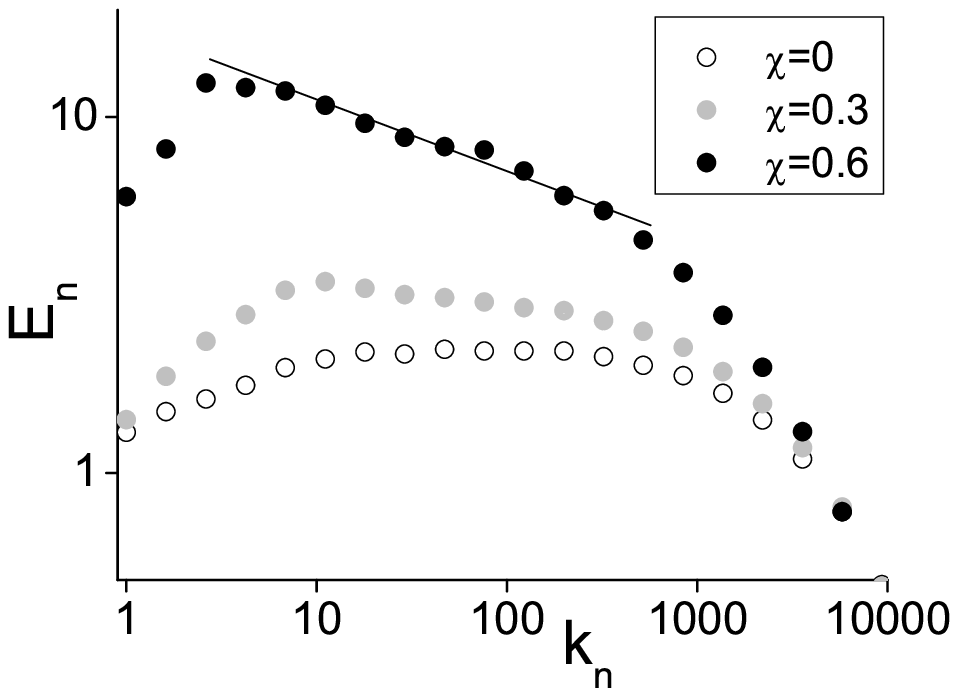}
\end{center}
\caption{Left panel: total energy of stationary forced
MHD turbulence versus the relative cross helicity injection rate $\chi/\epsilon$. The
time evolution of energy is shown in the inset, for $\chi=0$ (thick line),
$\chi=0.3$ (gray line), and $\chi=0.6$(thin line).
Right panel: total energy spectrum normalized by $k^{-2/3}$ for three different injection rates of cross helicity. The straight line corresponds to $E(k)\propto k^{-1.9}$.
Adapted from \citet{Mizeva2009}.}
\label{f1:fig2}
\end{figure}

With a stationary \textit{complex} forcing $\bF$ applied on one scale, the injection rate of two \textit{real} quantities, e.g. energy $\epsilon$ and cross helicity $\chi$, can be controlled.
However, with such a stationary forcing the solutions of Eqs.~(\ref{shell_NS}-\ref{shell_induction}) may reach an unphysical supercorrelated state $U_n = \pm B_n$,
with no fluxes and no inertial ranges \citep{Giuliani1998,Frick2000}. Instead using a forcing $F_n=f_n\e^{\i \phi_n(t)}$, where $\phi_n(t)$ is a random phase and $f_n$ is \textit{real}, leads to
long lasting dynamics of $\bU$ and $\bB$. However, using such random phase, the injection rate of only one quantity instead of two is then controlled.
For example, taking $\phi_n(t)$ constant during time interval $t_c$, but changing randomly from one time interval to the next, the injection rate of energy becomes
$\epsilon \approx f_n^2 t_c$ \citep{Plunian2007,Plunian2010} and the cross helicity injection is not controlled anymore.
In order to control the injection rate of two quantities again, the forcing must be split between two neighboring scales \citep{Mizeva2009}. In that case the forcing satisfies the following set of equations
\begin{eqnarray}
  \frac{1}{2}\sum_{n=n_F}^{n_F+1} U_{n}^* F_n+ U_{n} F_n^* &=& \epsilon, \label{forcing1}\\
  \frac{1}{2}\sum_{n=n_F}^{n_F+1} B_{n}^* F_n + B_{n} F_n^*  &=&  \chi. \label{forcing3}
\end{eqnarray}
\citet{Mizeva2009} used their helical shell model given in Eq.~(\ref{L1MHD})
with $\Pm=1$ to study the influence of cross helicity on MHD turbulence.
They found that the total energy of the system increases with the ratio $\chi/\epsilon$,
following the scaling law $E\propto (1-\chi/\epsilon)^{-4/3}$, as indicated by the full curve in Fig.~\ref{f1:fig2} (left panel). This corresponds to an increase in the non-linear transfer time $t_{NL}$ by a factor $\chi$. In other words, the energy transfer in the inertial range is depleted, implying energy accumulation in the large scales, and a steeper spectrum as shown in Fig.~\ref{f1:fig2} (right panel).
In the limit $\chi/\epsilon\rightarrow1$ the velocity and magnetic fields
are correlated, the non-linear terms in the MHD equations are
canceled, and the energy cascade transfer is blocked.

One may also want to control the injection rate of kinetic helicity. Depending on the model used this might require an additional forcing shell
\citep{Stepanov2006a} or forcing on two different helical modes \citep{Lessinnes2009a}.

\begin{figure}[ht]
\begin{center}
\includegraphics[width=0.45\textwidth]{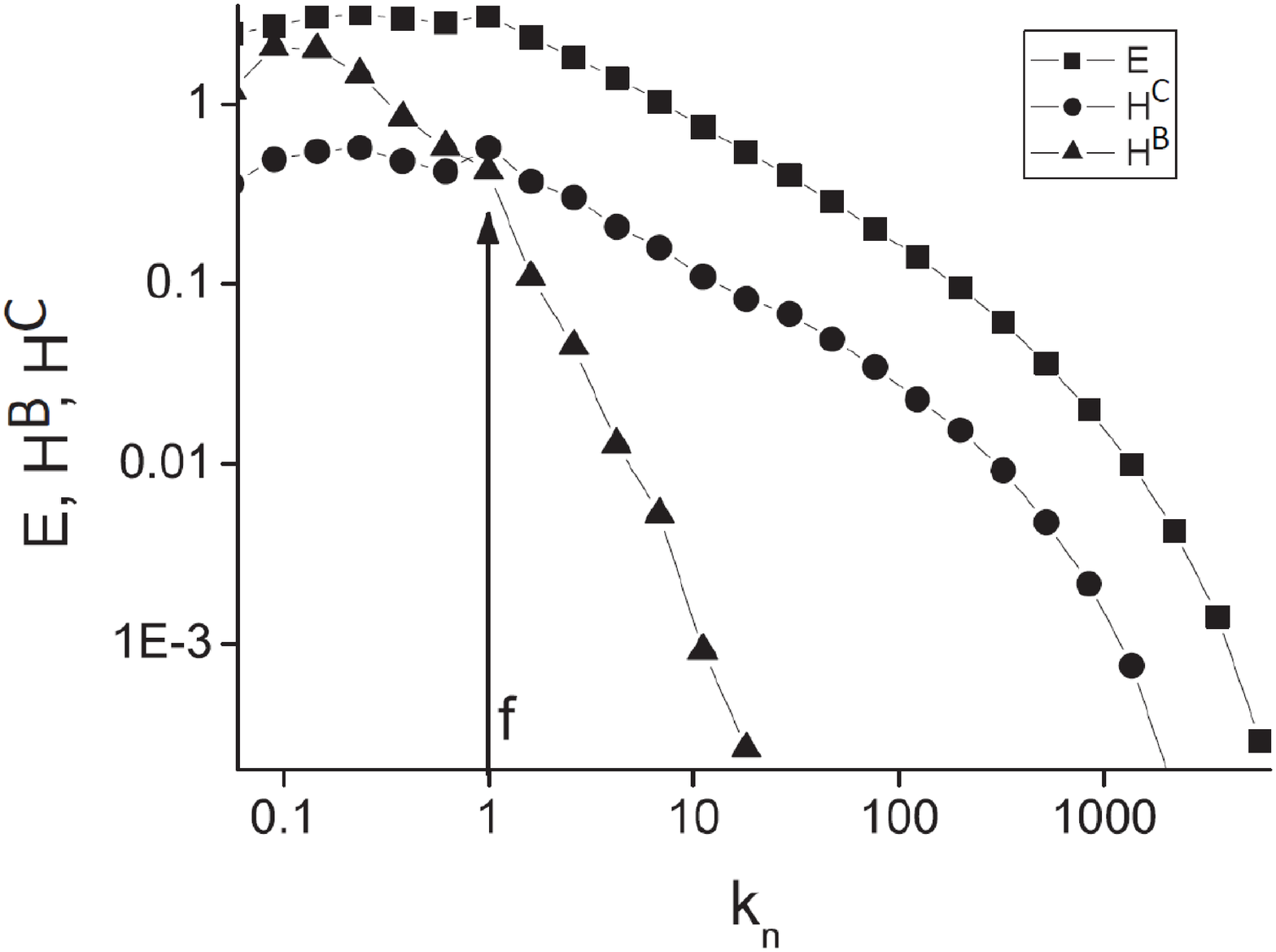}
\includegraphics[width=0.45\textwidth]{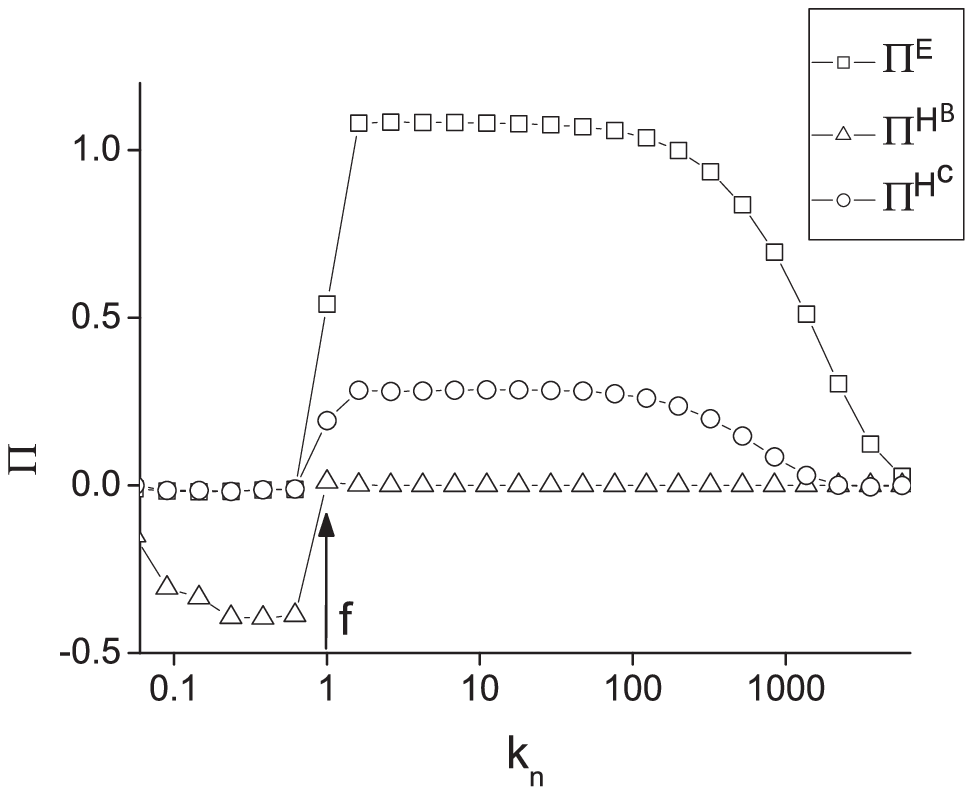}
\end{center}
\caption{Energy, cross helicity, magnetic helicity spectra (left panel)
and fluxes (right panel) of forced MHD turbulence with constant injected helicities
($\epsilon=1$, $\chi=0.3$, $\xi=0.4$).}
\label{figblin}
\end{figure}

One could also envisage applying some forcing $\bG$ in the time evolution equation for $\bB$. Such \textit{magnetic forcing} is naturally implemented in multi-scale dynamo models to account for the back-reaction of the large-scale fields onto small-scale turbulence \citep{Frick2006,Nigro2011}, and will be detailed in Sec.~\ref{s5}.
Note that \textit{magnetic forcing} can occur without injecting magnetic energy.
Let us take the example of constant
injection rates for kinetic energy $\epsilon$, cross helicity $\chi$ and magnetic-helicity $\xi$, in the absence of magnetic energy injection. Consequently we can take $\bF$ and $\bG$ to be constant in time (no random phase) within shell $n_F$. With the magnetic helicity defined by Eq.~(\ref{HB_helicalH1}),
$F_{n_F}$ and $G_{n_F}$ must satisfy
\begin{eqnarray}
  \frac{1}{2}\left(U_{n_F}^* F_{n_F}+ U_{n_F} F_{n_F}^*\right) &=& \epsilon, \\
  B_{n_F}^* G_{n_F}+ B_{n_F} G_{n_F}^* &=& 0, \\
  \frac{1}{2}\left(B_{n_F}^* F_{n_F} + B_{n_F} F_{n_F}^* + U_{n_F}^* G_{n_F} + U_{n_F} G_{n_F}^* \right)&=&  \chi,\\
  \i k_{n_F}^{-1}\left(B_{n_F}^* G_{n_F}^* - B_{n_F} G_{n_F}\right)&=&\xi.
\end{eqnarray}
The results obtained using such forcing applied at $k_{n_F}=1$ are given in Fig.~\ref{figblin}.  An inverse magnetic helicity cascade is found corresponding to a negative flux of magnetic helicity for $k<k_{n_F}$ (right panel).

\subsubsection{Small-scale dynamo action}
We speak of \textit{dynamo action} when the magnetic energy is sustained by the motion of an electroconducting fluid.
For the dynamo to work the magnetic Reynolds number, e.g. at the forcing scale of the system, must be higher than some threshold value $\Rm_c$.
Experiments and simulations show that
$\Rm_c$ depends on the presence of a mean flow and its characteristics, such as geometry and time dependency \citep{Peyrot2007,Peyrot2008,Ponty2011}, the electromagnetic boundaries, such as conductivity and permeability \citep{Frick02:MHD,Dobler03:PRE,Avalos2003,Avalos2005}, the presence of an external magnetic field, and global rotation.

\begin{figure}[ht]
\includegraphics[width=1\textwidth]{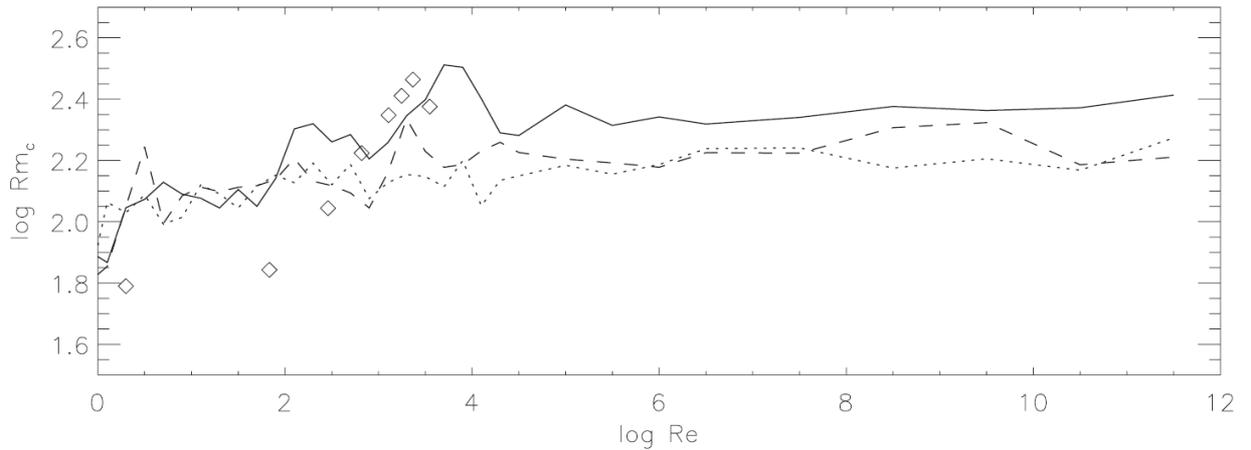}
\caption{Comparison of the dynamo threshold $\Rm_c$ versus $\Rey$, obtained from the N2-model given by Eqs.~(\ref{e3:sabra_u}-\ref{e3:sabra_b}) with different degrees of non-locality (the three curves) and from a DNS shown by the diamonds \citep{Iskakov2007}. From \citet{Buchlin2011}.}
\label{f4:Rmc}
\end{figure}

The dynamo threshold $\Rm_c$ can also be calculated assuming that dynamo action is produced only by the turbulence. Then dynamo is reached only if a sufficient amount of power $\epsilon_c$ is injected into the fluid, corresponding to $\Rm_c=\epsilon_c^{1/3}l_F^{4/3}/\eta$.
In Fig.~\ref{f4:Rmc} the dynamo threshold $\Rm_c$
is plotted versus $\Rey$ \citep{Buchlin2011}. The curves were obtained using the N2-model given by Eqs.~(\ref{e3:sabra_u}-\ref{e3:sabra_b}) and three degrees of non-locality: local $\gamma=-\infty$ (solid line), weakly non-local $\gamma=-5/2$ (dashed line) and strongly non-local $\gamma=-1$ (dotted line), where the parameter $\gamma$ is defined in Sec.~\ref{s3:N2models}. The results are similar to the ones obtained by \citet{Stepanov2006} using a 3D MHD GOY model, suggesting that the results are sensitive neither to the model (GOY or Sabra) nor to non-locality.
The diamonds correspond to $\Rm_c$ obtained from DNS \citep{Iskakov2007}.
The overall picture is that the dynamo threshold first increases with
$\Rey$, as found from DNS by \citet{Schekochihin2004}, and then decreases to reach a plateau for $\Rey \ge 10^4$. Similar results were obtained by \citet{Ponty2005} using Large Eddy Simulation techniques.

Using a helical shell model derived from a combination of the four H2-models given by Eq.~(\ref{helical_Upm}), \citet{Lessinnes2009a} compared the effect of maximum helical and non-helical forcing. They found that the dynamo threshold is lowered by at least a factor 1.5 when using a maximum helical forcing.\\

\begin{figure}[ht] \centering
\includegraphics[width=1\textwidth]{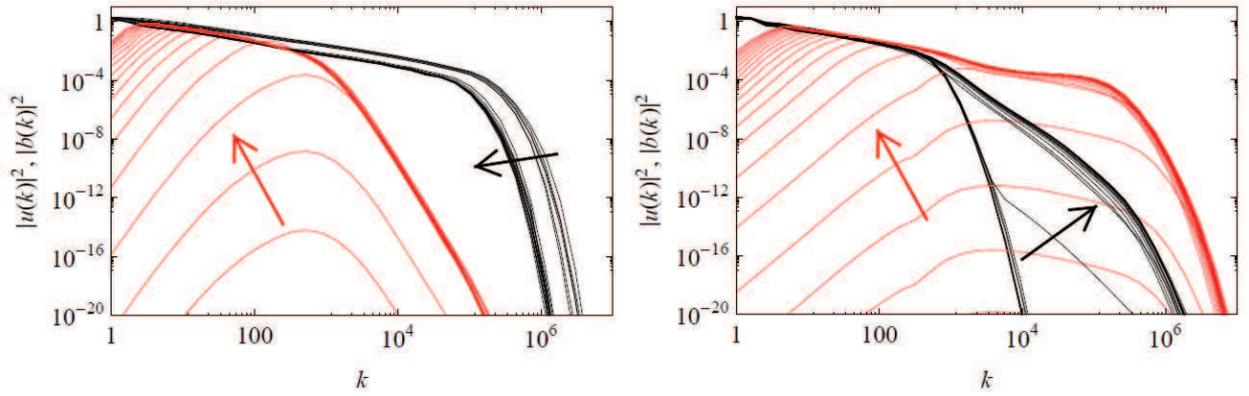}
\caption{Kinetic (black) and magnetic (red) energy spectra at different times for
$\Pm=10^{-3}$ (left-panel) and $\Pm=10^4$ (right-panel). The time interval between two samples is constant.
Along time, the
curve evolution is given by the arrows.
Adapted from \citet{Stepanov2008}.} \label{growth}
\end{figure}

Starting with a statistically stationary turbulent flow, the time evolution of the magnetic energy for $\Rm>\Rm_c$ can be split in two regimes, the kinematic regime during which the magnetic energy grows exponentially, and the saturated regime starting when the magnetic energy reaches some statistical stationary level. Other dynamics like on-off intermittency or chaotic reversals, which involve large-scale fields, are not captured with shell models unless large-scale equations are added (see Sec.~\ref{s5}). In Fig.~\ref{growth}, two examples are shown for
$\Pm\ll1$ (left-panel) and $\Pm\gg1$ (right-panel), and $\Rm\gg\Rm_c$ in both cases. The magnetic energy spectrum (in red) grows with time.  The end of the kinematic regime corresponds to the point at which the flow becomes sensitive to the magnetic field via the Lorentz forces.
For $\Pm\ll1$ (left-panel), a depletion of the kinetic energy spectrum occurs
due to the equipartition between magnetic and kinetic energies. A significant part of the injected power $\epsilon$ is then lost through magnetic dissipation, explaining why the viscous scale $l_{\nu}$ becomes larger ($k_{\nu}$ becomes smaller).
For $\Pm\gg1$ (right-panel) refilling of the kinetic energy spectrum occurs in a range of scales between $l_{\eta}$ and $l_{\nu}$, due to direct energy transfer from magnetic to kinetic.
The results of Fig.~\ref{growth} have been obtained with, again, the N2-model given by Eqs.~(\ref{e3:sabra_u}-\ref{e3:sabra_b}), with $\gamma=-5/2$ for $\Pm=10^{-3}$, and $\gamma=-1$ for $\Pm=10^4$.

For $\Pm \ll 1$ the results are consistent with
dynamo action occurring on a scale $k_{\rm kin}^{-1}$ of the same order of magnitude as the magnetic dissipation scale $k_{\eta}^{-1}$, and
a kinematic growth rate $\Gamma_{\rm kin}$ such that
\begin{equation}
\Gamma_{\rm kin}\propto \epsilon^{1/2}\eta^{-1/2}, \qquad	k_{\rm kin}\approx k_{\eta}\propto \epsilon^{1/4}\eta^{-3/4}.
\end{equation}
This corresponds to the dynamo action being produced by local energy transfers, the main physical mechanism being the action of flow shear against magnetic dissipation. This is also consistent with fast and small-scale kinematic dynamo action \citep{Childress1995}.
Taking other values of $\gamma \in\left[-\infty,-0.5\right]$ does not significantly change the results \citep{Plunian2007}.

For $\Pm\gg1$ the parameter $\gamma$ is taken to be equal to -1 in order
to have dynamo action occurring at the viscous scale $k_{\rm kin} \approx k_{\nu}$, where the flow shear is maximum \citep{Schekochihin2002a,Schekochihin2004b}. The
kinematic regime is then characterized by
\begin{equation}
	\Gamma_{\rm kin}=k_{\nu}u(k_{\nu})\propto \epsilon^{1/2}\nu^{-1/2}, \qquad k_{\rm kin} \approx k_{\nu}\propto\epsilon^{1/4}\nu^{-3/4}.
\end{equation}
The model is non-local and the kinematic dynamo is again fast and small-scale.

The transient regime leading towards saturation has also been studied in detail for $\Pm \ll 1$ and $\Pm \gg 1$.
An  inverse cascade of magnetic energy occurs towards large scales.
In the saturated state the energy ratio of magnetic to kinetic is found to be larger than unity (about 1.5). Such super-equipartition is spread over the whole inertial range.

\begin{figure}[ht] \centering
\includegraphics[width=0.55\textwidth]{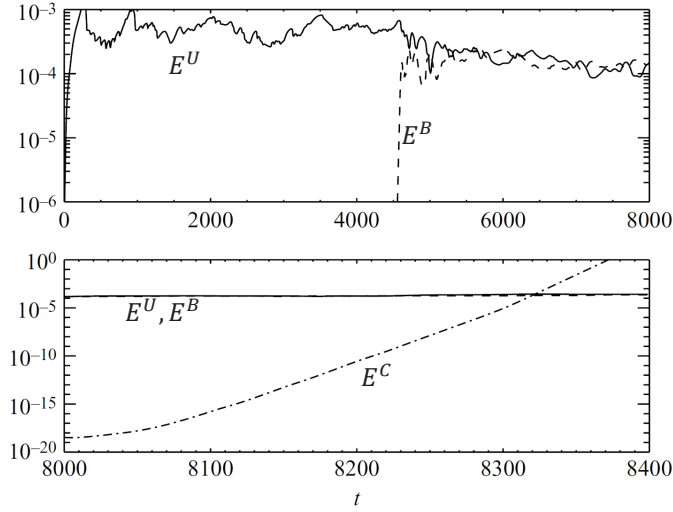}
\caption{Top panel: time series for the kinetic energy (solid line) and the magnetic energy (dashed line)
showing the dynamo evolving to a saturated state ($\nu=\eta=10^{-6}$). Bottom panel: time series for the kinetic energy (solid line), the magnetic energy (dashed line) and the energy of the passive field (dot-dashed line). Adapted from \citet{Cattaneo2009}.} \label{f4:CT}
\end{figure}

In such a saturated state, characterized by a statistical stationarity for both kinetic and magnetic energies, the time-average of the growth rate of the magnetic energy is thus equal to zero.
Does it imply that the saturated flow is unable to produce dynamo action anymore and can just compensating for dissipation? To investigate this question
\citet{Cattaneo2009} introduced a \textit{passive} vector $\bc$, satisfying a third equation identical to the induction equation:
\begin{eqnarray}
\left(\partial_t-\eta \nabla^2 \right) \bc &=&- (\bu\cdot \nabla) \bc + (\bc \cdot \nabla)  \bu ,    \qquad \nabla \cdot \bc =0.
\label{MHDC}
\end{eqnarray}
Here the vector $\bc$ is \textit{passive} in the sense that it does not back react onto the flow.
In other words, the system of Eqs.~(\ref{MHDU}-\ref{MHDB}) remains unchanged.
\citet{Cattaneo2009} solved the problem for $\Pm=1$ using two approaches, a DNS and the MHD GOY model given by Eqs.~(\ref{e3:goy_u}-\ref{e3:goy_b}). Both calculations lead to the same conclusion summarized in Fig.~\ref{f4:CT} for the GOY results. After a saturated regime has been reached (top figure)
the additional Eq.~(\ref{MHDC}) is switched on. The passive vector energy is found to grow exponentially (bottom), showing that the saturated flow keeps its small-scale dynamo characteristics,
presumably due to its chaotic dynamics.

\begin{figure}[ht]
\centering
\includegraphics[width=0.4\textwidth]{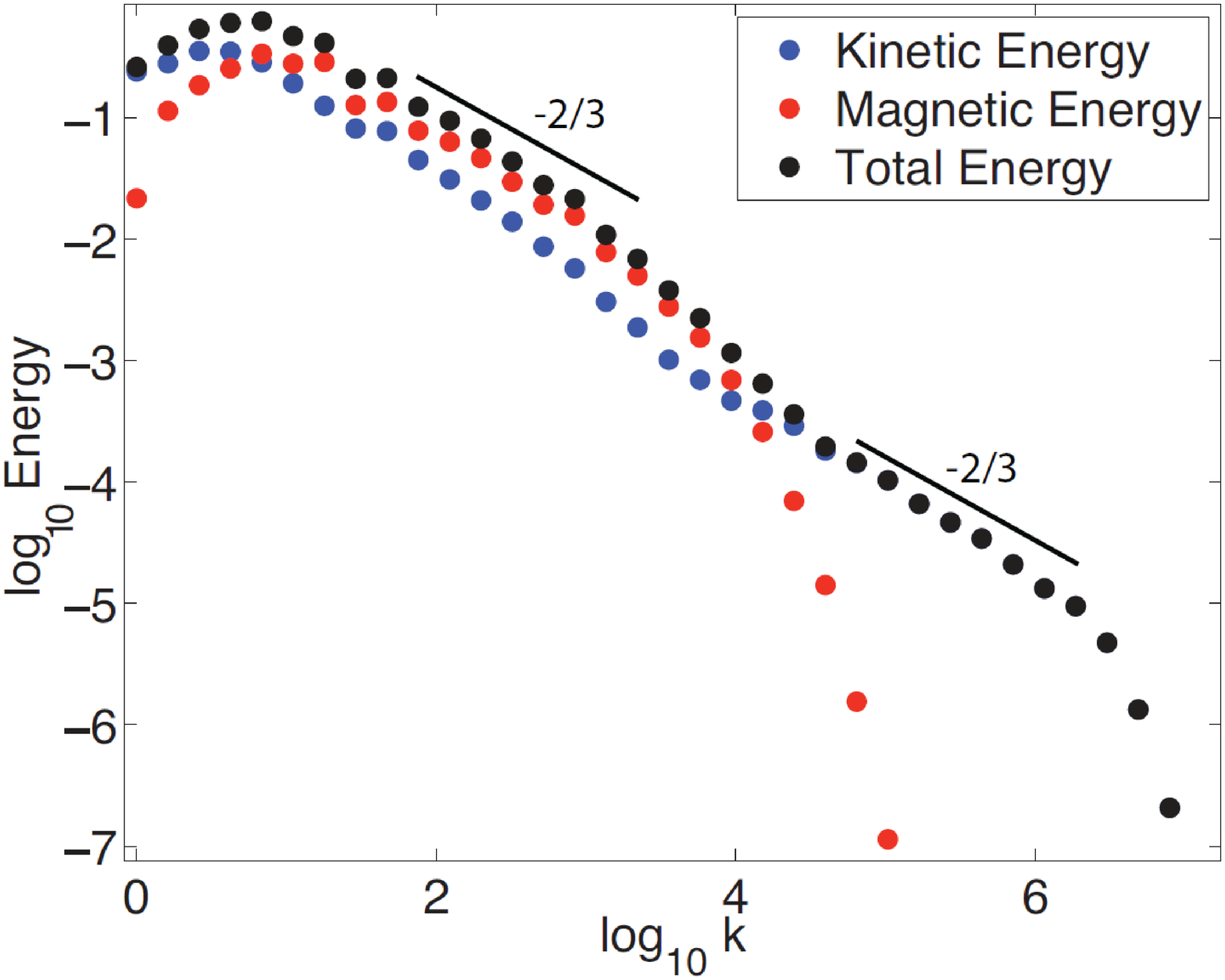}
\includegraphics[width=0.5\textwidth]{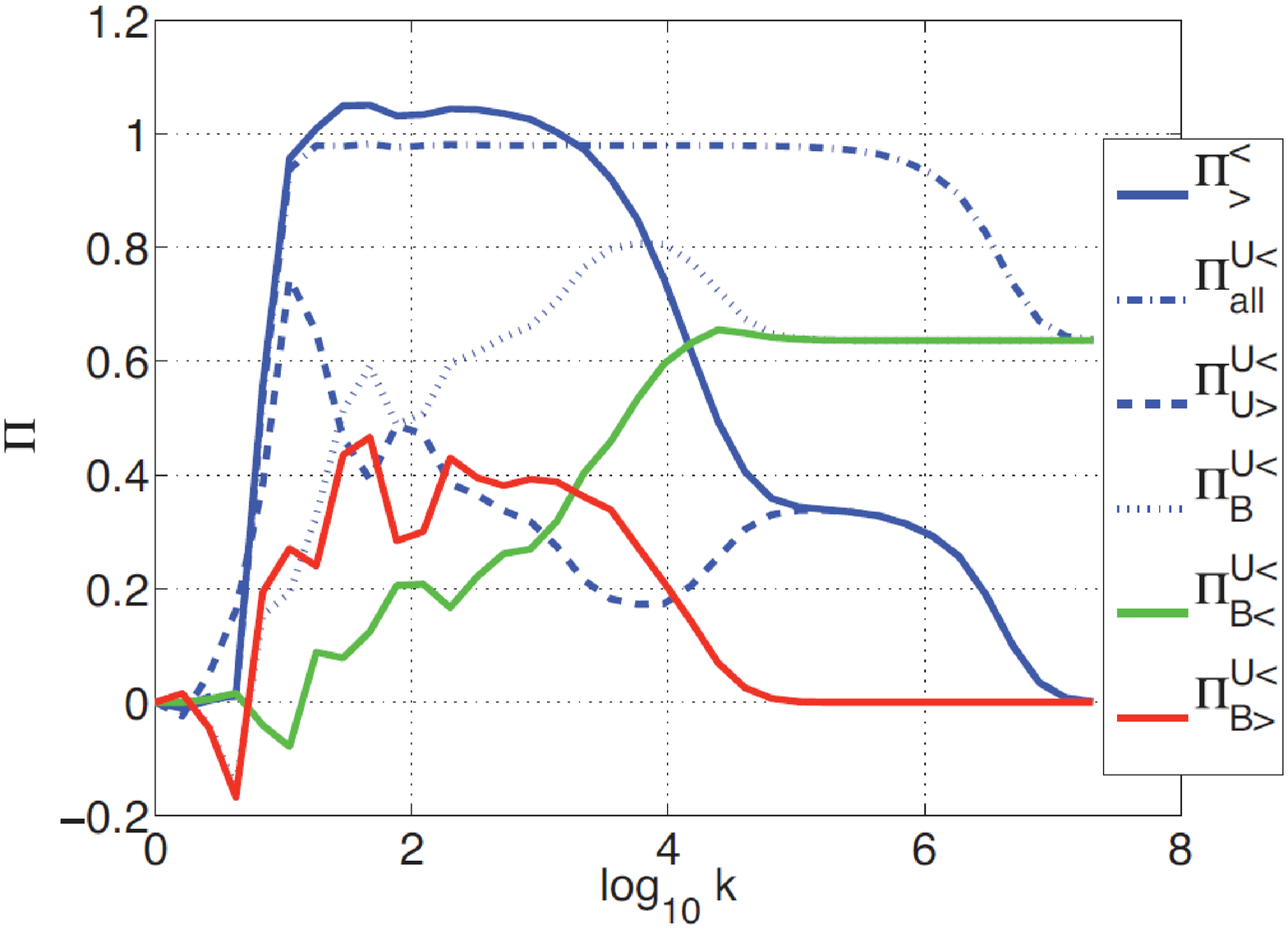}
\caption{Left panel: kinetic, magnetic and total energy spectra. Right panel:
energy fluxes versus $k$. Adapted from \citet{Lessinnes2009}.} \label{s4:Fluxes}
\end{figure}

Using again the MHD GOY model
\citet{Lessinnes2009} calculated a few of the fluxes introduced in Sec.~\ref{s3:fluxes} for the saturated regime. These are shown in Fig.~\ref{s4:Fluxes} for $\nu=10^{-9}$ and $\Pm=10^{-3}$, together with the
kinetic, magnetic and total energy spectra.
Rewriting Eqs.~(\ref{shell_NS32}-\ref{shell_induction32}) for $n> n_F$, in the form
\begin{eqnarray}
    d_t E_n^{U^{<}} + \Pi^{U^<}_{\text{all}} &=&  -  \nu \sum_{i=1}^{n} k_i^2U_i^2 + \epsilon, \label{flux_U}\\
    d_t E_n^{B^{<}} + \Pi^{B^<}_{\text{all}} &=&  -  \eta \sum_{i=1}^{n} k_i^2B_i^2, \label{flux_B}
\end{eqnarray}
we define
\begin{eqnarray}
\Pi^{U^<}_{\text{all}}&=&\Pi^{U^<}_{U^>}+\Pi^{U^<}_{B^<}+\Pi^{U^<}_{B^>},\\
\Pi^{B^<}_{\text{all}}&=&\Pi^{B^<}_{U^<}+\Pi^{B^<}_{U^>}+\Pi^{B^<}_{B^>},\\
\Pi^{<}_{>}&=&\Pi^{U^<}_{\text{all}}+\Pi^{B^<}_{\text{all}},\\
\Pi^{U^<}_{B}&=&\Pi^{U^<}_{B^<}+\Pi^{U^<}_{B^>}.
\end{eqnarray}
We stress again that fluxes are always $n$-dependent though not explicitly appearing in the above notation (see Sec.~\ref{s3:fluxes}).
The blue dot-dashed curve representing $\Pi^{U^<}_{\text{all}}$
in the right panel of Fig.~\ref{s4:Fluxes}
is constant in the range $k \in \left[k_F,k_{\nu}\right]$ with $\log_{10}k_F=1$ and $\log_{10}k_{\nu}\approx6$ where $\Pi^{U^<}_{\text{all}}=\epsilon \approx 0.98$.
Such a constant value necessarily leads to
a Kolmogorov spectral slope, as shown in the left panel of Fig.~\ref{s4:Fluxes}
\footnote{A -2/3 energy spectral slope corresponding to a -5/3 energy density spectral slope.}.
For scales smaller than the viscous scale ($k>k_{\nu}$), $\Pi^{U^<}_{\text{all}}$ drops by a value equal to the viscous dissipation $\epsilon_\nu\approx0.34$, leading to
$\Pi^{U^<}_{\text{all}}=\epsilon-\epsilon_\nu=\epsilon_{\eta}\approx 0.64$.
Note that the curves given by the blue-dashed and blue-dotted lines are symmetric
about a horizontal line for $k \in \left[k_F,k_{\nu}\right]$, implying that the relation $\Pi^{U^<}_{\text{all}}=\Pi^{U^<}_{U^>}+\Pi^{U^<}_{B}$ is indeed satisfied in this range of scales.
The blue-solid curve is above the blue-dot-dashed curve for $k \in [k_F,k_{\eta}]$,
with $\log_{10}k_{\eta} \approx 3.5$. This indicates that $\Pi^{B^<}_{\text{all}}>0$ for this range of scales.
Finally, the growth in the green-solid curve for $k \in [k_F,k_{\eta}]$
with $\Pi^{U^<}_{B^<}>0$ shows that dynamo action occurs at all scales in this range.

\begin{figure}[ht]
\centering
\includegraphics[width=0.5\textwidth]{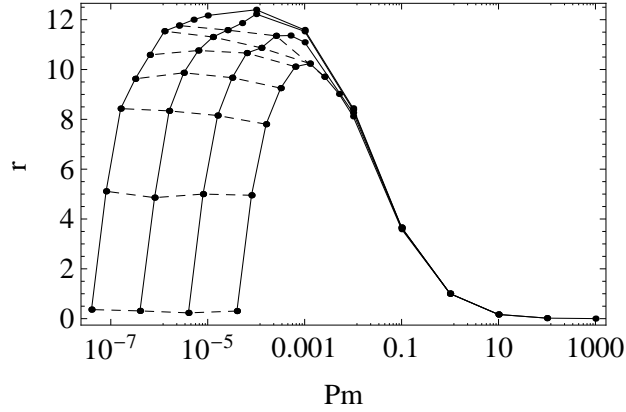}
\caption{Dissipation ratio versus $\Pm$.
The solid lines from right to left correspond to $\nu=10^{-5}$, $10^{-6}$, $10^{-7}$, $10^{-8}$. The dashed curves from bottom to top correspond to $\eta=1/4$, $1/8$, $1/16$, $1/32$, $1/64$, $1/128$, $1/256$. From \citet{Plunian2010}.}
\label{dissipation_ratio}
\end{figure}

As depicted in Fig.~\ref{s4:Fluxes} the power $\epsilon$ injected into the flow at the forcing scale is dissipated at smaller scales by Joule and viscous dissipation, according to
\begin{equation}
	\epsilon=\epsilon_{\eta}+\epsilon_{\nu}.
\end{equation}
On the other hand, there is no theoretical argument predicting how $\epsilon$ is distributed between $\epsilon_{\eta}$ and $\epsilon_{\nu}$.
\citet{Plunian2010} studied this problem for different values of $\Pm$, using
the helical shell model given by Eq.~(\ref{L1MHD}).
In Fig.~\ref{dissipation_ratio} the ratio $r=\epsilon_{\eta}/\epsilon_{\nu}$ versus $\Pm$ is plotted for different values of $\nu$ and $\eta$.
For a given value of $\nu$, in the limit $\Pm\rightarrow 0$, $\eta\rightarrow+\infty$ and dynamo action becomes impossible implying $r\rightarrow 0$.
For $\Pm = 1$ both kinetic and magnetic spectra are identical, implying $\epsilon_{\nu}=\epsilon_{\eta}=\epsilon/2$, and consequently $r=1$.
For a given $\nu$, there is always an intermediate value of $\Pm$ for which $r$ reaches a maximum.
This is related to a super-equipartition state in which the magnetic energy is higher than the kinetic energy at large scales.
For typical values of MHD turbulence, $\nu < 10^{-8}$ and $\Pm\approx 10^{-5}$, $r \ge 10$.

\subsection{Free-decaying MHD turbulence}
\label{s4:free decaying}
The decay properties of MHD turbulence have important applications in astrophysics. For example, the comparison of lifetime of a magnetic astrophysical object with
the decay time of its magnetic energy can help us decide whether the magnetic field is of primordial origin, or produced e.g. by dynamo action. In problems of this nature the magnetic helicity is known to play a crucial role.
The energy decay law satisfies $E(t)\propto t^{-1}$ or $E(t)\propto t^{-1/2}$,
depending on whether the magnetic helicity is respectively zero or maximal  \citep{Biskamp1999,Campanelli2004,Christensson2005}.
By solving the MHD GOY model given by Eqs.~(\ref{e3:goy_u}-\ref{e3:goy_b}), \citet{Antonov2001} and \citet{Antonov2001a}
found that cross helicity also plays a crucial role.

\begin{figure}[ht]
\includegraphics[width=0.478\textwidth]{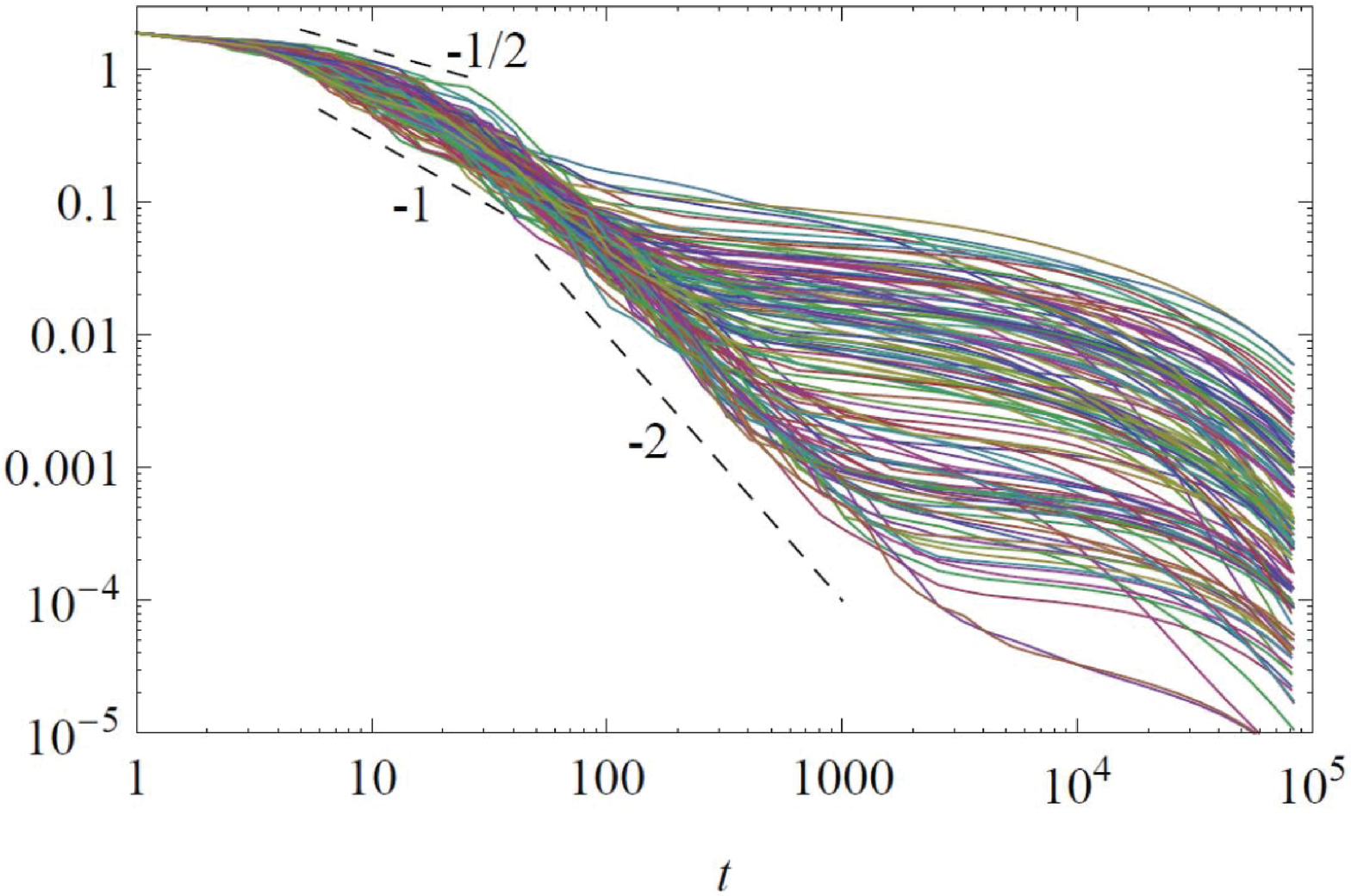}
\includegraphics[width=0.47\textwidth]{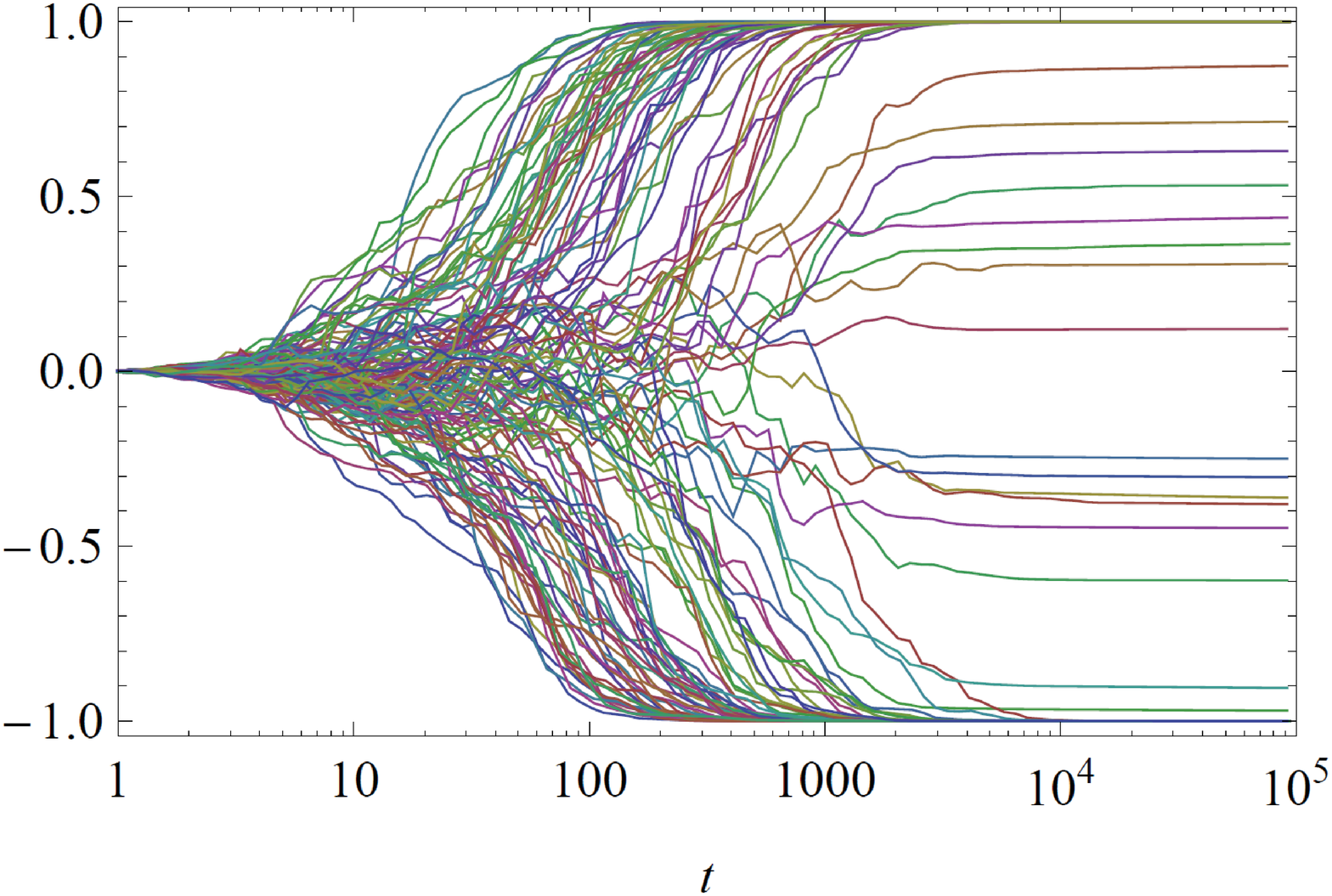}
\caption {Free decay of total energy $E$ (left panel) and normalized cross helicity $H^C/E$ (right panel) for $\Pm=1$. The 128 realizations differ by their initial conditions. From \citet{Frick2010}.}\label{f1:fig1}
\end{figure}

Using the helical shell model given by Eq.~(\ref{L1MHD}), \citet{Frick2010} explored the combined roles of cross helicity and magnetic helicity. Taking $\nu=\eta=10^{-5}$, they considered a set of 128 different initial conditions, with
both kinetic and magnetic energies concentrated in the first shell $n=0$, corresponding to $k_0=1$ ($E_0^U=E_0^B\approx 1$ and $U_{n>0}=B_{n>0}=0$). A very small amount of magnetic and cross helicities ($|H^B_0|< 10^{-4}$ and $|H^C_0|< 10^{-4}$) was injected.
In Fig.~\ref{f1:fig1} (left panel) the total energy versus time is plotted for the 128 initial conditions. At an early stage ($t<30$) we see that $E(t)\propto t^a$, with $a\in[-1,-1/2]$ (dashed lines), in good agreement with the DNS results and theoretical predictions. At a later time ($t>100$) many of the curves decrease with an exponential decay, in accord with pure dissipation.
In the right panel of Fig.~\ref{f1:fig1} the
ratio $H^C/E$ versus time is plotted. The curves on the left with an exponential decay correspond to the curves on the right with $H^C/E\rightarrow \pm1$. An $H^C/E \approx \pm1$ corresponds to $U_n\approx\pm B_n$, implying
again (see Sec.~\ref{s4:forcing}) a depletion of energy transfer in the inertial range and the accumulation of energy in the largest scales where the exponential decay occurs. Note that such a final state is reached at rather long times ($t>10^2$ to $10^3$), which may explain why this is not observed in DNS.
As shown in \citet{Frick2010}, cross helicity is generated at the dissipation scale, and then slowly cascades backwards towards the large scales.
This explains how a maximal cross helicity final state can be generated from a minimal cross helicity initial state ($|H^C_0/E_0|< 10^{-4}$).
Finally, in the left panel of Fig.~\ref{f1:fig1}, a few curves correspond to $E(t)\propto t^{-2}$ for $t>100$, with a limit
$H^C/E\ne \pm 1$.
The latter was observed previously by \citet{Brandenburg1996} and is presumably related to the concentration of magnetic helicity, rather than cross helicity, in the largest scales \citep{Frick2010}.

\begin{figure}[ht]\centering{
\includegraphics[width=0.49\textwidth]{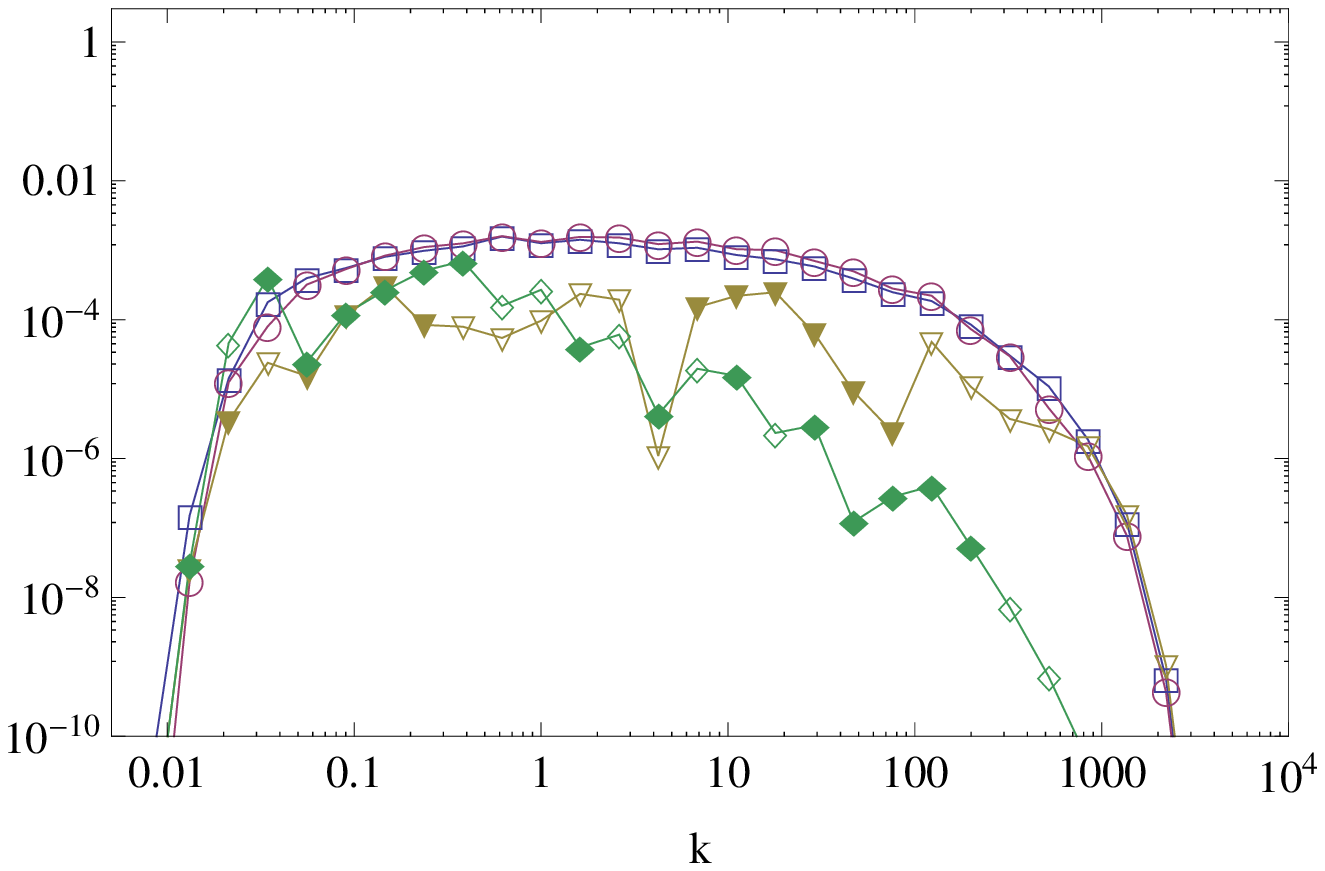}\raisebox{3cm}{\hspace{-0.6cm}(a)}
\includegraphics[width=0.49\textwidth]{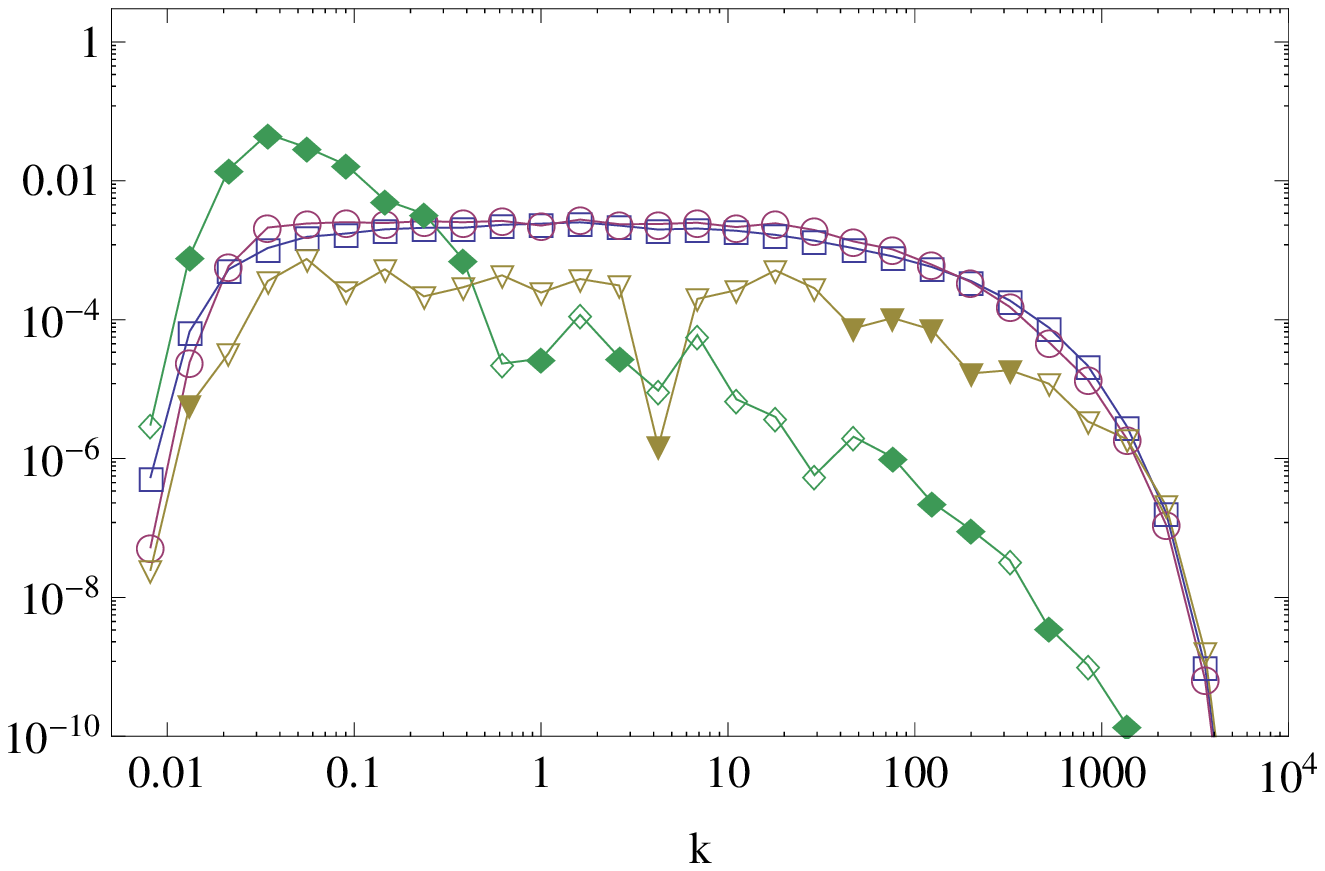}\raisebox{3cm}{\hspace{-0.6cm}(b)}
\includegraphics[width=0.49\textwidth]{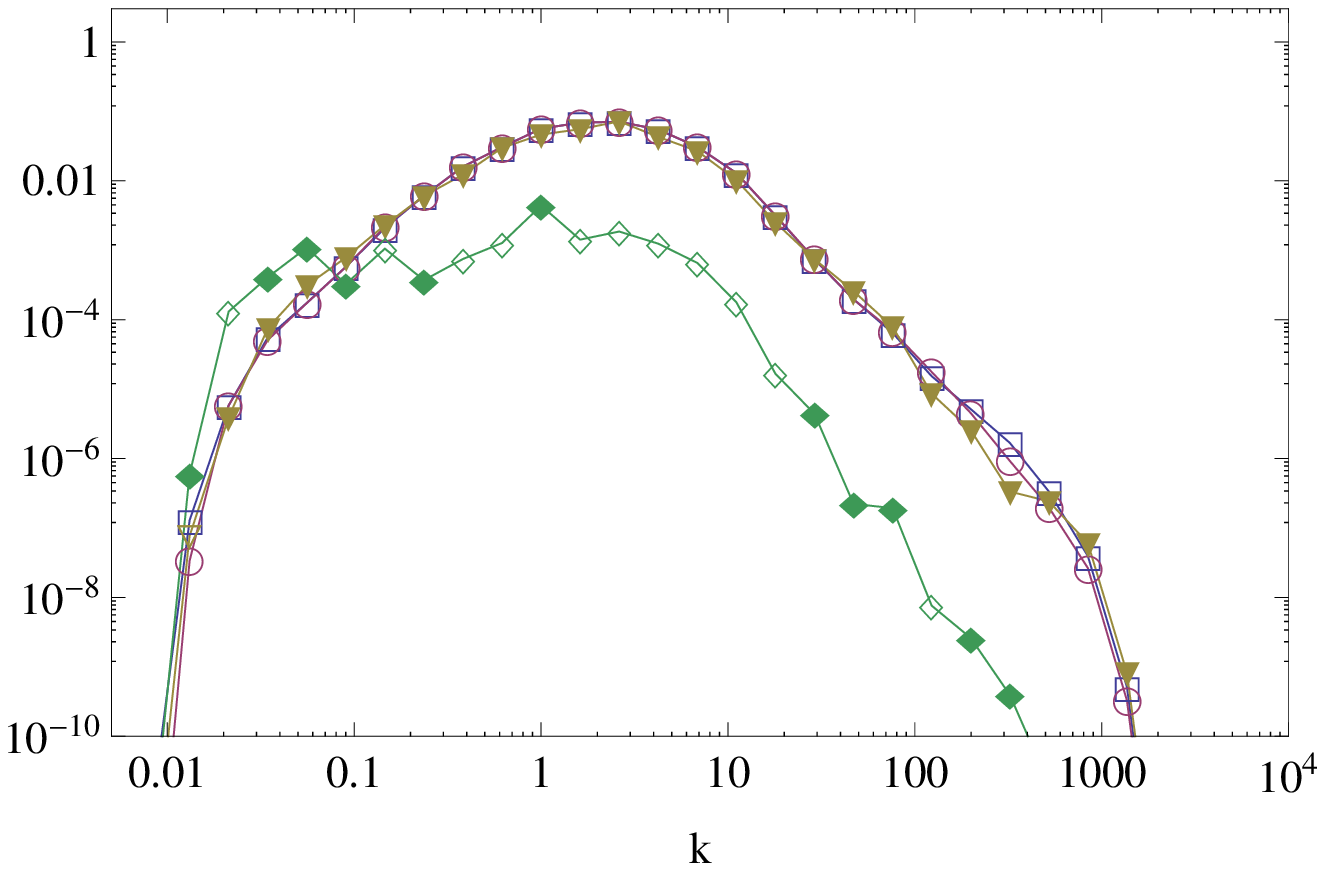}\raisebox{3cm}{\hspace{-0.6cm}(c)}
\includegraphics[width=0.49\textwidth]{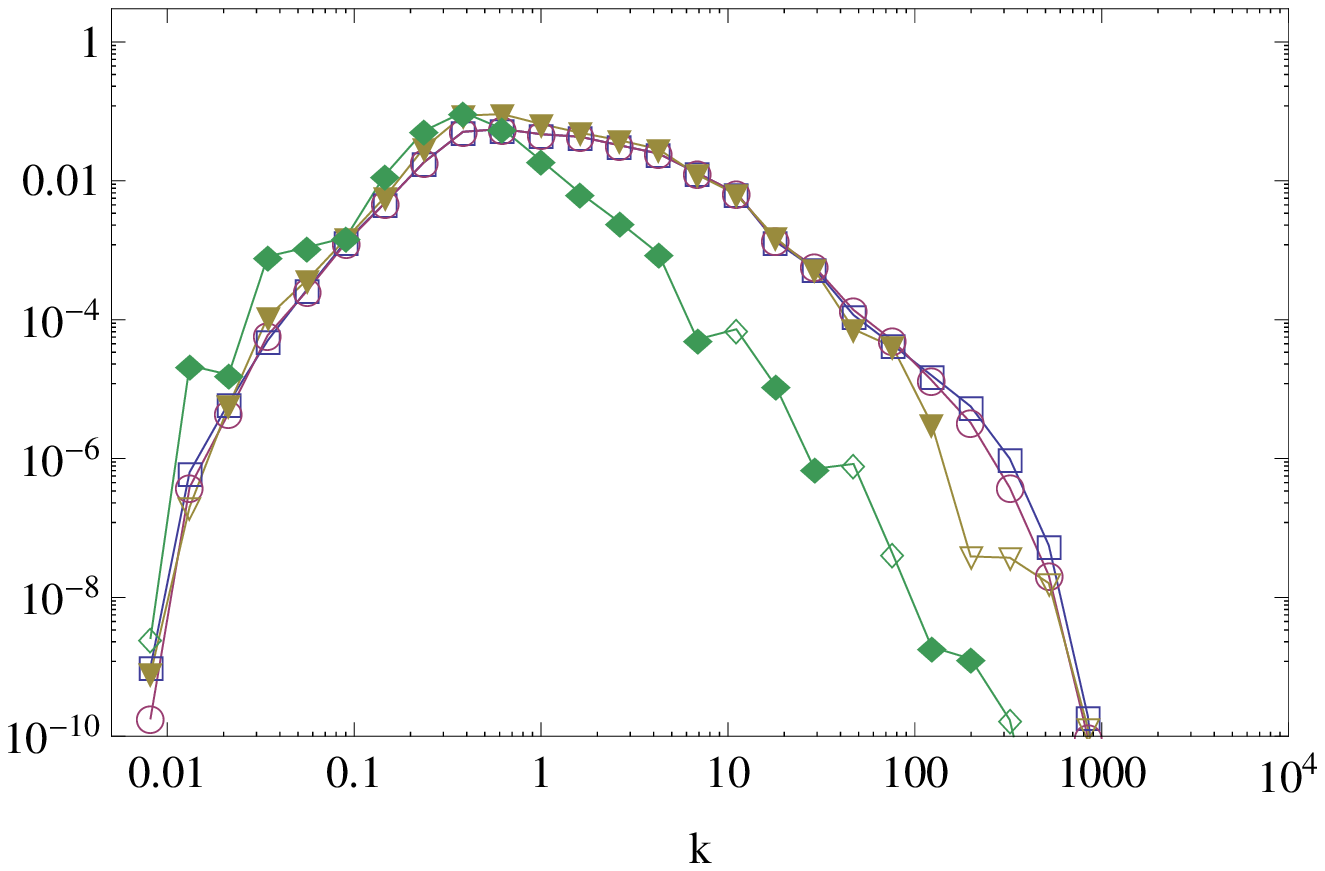}\raisebox{3cm}{\hspace{-0.6cm}(d)}}
\caption {Spectra of the kinetic energy (blue squares), the magnetic energy (magenta circles), cross helicity (brown triangles), magnetic helicity (green diamonds) obtained at time $t=1000$, from free-decaying MHD turbulence ($\nu=\eta=10^{-5}$) and normalized by $k^{-2/3}$. Each panel corresponds to different initial conditions with (a) $H^C_{t=0}=H^B_{t=0}=0$, (b) $H^C_{t=0}=0, H^B_{t=0}=1$, (c) $H^C_{t=0}=1, H^B_{t=0}=0$ (d) $H^C_{t=0}=H^B_{t=0}=1$. Filled (empty) symbols correspond to positive (negative) values of helicities.}
\label{fig02}
\end{figure}

To complete the study of free-decaying MHD turbulence we have calculated, using the same helical shell model given by Eq.~(\ref{L1MHD}),
the kinetic energy, magnetic energy, magnetic helicity and cross helicity spectra obtained at time $t=1000$, with four different combinations of initial conditions: maximal or zero for both cross and magnetic helicities. The results are presented in  Fig.~\ref{fig02} with
\begin{enumerate}[(a)]
	\item
$(H^C/E)_{t=0}=(H^B/E)_{t=0}=0$, the energy decays and helicities change their sign with $k$.
\item
$(H^C/E)_{t=0}=0$ and $(H^B/E)_{t=0}=1$, the magnetic energy cascade is accompanied by a simultaneous inverse cascade of magnetic helicity towards the largest scales. This leads to the accumulation of energy at the largest scale (the energy spectra are extended to the left).
\item
$(H^C/E)_{t=0}=1$ and $(H^B/E)_{t=0}=0$, the cross helicity
blocks the energy cascade, leading to steeper spectra.
\item
$(H^C/E)_{t=0}=(H^B/E)_{t=0}=1$, both effects are combined with essentially weak non-linear energy transfers.
\end{enumerate}
  

%% file: Applications.tex
\subsection{Multi-scale dynamos}
\label{s5}

Many MHD applications are primarily concerned with the properties of large-scale quantities, e.g. the chaotic reversals of the Earth's dipolar magnetic field, the dynamics of sunspots presumably due to underlying large-scale magnetic structures in the convection zone of the Sun and the zonal winds in Jupiter. Because of obvious numerical limitations it is not possible to solve the equations over all scales, ranging from the smallest dissipation scale to the size of the object itself. This is why we need a model for small scales (e.g. a shell model), so that only the large-scale equations remain to be solved (e.g. by DNS).
Such splitting, between large and small scales, makes it crucial to understand to what extent they are linked, namely how do
small-scale effects influence large-scale structures and vice-versa.
Beyond numerical issues, this problem is at the heart of the physical understanding of HD and MHD turbulence.

\subsubsection{Reynolds equations for MHD}
Both velocity and magnetic fields are split into large-scale and small-scale components
\begin{equation}
	\bu=\overline{\bu} + \bu', \qquad \bb=\overline{\bb} + \bb',
\end{equation}
e.g. after applying a large-scale filter.
On similarly splitting Eqs.~(\ref{MHDU}-\ref{MHDB}), we find that their large-scale component is
described by
\begin{eqnarray}
\left(\partial_t-\nu \nabla^2 \right)  \overline{\bu} &=&
- (\overline{\bu}\cdot \nabla) \overline{\bu} + (\overline{\bb} \cdot \nabla) \overline{\bb} - \overline{(\bu'\cdot \nabla) \bu'} + \overline{(\bb'\cdot \nabla) \bb'}
-\nabla \overline{p} + \overline{\bff}, \label{MHD<U>}\\
\left(\partial_t-\eta \nabla^2 \right) \overline{\bb} &=&
\nabla \times \left(\overline{\bu}\times \overline{\bb}\right) + \nabla \times \left(\overline{\bu'\times \bb'}\right),
\label{MHD<B>}
\end{eqnarray}
with, in addition, $\nabla \cdot \overline{\bu} =\nabla \cdot \overline{\bb}
=\nabla \cdot \bu' =\nabla \cdot \bb' =0$.
The effect of MHD turbulence on $\overline{\bu}$ is contained in the terms
$- \overline{(\bu'\cdot \nabla) \bu'} + \overline{(\bb'\cdot \nabla) \bb'}$, corresponding to the
\textit{Reynolds-Maxwell stress-tensor}
$R_{ij}=- \overline{u'_iu'_j-b'_ib'_j}$.
The effect of MHD turbulence on $\overline{\bb}$ is contained in the term
$\nabla \times \left(\overline{\bu'\times \bb'}\right)$, where $\overline{\bu'\times \bb'}$ is the so-called \textit{mean electromotive force} \citep{Steenbeck1966}.

\subsubsection{Reynolds-Maxwell stress-tensor}
\label{s5:transport coeff}
The first attempt to estimate the Reynolds-Maxwell stress-tensor using a shell model was made by  \citet{Geertsema1992}, in the context of a thin, differentially rotating, accretion disk.
A vectorial (instead of scalar) shell model was introduced.
Using a local system of cartesian coordinates ($x,y,z$) in the radial, azimuthal and vertical (normal to the disk plane) directions, they introduced three variables per shell for each field ,
$\bU_n=(U^x_n,U^y_n,U^z_n)$ and $\bB_n=(B^x_n,B^y_n,B^z_n)$.
The corresponding shell equations become
\begin{eqnarray}
\dot {\bf U}_n &=& \widetilde{\bW}_n({\bf a}, {\bf U},{\bf U})-\widetilde{\bW}_n({\bf a}, {\bf B},{\bf B}) - \nu k_n^2{\bf U}_n +
{\bf P}_n\left[\frac{3}{2}{\bf\Omega} {\bf U}^x_n {\bf e^y} - 2 {\bf\Omega}\times {\bf U}_n\right]  ,
\label{geer1}\\
\dot {\bf B}_n &=& \widetilde{\bW}_n({\bf a},{\bf U},{\bf B})-\widetilde{\bW}_n({\bf a}, {\bf B},{\bf U})  - \eta k_n^2 {\bf B}_n
- \frac{3}{2}{\bf\Omega} {\bf B}^x_n {\bf e^y}, \label{geer2}
\end{eqnarray}
where
\begin{equation}
\widetilde{\bW}_n({\bf a},{\bf X},{\bf Y})=k_n C_1\left[({\bf a}_n \cdot{\bf X}_{n - 1}){\bf Y}_{n-1}
- \lambda ({\bf a}_{n+1}\cdot{\bf X}_{n}){\bf Y}_{n+1} \right]+
k_n C_2 \left[({\bf a}_{n-1}\cdot{\bf  X}_{n}){\bf Y}_{n - 1} - \lambda( {\bf a}_{n}\cdot{\bf X}_{n+1}) {\bf Y}_{n +1}\right]
\end{equation}
is a vectorial generalization of model (\ref{gloagen}), and ${\bf a}_n$ is a set of arbitrary unit vectors.
In Eq.~(\ref{geer1}), ${\bf P}_n$ is an analog of the projection tensor
of Eqs.~(\ref{Fourieru}-\ref{Fourierb})
though here it applies to linear terms only \citep{Geertsema1992}. The Coriolis forces $- 2 {\bf\Omega}\times {\bf U}_n$ are also included. The terms $\frac{3}{2}{\bf\Omega} {\bf U}^x_n {\bf e^y}$ and $- \frac{3}{2}{\bf\Omega} {\bf B}^x_n {\bf e^y}$ are the contributions from Keplerian differential rotation.
Numerical solutions show that the turbulent shear-stress $\sum\limits_n \left(B_n^xB_n^y-U_n^xU_n^y\right)$ can supply the strong effective dissipation needed to explain the dynamics of  accretion disks as originally suggested by \citet{Shakura1973}.

\subsubsection{A subgrid shell model}
\begin{figure}[ht]
\centering
  \includegraphics[width=0.4\textwidth]{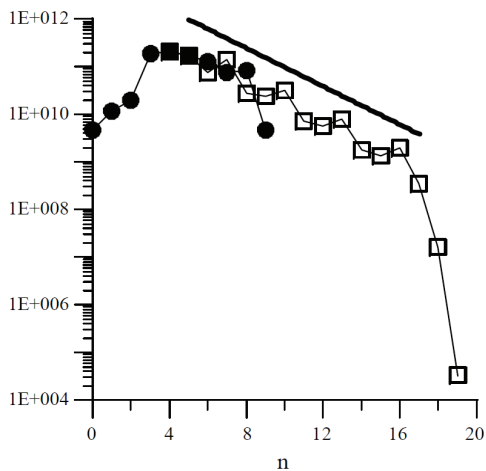}
  \caption{Second order structure function versus shell number, from DNS (filled circles) and shell model (empty squares). The straight line indicates a Kolmogorov slope of -2/3. The two common shells correspond to the filled squares. From \cite{Frick2002}.}
\label{fig:gridshell}
\end{figure}

A subgrid shell model was introduced by \citet{Frick2002} for HD convection in a rotating spherical layer heated from below, with application to the Earth's core in mind.
The large-scale flow $\overline{\bu}$ and temperature $\overline{\theta}$
satisfy
\begin{eqnarray}
\label{convection1}
\partial _t \overline{\bu} + (\overline{\bu}\cdot{\bf\nabla})\overline{\bu} &=&
 - {\bf\nabla} \overline{p} + (\nu+\nu_t) \nabla^2\overline{\bu}+ \overline{\bff},\\
\label{convection2}
\partial_t\overline{\theta}+\overline{\bu}\cdot\nabla(\overline{\theta}+\overline{\theta}_0)&=&(\kappa+\kappa_t) \nabla^2 \overline{\theta},
\end{eqnarray}
where $\overline{\bff}$ includes the Coriolis and Archimedean forces, and $\overline{\theta}_0$ is the  temperature profile prescribed throughout the layer. In addition, appropriate boundary conditions are applied.
The system of Eqs.~(\ref{convection1}-\ref{convection2}) is solved by DNS with the resolution given by the grid size.
The effect of subgrid turbulence on the scales larger than the grid size will be modeled by
the turbulent transport coefficients $\nu_t$ and $\kappa_t$.
For scales smaller than the grid size the GOY model given by Eq.~(\ref{GOY_epsilon}) for $\varepsilon=1-\lambda^{-1}$ (3D turbulence) is solved.
In order to provide the correct linkage between the DNS and the shell model, the DNS kinetic energy is calculated in different Fourier shells as introduced in Sec.~\ref{s2:transfers}, except that here the sequence of shells is geometric. This provides the velocity $U_n$ in the two first shells of the shell model, in which no other forcing is applied.
The dissipation rate $\epsilon$ in the shell model is the total energy dissipated per unit of time. It leads to numerical values for $\nu_t$ and $\kappa_t$, estimated as $\nu_t\approx \kappa_t \approx 0.1 (l_c^4 \epsilon)^{1/3}$, where $l_c$ is the mean scale corresponding to the two common shells between the DNS and the shell model.
In Fig.~\ref{fig:gridshell} the kinetic energy is plotted versus the shell number.
The two common shells are $n=4$ and $n=5$ indicated by black squares.
Such an example demonstrates the feasibility of using a shell model as a sub-grid model.
Combined with the previous example \citep{Geertsema1992} it could be used for the creation of a vectorial subgrid MHD shell model.

\subsubsection{The mean electromotive force}
The theory of astrophysical dynamos has been developed mainly using the
framework of the mean-field approach \citep{Krause1980,Radler2007a,Radler2007b}.
The latter yields various models
for large-scale magnetic fields of celestial bodies,
such as galaxies, stars and planets, reproducing to some
extent the available observations \citep{Ruzmaikin1988,Beck1996,Brandenburg2005}.
This approach has recently been challenged \citep{Vishniac2003,Cattaneo2009b} and its compatibility with small-scale dynamo action questioned \citep{Cattaneo2001}.
At the heart of the controversy is (i) how the mean electromotive force $\overline{\bu'\times \bb'}$ (the mean e.m.f.),
is estimated \citep{Courvoisier2010} and (ii) if it is at all relevant to the existence of large-scale magnetic fields.

Here we assume that the large-scale magnetic field is generated only by the
mean e.m.f. ($\overline{\bu}=0$). The large-scale magnetic field $\overline{\bb}$ is decomposed into its poloidal and toroidal parts, respectively related to the scalar quantities
$\bar{b}_P$ and $\bar{b}_T$  \citep{Moffatt1978}. The latter satisfies

\begin{eqnarray}
	\left(d_t+\eta k_L^2 \right) \bar{b}_P &=&
\left\{\nabla \times \left(\overline{\bu'\times \bb'}\right)\right\}_P,\label{emfP}\\
	\left(d_t+\eta k_L^2 \right) \bar{b}_T &=&
\left\{\nabla \times \left(\overline{\bu'\times \bb'}\right)\right\}_T,\label{emfT}
\end{eqnarray}
where $k_L$ is the wave number of the large-scale magnetic field,
and the terms on the right hand side (r.h.s.) are the scalar quantities related to the
poloidal and toroidal parts of the mean e.m.f.
The idea of estimating the r.h.s. terms using a shell model
started with \citet{Sokoloff2003}. The r.h.s. terms have been deduced either using some parametrization derived from the mean-field approach \citep{Sokoloff2003,Frick2006}, or directly from the shell model itself \citep{Nigro2011}.
As for the subgrid shell model, the essential point is to provide the correct linkage between the large-scale equations and the shell model.

In the mean-field approach, the terms on the r.h.s. of Eqs.~(\ref{emfP}-\ref{emfT}) are parametrized with the so-called $\alpha$
and $\beta$ effects, such that
\begin{equation}
	\left\{\nabla \times \left(\overline{\bu'\times \bb'}\right)\right\}_P
	\approx \i k_L \alpha \bar{b}_T - \beta k_L^2 \bar{b}_P, \qquad
	\left\{\nabla \times \left(\overline{\bu'\times \bb'}\right)\right\}_T
	\approx - \i k_L \alpha \bar{b}_P - \beta k_L^2 \bar{b}_T,
\end{equation}
where, using approximations, the $\alpha$ and $\beta$ parameters are estimated to be
\begin{equation}
\alpha=\alpha^u+\alpha^b, \quad \alpha^u= -\frac{\tau}{3}\left\langle \bu'\cdot\nabla\times\bu'\right\rangle, \quad \alpha^b\approx \frac{\tau}{3} \left\langle \bb'\cdot\nabla\times\bb'\right\rangle,
\quad
\beta \approx \frac{\tau}{3}\left\langle \bu'^2\right\rangle,
\label{alpha-beta3}
\end{equation}
$\tau$ being some characteristic time of turbulence.
The term $\alpha^u$, is the usual kinematic $\alpha$-effect and acts in favor of dynamo action. The term $\alpha^b$ is related to the feedback effect through the Lorentz forces \citep{Frisch1987} and acts against the dynamo. The term $\beta$ is the so-called \textit{turbulent magnetic diffusivity}.
With a shell model of L2-type (GOY or Sabra), and identifying $\tau$ with the turn-over time, these terms can be written in the form \citep{Frick2006}
\begin{equation}
\alpha^u = -\frac{1}{3}\sum_n (-1)^n |U_n|, \quad \alpha^b = \frac{1}{3}\sum_n (-1)^n |B_n|, \quad \beta = \frac{1}{3} \sum_n k_n^{-1} |U_n| .
\label{alpha-bet}
\end{equation}
\citet{Sokoloff2003} took $\alpha^b=0$, but included other feedback effects related to Alfv\'en waves.
In \citet{Nigro2011}, the terms on the r.h.s. of Eqs.~(\ref{emfP}-\ref{emfT}) are estimated
as $\i k_L\sum_n \left(U_n^*B_n-U_nB_n^*\right)$,
where $U_n$ and $B_n$ are obtained from a shell model.

Whatever the expression of the mean e.m.f. in terms of $U_n$ and $B_n$,
when considering the full model composed of Eqs.~(\ref{emfP}-\ref{emfT}) plus the shell model equations, at least the total energy should be conserved in the absence of viscosity and diffusivity ($\nu=\eta=0$).
This leads necessarily to additional terms in the shell model equations. For details we refer the interested reader to the previously mentioned papers.

\begin{figure}[ht]
\centering
  \includegraphics[width=\textwidth]{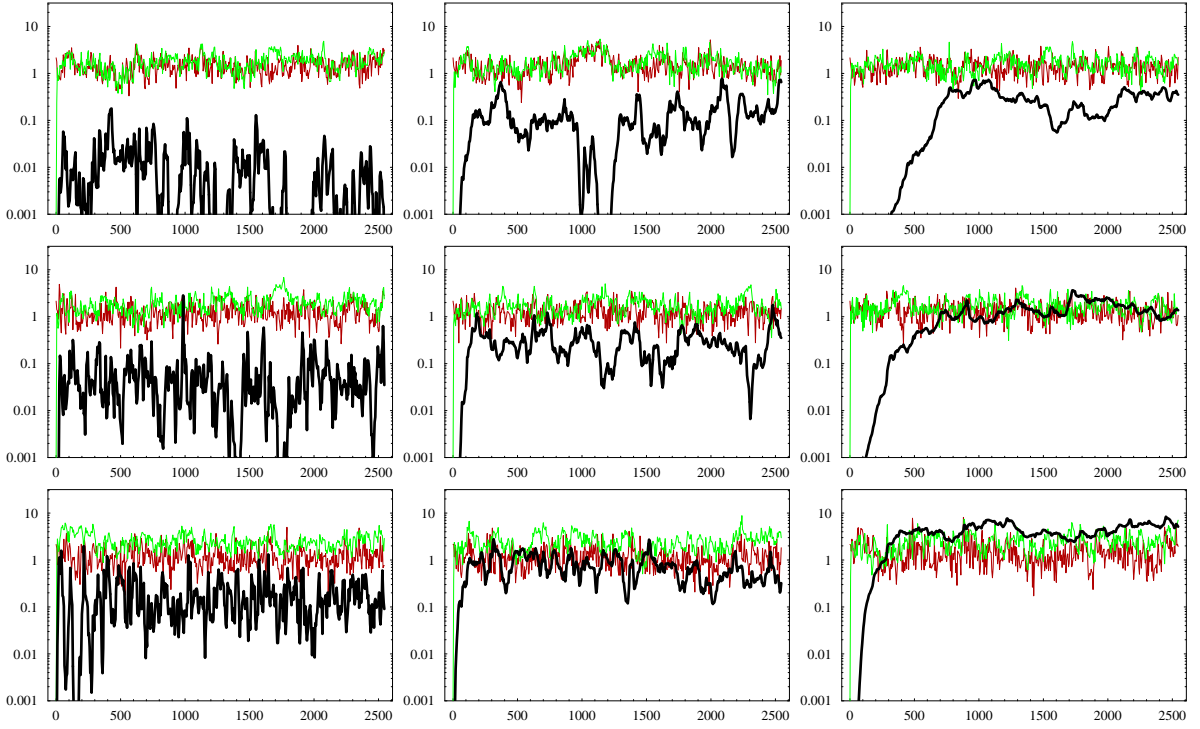}
  \caption{Time series for $E^{\bar{b}}$ (thick black curve),
  $E^B$ (green curve) and $E^U$ (red curve). From top to bottom, the injection rate of kinetic helicity is increased, $\zeta= 0; 0.04; 0.16$.
  From left to right, the scale separation is increased, $k_L = 1/2; 1/8; 1/32$.
  From \cite{Frick2006}. }
\label{evol3}
\end{figure}

In their shell model \cite{Frick2006} used a random forcing on two scales to control both the injection of energy and kinetic helicity. The rate of energy injection was constant $\epsilon=1$ and the
rate of kinetic helicity injection was varied $\zeta = 0; 0.04; 0.16$. In
Fig.~\ref{evol3} the kinetic and magnetic energies $E^U$ and $E^B$ are plotted versus time, together with the large-scale magnetic energy $E^{\bar{b}}=\frac{1}{2}(\bar{b}^2_P+\bar{b}^2_T)$. For $\nu=\eta=10^{-6}$ the results are given from top to bottom for increasing values of $\zeta$, and from left to right for decreasing values of $k_L$.

The first observation is that the time span corresponding to the kinematic growth is always roughly the same for small-scale magnetic energy (green curve), whereas for the
large-scale magnetic energy (black curve) it decreases with $k_L$ and $\zeta$. This is directly related to the growth rate of the large-scale magnetic energy which can be estimated as $\Gamma_{\rm kin}\approx\alpha k_L - (\beta + \eta)k_L^2$.
In the kinematic range, $\alpha\approx\alpha^{u} \propto \zeta$. Then for small $k_L$, $\Gamma_{\rm kin}\approx\zeta k_L$, implying that $\Gamma_{\rm kin}$ increases with $\zeta$ and $k_L$.
The second observation is that the level of saturation of the large-scale magnetic energy (black curve)
increases with both $\zeta$ and $k_L^{-1}$. This is related to (i) the definition
of $\alpha^u$ given by Eq.~(\ref{alpha-beta3}), which increases with $\zeta$ and (ii) to the magnetic dissipation
which is proportional to $k_L^2$.

Taking $\alpha^b=0$, \cite{Frick2006}
calculated the cross-correlation between
$\alpha^u$ and the large-scale magnetic energy $E^{\bar{b}}$ in the saturated state. They found a systematic delay between $\alpha^u$-quenching due to the growth of $E^{\bar{b}}$, and the growth of $E^{\bar{b}}$  due to $\alpha^u$. Such dynamics excludes any simple algebraic relation between $\alpha^u$ and $E^{\bar{b}}$, in contrast to what is usually assumed in the mean-field approach.

\begin{figure}[ht]
\begin{center}
  \includegraphics[width=.55\textwidth]{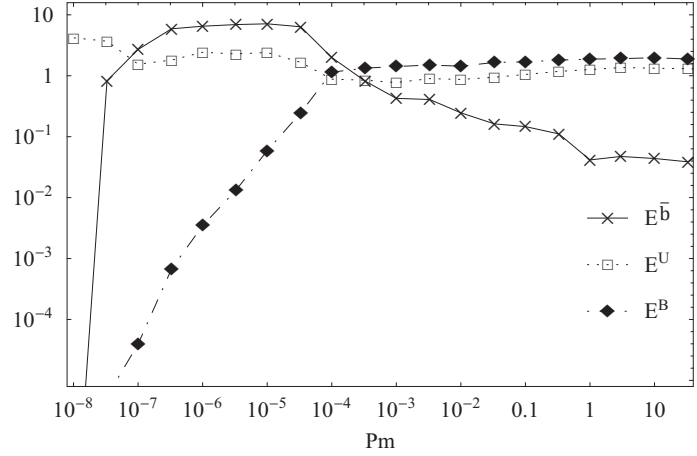}
  \end{center}
  \caption{$E^U$ (white squares), $E^B$  (black diamonds) and $E^{\bar{b}}$ (crosses) 
  versus $\Pm$, for  $\nu=10^{-6}$, $\zeta=0.08$ and $k_L=1/16$. From \cite{Frick2006}.}
  \label{prmE}
\end{figure}

In Fig.~\ref{prmE} the mean values of the three energies $E^U$, $E^B$ and $E^{\bar{b}}$ versus $\Pm$
are plotted for a given value of the viscosity $\nu=10^{-6}$.
For $\Pm \le 10^{-4}$, the small-scale dynamo works poorly and even stops for $\Pm \le 10^{-7}$ because the magnetic Reynolds number becomes too low.
Thus for $\Pm \ll 10^{-4}$, $|\alpha^b|\ll1$, implying that the large-scale magnetic field is generated by the $\alpha^u$-effect alone. Note that the level of large-scale magnetic energy is even greater than that of the kinetic energy. Now as $\nu$ is fixed, increasing $\Pm$ corresponds to raising $\Rm$ thus favoring the small-scale dynamo. If so then $|\alpha^b|$ also increases, implying quenching of the large-scale dynamo
as depicted by the negative slope of $E^{\bar{b}}$ versus $\Pm$ in Fig.~\ref{prmE}. Similar results were obtained by DNS \citep{Ponty2011}.
The former results for $\Pm < 10^{-4}$ would presumably drastically change
if $\Pm$ rather than $\nu$ was kept fixed. Indeed the reason why the small-scale dynamo shrinks for $\Pm\le 10^{-4}$ is that $\Rm$ becomes low.
Increasing $\Rm$, while keeping $\Pm$ fixed at a low value, might lead to a stronger small-scale dynamo
and to quenching of the large-scale dynamo due to an increase of $|\alpha^b|$.

\begin{figure}[ht]
\begin{center}
  \includegraphics[width=0.55\textwidth]{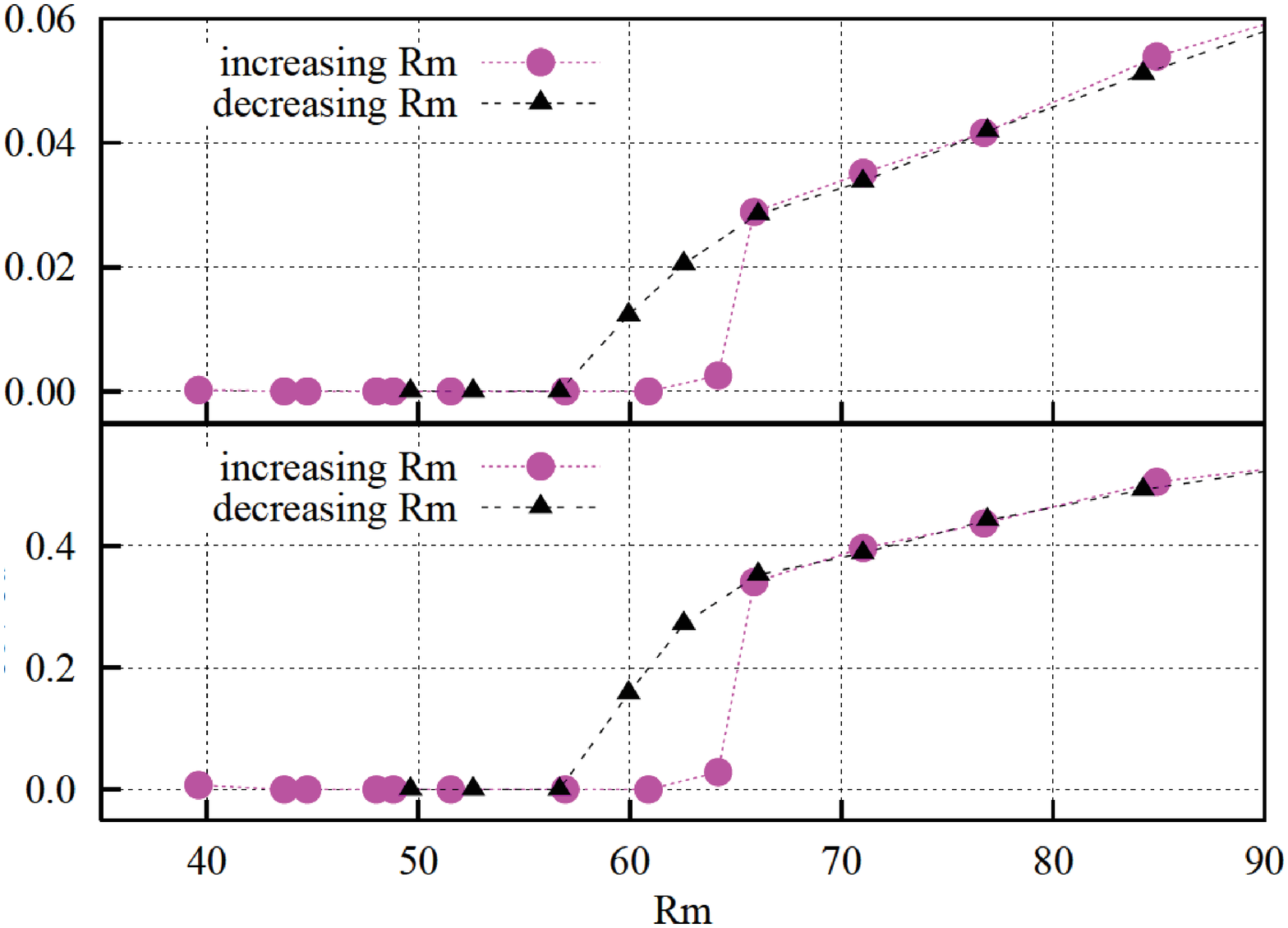}
  \end{center}
  \caption{Ratio $E^{\bar{b}}/E^U$ (top) and $E^B/E^U$ (bottom) versus $\Rm$
  for $\nu=10^{-5}$. Adapted from \cite{Nigro2011}.}
  \label{hysterisis}
\end{figure}

For a given viscosity $\nu=10^{-5}$, \citet{Nigro2011} studied the dynamics of their multi-scale dynamo introduced above. In particular, they found a hysteresis cycle represented in Fig.~\ref{hysterisis} for both ratios $E^{\bar{b}}/E^U$ and $E^B/E^U$ versus $\Rm$. On increasing $\Rm$, the dynamo starts for $\Rm\approx 65$. On decreasing $\Rm$,
the dynamo remains for $\Rm<60$. Such subcritical dynamos are also observed in DNS, e.g. for rotating convective dynamos \citep{Morin2009}.

\begin{figure}[ht]
  \centering
\includegraphics[width=0.7\textwidth]{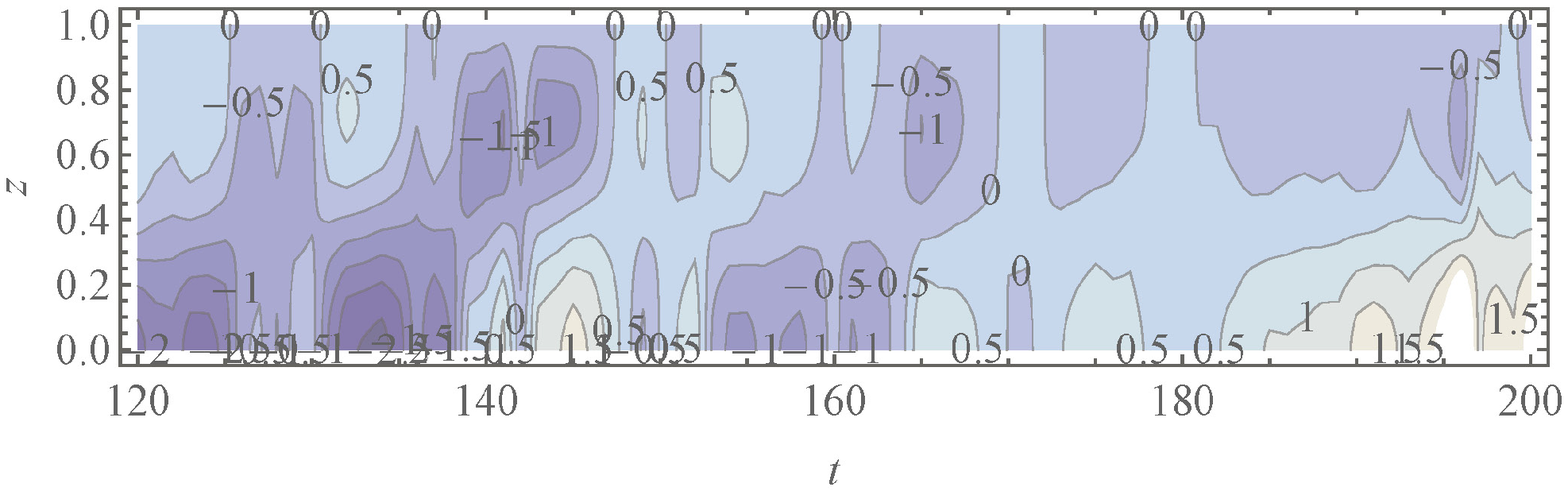}
\includegraphics[width=0.7\textwidth]{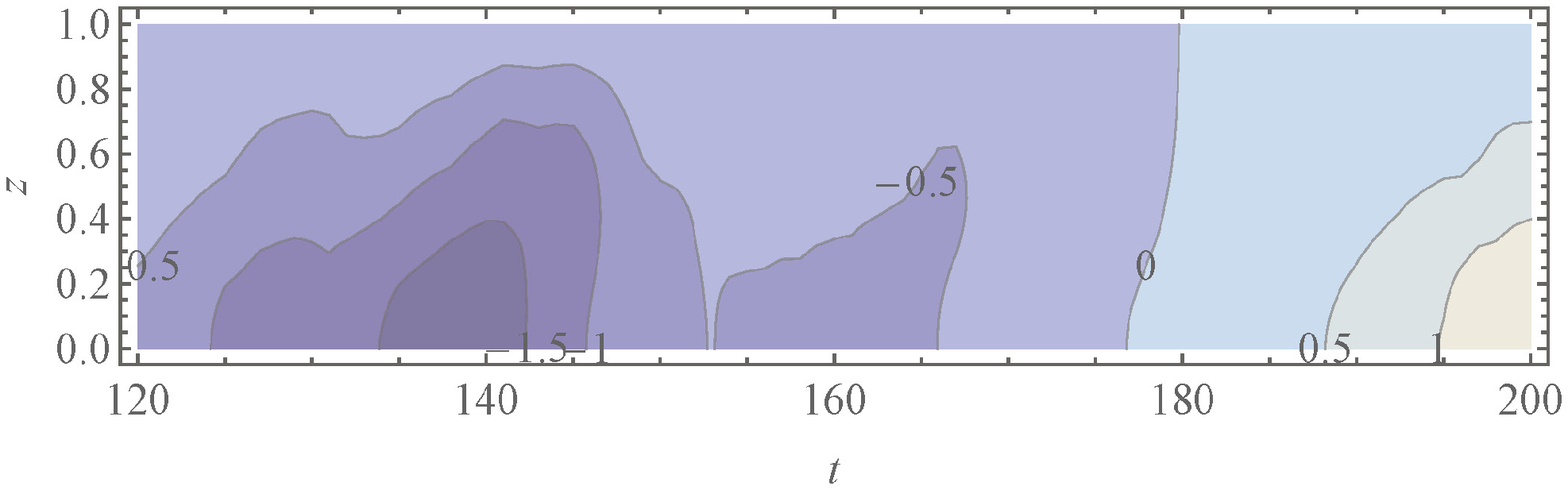}
\caption{Isolines of poloidal (top) and toroidal (bottom) components of the large-scale magnetic field versus $z$ (vertical axis) and time (horizontal axis). From \citet{Stepanov2008a}.}
\label{figdisk}
\end{figure}

\citet{Stepanov2006}, with galactic disks in mind, elaborated a multi-scale dynamo model
in which the toroidal part of the large-scale magnetic field is generated by differential rotation rather than by means of mean e.m.f.. In the mean-field terminology the dynamo is denoted as an $\alpha \omega$-dynamo, in contrast to the previous models denoted as $\alpha^2$-dynamos. The large-scale magnetic field equations are
 \begin{eqnarray}\label{mfa}
 \partial_t \bar{b}_P &=& \alpha k_L \bar{b}_T + \beta k^2_L
 \partial^2_{z} \bar{b}_P,\\ \label{mfb}
 \partial_t \bar{b}_T &=& - G  \partial_z \bar{b}_P + \beta k^2_L
 \partial^2_{z} \bar{b}_T \label{mean}
 \end{eqnarray}
 where $\bar{b}_P$ and $\bar{b}_T$ not only depend on time but also on the coordinate $z$ perpendicular to the
 plane of the galactic disk, with $-1 \le z \le 1$. Appropriate boundary conditions in $z$ are applied to
 $\bar{b}_P$ and $\bar{b}_T$ and their first $z$-derivatives.
 The parameter $G$ is  related to the intensity of the differential rotation.
 The parameter $\alpha$ also depends on $z$ as $\alpha=(\alpha^u+\alpha^b)\sin(\pi z)$.
The linkage between the shell model and the large-scale model is similar to that in \citet{Frick2006}.
In contrast to the previous models this model has to be integrated in $z$.
In Fig.\ref{figdisk} the $z$-profile of the poloidal and toroidal magnetic field components are plotted versus time. The poloidal component exhibits a much stronger dependence on $z$ (smaller scales) and varies faster than the toroidal component. The toroidal component exhibits one reversal.

\subsubsection{Reversals of large-scale magnetic field}
The system of Eqs.(\ref{MHDU}-\ref{MHDB})
is invariant on changing $\bb$ to $-\bb$, allowing for opposite magnetic field polarities
with the same velocity field $\bu$. Magnetic reversals are indeed observed, e.g. the Earth geomagnetic dipole fluctuates and reverses chaotically \citep{Merrill1983}, while the solar magnetic field reverses with a rather stable periodicity \citep{Weiss2009}.
The experimental results of \citet{Ravelet2008} showing chaotic magnetic reversals have recently
emphasized the importance of small-scale turbulence in triggering the reversals of the large-scale magnetic field \citep{Petrelis2008,Petrelis2009}. It is then rather natural to expect chaotic reversals in multi-scale shell models. This is visible in Fig.~\ref{evol3} where each time the black curve, corresponding to $E^{\bar{b}}$, crosses the horizontal axis, $\bar{b}$ is changed to $-\bar{b}$.

\citet{Ryan2007,Ryan2011} considered a large-scale $\alpha\omega$-dynamo model in which  $\alpha$ is calculated from the 3D HD GOY model (no small-scale magnetic field, no back-reaction from the large-scale magnetic field onto the turbulence). They find that the reversals fit a log-normal distribution as do the paleomagnetic data, stressing the importance of multiplicative noise in the underlying geodynamo.

\begin{figure}[ht]
  \centering
\includegraphics[width=0.98\textwidth]{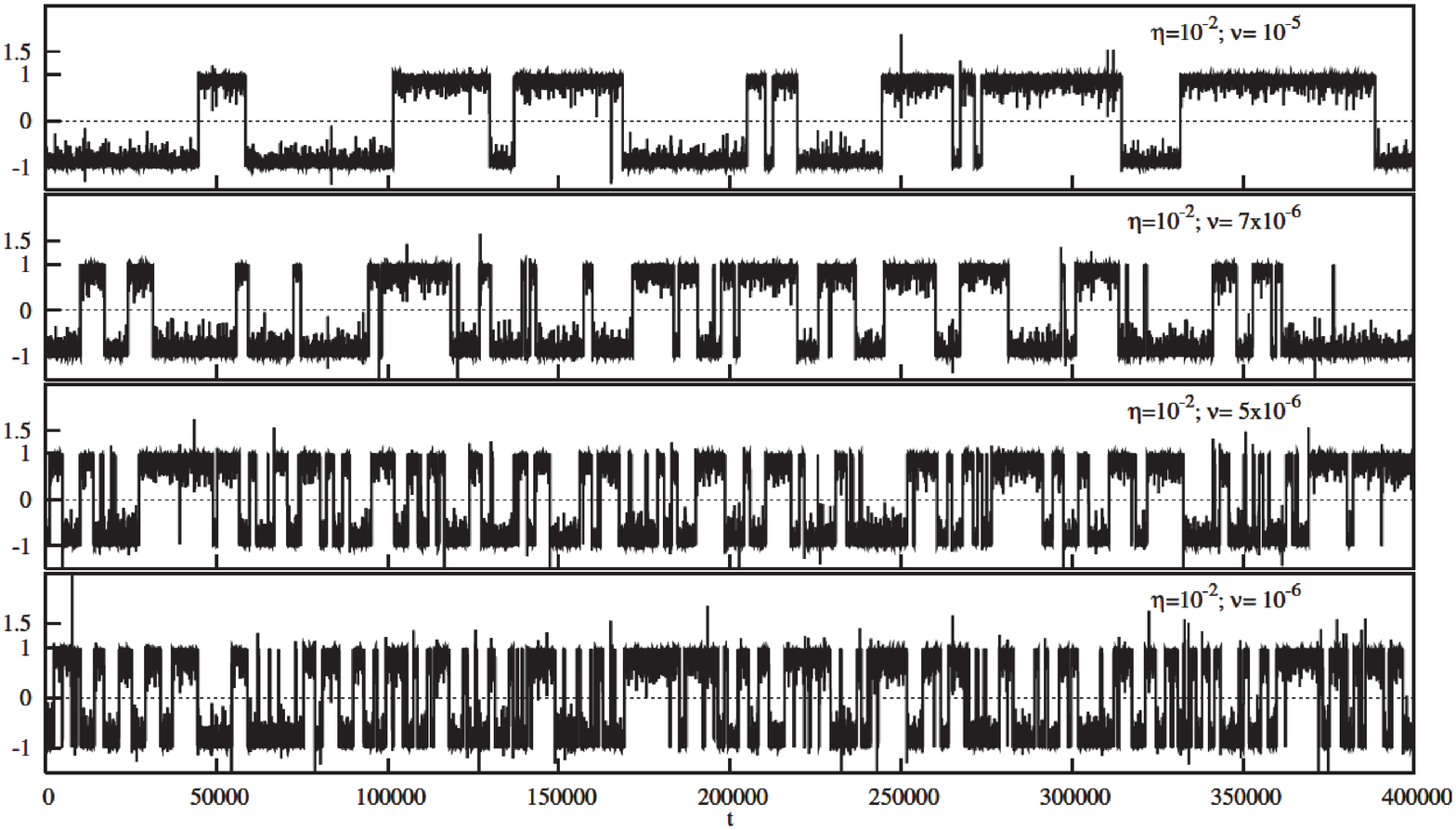}
\caption{Large-scale magnetic field versus time. From top to bottom the viscosity is decreased.
Adapted from \citet{Nigro2010}.}
\label{fig:reversal}
\end{figure}

Considering the MHD Sabra model given by Eqs.~(\ref{e3:sabraL2_u}-\ref{e3:sabraL2_b}),
 \citet{Benzi2010a} and \citet{Nigro2010}
added a term to the magnetic field equation for one of the largest shells, say $n=1$,
such that the new equation for $B_1$ becomes
\begin{equation}
d_t B_1 =  \tilde{W}_1(U,B)-\tilde{W}_1(B,U) - M(B_1).
\end{equation}
They chose
$M(B_1)\propto B_1$ or $M(B_1)\propto B_1^3$ \citep{Benzi2010a}, or a combination of both \citep{Nigro2010}.
Of course energy conservation is thus violated, but this was the simplest way of breaking the symmetry of the system and have reversals for $B_1$ with periods of constant sign. They found that the mean period between reversals increases with the magnetic diffusivity \citep{Benzi2010a} and the viscosity \citep{Nigro2010}.
In Fig.~\ref{fig:reversal} an example is given for different values of viscosity.
From top to bottom the viscosity is decreased by a factor 10, while the mean period between
two reversals clearly becomes shorter.

\subsection{Alfv\'en waves and rotation}
\label{s6:rotalfven}
Assuming that MHD turbulence is embedded in a homogeneous magnetic field $\bb_0$, and in a rotating frame with angular velocity $\bOm$, Eqs.~(\ref{MHDU}-\ref{MHDB}) are changed into
\begin{eqnarray}
\left(\partial_t-\nu \nabla^2 \right)  \bu &=&- (\bu\cdot \nabla) \bu + (\bb \cdot \nabla) \bb + (\bb_0 \cdot \nabla) \bb -2 \bOm \times \bu -\nabla p + \bff, \qquad \; \nabla \cdot \bu = 0,  \label{AlfvenU}\\
\left(\partial_t-\eta \nabla^2 \right) \bb &=&- (\bu\cdot \nabla) \bb + (\bb \cdot \nabla)  \bu + (\bb_0 \cdot \nabla)  \bu,    \qquad \qquad \qquad \qquad \qquad\nabla \cdot \bb =0,
\label{AlfvenB}
\end{eqnarray}
As explained in Sec.~\ref{s2:MHDequations}, Alfv\'en waves may propagate along $\bb_0$ in both directions provided that dissipation is not too strong. This latter statement is actually what makes experimental evidence difficult to obtain, especially in a liquid metal \citep{Alboussiere2011}.
For a sufficiently strong $\bb_0$ or $\bOm$ the MHD turbulence becomes anisotropic, making the use of isotropic 3D shell models questionable. In order to account for such anisotropy, other shell models have been elaborated as presented below.
Mainly three lines of research have been followed so far.
\begin{itemize}
	\item
	Two shell models of Alf\'enic turbulence have been elaborated by \citet{Carbone1989,Carbone1990} within an isotropic
	and an anisotropic framework. In the latter case they introduced an angular dependence with respect to the direction of anisotropy (the direction of $\bb_0$). The model coefficients were estimated using a standard statistical approach (Direct interaction approximation, Markovian approximation and random phase hypothesis). The 
interested reader should refer to the above mentioned papers.
	\item
Another line of development (Sec.~\ref{s6:rotation}) consists in
keeping an isotropic MHD shell model and simply adding
a term corresponding to $\bb_0$ \citep{Biskamp1994,Hattori2001}, or $\bOm$ for HD turbulence \citep{Hattori2004,Chakraborty2010a}, or both \citep{Plunian2010}. Such a model is, of course, well suited to isotropic phenomenology like IK for Alf\'en waves.
\item
The usual way of dealing in the presence of an applied magnetic field $\bb_0$ is to write the MHD equations in the plane perpendicular to the direction of $\bb_0$, leading to the so-called \textit{reduced} MHD equations (RMHD). The first RMHD shell model was developed in the context of intermittent heating in solar coronal loops \citep{Nigro2004}. Its statistical properties compare successfully with observation
\citep{Nigro2005,Buchlin2005,Buchlin2007,Buchlin2007a}. In the context of the Alfv\'enic solar wind the RMHD shell model gives a good description of the transition between weak and strong MHD turbulence \citep{Verdini2012b}.
It also provides a mechanism for the presence of the low frequency magnetic spectrum observed inside the sub-Alfv\'enic solar wind \citep{Verdini2009,Verdini2012a}.
The model and a few results are summarized in Sec.~\ref{S4:Alfven}.
\end{itemize}

\subsubsection{Isotropic shell models}
\label{s6:rotation}
A shell model with an externally applied magnetic field and global rotation of intensities $b_0$ and $\Omega$, can be written in the form
\begin{eqnarray}
   d_t U_n &=& \widetilde{W}_n(\bU,\bU)-\widetilde{W}_n(\bB,\bB) + \i k_n b_0 B_n + \i \Omega U_n -   \nu k_n^2U_n + F_n, \label{AlfvenNS}\\
   d_t B_n &=& \widetilde{W}_n(\bU,\bB)-\widetilde{W}_n(\bB,\bU) + \i k_n b_0 U_n -   \eta k_n^2B_n.   \label{Alfveninduction}
\end{eqnarray}
Obviously, taking $b_0\ne 0$ and $\Omega \ne 0$ implies the failure of magnetic and cross helicity conservation respectively, in agreement with the original equations.
In addition to the eddy turn-over time $t_{NL}=l/ u_l$,
the introduction of
$b_0$ and $\Omega$ leads to two other time-scales, $t_A=l/b_0$ and $t_{\Omega}=\Omega^{-1}$, over which energy transfers may occur.
Depending on which is smallest, $t_{NL}$, $t_A$ or $t_{\Omega}$, one expects different types of turbulence characterized by different slopes for the energy density spectrum within the inertial range, respectively
$E(k)\propto k^{-5/3}$ (type K), $E(k)\propto k^{-3/2}$ (type A) or $E(k)\propto k^{-2}$ (type R). As $t_{NL}$ and $t_A$
depend on scale $l$, an inertial range with several slopes generally occurs.

\begin{figure}[ht]
\begin{center}
\includegraphics[width=0.45\textwidth]{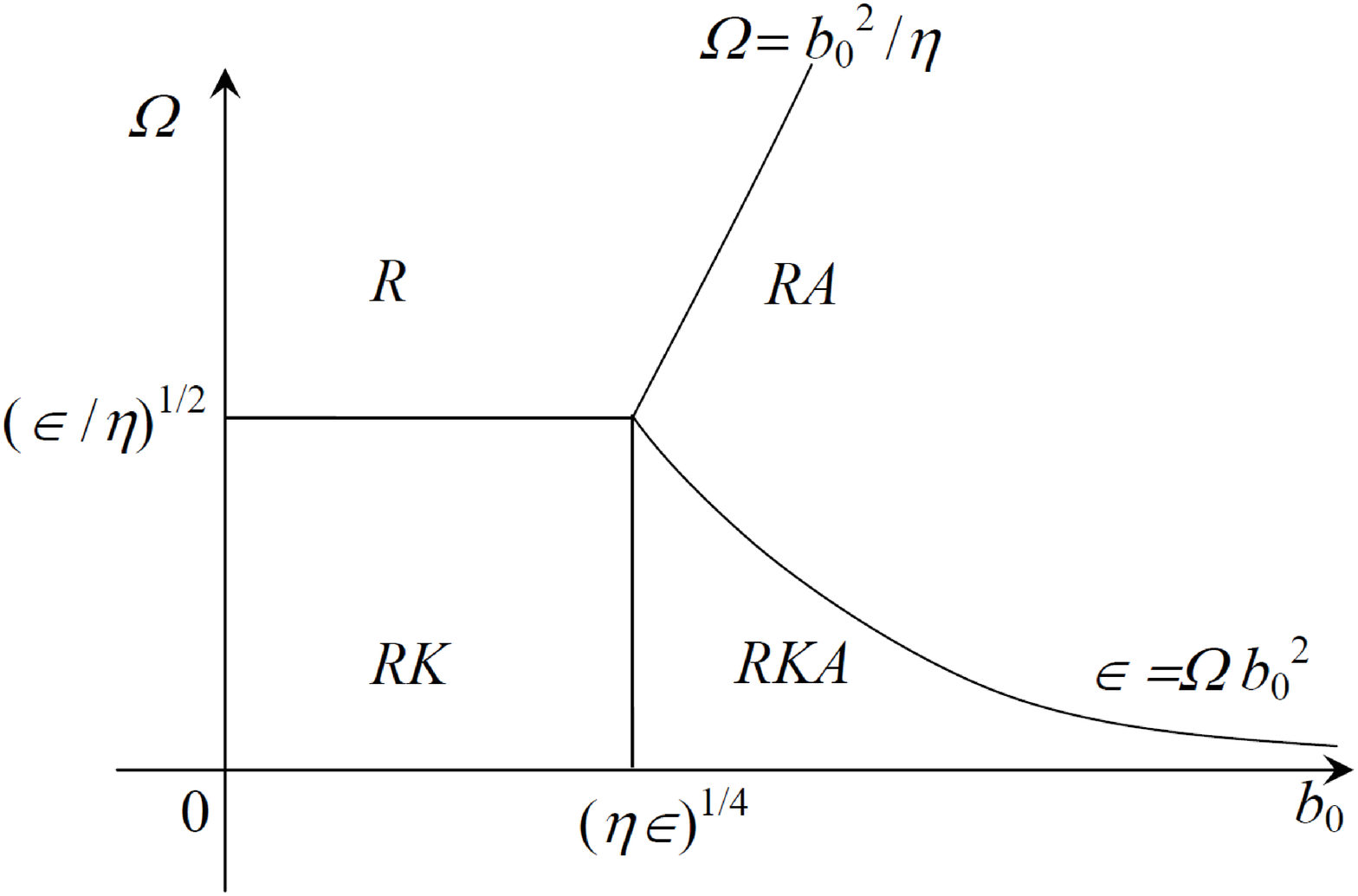}
\includegraphics[width=0.45\textwidth]{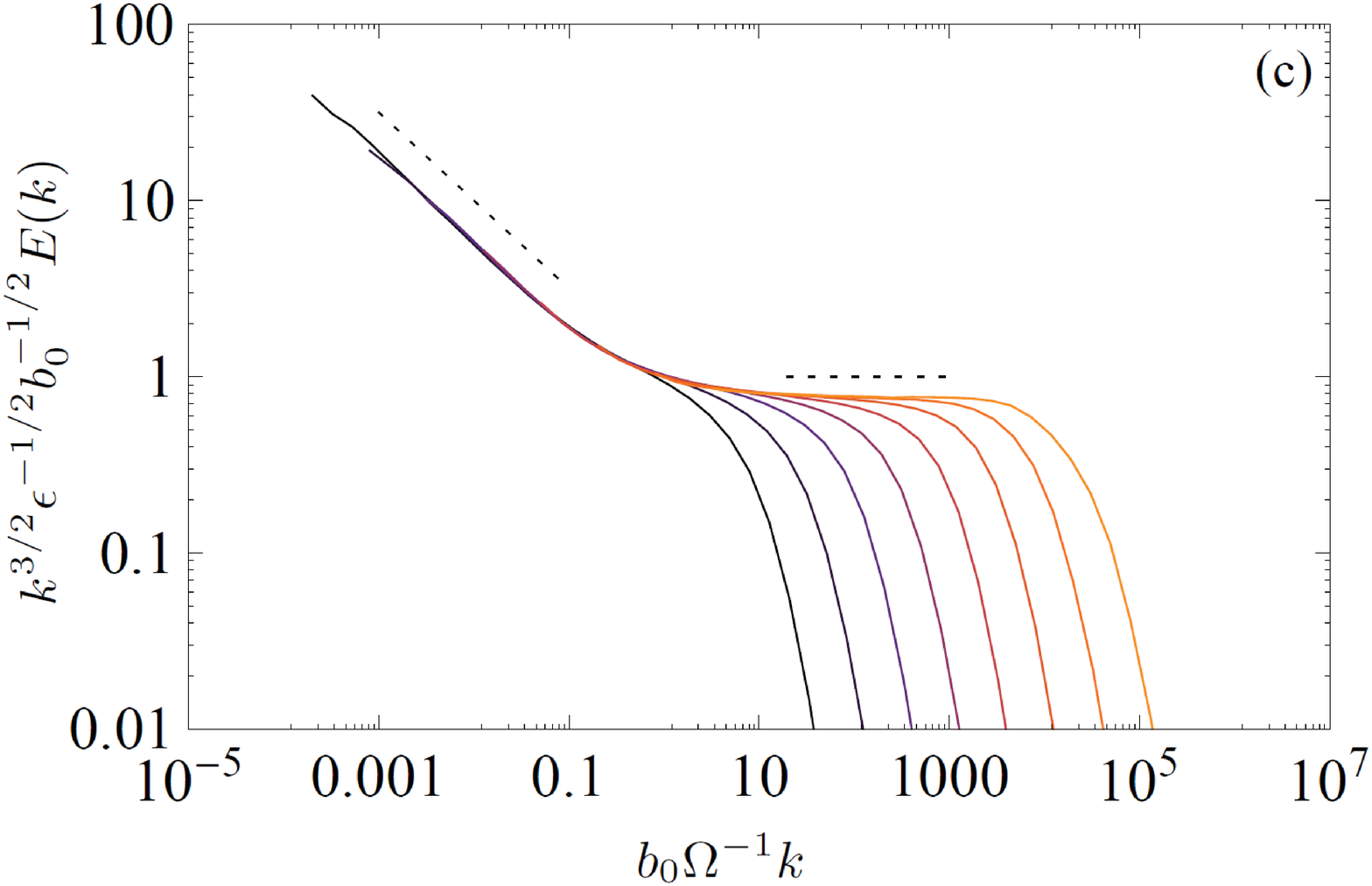}
\includegraphics[width=0.45\textwidth]{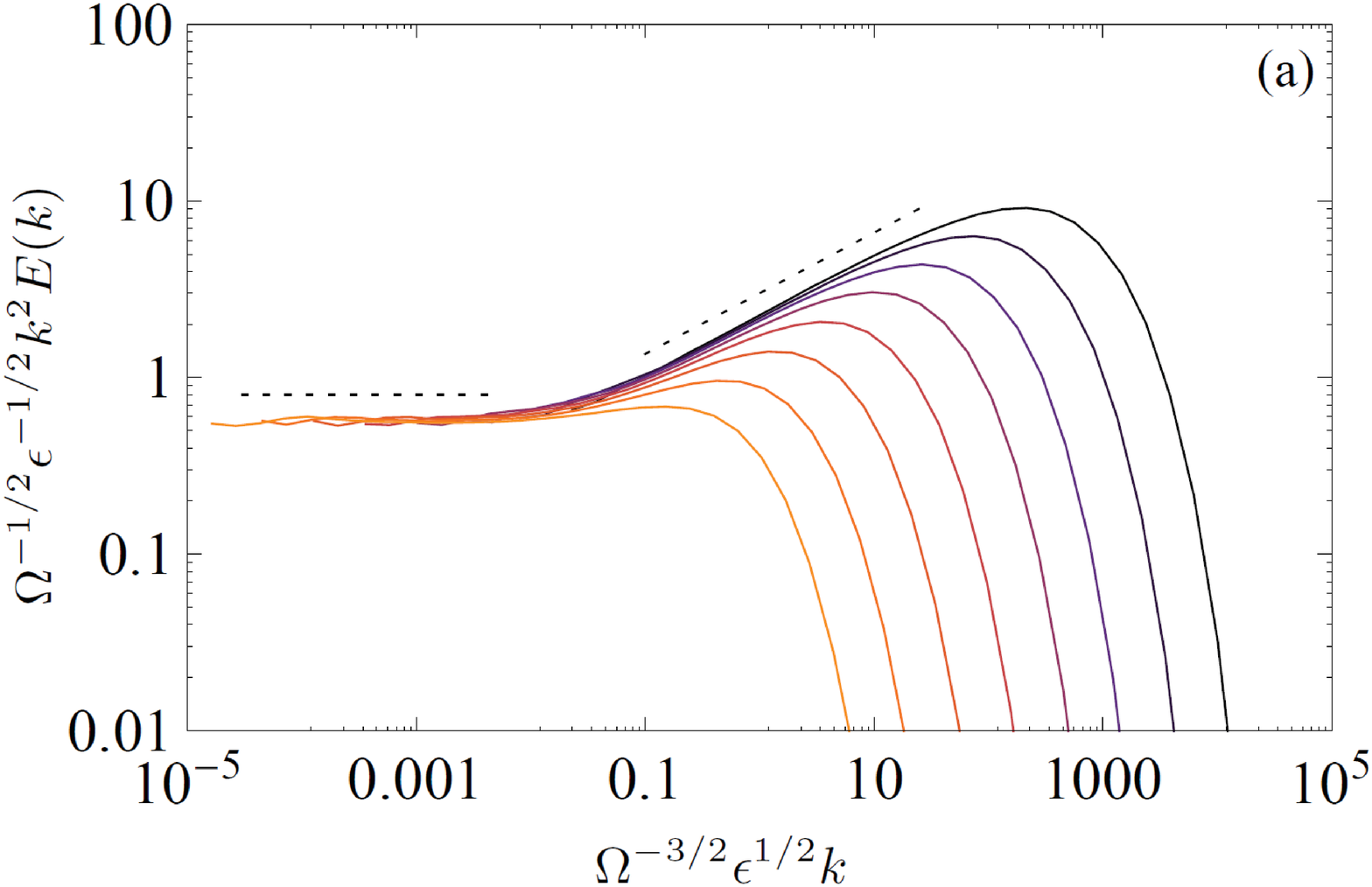} \includegraphics[width=0.45\textwidth]{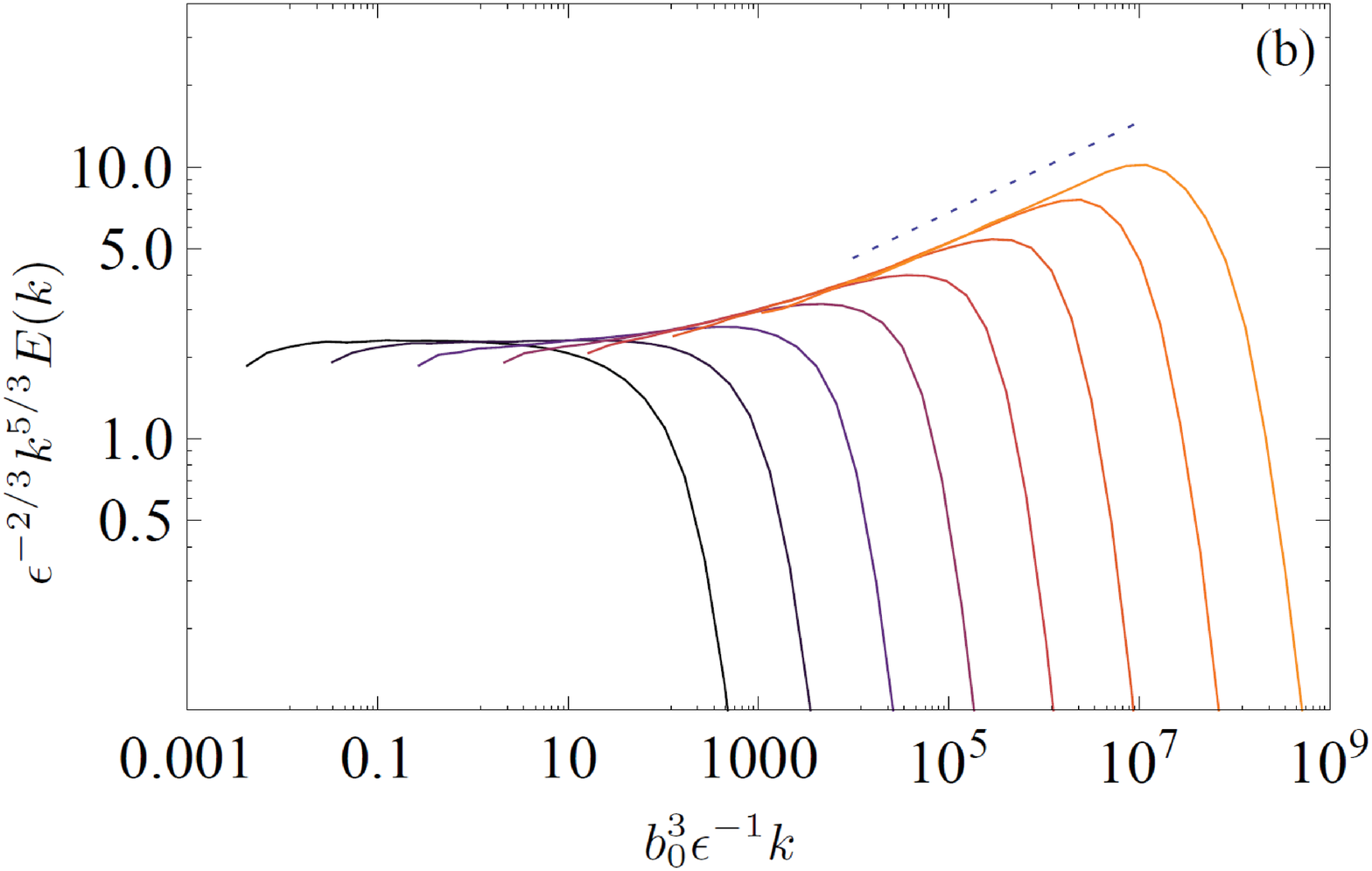}
  \end{center}
\caption{Top-left panel: all possible non-linear regimes in the map $(b_0,\Omega)$.
Other panels: normalized energy density spectrum versus normalized wave number. They correspond to regimes RA (top-right), RK (bottom-left) and KA (bottom-right).
The different curves correspond to different values of $b_0$, $\Omega$ or both.
Adapted from \citet{Plunian2010}.}
\label{figslopes}
\end{figure}

In the top-left panel of Fig.~\ref{figslopes}, we summarize the range of values of parameters $b_0$ and $\Omega$
for which the four following regimes are possible, provided $\Pm=1$, Rotation (R), Rotation-Kolmogorov (RK), Rotation-Kolmogorov-Alfv\'en (RKA) and Rotation-Alfv\'en (RA).
Each regime is characterized by different inertial ranges with, going from small to large $k$, slope $k^{-2}$ for (R),
slopes $k^{-2}$ and $k^{-5/3}$ for (RK), slopes $k^{-2}$, $k^{-5/3}$ and $k^{-3/2}$  for (RKA), and slopes $k^{-2}$ and $k^{-3/2}$ for (RA).
In addition, two other regimes are also possible for $\Omega=0$, Kolmogorov (K) and Kolmogorov-Alfv\'en (KA) depending on whether $b_0$ is weaker than $(\eta\epsilon)^{1/4}$ or not.
In the other panels of Fig.~\ref{figslopes} sets of results are shown for the cases (RA), (RK) and (KA) with $\nu=10^{-7}$ and $\Pm=1$, using the helical L1 shell model given by Eq.~(\ref{L1MHD}).
We find excellent agreement between the results obtained with the shell model and isotropic phenomenological predictions.

For $\Pm \ll 1$, the characterization of the different regimes is more complex, but still tractable analytically. However, the results obtained with the shell model clearly show that the distinction between the different regimes is in most cases no longer possible.
This is because the magnetic dissipation scale for $\Pm < 1$ is larger than the Kolmogorov scale implying that the spectral slopes are not clear enough.

Finally, note that in \citet{Plunian2010} it was necessary to introduce correlation times
for $b_0$ and $\Omega$ in order to avoid spurious supercorrelation
between $U_n$ and $B_n$, as previously explained (Sec.~\ref{s4:forcing}).
This is presumably related to the number of degrees of freedom in the helical L1-model (\ref{L1MHD}).
For HD turbulence with a  constant $\Omega$ we verified that such a problem disappears using
the helical L2-model given by Eq.~(\ref{HelMHDSABRA}).

\subsubsection{RMHD shell models}
\label{S4:Alfven}
The Reduced MHD equations accounting for anisotropy due to a strong externally applied magnetic field  $\bb_0$, are obtained by restricting the initial MHD Eq.~(\ref{elsasser2}) to field-perpendicular equations \citep{Strauss1976}
\begin{equation}
	\partial_t \bz_{\perp}^{\pm} \mp b_0 \partial_x \bz_{\perp}^{\pm}  + (\bz_{\perp}^{\mp}\cdot \nabla_{\perp})\bz_{\perp}^{\pm} + \nabla_{\perp} p =
	r^+ \nabla_{\perp}^2\bz^{\pm} + r^- \nabla_{\perp}^2\bz^{\mp}, \qquad \nabla_{\perp} \cdot \bz_{\perp}^{\pm}=0,
	\label{RMHD}
\end{equation}
where $x$ is the coordinate along the direction of $\bb_0$,
$\bz_{\perp}^{\pm}$ is the projection of the Els\"{a}sser variables onto the plane perpendicular to $\bb_0$,
and $r^{\pm}=\frac{1}{2}(\nu \pm \eta)$.

\cite{Nigro2004} introduced a RMHD shell model for Eq.~(\ref{RMHD})
keeping the $x$-dependency along the direction of $\bb_0$ (with appropriate boundary conditions)
while using a 2D shell model in the perpendicular direction to account for 2D MHD turbulence. The spectral space is thus divided into cylindrical shells, each shell $n$ being characterized by $k_{n}\le |\bk_{\perp}|< k_{n+1}$, where $\bk_{\perp}$ is the component of the wave number perpendicular to $\bb_0$. The shell variables $U_n$ and $B_n$ depend on both time and $x$.
Such a \textit{hybrid} shell model written in Els\"{a}sser variables $Z_n^{\pm}=U_n\pm B_n$, is given by
\begin{equation}
	\left(\partial_t \mp b_0\partial_x\right)Z_n^{\pm} = \widetilde{W}_n(Z_n^{\mp},Z_n^{\pm})-k_n^2 \left(r^+ Z_n^{\pm} + r^- Z_n^{\mp}\right),
	\label{anisotropicMHD}
\end{equation}
\citet{Nigro2004} used the MHD GOY model (\ref{GOY_general2}) with $\lambda=2$ and $\varepsilon=5/4$
to obtain
\begin{equation}
	\widetilde{W}_n(Z_n^{\mp},Z_n^{\pm}) =
	+\i k_n\left(\frac{13}{24}Z_{n+2}^{\pm}Z_{n+1}^{\mp}+
	\frac{11}{24}Z_{n+2}^{\mp}Z_{n+1}^{\pm}-
	\frac{19}{48}Z_{n+1}^{\pm}Z_{n-1}^{\mp}-
	\frac{11}{48}Z_{n+1}^{\mp}Z_{n-1}^{\pm}+
	\frac{19}{96}Z_{n-1}^{\pm}Z_{n-2}^{\mp}-
	\frac{13}{96}Z_{n-1}^{\mp}Z_{n-2}^{\pm}
	\right)^*.
	\label{aniMHD2}
\end{equation}

\begin{figure}[ht]
\begin{center}
\includegraphics[width=0.39\textwidth]{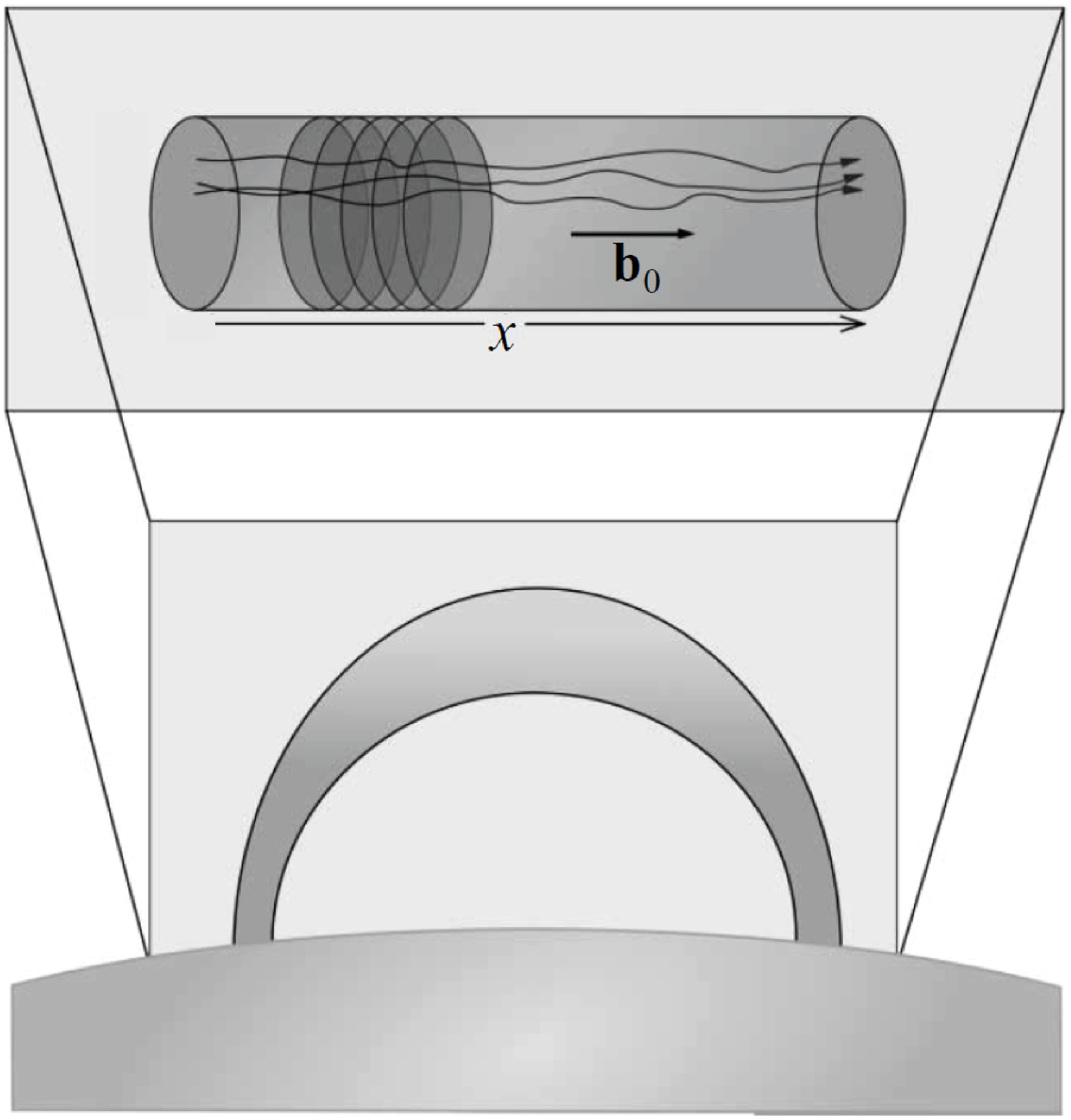}
\includegraphics[width=0.47\textwidth]{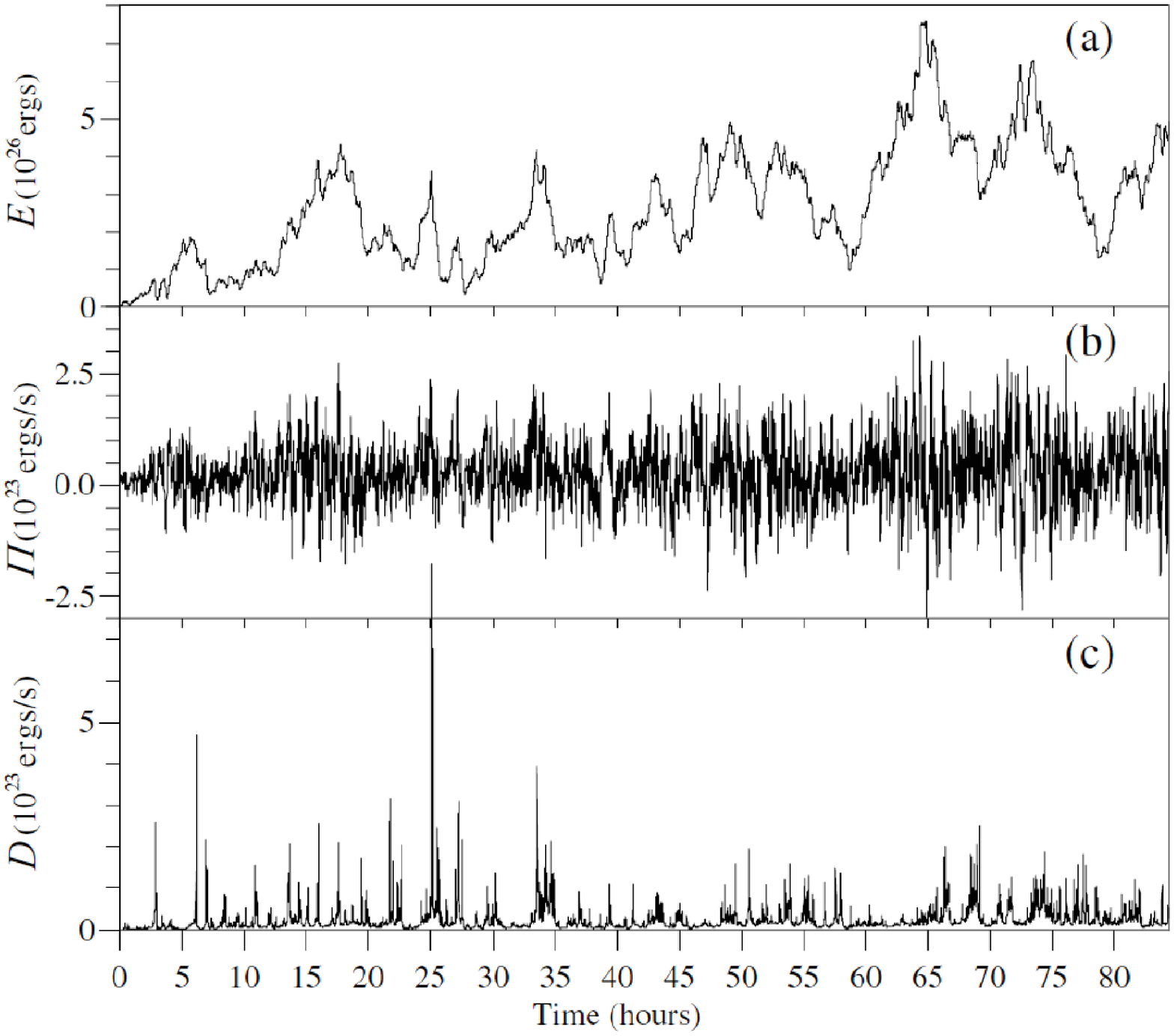}
\end{center}
\caption{Left panel: coronal loop in the direction of $\bb_0$. The 2D shell models in planes perpendicular to $\bb_0$ are piled up along $\bb_0$. Adapted from \citet{Buchlin2007a}.
Right panel: time evolution of (a) energy, (b) energy flux, and (c) dissipated power. Adapted from \citet{Nigro2004}.
}
\label{figcoronaldiss}
\end{figure}
To model a coronal loop, such as the one represented in the left-panel of Fig.~\ref{figcoronaldiss},
kinetic energy is injected at one foot $x=0$ of the coronal loop,
with no motion in the other foot $x=1$. At $x=0$ the motion is non-zero only in the first three shells ($n=0,1,2$) and has a Gaussian time distribution.
With appropriate values for the different dimensional parameters (motion at $x=0$, length and radius of the coronal loop, Alfv\'en wave velocity), and for $\nu=\eta=10^{-7}$, \citet{Nigro2004} estimated the time evolution of the total energy, net incoming energy flux and dissipated power (right panel in Fig.~\ref{figcoronaldiss}). This study strongly supports the idea that coronal nanoflares are due to intermittent events produced by Alfv\'en wave turbulence. About 60\% of the energy entering the system is in fact dissipated. The dissipation compares well with the energy involved in nanoflares and, from Fig.~\ref{figcoronaldiss}, clearly displays intermittency with statistics comparable to those of nanoflare emissions. Introducing a more complex model with several layers in $x$, \citet{Verdini2012c}
studied the combined effects of turbulence and energy leakage on the coronal heating.

\begin{figure}[ht]
\begin{center}
\includegraphics[width=0.58\textwidth]{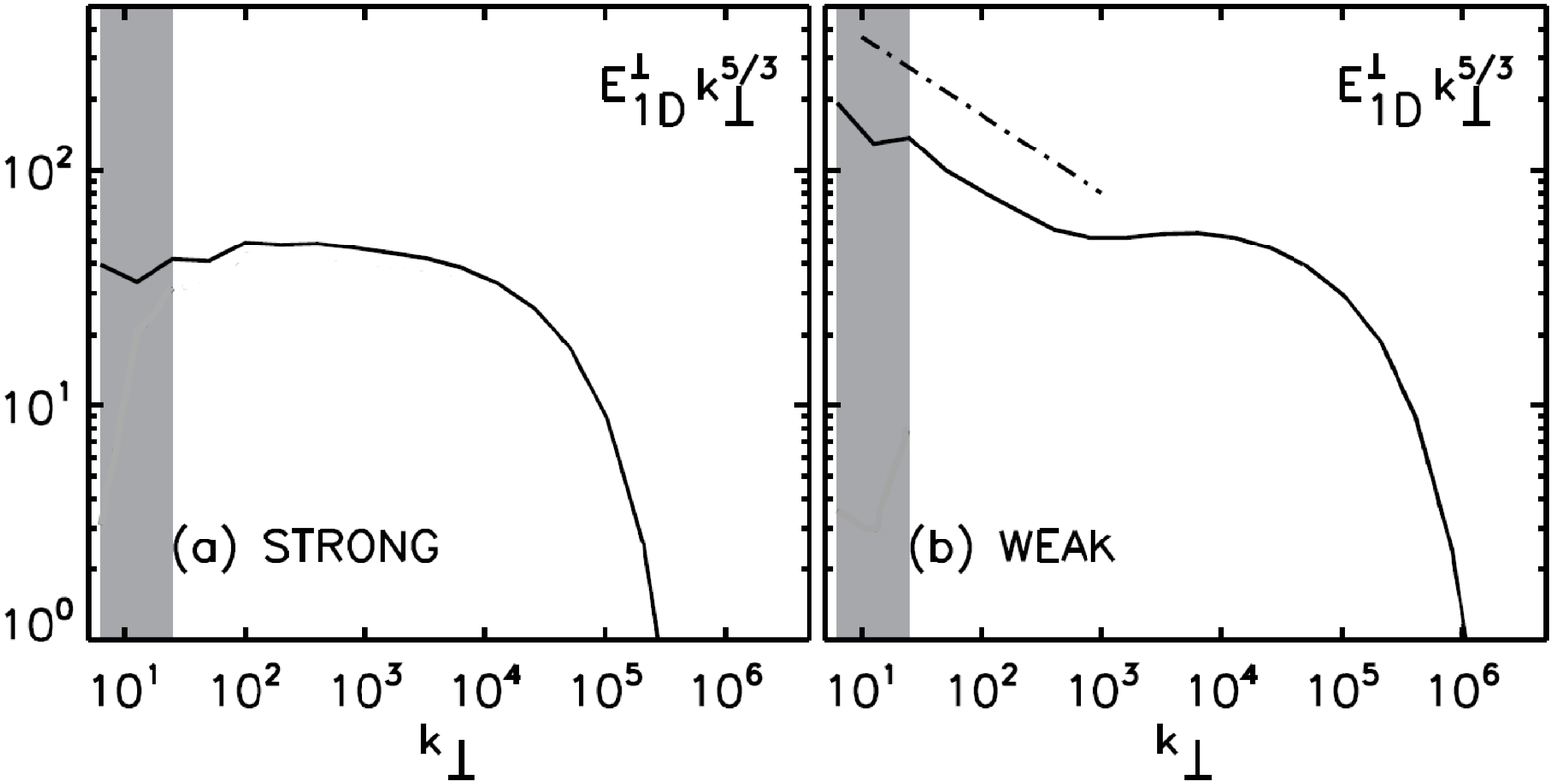}
\includegraphics[width=0.4\textwidth]{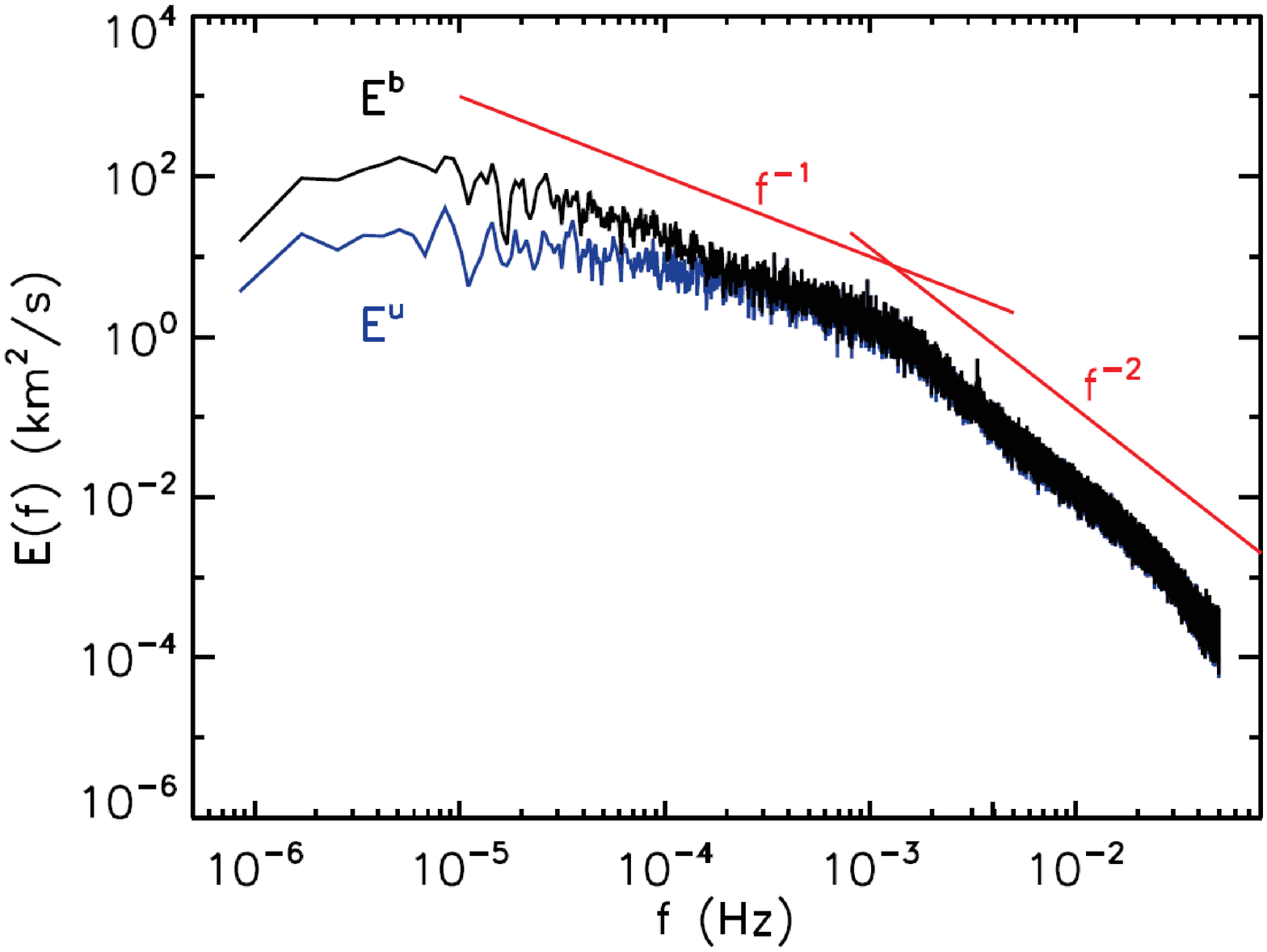}
\end{center}
\caption{Left panel: reduced perpendicular energy spectra normalized by $k^{-5/3}$ (solid lines).
The dot-dashed line is the $k_{\perp}^{-2}$ scaling.
Right panel: reduced frequency spectra for $E^{u,b}$ at $r_{top}$ (solid lines). Adapted from \citet{Verdini2012b} and \citet{Verdini2012a}.}
\label{figweak-strong}
\end{figure}

The same RMHD shell model given by Eqs.(\ref{anisotropicMHD}-\ref{aniMHD2}) was also used by \citet{Verdini2012b} to investigate the transition from weak to strong turbulence,
depending whether $t_A \ll t_{NL}$, or $t_A \approx t_{NL}$,  where $t_A\propto(k_{\parallel}b_0)^{-1}$ is the characteristic time responsible for Alfv\'en waves propagation, and $t_{NL}\propto (k_{\perp}b(k_{\perp}))^{-1}$ is the eddy turn-over time (see Sec.~\ref{s2:spectra}). In order to prescribe the ratio $t_A / t_{NL}$ , they added
a forcing term $f_n^{\pm}(x,t)$ in Eq.~(\ref{anisotropicMHD}) leading to an estimation of $t_A$ and $t_{NL}$ at the forcing scale. Typical perpendicular spectra are shown in the left-panel of Fig.~\ref{figweak-strong} for $t_A / t_{NL}\approx 1$ and $\approx 1/32$. They found that
(a) only one slope $k^{-5/3}$ is obtained for strong turbulence, whereas
(b) two slopes $k^{-2}$ and $k^{-5/3}$ are found for weak turbulence.
In the notation introduced in Sec.~\ref{s6:rotation}, the weak regime is of (AK) type, in striking opposition with the (KA) isotropic case.

In the context of solar wind MHD turbulence,
\citet{Verdini2009} introduced a modified RMHD shell model in the form
\begin{equation}
	\frac{\partial Z_n^{\pm}}{\partial t}  +(U \pm b_0)\frac{\partial Z_n^{\pm}}{\partial x} -\frac{1}{4}(U \mp b_0)\left(\frac{1}{\rho}\frac{d\rho}{dx}\right)Z_n^{\pm}
	+ \frac{1}{4}(U \mp b_0)\left(\frac{1}{\rho}\frac{d\rho}{dx}+2\frac{1}{R}\frac{dR}{dx}\right)Z_n^{\mp}= \widetilde{W}_n(Z_n^{\mp},Z_n^{\pm})-k_n^2 \left(r^+ Z_n^{\pm} + r^- Z_n^{\mp}\right),
	\label{RMHD2}
\end{equation}
where $U$, $\rho$ and $R$ stand for the background solar wind,
the fluid density and the radius of the magnetic flux tube respectively.
The profile of the latter quantities as a function of $x$ was assigned
along with $b_0$.
The modulus and correlation time of $Z_n^+$ was prescribed at $x=0$ and only in the three first shells. This corresponds to the injection of high frequency Alfv\'en waves at the base of the chromosphere.
By solving Eq.~(\ref{RMHD2}) \citet{Verdini2012a} found
a double power law for the magnetic frequency spectrum that is reproduced in the right panel of Fig.~\ref{figweak-strong}. At low frequency it compares well with the $1/f$ magnetic spectrum of the sub-Alfv\'enic solar wind represented in Fig.~\ref{MHDspectra-exp-sol}. For higher frequencies \citet{Verdini2012a} argue that the -2 slope that they find in their model is masked in the solar wind by the more energetic perpendicular spectrum which has a $-5/3$ slope.

\subsection{Hall-effect}
\label{s7}
In strongly magnetized conducting media,
e.g. in weakly
ionized accretion disks, white dwarfs and
neutron stars,
the Hall drift of the magnetic field can operate on a much shorter time-scale than the other transport processes. Though non-dissipative, the Hall current redistributes the magnetic energy from large to small scales and thence enhances the dissipation rate of the magnetic field \citep{1999MNRAS.304..451U}.
The rate of magnetic field decay is important to understanding the history
of such astrophysical objects.
The Hall
effect is probably most pronounced in young neutron stars, called magnetars because of their strong magnetic field of up to $10^{15}$ G. 

In the absence of motion and ambipolar diffusion, the evolution of the magnetic field due to the Hall effect can be written in its dimensionless form as \citep{GR92}
\begin{equation}
\frac{\partial\bf{b}}{\partial t} = {\rm R}_{\rm H}^{-1} \nabla^2\bf{b} -
\bf{\nabla}\times\left[(\bf{\nabla}\times\bf{b}) \times \bf{b}
\right], \label{hall}
\end{equation}
where $\Rh=(e b_0 t_e)/(m_e^* c)$, $b_0$ the
characteristic magnetic field strength at the largest scale,
$e$ the elementary charge, $c$ the speed of light, $m_e^*$ the effective
mass of an electron and $t_e$ the electron relaxation time.

The last term of Eq.~(\ref{hall}) describes the advection of the magnetic field
by the Hall drift and is expected to produce a non-linear cascade from large to small magnetic scales.
Using the assumption of strong turbulence, \citet{V73} derived 
a $k^{-7/3}$ scaling law for the magnetic energy density spectrum.
However, considering weakly
interacting waves \citet{GR92} found an alternative
$k^{-2}$ scaling law.
In both cases the spectra are steeper than for Kolmogorov turbulence.

The possibility of a cascade is of course appealing for shell models.
In addition, for $\Rh \rightarrow \infty$ Eq.~(\ref{hall}) gives rise to two quadratic invariants,
the energy $E^b$ and the magnetic helicity $H^b$, which are sufficient for the derivation of a shell model including the Hall effect. The difference with MHD shell models is that the non-linear term is now of the form
$k_n Q_n(\bX)$.

\citet{Frick2003} introduced a Hall shell model in the form
\begin{equation}
(d_t + {\rm R}_{\rm H}^{-1} k_n^2)B_n = k_n^2 \bigg( B_{n+2} B_{n+1} - \frac{\varepsilon}{\lambda^2} B_{n+1} B_{n-1}
   - \frac{1-\varepsilon}{\lambda^4} B_{n-1} B_{n-2} \bigg)+F_n \;,
\label{mod_hall1}
\end{equation}
with the two quadratic invariants
\begin{equation}
	E^B=\frac{1}{2}\sum_n B_n^2, \qquad H=\sum_n (\varepsilon-1)^{-n} B_n^2.
	\label{maghel1}
\end{equation}
In Eq.(\ref{mod_hall1})
$F_n$ is an externally applied electromotive force and $B_n$ is \textit{real}.
Indeed the \textit{real} and \textit{imaginary} parts
of \textit{complex} variables $B_n$ would not couple, as can be seen from Eq.~(\ref{hall}) written in
Fourier space.
We take $\varepsilon=1-\lambda$ in order to impose the conservation of magnetic helicity
\begin{equation}
	H^B=\sum_n (-1)^n k_n^{-1} B_n^2.
\end{equation}
Solutions of Eq.~(\ref{mod_hall1}) for $F_n=0$ (free-decaying Hall-turbulence) show that the initial magnetic energy concentrated in the large scales triggers the cascade process. This results in a magnetic inertial range of the form $E^B(k_n)\sim k_n^{-4/3}$, corresponding to the \citet{V73} energy density $E^b(k)\sim k^{-7/3}$. However, the energy spectrum steepens rapidly in time, unless stochastic forcing is applied \citep{Frick2007}.

On representing the magnetic field as the sum of poloidal $\bb_\tp$ and toroidal $\bb_\ttor$ components,
Eq.~(\ref{hall}) takes the form \citep{1999MNRAS.304..451U}
\begin{eqnarray}
\frac{\partial{\bf b}_\tp}{\partial t} &=& {\rm R}_{\rm H}^{-1} \nabla^2{\bf
b}_\tp - {\bf \nabla}\times\left[({\bf \nabla}\times{\bf b}_\ttor)
\times{ \bf b}_\tp \right], \label{pol} \\ 
\frac{\partial{\bf
b}_\ttor}{\partial t} &=& {\rm R}_{\rm H}^{-1} \nabla^2{\bf b}_\ttor - {\bf
\nabla}\times\left[({\bf \nabla}\times{\bf b}_\ttor) \times{ \bf b}_\ttor
+ ({\bf \nabla}\times{\bf b}_\tp) \times{ \bf b}_\tp\right].
\label{tor}
\end{eqnarray}
From Eqs.~(\ref{pol}-\ref{tor}), we clearly see that the two components $\bb_\tp$ and $\bb_\ttor$ are coupled. However, coupling is not symmetric. An initial poloidal configuration
can generate a toroidal component with the term $-\nabla \times \left[({\bf \nabla}\times{\bf b}_\tp) \times{ \bf b}_\tp\right]$. On the other hand, a poloidal component cannot be generated if the initial magnetic configuration is purely toroidal unless some transient dynamo type instability occurs for $\bb_\tp$.
Note that, in the presence of a background magnetic field, such a mechanism is at the heart of the so-called Hall drift instability \citep{2002PhRvL..88j1103R}. In the context of neutron stars, this instability
leads to non-local energy transfer to small scales and then
to enhanced crustal field dissipation.

\citet{Frick2003} derived the following shell model for both poloidal $P$ and toroidal $T$ components
of the magnetic field
\begin{eqnarray}
d_t {P}_n + {\rm R}_{\rm H}^{-1} k_n^2 P_n  = k_n^2 \left(P_{n+2}
T_{n+1} - \frac{\lambda+1}{\lambda^2} P_{n+1} T_{n-1} +
\frac{1}{\lambda^3} P_{n-1} T_{n-2} \right)+
k_n^2 \left(T_{n+2} P_{n+1} - \frac{\lambda+1} {\lambda^2} T_{n+1} P_{n-1} +
\frac{1}{\lambda^3} T_{n-1}
P_{n-2} \right), \label{mod21}\\
 d_t T_n + {\rm R}_{\rm H}^{-1} k_n^2
T_n
=  k_n^2 \left(P_{n+2} P_{n+1} - \frac{\lambda+1}{\lambda^2} P_{n+1} P_{n-1} + \frac{1}{\lambda^3} P_{n-1}
P_{n-2} \right)+ k_n^2 \left(T_{n+2} T_{n+1}
- \frac{\lambda+1}{\lambda^2} T_{n+1} T_{n-1} + \frac{1}{\lambda^3} T_{n-1} T_{n-2} \right), \label{mod22}
\end{eqnarray}
with magnetic energy and magnetic helicity defined by
\begin{equation}
    E^B = \sum_n (P_n^2+T_n^2), \qquad
    H^B = \sum_n k_n^{-1} P_n T_n.
    \label{maghel2}
\end{equation}
As in helical models a remarkable advantage of definition (\ref{maghel2}) is that, contrary to (\ref{maghel1}), 
magnetic helicity and magnetic energy are not shell-by-shell correlated.
Another advantage is
that it is possible to describe the energy exchange between both
poloidal and toroidal components of the magnetic field. In
particular, it was shown that a transient poloidal field
component can develop from an initial purely toroidal configuration \citep{Frick2003}.
With again stochastic forcing incorporated into Eqs.(\ref{mod21}-\ref{mod22}),
\citet{Frick2007} found 
a stable inertial range characterized by the \citet{V73} energy spectrum slope $E^b(k) \sim k^{-7/3}$, and ${ H^b}(k)\sim k^{-1} \, {E^b}(k)$.

With the application to the solar wind in mind, \citet{hori2005} and \citet{Galtier2007} 
included the Hall effect in the incompressible MHD Eqs.~(\ref{MHDU}-\ref{MHDB})
\begin{eqnarray}
\left({\partial}_t-\nu \nabla^2 \right)  \bu &=& - (\bu\cdot \nabla) \bu + (\bb \cdot \nabla) \bb -\nabla p,
\qquad  \nabla \cdot \bu = 0, \label{HMHDU}\\
\left(\partial_t-\eta \nabla^2 \right) \bb &=&- (\bu\cdot \nabla) \bb + (\bb \cdot \nabla)  \bu  - \Rh \bf{\nabla}\times\left[(\bf{\nabla}\times\bb) \times \bb \right], \quad \nabla \cdot \bb =0. \label{HMHDB}
\end{eqnarray}
In the limit $(\nu,\eta)\rightarrow (0,0)$, the system of Eqs.(\ref{HMHDU}-\ref{HMHDB}) contains three quadratic invariants, the total energy $E$ and magnetic helicity $H^b$ defined in Eqs.(\ref{totalenergy}-\ref{defhelicities}), and a third (new) quantity
\begin{equation}
H^i=\int_V ( \ba+\Rh\bu)\cdot (\bb+\Rh \nabla\times\bu)dV,
\label{hall3}
\end{equation}
called the ion helicity. 
Note that $H^i=H^b+2\Rh H^c + \Rh^2 H^u$.
In the limit $\Rh\rightarrow 0$, $H^i$ reduces to magnetic helicity, while $(H^i-H^b)/2\Rh$, also a conserved quantity, reduces to cross helicity $H^c$.

In the spirit of the \textit{complex} GOY shell model, the Hall MHD shell model for $\lambda=2$, takes the form \citep{hori2005,Galtier2007}
 \begin{eqnarray}
(d_t + \nu k_n^2)U_n &=& \i k_n \Bigl \lbrace
(U_{n+1}U_{n+2} - B_{n+1}B_{n+2} ) 
- \frac{1}{4} (U_{n-1}U_{n+1} -
 B_{n-1}B_{n+1} )
-\frac{1}{8} (U_{n-2}U_{n-1}
- B_{n-2}B_{n-1} )\Bigl \rbrace^* +  F_n.\nonumber\\
\label{hgoy_u}\\
(d_t + \eta k_n^2)B_n &=&
\frac{\i k_n}{6} \Bigl \lbrace U_{n+1}B_{n+2} - B_{n+1}U_{n+2}  
+  U_{n-1}B_{n+1} -  B_{n-1}U_{n+1} +  U_{n-2}B_{n-1} - B_{n-2}U_{n-1} \Bigl \rbrace^* \nonumber\\
&+& (-1)^n \i \Rh k_n^2 \Bigl \lbrace B_{n+2} B_{n+1} - \frac{1}{4} B_{n+1} B_{n-1}
   - \frac{1}{8} B_{n-1} B_{n-2} \Bigl \rbrace^*.
\label{hgoy_b}
\end{eqnarray}
In the limit $(\nu,\eta)\rightarrow (0,0)$ the following quantities are conserved
\begin{equation}
	E^U+E^B=\frac{1}{2}\sum_n |U_n|^2+|B_n|^2, \quad H^B=\frac{1}{2}\sum_n (-1)^n k_n^{-1}|B_n|^2, \quad 
	H^I=H^B+2\Rh H^C + \Rh^2 H^U
\end{equation}
with
\begin{equation}
	H^C=\frac{1}{2}\sum_n U_nB_n^* + U_n^* B_n, \qquad H^U= \frac{1}{2}\sum_n (-1)^n k_n |U_n|^2.
\end{equation}
Note that in Eq.~(\ref{hgoy_b}) the term accounting for the Hall drift differs
from Eq.~(\ref{mod_hall1}) because here the variables are \textit{complex}.

\citet{Galtier2007} and \citet{hori2008} 
found that the large-scale magnetic energy density follows a $k^{-5/3}$
spectrum that steepens to $k^{-7/3}$ at scales smaller than $\Rh$
 if the magnetic energy overtakes the kinetic energy or to $k^{-11/3}$ in the inverse case. This might explain why the magnetic spectrum of the solar wind
 steepens at high frequencies ($f>1Hz$).

%% file: Conclusion.tex
\section{Summary and outlook}
\label{s9}
Table \ref{summary} summarizes the MHD shell models that have been discussed in the present review. The L2 GOY model given by Eqs.~(\ref{GOY_general2}-\ref{e3:goy_b}) is without doubt the most often used, in its 2D and 3D forms. It gives excellent results in terms of spectra, intermittency, and energy transfer and flux. However, we saw that such a model can be misleading when dealing with kinetic helicity in 3D HD turbulence.
This is also true for its cousins the L2 and N2 models given respectively by Eqs.~(\ref{e3:sabraL2_u}-\ref{e3:sabraL2_b}) and Eqs.~(\ref{SabraW}-\ref{e3:sabra_b}). 
On the other hand helical models do not suffer from this drawback. So far only the H1 local model given by Eq.~(\ref{L1MHD}) has been used for MHD turbulence. The other helical models given by Eq.~(\ref{HelMHDGOY}) or Eq.~(\ref{HelMHDSABRA}) are given here for completeness, but have not been tested yet.

As shown in Appendix, the models depicted in Table \ref{summary} represent only a subset of all possible models that satisfy total energy and cross helicity conservation. In addition any combination of models will again satisfy both conservation laws. In Table \ref{summary} 
the models which depend on \textit{one free parameter} are the result of the combination of two other models. 
Regarding the \textit{non-local parameter} in Table \ref{summary},
it controls the degree of non-locality in 
non-local models. We saw that this parameter can be estimated with the help of a hierarchical approach or from phenomenological arguments (Sec.~\ref{s3:nonlocal}).

In addition to automatically satisfying the conservation of total energy and cross helicity, the general formulation given in Eqs.~(\ref{shell_NS2}-\ref{relation1}) has the advantage of offering a mathematical framework for the definition of energy transfer and flux. This general formulation has also been  used to define transfer and flux of kinetic helicity in HD turbulence. No doubt it can also be applied to transfer and flux of cross and magnetic helicities in 3D MHD turbulence.

The dimension of MHD turbulence, 2D or 3D, can be changed imposing either the square of magnetic potential or the magnetic helicity as the third quadratic invariant. Though this is possible to achieve for any models, except L1 for
which magnetic helicity cannot be defined, in Table \ref{summary} we denote by 2D, 3D or both
the corresponding version that has been developed so far.

Finally modern shell models always use \textit{complex} instead of \textit{real} variables mainly because it increases the degree of freedom of the system, leading to more realistic dynamics.

The applications presented in Sec.~\ref{s4} correspond to a large panel of situations requiring high values for the kinetic and magnetic Reynolds numbers: small-scale and multi-scale dynamos, free-decaying MHD turbulence, Alfv\'en wave turbulence and Hall-effect turbulence. In order to deal with anisotropy \textit{hybrid} models have been developed by mixing 2D MHD turbulence in a plane perpendicular to the direction of anisotropy with direct integration in the parallel direction. 
A few attempts have been made to develop subgrid shell models, which are indeed an interesting and promising application of shell models. 

Shell model simulations require the same time-stepping as direct numerical simulations. However, there is a significant gain in using shell models, essentially because the number of grid points is much lower than in DNS, and partial derivatives are absent. Shell models are user-friendly tools that guide our intuition in realistic parameter regimes still inaccessible to DNS.  

\begin{table}[ht]
\begin{center}
\begin{tabular}{|@{\hspace{0.cm}}c|@{\hspace{0.2cm}}l|@{\hspace{0.2cm}}c|@{\hspace{0.2cm}}l|@{\hspace{0.2cm}}l|@{\hspace{0.2cm}}l|@{}}
\hline
$\text{Model}$	&		\text{Equations}	                      &	\text{Variables}	&	\text{Conservative quantities}                                                      & Comments                  &	Reference\\*[0cm] \hline
\text{L1}       & (\ref{gloagen}-\ref{e3:gloag_b})		      &	\text{real}	      &	$E=\frac{1}{2}\sum(U_n^2+B_n^2)$                                                  &One free                   &	\citet{Gloaguen1985}	\\*[0cm]
                &    		                                    &	$U_n$, $B_n$	    &	$H^C=\sum U_nB_n$		                                                              &parameter                  &		\\*[0cm] \hline
\text{L1}       & (\ref{gloagencomplex}-\ref{e3:bisk_b})		&	\text{complex}	  &	$E=\frac{1}{2}\sum(|U_n|^2+|B_n|^2)$		                                          &One free                   &	\citet{Biskamp1994}	\\*[0cm]
                &    	                                      &	$U_n$, $B_n$	    &	$H^C=\frac{1}{2}\sum(U_nB_n^*+U_n^*B_n)$                                          &parameter                  &		\\*[0cm]\hline
\text{L2}       & (\ref{GOY_general2}-\ref{e3:goy_b})		    &	\text{complex}	  &	$E=\frac{1}{2}\sum(|U_n|^2+|B_n|^2)$		                                          &                           &\citet{Brandenburg1996}		\\*[0cm]
  GOY           &   		                                    &	$U_n$, $B_n$	    &	$H^C=\frac{1}{2}\sum(U_nB_n^*+U_n^*B_n)$                                          &                           & \citet{Basu1998}		\\*[0cm]
                &    		                                    &		                &	$H^B=\frac{1}{2}\sum(-1)^nk_n^{-1}|B_n|^2$ &$\Rightarrow$ \text{3D MHD}&	\citet{Frick1998}	\\*[0cm]
                & 		                                      &		                &	$A=\frac{1}{2}\sum k_n^{-2}|B_n|^2$         &$\Rightarrow$ 2D MHD&		\\*[0cm]\hline
\text{L2}       & (\ref{e3:sabraL2_u}-\ref{e3:sabraL2_b})		&	\text{complex}	  &	$E=\frac{1}{2}\sum(|U_n|^2+|B_n|^2)$		                                          &&		\\*[0cm]
  Sabra         &    		                                    &	$U_n$, $B_n$	    &	$H^C=\frac{1}{2}\sum(U_nB_n^*+U_n^*B_n)$		                                      && 	\citet{Plunian2007}	\\*[0cm]
                &                                           & 				          &	$H^B=\frac{1}{2}\sum(-1)^nk_n^{-1}|B_n|^2$                                        & 3D MHD&		\\*[0cm]\hline
\text{N2}       & (\ref{N2hierachical}),                     &\text{real}	      &	$E=\frac{1}{2}\sum(U_n^2+B_n^2)$		                                              &&		\\*[0cm]
  Hierar-       & (\ref{frick21}-\ref{frick22})		          &	$U_n$, $B_n$	    &	$H^C=\sum U_nB_n$		                                                              && 	\citet{Frick1984}	\\*[0cm]
  -chical       &                                           & 				          &	$A=\frac{1}{2}\sum k_n^{-2}B_n^2$                                                  & 2D MHD&		\\*[0cm]\hline
\text{N2}       & (\ref{SabraW}-\ref{e3:sabra_b})		        &	\text{complex}	  &	$E=\frac{1}{2}\sum(|U_n|^2+|B_n|^2)$		                                          &One non-local&		\\*[0cm]
           &    		                                    &	$U_n$, $B_n$	    &	$H^C=\frac{1}{2}\sum(U_nB_n^*+U_n^*B_n)$		                                      &parameter& 	\citet{Plunian2007}	\\*[0cm]
                &                                           & 				          &	$H^B=\frac{1}{2}\sum(-1)^nk_n^{-1}|B_n|^2$                                        & 3D MHD&		\\*[0cm]\hline
\text{H1}      & (\ref{L1MHD})		                          &	\text{complex}	  &	$E=\frac{1}{2}\sum(|U_n|^2+|B_n|^2)$		                                    &One free&		\\*[0cm]
                &     		                                    &	$U_n$, $B_n$	    &	$H^C=\frac{1}{2}\sum(U_nB_n^*+U_n^*B_n)$		              &parameter& \citet{Mizeva2009}		\\*[0cm]
                &    	                                     	&		                &	$H^B=\frac{i}{2}\sum k_n^{-1}({B_n^*}^2-B_n^2)$                                    & 3D MHD&		\\*[0cm]\hline
\text{H2}      & 	 (\ref{HelMHDGOY})                                     	&	\text{complex}	  &	$E=\frac{1}{2}\sum(|U_n^+|^2+|U_n^-|^2+|B_n^+|^2+|B_n^-|^2)$		                  &&		\\*[0cm]
             &  (\ref{HelMHDSABRA})	                    	&$U_n^{\pm}$, $B_n^{\pm}$&	$H^C=\frac{1}{2}\sum(U_n^+{B_n^+}^*+U_n^-{B_n^-}^*+c.c.)$		                && 	This paper	\\*[0cm]
           &    	                	&		                &	$H^B=\sum k_n^{-1}(|B_n^+|^2-|B_n^-|^2)$                            & 3D MHD& 	\\*[0cm]
\hline
	\end{tabular}	
	\caption{Summary of main MHD shell models.}
	\label{summary}
	\end{center}
\end{table}

\section*{Acknowledgements}
This collaboration  benefited from the International
Research Group Program supported by Perm region Government.
R.S. acknowledges the grant YD-4471.2011.1 from the Council of the President of the Russian
Federation and the RFBR grant 11-01-96031-ural.
P.F. acknowledges the RFBR grant  11-01-00423.
We warmly acknowledge J. Torbet for great improvement of the writing.

%% file: Appendix.tex
\appendix
\label{Appendix}
\section{L1-models}
\label{s7:L1}
The complex L1-models have the form
\begin{equation}
	{\widetilde W}_n(\bX,\bY)=k_n \sum_{i,j=-1,|i-j|\leq1}^{+1} a_{ij}^{{\widetilde W}} X_{n+i} Y_{n+j} + b_{ij}^{{\widetilde W}} X_{n+i}^* Y_{n+j} + c_{ij}^{{\widetilde W}} X_{n+i} Y_{n+j}^* + d_{ij}^{{\widetilde W}} X_{n+i}^* Y_{n+j}^*, \qquad ,
	\label{Wntildeform}
\end{equation}
involving 28 complex coefficients.
The number of complex coefficients reduces to 11 on applying the property (\ref{relation1}).
Then the general shape of complex L1-models becomes
\begin{eqnarray}
\widetilde{W}_n(\bX,\bY)=k_n \left[ \right.
&&  A_1 (X^*_{n-1}    Y^*_{n-1}  -  \lambda X^*_{n}   Y^*_{n+1} )
 +  A_2 (X^*_{n}      Y^*_{n-1}   -  \lambda X^*_{n+1} Y^*_{n+1} )\nonumber\\
&+& A_3 (X_{n-1}      Y^*_{n-1}   -  \lambda X_{n}     Y^*_{n+1} )
 +  A_4 (X_{n}        Y^*_{n-1}   -  \lambda X_{n+1}   Y^*_{n+1} )\nonumber\\
&+& A_5 X^*_{n-1}    Y_{n-1}     -  \lambda A_5^* X_{n}     Y_{n+1}
 +  A_6 X^*_{n}      Y_{n-1}     -  \lambda A_6^* X_{n+1}   Y_{n+1}   \nonumber\\
&+& A_7 X_{n-1}      Y_{n-1}     -  \lambda A_7^* X^*_{n}   Y_{n+1}
 +  A_8 X_{n}        Y_{n-1}   -    \lambda A_8^*X^*_{n+1} Y_{n+1}   \nonumber\\
&+& (A_9 X_{n-1}-  A_9^* X_{n-1}^*
 + A_{10} X_{n}-   A_{10}^* X_{n}^*
 + A_{11} X_{n+1}- A_{11}^* X_{n+1}^*)Y_n \left. \right]
.
\label{L1complex}
\end{eqnarray}
where the $A_i$ are complex parameters. 
After our definition of L1-models given in Sec.~\ref{s3:general formalism} shell $n$ cannot interact with itself only, implying $A_{10}=0$.
A similar general shape for complex L1-models has been introduced in the seminal paper by \citet{Gloaguen1985} but using Els\"asser variables $\bZ^{\pm}$.
Liouville's theorem yields additional constraints on the possible choice for the $A_i$. The model given by Eq.~(\ref{gloagencomplex}) investigated by \citet{Biskamp1994} corresponds with taking $A_i=0$ for $i=3,\cdots,11$.

\section{L2-models}
\label{s7:L2}
Any complex L2-model has the form
\begin{equation}
	{\widetilde W}_n(\bX,\bY)=k_n \sum_{\substack{i,j=-2,\\|i+j|=3 \text{ or } i=-j=\pm1}}^{+2}
	a_{ij}^{{\widetilde W}} X_{n+i} Y_{n+j} + b_{ij}^{{\widetilde W}} X_{n+i}^* Y_{n+j} + c_{ij}^{{\widetilde W}} X_{n+i} Y_{n+j}^* + d_{ij}^{{\widetilde W}} X_{n+i}^* Y_{n+j}^*.
	\label{WntildeformL2}
\end{equation}
A total of 24 complex coefficients are involved.
On applying the property (\ref{relation1}) the number of complex coefficients is reduced to 12
and the general shape of complex L2-models becomes
\begin{eqnarray}
\widetilde{W}_n(\bX,\bY)&=&k_n \left[ \right.
    C_1 (X^*_{n-2}    Y^*_{n-1}  -  \lambda X^*_{n-1}   Y^*_{n+1} )
 +  C_2 (X^*_{n-1}      Y^*_{n-2}   -  \lambda^2 X^*_{n+1} Y^*_{n+2} )
 +  C_3 (X^*_{n+1}      Y^*_{n-1}   -  \lambda X^*_{n+2} Y^*_{n+1} )
 \nonumber\\
&+& C_4 (X_{n-2}      Y^*_{n-1}   -  \lambda X_{n-1}     Y^*_{n+1} )
 +  C_5 X^*_{n-1}      Y_{n-2}     -  \lambda^2 C_5^* X_{n+1}   Y_{n+2}
 +  C_6 X^*_{n+1}      Y_{n-1}     -  \lambda C_6^* X_{n+2}   Y_{n+1}
 \nonumber\\
&+& C_7 X^*_{n-2}    Y_{n-1}     -  \lambda C_7^* X_{n-1}     Y_{n+1}
 +  C_8 (X_{n-1}        Y^*_{n-2}   -  \lambda^2 X_{n+1}   Y^*_{n+2} )
 +  C_{9} X_{n+1}        Y_{n-1}   -    \lambda C_{9}^*X^*_{n+2} Y_{n+1}
 \nonumber\\
&+& C_{10} X_{n-2}      Y_{n-1}     -  \lambda C_{10}^* X^*_{n-1}   Y_{n+1}
 +  C_{11} X_{n-1}        Y_{n-2}   -    \lambda^2 C_{11}^*X^*_{n+1} Y_{n+2}
 +  C_{12} (X_{n+1}        Y^*_{n-1}   -  \lambda X_{n+2}   Y^*_{n+1} )
\left. \right],\nonumber\\
\label{L2complex}
\end{eqnarray}
where the $C_i$ are again complex parameters.
Liouville's theorem is automatically satisfied as $\widetilde{W}_n(\bX,\bY)$ does not depend on
$X_n$, $X_n^*$, $Y_n$ nor $Y_n^*$.
The GOY model is obtained by setting all $C_i$ coefficients to zero, except $C_1, C_2$ and $C_3$.
The Sabra model is obtained by setting all $C_i$ coefficients to zero, except $C_{11}, C_{12}$ and $C_{13}$.

\section{N1-models}
\label{s7:N1}
The N1-models are obtained from the L1-model (\ref{L1complex}), including the non-local interactions.
Their general shape is
\begin{eqnarray}
\widetilde{W}_n(\bX,\bY)&=&k_n \sum_{m\ge 1}\left[ \right.
  A_1^1 (X^*_{n-m}    Y^*_{n-1}  -  \lambda X^*_{n-m+1}   Y^*_{n+1} )
 +  A_1^2 (X^*_{n-1}    Y^*_{n-m}  -  \lambda^m X^*_{n+m-1}   Y^*_{n+m} )
 +  A_2 (X^*_{n}      Y^*_{n-m}   -  \lambda^m X^*_{n+m} Y^*_{n+m} )\nonumber\\
&+& A_3^1 (X_{n-m}      Y^*_{n-1}   -  \lambda X_{n-m+1}     Y^*_{n+1} )
 +  A_3^2 (X_{n-1}      Y^*_{n-m}   -  \lambda^m X_{n+m-1}     Y^*_{n+m} )
 +  A_4 (X_{n}        Y^*_{n-m}   -  \lambda^m X_{n+m}   Y^*_{n+m} )\nonumber\\
&+& A_5^1 X^*_{n-m}    Y_{n-1}     -  \lambda {A_5^1}^* X_{n-m+1}     Y_{n+1}
 +  A_5^2 X^*_{n-1}    Y_{n-m}     -  \lambda^m {A_5^2}^* X_{n+m-1}     Y_{n+m}
 +  A_6 X^*_{n}      Y_{n-m}     -  \lambda^m A_6^* X_{n+m}   Y_{n+m}   \nonumber\\
&+& A_7^1 X_{n-m}      Y_{n-1}     -  \lambda {A_7^1}^* X^*_{n-m+1}   Y_{n+1}
 +  A_7^2 X_{n-1}      Y_{n-m}     -  \lambda^m {A_7^2}^* X^*_{n+m-1}   Y_{n+m}
 +  A_8 X_{n}        Y_{n-m}   -    \lambda^m A_8^*X^*_{n+m} Y_{n+m}   \nonumber\\
&+& (A_9 X_{n-m+2}-  A_9^* X_{n-m+2}^*)Y_n \left. \right]
.
\label{N1complex}
\end{eqnarray}
where the $A_i$ and $A_i^j$ are complex parameters depending on $m$. The non-local version of the model (\ref{gloagen}) by \citet{Gloaguen1985} becomes
\begin{equation}
\widetilde{W}_n(\bX,\bY)=k_n \left[C_1^1(X_{n - m}Y_{n-1} - \lambda X_{n-m+1}Y_{n +1} )+C_1^2(X_{n - 1}Y_{n-m} - \lambda^m X_{n+m-1}Y_{n +m} )+C_2 (X_{n}Y_{n - m} - \lambda^m X_{n+m} Y_{n +m})\right], \label{gloagenNL}
\end{equation}
where $C_1^1$, $C_1^2$ and $C_2$ are real parameters, and $\bX$ and $\bY$ are real variables.

\section{N2-models}
\label{s7:N2}
The N2-models are obtained from the L2-model (\ref{L2complex}), including the non-local interactions.
Their general shape is
\begin{eqnarray}
\widetilde{W}_n(\bX,\bY)&=&k_n \sum_{m\ge 1}\left[ \right.
    C_{1} (X^*_{n-m-1}    Y^*_{n-1}  -  \lambda X^*_{n-m}   Y^*_{n+1} )
 +  C_{2} (X^*_{n-1}      Y^*_{n-m-1}   -  \lambda^{m+1} X^*_{n+m} Y^*_{n+m+1} )
 +  C_{3} (X^*_{n+1}      Y^*_{n-m}   -  \lambda^m X^*_{n+m+1} Y^*_{n+m} )
 \nonumber\\
&+& C_{4} (X_{n-m-1}      Y^*_{n-1}   -  \lambda X_{n-m}     Y^*_{n+1} )
 +  C_{5} X^*_{n-1}      Y_{n-m-1}     -  \lambda^{m+1} C_{5}^* X_{n+m}   Y_{n+m+1}
 +  C_{6} X^*_{n+1}      Y_{n-m}     -  \lambda^m C_{6}^* X_{n+m+1}   Y_{n+m}
 \nonumber\\
&+& C_{7} X^*_{n-m-1}    Y_{n-1}     -  \lambda C_{7}^* X_{n-m}     Y_{n+1}
 +  C_{8} (X_{n-1}        Y^*_{n-m-1}   -  \lambda^{m+1} X_{n+m}   Y^*_{n+m+1} )
 +  C_{9} X_{n+1}        Y_{n-m}   -    \lambda^m C_{9}^*X^*_{n+m+1} Y_{n+m}
 \nonumber\\
&+& C_{10} X_{n-m-1}      Y_{n-1}     -  \lambda C_{10}^* X^*_{n-m}   Y_{n+1}
 +  C_{11} X_{n-1}        Y_{n-m-1}   -    \lambda^{m+1} C_{11}^*X^*_{n+m} Y_{n+m+1}
 +  C_{12} (X_{n+1}        Y^*_{n-m}   -  \lambda^m X_{n+m+1}   Y^*_{n+m} )
\left. \right],\nonumber\\
\label{N2complex}
\end{eqnarray}
where the $C_i$ are again complex parameters depending on $m$.
Liouville's theorem is again automatically satisfied as $\widetilde{W}_n(\bX,\bY)$ does not depend on
$X_n$, $X_n^*$, $Y_n$ nor $Y_n^*$. Non-local version of GOY and Sabra models is obtained by setting all $C_i$ coefficients to zero, except $C_1, C_2$ and $C_3$ for GOY, and $C_{11}, C_{12}$ and $C_{13}$ for Sabra.

\section{Numerical aspects}
\label{s8}

\subsection*{Computational gain using a shell model}
The computational gain using a shell model rather than a DNS can be estimated.
The cost of a simulation is proportional to MN, where M is the number of time steps and N the number of grid points.
First we consider HD turbulence.

In each direction the number of grid points can be estimated as the ratio between the scale at which energy is injected and the scale at which energy is dissipated.
Defining the Reynolds number at the forcing scale by $\Rey = l_F u_{l_F}  /\nu$, with, from Eqs.~(\ref{Scalinglaw}) and (\ref{eqlnu}), $u_{l_F} \propto (\epsilon l_F)^{1/3}$ and $l_{\nu} \propto \epsilon^{-1/4} \nu^{3/4}$, we find
\begin{equation}
	l_F/l_\nu \approx \Rey^{3/4}.
	\label{estimate}
\end{equation}
For a DNS in 3D this leads to
\begin{equation}
	N^{\rm DNS}_{\rm HD} \approx (l_F/l_\nu)^3  \approx \Rey^{9/4}.
\end{equation}
For an isotropic shell model not only is there just one direction, but the sequence of wave numbers is simply geometric. Replacing $k_{\nu}=\lambda^{n_{\nu}}$ and $k_{F}=\lambda^{n_{F}}$ in Eq.~(\ref{estimate}) gives
\begin{equation}
	N^{\rm Shell}_{\rm HD}=n_\nu-n_F\approx\ln\Rey.
\end{equation}
Therefore the number of grid points in a shell model is about $\Rey^{9/4}$ less
than in a 3D DNS. Note also that the number of arithmetic operations per variable is about 10 times smaller for a shell model than for a DNS where finite differences are calculated in three directions.

An estimate of the number of time steps $M_{\rm HD}$ is given by the ratio $t_F/t_{\nu}$ where $t_F$ and $t_{\nu}$
are the turn-over times $l/u_l$ at the forcing and dissipation scales. Assuming Kolmogorov turbulence $u_l\propto \epsilon^{1/3}l^{1/3}$, we find $t_F/t_{\nu} \approx (l_F/l_\nu)^{2/3}$.
Applying Eq.~(\ref{estimate}) we find
\begin{equation}
	M_{\rm HD}=t_F/t_{\nu} \approx \Rey^{1/2},
	\label{estimate2a}
\end{equation}
which is the same for DNS and shell models.

In MHD, for $\Pm \le 1$ the previous estimates hold. However, for $\Pm > 1$ the number of grid points
in each direction increases by a factor $l_\nu / l_\eta \approx \Pm^{1/2}$, leading to
\begin{equation}
	l_F/l_\eta \approx \Rey^{3/4} \Pm^{1/2},
	\label{estimate2}
\end{equation}
and then to
\begin{equation}
	N^{\rm DNS}_{\rm MHD} \approx (l_F/l_\eta)^3  \approx \Rey^{9/4} \Pm^{3/2}.
\end{equation}
For an isotropic MHD shell model, replacing $k_{\eta}$ by $\lambda^{n_{\eta}}$ and $k_{F}$ by $\lambda^{n_{F}}$ in
Eq.~(\ref{estimate2}) leads to
\begin{equation}
	N^{\rm Shell}_{\rm MHD}=n_\nu-n_F\approx\frac{3}{4}\ln\Rey + \frac{1}{2}\ln\Pm.
\end{equation}
So for $\Pm>1$, the number of grid points in a shell model is about $\Rey^{9/4}\Pm^{1/2}$ less
than in a 3D DNS.

An estimate of the number of time steps $M_{\rm MHD}$ is now given by the ratio $t_F/t_{\eta}$ where $t_{\eta}$
is the magnetic dissipation time. By definition the latter is given by $t_{\eta}=l_{\eta}^{2}/\eta$. With
$t_{\nu}=l_{\nu}^{2}/\nu$ and again $l_\nu / l_\eta \approx \Pm^{1/2}$ then $t_{\eta}=t_{\nu}$. 
From Eq.~(\ref{estimate2a}) we find
\begin{equation}
	M_{\rm MHD}=t_F/t_{\eta} \approx \Rey^{1/2},
\end{equation}
which again is the same for a DNS and a shell model.
Note that the number of time steps is approximately the same in HD and MHD, and does not depend on $\Pm$.

\subsection*{Numerical integration}

A system of ODEs can be integrated with standard numerical methods like Runge--Kutta. However, special care must be taken because the shell model system of equations is stiff. Indeed as shown before the characteristic times vary considerably between large and small scales by a factor $\Rey^{1/2}$ if $\Pm\le1$,
or $\Rey^{1/2}\Pm^{1/3}$ if $\Pm\ge1$.

One could think of using a constant time-step equal to the smallest characteristic time of the problem,
corresponding here to the dissipation time. However, because the dissipation rate fluctuates strongly, this can lead to a numerical effect of negative viscosity. This is why it is much better to use an adaptive time step, and in addition keeping the latter at least one order of magnitude smaller than
the dissipation time.

Another possibility is to split the time-step into two parts, each part solving different physics. During the first part only the non-linear terms are integrated, leading to a value for
$U_n$ (in HD). During the second part the exact solution for the dissipative term is calculated, replacing $U_n$ by $U_n\exp(-\nu k_n^2 {\bigtriangleup t})$.
This helps to avoid the negative viscosity effect previously mentioned, but it reduces accuracy to a first order approximation  $O(\bigtriangleup t)$.

\citet{Stepanov2002} found another way to increase the accuracy of the method.
To explain the method, we write the ODEs system for a HD shell model in the following form
\begin{equation}
\dot{U}_n(t)=F_n(U_m(t),U_q(t),t) - p_n U_n(t), \;\; n=1,N, \label{mod}
\end{equation}
where $p_n=\nu k_n^2$, and $F_n(U_m(t),U_q(t),t)$ is the quadratic function describing the interactions between shells $m$ and $q$ with shell $n$.
Introducing
\begin{equation}\label{sub}
  V_n(t) = U_n(t)\exp(p_n t).
\end{equation}
the system (\ref{mod}) becomes
\begin{equation}\label{mod1}
\dot{V}_n(t)=F_n(V_m(t)\exp(-p_mt),V_q(t)\exp(-p_qt),t) \exp(p_n t).
\end{equation}
Starting with $U_n$ at $t=t_0$ we want to calculate $U_n$ at $t=t_0+\bigtriangleup t$. 
Setting $V_n(t_0)=U_n(t_0)$, we calculate
$V_n(t_0+{\bigtriangleup t})$ using Eq.~(\ref{mod1}). Then using Eq.~(\ref{sub}) we find
$U_n(t_0+{\bigtriangleup t})=V_n(t_0+{\bigtriangleup t})\exp(-p_n {\bigtriangleup t})$.
Now when calculating $V_n(t_0+{\bigtriangleup t})$, the product of coefficients $\exp(-p_m t)$ and
$\exp(-p_q t)$ with $\exp(p_n t)$ in Eq.~(\ref{mod1}) can lead to a significant residual error especially
at small scales where $p_i {\bigtriangleup t} $ are much larger than unity.
Therefore the numerical integration of Eq.~(\ref{mod1}) needs to be elaborated a little.
Starting with a 4th order Runge-Kutta method, we calculate the explicit quadratic form for $F_n$ in order to avoid $\exp(p_n {\bigtriangleup t})$ in the numerical integration of Eq.~(\ref{mod1}). This leads to the following numerical scheme
\begin{eqnarray}
 K_n^1&=&{\bigtriangleup t} F_n(U_m(t_0),U_q(t_0),t_0), \nonumber \\
 K_n^2&=&{\bigtriangleup t} F_n(U_m(t_0) \exp(-p_m {\bigtriangleup t}/2)+\exp(-p_m {\bigtriangleup t}/2) K_m^1/2, U_q(t_0) \exp(-p_q {\bigtriangleup t}/2)+\exp(-p_q {\bigtriangleup t}/2) K_q^1/2 ,t_0+{\bigtriangleup t}/2), \nonumber \\
 K_n^3&=&{\bigtriangleup t} F_n(U_m(t_0)\exp(-p_m {\bigtriangleup t}/2)+K_m^2/2,U_q(t_0)\exp(-p_q {\bigtriangleup t}/2)+K_q^2/2,t_0+{\bigtriangleup t}/2), \label{shem1} \\
 K_n^4&=&{\bigtriangleup t} F_n(U_m(t_0)\exp(-p_m {\bigtriangleup t})+\exp(-p_m {\bigtriangleup t}/2) K_m^3,U_q(t_0)\exp(-p_q {\bigtriangleup t})+\exp(-p_q {\bigtriangleup t}/2) K_q^3,t_0+{\bigtriangleup t}), \nonumber
 \\ \nonumber \\
 U_n(t_0+{\bigtriangleup t})&=& U_n(t_0)\exp(-p_n {\bigtriangleup t}) + (K_n^1 \exp(-p_n {\bigtriangleup t}) + 2 K_n^2 \exp(-p_n {\bigtriangleup t}/2)+
 2 K_n^3 \exp(-p_n {\bigtriangleup t}/2) + K_n^4)/6 + O({\bigtriangleup t}^5). \label{shem2}
\end{eqnarray}
The factors $\exp(-p_i {\bigtriangleup t})$ and $\exp(-p_i {\bigtriangleup t}/2)$ with $i\in\left[1,N\right]$ can be precalculated only once.
The negative viscosity effect is eliminated and the accuracy is up to a fifth order approximation $O({\bigtriangleup t}^5)$.

More complex methods can be used to deal with stiffness. The VODE time-stepping scheme  \citep{Brown1998} can be used for shell model simulations. The integration method is a variable-coefficient form of Backward Differentiation Formula methods. It requires computing the Jacobian of ODEs. For a local shell model the Jacobian is a sparse matrix. However, for non-local models the Jacobian is a full matrix and the method loses efficiency.  

\subsection*{Statistics}
In turbulence, the accuracy of e.g. spectral laws and scaling exponents is directly related to the statistics. High accuracy requires to record data over a sufficiently long time, thus again increasing the cost of simulation. Compared to DNS, another advantage of shell models is the possibility to reach accurate statistics. 

This was well demonstrated in the study of helicity cascade by \citet{Lessinnes2011}. 
The challenge was to calculate the helicity spectrum $H(k)$ resulting from the sum of two quantities $H^{\pm}(k)= \pm A k^{-2/3} + B k^{-5/3}$, that is zero at leading order and non-zero at next order.
The ratio $|H(k)|/|H^{\pm}(k)| \propto k^{-1}$ was as small as $10^{-5}$. A relative error of $10^{-6}$ has been taken in the VODE time scheme.
In addition the time fluctuations of $|H^{\pm}(k)|$ were larger than those of energy by a factor $k$,
requiring again more data.

\begin{figure}
\begin{center}
    \includegraphics[width=0.49\textwidth]{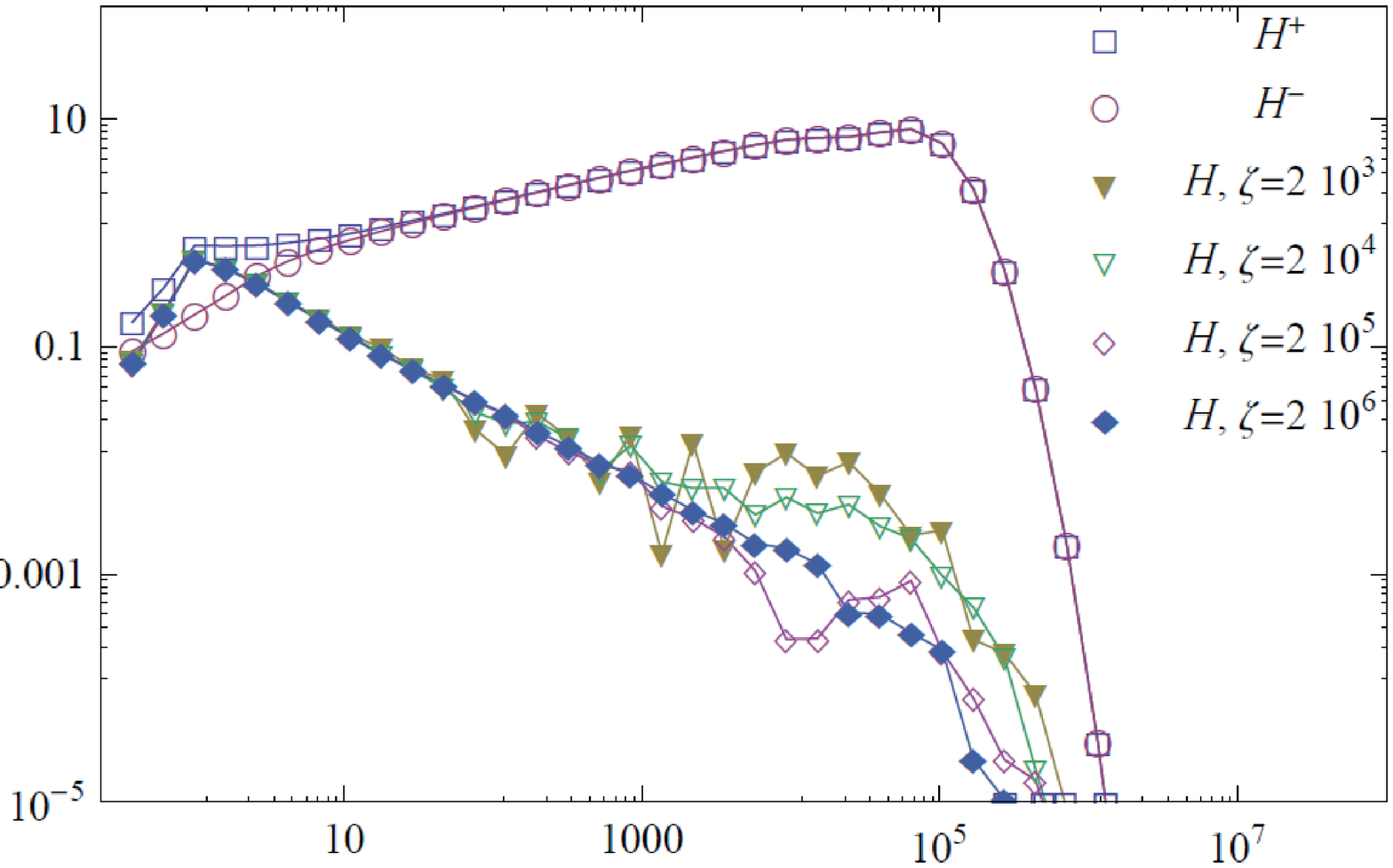}
    \includegraphics[width=0.49\textwidth]{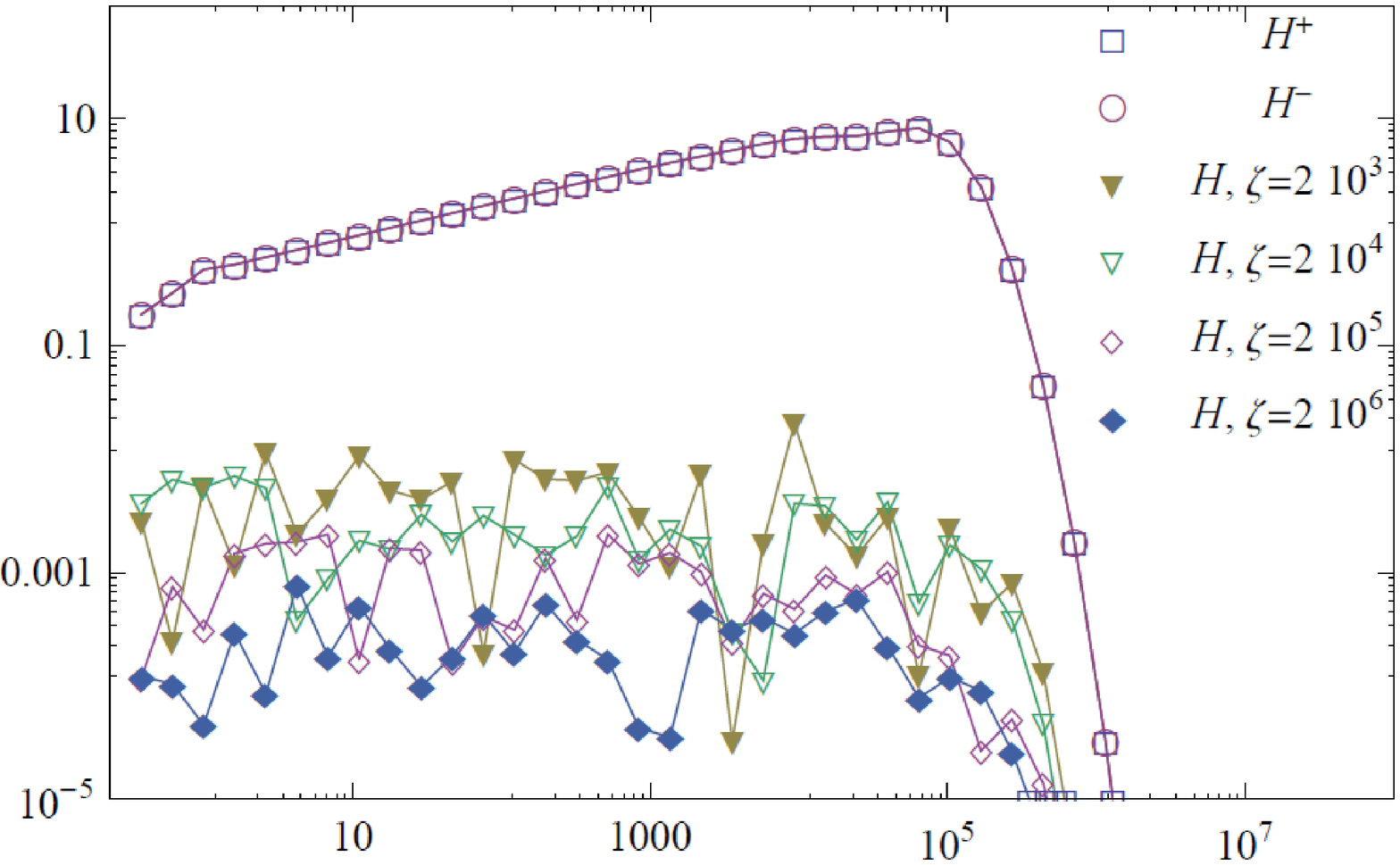}
    \centering{$k$\hspace{7cm}$k$}
    \end{center}
\caption{Spectra of $|H^+|, |H^-|$ and $|H|$, for different amounts of data measured by the parameter $\zeta$, and for two different forcings, helical (left) and non-helical (right).}
\label{fig:stat}
\end{figure}

In Fig.~\ref{fig:stat}
the results are illustrated for different amounts of data $\zeta=Q T$, where $Q$ is the number of independent runs, each run being performed during time $T$.
For a helical forcing (Fig.~\ref{fig:stat}, left panel) we clearly see that $H(k)$ does converge to a well-defined spectrum with increasing $\zeta$, while $H^{\pm}(k)$ spectra do not change.
In the right panel of Fig.~\ref{fig:stat}, $H(k)$ is getting smaller with increasing $\zeta$
as expected for a non-helical forcing.

The degree of convergence follows the power law $\zeta^{-1/2}$ in agreement with the following formula
\begin{equation}
\sigma_{\langle a\rangle} = \frac{\sigma_a}{\sqrt{N}},
\end{equation}
derived for a Gaussian process $a$, where $\sigma_{\langle a\rangle}$ is the standard deviation due to finite sampling of the mean value of $a$,
$\sigma_a$ the standard deviation of $a$ and $N$ the number of independent
values of $a$ within the sampling.
